\documentstyle[12pt,psfig,epsfig,amssymb,longtable]{thesis_prov}

\input{epsf}
\input{psfig}

\newcommand{\pp} {\mbox{$p+p$}}

\newcommand{\ee} {\mbox{$e^+e^-$}}

\newcommand{\Au} {\mbox{$Au+Au$}}

\newcommand{\dA} {\mbox{$d+Au$}}
\newcommand{\pt} {\mbox{$p_T$}}

\newcommand{\sqs} {\mbox{$\sqrt{s}$}}
\newcommand{\scc} {\mbox{$\sigma_{c\overline{c}}$}}
\newcommand{\sbb} {\mbox{$\sigma_{b\overline{b}}$}}
\newcommand{\QQ} {\mbox{$Q\overline{Q}$}}
\newcommand{\qq} {\mbox{$q\overline{q}$}}

\newcommand{\kt}  {\mbox{$\langle k_{T} \rangle$}}

\newcommand{\cunit} {\mbox{$[mb\cdot c^3/GeV^2]$}}

\def\Journal#1#2#3#4{{#1}{\bf #2}, #3 (#4)}

\def\IJMPA{{Int. J. Mod. Phys.}~{\bf A}}
\def\EPJ{{Eur. Phys. J.}~{\bf C}}
\def\JHEP{{J. High Energy Phys.}~}
\def\JPG{{J. Phys}~{\bf G}}
\def\NIM{Nucl. Instrum. Methods }
\def\NIMA{{Nucl. Instrum. Methods }~{\bf A}}

\def\NPB{{Nucl. Phys.}~{\bf B}}
\def\PLB{{Phys. Lett.}~{\bf B}}
\def\PL{{Phys. Lett.}}

\def\PRL{Phys. Rev. Lett.\ }
\def\PRD{{Phys. Rev.}~{\bf D}}
\def\PRC{{Phys. Rev.}~{\bf C}}
\def\PRB{{Phys. Rev.}~{\bf B}}
\def\CPC{{Comp. Phys. Comm. }}
\def\EPJ{{Eur. Phys. J.}~{\bf C}}
\def\HIP{{Heavy Ion Phys. }}
\def\ZPC{{Z. Phys.}~{\bf C}}

\def\RMP{Rev. Mod. Phys.\ }

\def\PR{Phys. Rep.\ }

\begin{document}
\pagestyle{prelim}            
%
%
\dissertation
\singlespace
%
\author{Sergey Butsyk}

%
\degree{Doctor of Philosophy}

%
\department{Physics}

%
\title{Study of Open Charm Production in p+p Collisions at $\sqs$ = 200 GeV}

%
\month{May} \year{2005}

%
\maketitle

\makecopyright
%
\begin{approval}
 \member{Thomas Hemmick (Advisor)\\Professor, Department of Physics and Astronomy}
 \member{Edward Shuryak (Chair)\\Professor, Department of Astronomy and Physics}
 \member{Michael Marx\\Professor, Department of Astronomy and Physics}
 \member{Ralf Averbeck\\Assistant Professor, Department of Astronomy and Physics}
 \member{Yasuyuki Akiba\\ Senior Scientist, RIKEN Institution}
\end{approval}

%
\begin{abstract}
The PHENIX experiment at the Relativistic Heavy Ion Collider
(RHIC) with its unique electron identification system enables us
to perform high precision measurements of electron yields. By
measuring electron production at high transverse momentum, we can
disentangle the contribution of electrons originating from
semi-leptonic decays of heavy quarks (charm or bottom) from the
less interesting ``photonic'' decay modes of light mesons. $D$/$B$
mesons carry single heavy valence quarks and are usually referred
to as ``Open Charm'' and ``Open Bottom'' particles,
differentiating them from Closed Flavor particles such as $J/\psi$
and $Y$ mesons. Due to the large mass of the heavy quarks, their
production mechanisms can be adequately explained by perturbative
QCD (pQCD) theory.

This dissertation presents the measurement of electrons from heavy
flavor decays in proton + proton collisions at RHIC at collision
energy $\sqs = 200$ GeV over a wide range of transverse moment
($0.4 < p_T <5$ GeV/c). Two independent analysis techniques of
signal extraction were performed. The ``Cocktail'' subtraction is
based on the calculation and subtraction of the expected ``photon
- related'' electron background based upon measured yields of
light mesons. The ``Converter'' subtraction is based upon a direct
measurement of photon yields achieved introducing additional
material in the PHENIX acceptance and deducing the photon
abundance by measuring the increase in electron yield. This is the
first measurement of the Open Charm crossection at this collision
energy and it is an important baseline measurement for comparison
with nucleus + nucleus collisions. The modification of Open Charm
production in heavy ion collisions compared to the presented $\pp$
result can be used to study the final state interaction of the
heavy quarks with hot dense matter inside the collisions. The
results of the Open Charm measurements are compared to current
pQCD predictions both in Leading Order (LO) $O(\alpha_s^2)$ and
Next-to-Leading Order (NLO) $O(\alpha_s^3)$. The final result for
the total Open Charm crossection is $\sigma_{c\bar{c}}= (0.920 \pm
0.148(stat)\pm 0.524(sys))$ mb, in good agreement with the
published STAR experiment measurements of Open Charm crossection
in $\dA$ colliding system.

\end{abstract}

\begin{dedication}
To my mom, my dad and Julia
\end{dedication}

\tableofcontents \listoffigures \listoftables

\begin{acknowledgements}

    First of all I would like to express my deepest gratitude to my
colleagues from SUNYSB and PNPI for letting me to take part in the
creation and running of one of the most pioneering physics devices
in high energy field - PHENIX detector. Especially I would like to
thank Professor Thomas Hemmick, for teaching me basically all the
aspects of modern experimental physics, for trusting in me and his
unique ability to strive for perfection. I want to thank Dr.
Federica Messer and Professor Axel Drees for giving me a chance to
participate in an exciting and challenging world of PHENIX
software, all the members of the PHENIX electron working group
(especially Dr. Yasuyuki Akiba and Dr. Ralf Averbeck) for an
encouraging results in the development of the signal extraction
technique, used in this work and a significant contribution to the
analysis. Needless to say that without the friendly team of
graduate students at Nuclear and Heavy Ion group (Felice
Matathias, Jianyoung Jia, Anuj Purwar, Jamil Egdemir) this mission
would never have been accomplished. Thank you very much for
fruitful and thought-provoking discussions and support.

\end{acknowledgements}
\pagestyle{body}              
\chapter{Introduction}\label{sec:ch1}

Heavy quarks are a unique tool in High Energy and Nuclear Physics.
Discovered in 1974, the charm quark has attracted the attention of
both experimentalists and theoreticians. The production of particles
with Open Charm (carrying a single $c$ or $\bar{c}$ quark, such as
$D$-mesons and $\Lambda_c$ baryons) and Closed Charm ($c\bar{c}$
mesons) is a rare process purely from energy considerations, because
the charm mass, $m_c \approx 1.2 - 1.5$ GeV, is so high. This energy
scale allows one to use (with some caution, of course) perturbative Quantum
Chromodynamics (pQCD) - a theoretical model of the strong color field
interaction to describe the production mechanisms and rates of Open Charm.

Open Charm particles are produced through the fragmentation of
$c\bar{c}$ pairs. Creation of those pairs is intrinsically sensitive
to the initial state of the partonic system. pQCD predicts that the
primary production mechanism in Leading Order approximation is "gluon
fusion". Thus, the production rate of charm pairs is proportional to
the initial gluon density inside the collision. Open Charm production
in in heavy ion collisions at the Relativistic Heavy Ion Collider
(RHIC) can thereby lead us to understanding how the initial properties
of the matter change with the size of the colliding system. RHIC
raised the High Energy Nuclear physics to the next level, enabling us to study
with the same detector setup the full variety of collisions - $\pp$,
$\dA$, $\Au$ at $\sqs = 200$ GeV. Detailed comparison of Open Charm
production in those collisions will help us to answer the fundamental
questions:
\begin{itemize}
    \item What are the mechanisms governing charm production?\\
    The $\pp$ collision is a reference measurement, which should be
    exactly explainable by perturbative QCD.  Production of Open Charm
    at lower energies has always been a serious confirmation of pQCD's
    applicability for heavy quarks.  So far, exact calculations from
    pQCD exist to second order
    (NLO)~\cite{nlo_charm,qcd_hq1,qcd_hq2,nlo1,nlo2} and describe the
    lower energy data within the theoretical uncertainties.  It may
    well be that at our high energy, the theory cannot neglect these
    higher order partonic interaction processes in order to
    describe the data. The theory of hadronization of the $c$ and
    $\bar{c}$ is presently at a rudimentary stage and in order to
    constrain the different theoretical predictions, we need to study
    $\pp$ collisions at different rapidity ranges.

    \item How does the production sensitive to additional "cold" matter
    in the collision?\\
    A comparison of $\pp$ production with $\dA$ will measure the
    modification of the gluon structure function in "cold" hadronic
    matter. In this case, the ability to perform rapidity scan is
    essential to study the $x$ dependence of the gluon
    modification factor that was clearly observed in case of light
    quark mesons.

    \item What happens with the charm quark production in "hot"
    dense matter of $\Au$ collisions?\\
    Direct comparison of charm production crossection in $\Au$ with $\dA$
    and $\pp$ will answer this question. So far there are a lot of different
    predictions, expecting both enhancement due to the initial
    state thermal production and suppression due to the media
    induced gluon radiation. So far, we observed a significant
    suppression for the light quarks, in case of the
    heavy quark this effect may be significantly reduced due to
    the suppression of gluon radiation at small angles for the
    charm quark.
\end{itemize}

There is a lot of very interesting new physics that can be
discovered if this programm is accomplished.  In this thesis we
begin the program by accomplishing the baseline measurement of
Open Charm production in $\pp$ collisions at mid rapidity.

There are two experiments at RHIC capable of performing this task,
each with certain advantages and disadvantages - PHENIX and STAR.

\begin{figure}[t]
\begin{tabular}{lr}
\begin{minipage}{0.5\linewidth}\epsfig{figure=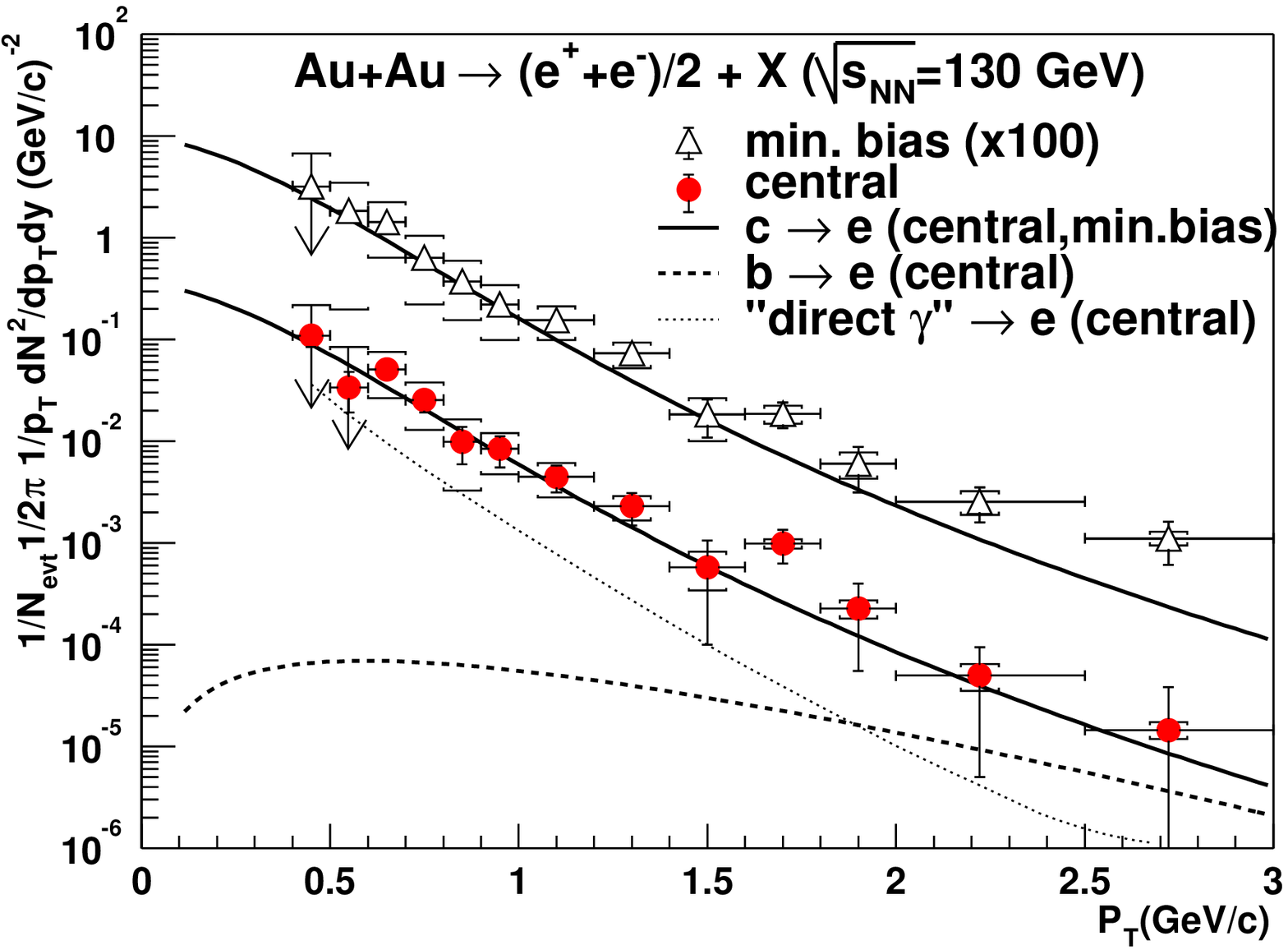,width=1\linewidth,clip}
\caption{\label{fig:ch1.se_130} Invariant crossection for
"Non-photonic" electrons at $\sqs =130$ GeV at
PHENIX~\cite{ppg011}.}
\end{minipage}
&
\begin{minipage}{0.5\linewidth} \centering
\epsfig{figure=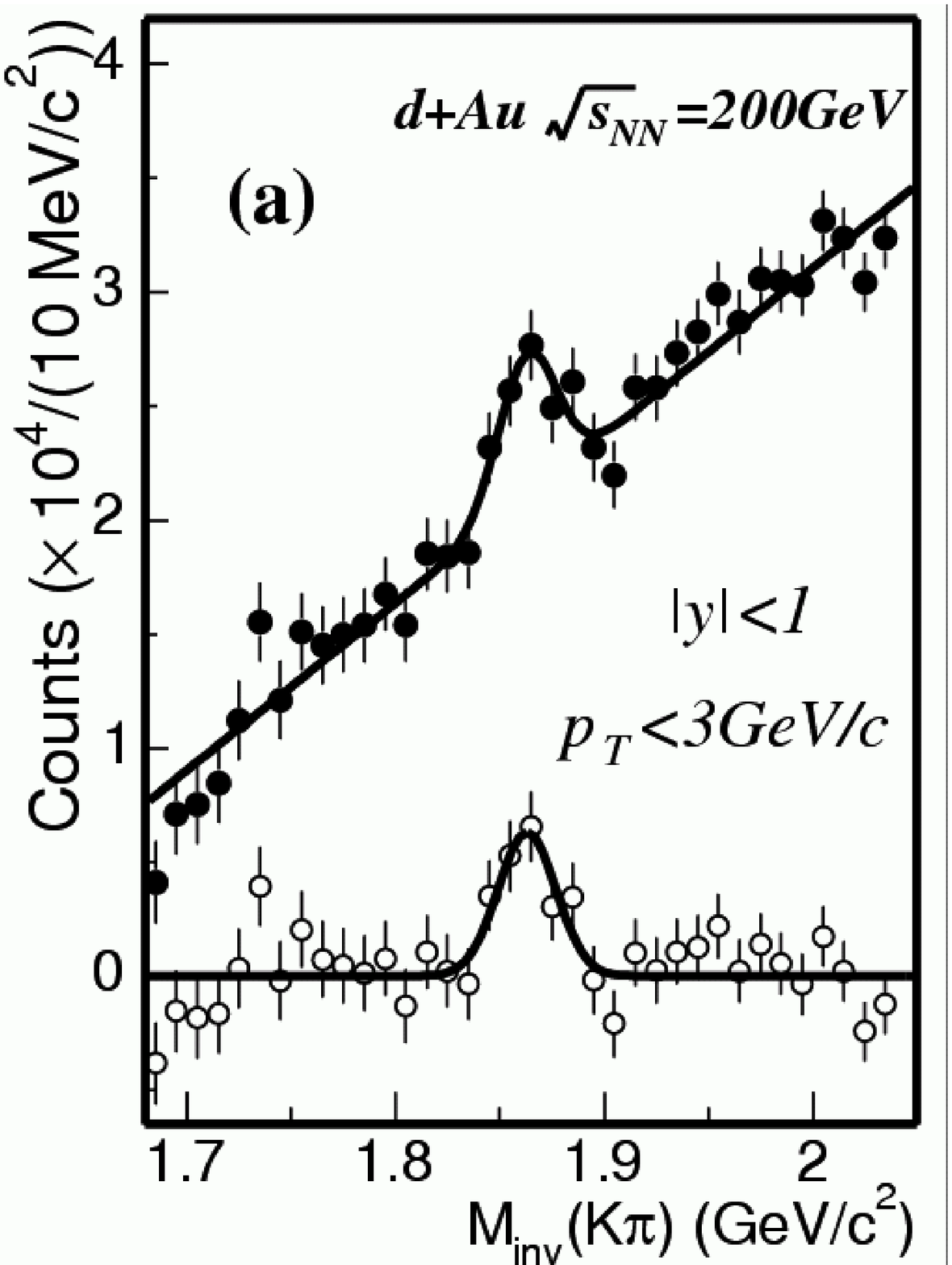,width=0.55\linewidth,clip}
 \caption{\label{fig:ch1.star_d} Mass spectrum of $K\pi$ pairs in $\dA$ collisions
  with clear $D^0$ peak~\cite{STAR_se_pp}}
\end{minipage}
\end{tabular}
\end{figure}

PHENIX was a pioneer in Open Charm measurements at RHIC. Built
with a strong accent towards the electron and muon identification,
PHENIX successfully measured Heavy Flavor production at $\sqs =
130$ GeV~\cite{ppg011}. Fig.~\ref{fig:ch1.se_130} shows the
invariant multiplicity for the electrons, originated from the
semi-leptonic decay of the charm and bottom quarks
$c(b)\rightarrow eX$.

STAR also has elaborated electron identification using
Electromagnetic Calorimeter, Time-of-Flight Detector and Time
Projection Chamber. They were able to perform similar
semi-leptonic measurement in $\pp$ and $\dA$ collisions at $\sqs =
200$ GeV~\cite{STAR_se_pp}. But STAR has also been able to
directly measure decays of $D$ mesons into hadronic channels. This
is a more direct measurement of the Open Charm production then
measuring decay leptons. Fig.~\ref{fig:ch1.star_d} shows the
reconstruction of $D^0\rightarrow K\pi$ decay channel by STAR
collaboration. It is doubtful that these measurements can be also
done in case of $\Au$ collisions. Direct measurement of hadronic
decays of Open Charm mesons in PHENIX is very hard due to the
limited acceptance for these decays.

My dissertation sets the cornerstone to the foundation of the
Heavy Flavor measurements at energy $\sqs = 200$ GeV - I will
present the experimental measurement of Open Charm production at
mid-rapidity in PHENIX at RHIC from single electron channel
($c\rightarrow eX$). The paper is organized in the following way:
Chapter~\ref{sec:ch2} presents the current theoretical aspects of
the heavy quark production and fragmentation. The experimental
setup, used in the analysis is described in Chapter~\ref{sec:ch3}.
Chapter~\ref{sec:ch4} is dedicated to data reduction towards the
final Open Charm electron crossection. A discussion and
theoretical comparison is summarized in Chapter~\ref{sec:ch5} and
final conclusions and a future outlook are presented in
Chapter~\ref{sec:ch6}.

\chapter{Theory}\label{sec:ch2}

\section{Quantum Chromodynamics}

Quantum ChromoDynamics (QCD) is the set of physical laws,
describing our current understanding of strong interactions
between the quarks and gluons. QCD was developed as an extension
of Quantum Electrodynamics via the imposition of a local $SU(3)$
symmetry of rotation in color space.  The Standard Model couples
the $SU(3)$ strong interaction with the unified electro-week
interaction making the overall symmetry group $SU(3)\times SU(2)
\times U(1)_Y$. The Standard Model describes the large variety of
mesonic and baryonic states by the existence of the deeper level
of elementary constituents of matter:
\textbf{quarks}~\cite{gell_mann}. All the variety of particles
that can be found in Particle Data Tables~\cite{PDG} can be
explained in terms of six spin-$\frac{1}{2}$ quark
\textbf{flavors}:

\begin{tabular}{|c|c|}
\hline
  $Q = +\frac{2}{3}$ & u c t \\
  \hline
  $Q = -\frac{1}{3}$ & d s b \\
  \hline
\end{tabular}

The meson state in this case consists of quark-antiquark
($q\overline{q}$) and baryonic state is a combination of three
quarks ($qqq$) or antiquarks ($\bar{q}\bar{q}\bar{q}$), this model
gives us a unique correspondence between the observed particle and
its state.

However, the early quark picture faced a problems with Fermi-Dirac
statistics, for example $\Lambda^{++}$ baryon ($J = \frac{3}{2}$)
should have been explained as a combination of $u\uparrow
u\uparrow u\uparrow$ quarks which contradicts Fermi statistics of
spin-$\frac{1}{2}$ particles. This problem was solved by
introducing a new quantum number~\cite{han}, \textbf{color}, such
that each species of quark can have $N_C = 3$ different colors. In
this picture mesons and baryons are described as color-singlet
combinations of colored quarks.
\begin{equation}
B =
\frac{1}{\sqrt{6}}\epsilon^{\alpha\beta\gamma}|q_{\alpha}q_{\beta}q_{\gamma}\rangle
\ \ \ \ \ M =
\frac{1}{\sqrt{3}}\epsilon^{\alpha\beta}|q_{\alpha}\overline{q}_{\beta}\rangle
\end{equation}
States, with color content not equal to zero were never observed,
which led the theorists to postulate the hypothesis that all
states are color-singlets. This statement is known as
\textbf{``confinement hypotheses''} implying directly the
non-existence of free quark colored states.

\begin{figure}[t]
\begin{center}
\epsfig{figure=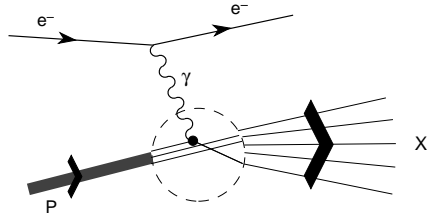,width=0.4\linewidth}
\caption{\label{fig:ch2.DIS} Deep Inelastic $e^-p\to e^-X$
scattering.~\cite{asp_qcd}}
\end{center}
\end{figure}

The next valuable piece to the construction of the QCD theory was
obtained from \textbf{"deep inelastic scattering"} (DIS) of high
energy electrons on the proton.  Fig.~\ref{fig:ch2.DIS} shows the
interaction of of the electron with the content of the proton.

At higher energies, the virtual photon is sensitive to smaller
distances and we are thereby able to resolve the structure of the
proton by increasing the energy of the electron beam. Denoting the
photon momentum transfer as $Q^2 = 4EE'sin^2(\theta /2)$, where
$E$ and $E'$ are incident and scattered electron energies.

From the studies of scattering crossections, it was found that at
large values of $Q^2$ a sizable continuum contribution to the
crossection still persists, suggesting an existence of pointlike
objects inside the proton. Those pointlike spin-$\frac{1}{2}$
constituents were called \textit{partons}. It was also found that
proton structure functions only depends on the transferred
momentum $Q^2$ and the relative contribution of parton momentum to
the total proton momentum $x$ ~\cite{Bjorken}. This surprising
discovery of pointlike partons inside the proton, initially
contradicting previously explained confinement of the quarks by a
strong color force leads to an fundamental property of strong
interactions known as \linebreak "\textit{asymptotic freedom}" -
the interaction between quarks should become weaker at short
distances, so that at high $Q^2$ transfer, quarks behave as free
particles. The interaction in $q\overline{q}$ system grows as we
try to separate the quarks. At some point the energy of the string
become more then $2m_{Q}$, where $m_{Q}$ is the mass of another
quark. It becomes energetically favorable to split the string into
two $q\overline{Q}$ and $Q\overline{q}$ mesons. By increasing the
energy, we can create more and more colorless meson states.
Otherwise, if we try to approach two quarks, the potential loses
strength and the quarks start behave like free particles.

 The field carriers in QED theory are photons, and the QED
Lagrangian describes the interaction of a Dirac fermion with the
electromagnetic field $A_{\mu}$. By analogy, the QCD Lagrangian
can be constructed. This Lagrangian will describe the interaction
of $q^{\alpha}_f\overline{q^{\alpha}_f}$ quarks of color $\alpha$
and flavor $f$. The free Lagrangian is given by
Eq.~\ref{eq:ch2.QCD_free_L}~\cite{asp_qcd}.

\begin{equation}
L_0 = \sum_f\,\overline{q}_f \,( i\gamma^\mu\partial_\mu - m_f)
q_f \label{eq:ch2.QCD_free_L}
\end{equation}

Propagation of the quark wave function requires the Lagrangian to be
invariant under local $SU(3)_C$ color space transformation. This
invariance introduce $3^2-1=8$ component gauge bosons $G_a^{\mu}(x)$,
those particles are called \textbf{gluons}. The total invariant
Lagrangian of QCD can be written as follows~\cite{asp_qcd}:

\begin{eqnarray}
\label{eq:L_QCD_pieces} L_{QCD} & = &
 -{1\over 4}\, (\partial^\mu G^\nu_a - \partial^\nu G^\mu_a)
  (\partial_\mu G_\nu^a - \partial_\nu G_\mu^a)
 + \sum_f\,\bar q^\alpha_f \, \left( i\gamma^\mu\partial_\mu - m_f\right)\,
q^\alpha_f \qquad\nonumber\\ && \mbox{}
 + g_s\, G^\mu_a\,\sum_f\,  \bar q^\alpha_f \gamma_\mu
 \left({\lambda^a\over 2}\right)_{\!\alpha\beta} q^\beta_f
\\ && \mbox{}
 - {g_s\over 2}\, f^{abc}\, (\partial^\mu G^\nu_a - \partial^\nu G^\mu_a) \,
  G_\mu^b G_\nu^c \, - \,
{g_s^2\over 4} \, f^{abc} f_{ade} \, G^\mu_b G^\nu_c G_\mu^d
G_\nu^e \,  \nonumber
\end{eqnarray}

\begin{figure}[h]
\centering \epsfig{figure=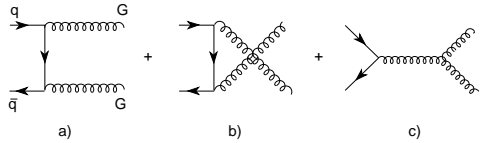,width=0.45\linewidth}
\caption{\label{fig:ch2.qqgg}Tree-level Feynman diagrams
contributing to $q\bar q\to gg$.~\cite{asp_qcd}}
\end{figure}

The first line presents the kinetic energy of quark and gluon
field. The second line describes the color interaction between
gluon and quark, where $\lambda^a$ are color field $SU(3)_C$
matrices. The last line shows gluon self-interactions, exhibiting
non-abelian properties of the color field\footnote{Gluon field
self-interaction is a striking difference of QCD, compared to QED,
where photon can not split into two photons}. The coupling
constant $g_s$ is the universal \textit{strong coupling constant}.
Examples of $q\overline{q}\rightarrow gg$ scattering Feynman
Diagrams allowed in QCD are shown in
Fig.~\ref{fig:ch2.qqgg}~\cite{asp_qcd}.

\section{Renormalization of QCD}

\begin{figure}[b]
\centering \epsfig{figure=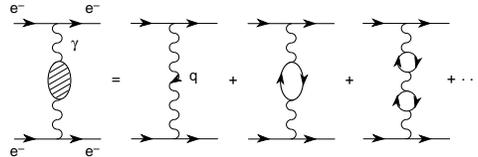,width=0.45\linewidth}
\caption{\label{fig:ch2.qed_loops}Photon self-energy contribution
to $e^-e^-$ scattering.~\cite{asp_qcd}}
\end{figure}

Renormalization techniques have been widely developed in QED as a
useful mechanism to absorb the loop diagrams that cause
logarithmic divergences. Ultraviolet logarithmic divergences
appear in QED if the photon polarizes the vacuum, creating a
virtual \ee pair. The calculated crossection for this loop process
produces an infinity which is "non-physical". In order to solve
this problem, we assume (see Fig.~\ref{fig:ch2.qed_loops}) that
for \ee interactions processes, the effective coupling constant
that we measure is changing due to vacuum polarization. The
``bare'' coupling constant in QED has been known since Thompson
experiments as $\alpha_0 = e^2/(4\pi)\approx 1/137$. Assuming that
the coupling may vary as a function of photon transferred energy
$Q^2$ due to appearance of the loops leads to very powerful
conclusion that the we can actually measure \textbf{renormalized
running coupling} $\alpha^{QED}(Q^2)$ which can be written as:

\begin{equation}
\label{eq:ch2.alpha_run} \alpha^{QED}(Q^2) \, = \,
{\alpha^{QED}(Q_0^2)\over 1 - {\beta_0\cdot
\alpha^{QED}(Q_0^2)\over 2\pi} \ln{(Q^2/Q_0^2)}}
\end{equation}

where $\beta_0$ is a solution of differential equation
(\ref{eq:ch2.beta}) and for one loop contribution is equal to
$\beta^{QED}_0 = \frac{2}{3}$.

\begin{equation}
\label{eq:ch2.beta} Q^2\, {d\alpha\over dQ^2} \,\equiv\, \alpha
\,\beta(\alpha) \, ; \qquad\qquad \beta(\alpha)\, =\, \beta_0
{\alpha\over \pi} + \beta_1 \left({\alpha\over\pi}\right)^2 +
\cdots \end{equation}

$Q_0^2$ is an arbitrary scale, at which we can neglect the
contributions from the radiative corrections. We can take it as
$Q_0^2 \rightarrow 0$ which produces classical value for the
coupling constant $\alpha(Q_0^2) = \alpha_0 = 1/137$.

The fact that $\beta_0 >0$ leads to the conclusion that alpha is
growing with the increase of the energy transferred and, i.e. the
effective charge of the electron is decreasing with the distances
between the charges. This is intuitively understandable as the
produced virtual \ee pairs acts like a screen around the charge in
the polarized dielectric medium.

\pagebreak

\begin{figure}[t]
\centering \epsfig{figure=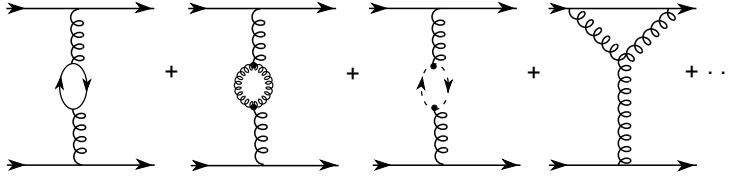,width=0.7\linewidth}
\caption{\label{fig:ch2.qcd_loops}Feynman diagrams contributing to
the renormalization of the strong coupling.~\cite{asp_qcd}}
\end{figure}

A similar renormalization approach can be applied to the QCD
theory, but in the case of QCD we need to take into account gluon
self-interaction diagrams. Processes, contributing to the
renormalization of the scale are shown in
Fig.~\ref{fig:ch2.qcd_loops}. The final equation for the QCD
running coupling gives identical result to QED
Eq.~\ref{eq:ch2.alpha_run}, but the $\beta$ function will be
drastically different~\cite{Gross}.

\begin{equation}
\label{eq:QCD_beta} \beta_0 \, = \, {2\, N_f - 11\, N_C \over 6}
\end{equation}

where $N_f$ is a number of flavors, $N_C$ is the number of colors.
The first term is due to the creation of $q\overline{q}$ loops of
different flavor and is similar to the QED value. The second,
negative contribution in Eq.~\ref{eq:QCD_beta} corresponds to
gluonic self-interaction loops and is proportional to number of
colors. The existence of the second term makes the factor
$\beta_0$ negative for $N_f<16$ this leads to the important result
that that at short distances (large $Q^2$) $\alpha_s(Q^2)
\rightarrow 0$. This is direct confirmation of \textsl{asymptotic
freedom }.

In the one loop approximation we can write the strong interaction
running coupling as (solving Eq.~\ref{eq:ch2.beta} for $\alpha$):
\begin{equation}
\label{eq:ch2.alpha_0} \alpha_s^{(0)}(Q^2) \, = \, {2\pi\over
-\beta_0\ \ln{(Q^2/\Lambda_{QCD}^2)}}
\end{equation}

where $\Lambda_{QCD}$ comes as an integration parameter at an
energy scale at which the QCD coupling constant starts to exhibit
non-perturbative properties. If the momentum transfer $Q$ is much
larger then $\Lambda_{QCD}$, we are in asymptotic freedom regime
and $\alpha_s(Q^2) \rightarrow 0$ allows us to use the
\textbf{perturbative QCD} (pQCD) technique. In the case of energy
transfer, comparable to $\Lambda_{QCD}$, Eq.\ref{eq:ch2.alpha_0}
blows-up the strong coupling constant making all orders diagrams
to be of similar contribution and the pQCD approach cannot be
used.

\pagebreak

It is necessary to add that there are several theoretical
approaches to pQCD renormalization (referred to as the
\textit{renormalization scheme}). Currently, the most adopted and
cited approach is use of $\overline{MS}$ - modified minimal
subtraction scheme~\cite{MS_scheme}.

In $\overline{MS}$ case the next orders of two loop corrections we
have following expression for next term of $\beta$ function
deconvolution Eq.~\ref{eq:ch2.beta}~\cite{beta_dec}.

\begin{equation}
\label{eq:beta_hl} \beta_1 = -{51\over 4} + {19\over 12}\, N_f
\end{equation}

\begin{figure}[b]
\centering \epsfig{figure=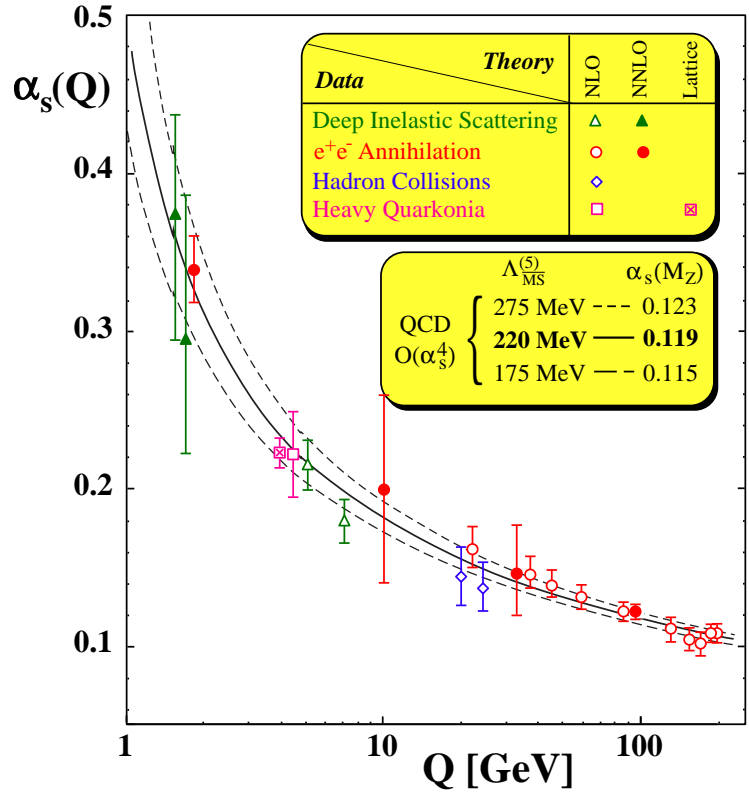,width=0.7\linewidth}
\caption{\label{fig:ch2.alpha_qcd}Measurements of $\alpha_s$ as a
function of energy scale~\cite{bethke}.}
\end{figure}

For $N_f \le8$, $\beta_1 <0$ which even stronger justifies the
use of asymptotic freedom. Fig.~\ref{fig:ch2.alpha_qcd} shows
the dependence of $\alpha_s$ on the energy scale $Q$ for the
experimental data comparison with pQCD theory~\cite{bethke}. It is
also clearly seen from this figure that $\Lambda_{QCD}$ is on the
order of $\sim220$ MeV and should be calculated for the expected
number of flavors ($\Lambda^{(5)}_{QCD}$ assumes $N_f = 5$).

\section{Heavy quark production in QCD}\label{sec:HQ_prod}

In this section I discuss the production mechanisms of heavy quark
particles in $\pp$ collisions using the perturbative QCD
formalism, described in previous Section.

The main advantage of the heavy quark measurements is that the
initial content of the heavy flavor in the proton is negligibly small
(there are some papers though that assume the intrinsic heavy
quark content of the nucleon to be
non-zero~\cite{intrinsic_flavor,Vogt_intrinsic} but we will not
include this aspect into our considerations). Then, the heavy quarks
should be produced in hard partonic scattering. The big advantage
of the heavy quark production is that the energy scale for the
production of charm quarks $Q^2 \sim m_c^2$ is significantly
higher then $\Lambda_{QCD}$. This gives us a coupling constant of
the order of $\alpha_{s}\sim(0.3-0.5)$ (see
Fig.~\ref{fig:ch2.qcd_loops}), which is small enough to utilize
perturbative theory. pQCD should give even more accurate results
for the higher mass bottom quark.

\section{Hard scattering processes}

\begin{figure}[h]
\centering
\epsfig{figure=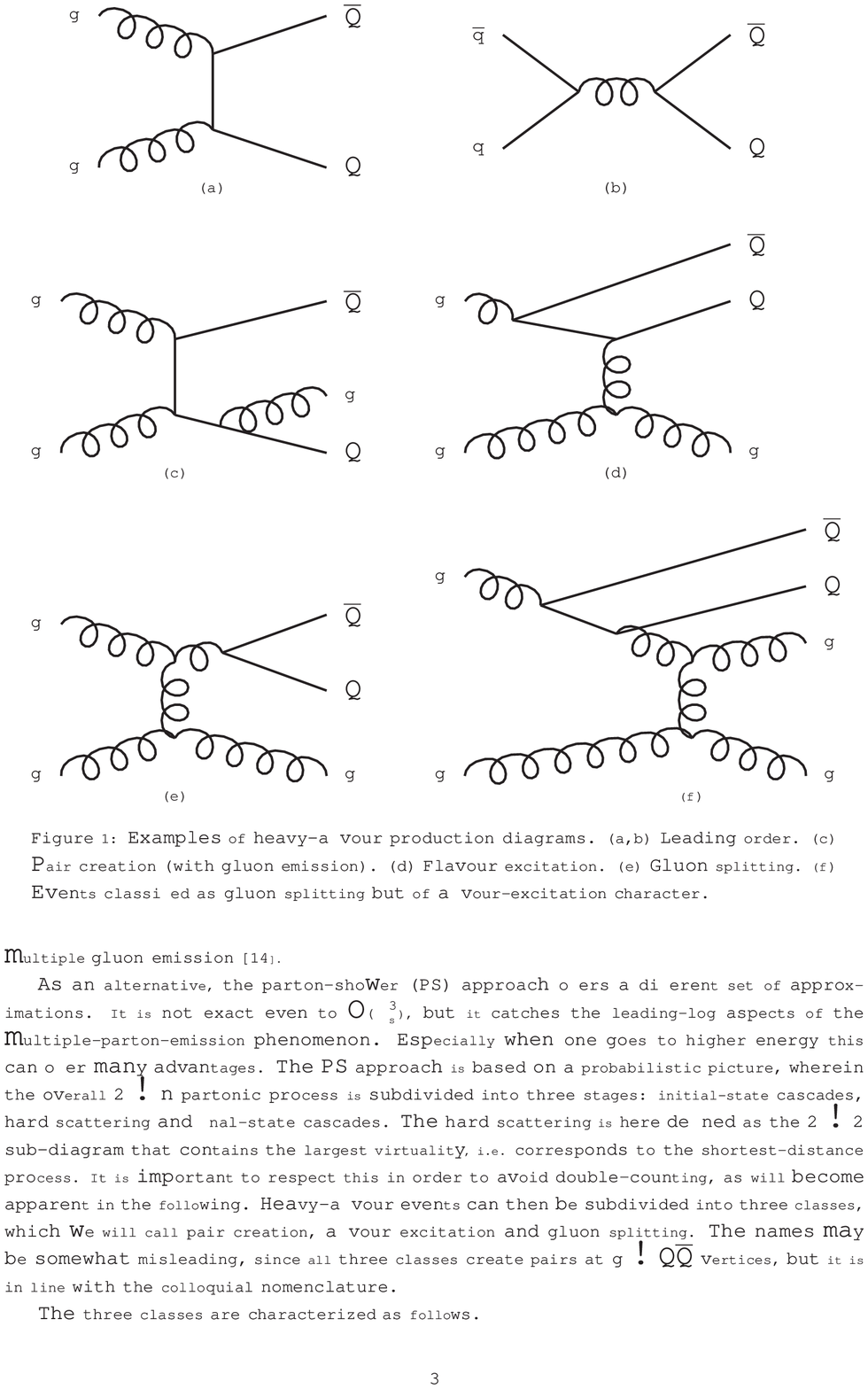,width=0.8\linewidth,clip}
\caption{\label{fig:ch2.QQ_diag}LO and most important NLO heavy
quark production diagrams. LO - a) "gluon fusion" b)
"quark-antiquark annihilation"  NLO - c) Pair creation with gluon
emission in output channel d) "flavor excitation" e) "gluon
splitting" f) "gluon splitting but of "flavor excitation"
character ~\cite{PShower}}
\end{figure}

In the partonic model we can assume that the colliding protons
consist of a collection of partons (quarks, antiquarks, gluons)
that interact with the content of the other proton. In Leading
Order pQCD, hard scattering of the partons can create heavy
quark-antiquark pairs $\QQ$\footnote{$Q$ index refers to heavy
quarks} by the diagrams, shown in Fig.~\ref{fig:ch2.QQ_diag}a,b).
LO sub-processes are usually called "\textbf{gluon fusion}"
($gg\rightarrow \QQ$) and "\textbf{quark-antiquark annihilation}"
($\qq \rightarrow \QQ$).

The general perturbative extension for the total crossection of
quark pair production on the partonic level can be expressed by
the following equation~\cite{whatK}.

\begin{eqnarray}
\label{eq:ch2.partonic_crossection}
\sigma_{ij}(\hat{s},m_Q^2,\mu_R^2) =
\frac{\alpha^2_s(\mu_R^2)}{m_Q^2} \sum\limits_{k=0}^{\infty} \,\,
\left( 4 \pi \alpha_s(\mu_R^2) \right)^k \sum\limits_{l=0}^k \,\,
f^{(k,l)}_{ij}(\eta) \,\, \ln^l\left(\frac{\mu_R^2}{m_Q^2}\right)
\end{eqnarray}

where $\hat{s} = x_1x_2s\,$ is the partonic energy in center of
mass. The useful dimensionless parameter $\eta =
\hat{s}/4m_Q^2-1$~\cite{Vogt_eta} reflects the threshold
production of the heavy quark ($\sqrt{\hat{s}}$ should be at least
$2m_Q$ to create a quark-antiquark pair). $i$ and $j$ are the
partonic indexes corresponding to each particular Feynman diagram
on the Fig.\ref{fig:ch2.QQ_diag}. $f^{(k,l)}_{ij}(\eta)$ is a
dimensionless scaling function representing the amplitude of a
given partonic scattering diagram. Index $k$ shows the order of
the process diagram, $k=0$ corresponds to LO, one can see that in
the leading order the total crossection is a function of
$\alpha_s^2$. Next order of perturbation $k=1$ corresponds to
Next-Leading-Order (NLO) contributions (some of the NLO diagrams
are shown in Fig.~\ref{fig:ch2.QQ_diag}c-d)). The NLO crossection
is of the order of $\alpha_s^3$ and in principle are of the next
order of smallness, unfortunately, the NLO crossection develops a
logarithms term for $l=1$ which can $ln(\frac{\mu_R^2}{m_Q^2})$ be
make NLO crossection of the order and even higher then the LO
contribution. Case of $k=2$ is usually referred to as Next-to-Next
Leading Order (NNLO) so far the exact crossection for NNLO heavy
quark production has not been calculated because of a large number
of contributing diagrams and mathematical complications in
resummation procedure. There have been theoretical attempts to
calculate the NNLO crossection contribution near production
threshold~\cite{Vogt_eta}.

$Q^2$ in Eq.~\ref{eq:ch2.partonic_crossection} is called
\textbf{renormalization scale} and usually assumed to be
proportional to $m_Q^2$ or $m_Q^2+p_{T\, Q}^2$ depending on the
model. The proportionality coefficient $sc_R$ in $\mu_R^2 =
(sc_R\cdot m_Q)^2$ is a variable parameter in QCD and usually
assumed to be in the range of $sc_R\in (\frac{1}{2} ; 2)$. Low
values of $sc_R = \frac{1}{2}$ for charm quark creation produces
scales on the order of $m_c^2/4 \le 1\ GeV^2$ and from
Fig.~\ref{fig:ch2.alpha_qcd} one can see that $\alpha_s$ becomes
quite large to utilize perturbative technique.

The next general step in QCD is to assume that we know the
distribution of the partons inside of the proton. This
distribution is called \textbf{parton distribution function}
(PDF)~\cite{PDFLIB,CTEQ5} $f_i^p(x,\mu_F^2)$ described in terms of
Feynman variable $x$ and the momentum transfer scale $\mu_F^2$.
$x$ is a relative fraction of partonic momentum to the total
momentum of the hadron $x_i=\frac{p_{z\, i}}{p_{z\, max}}$. Scale
dependence of the PDF is described by DGLAP
equations~\cite{gribov,altarelli,dokshitzer} and the shape of the
parton distribution is derived from comparison of pQCD predictions
to experimental measurements.

\begin{figure}[h]
\centering
\epsfig{figure=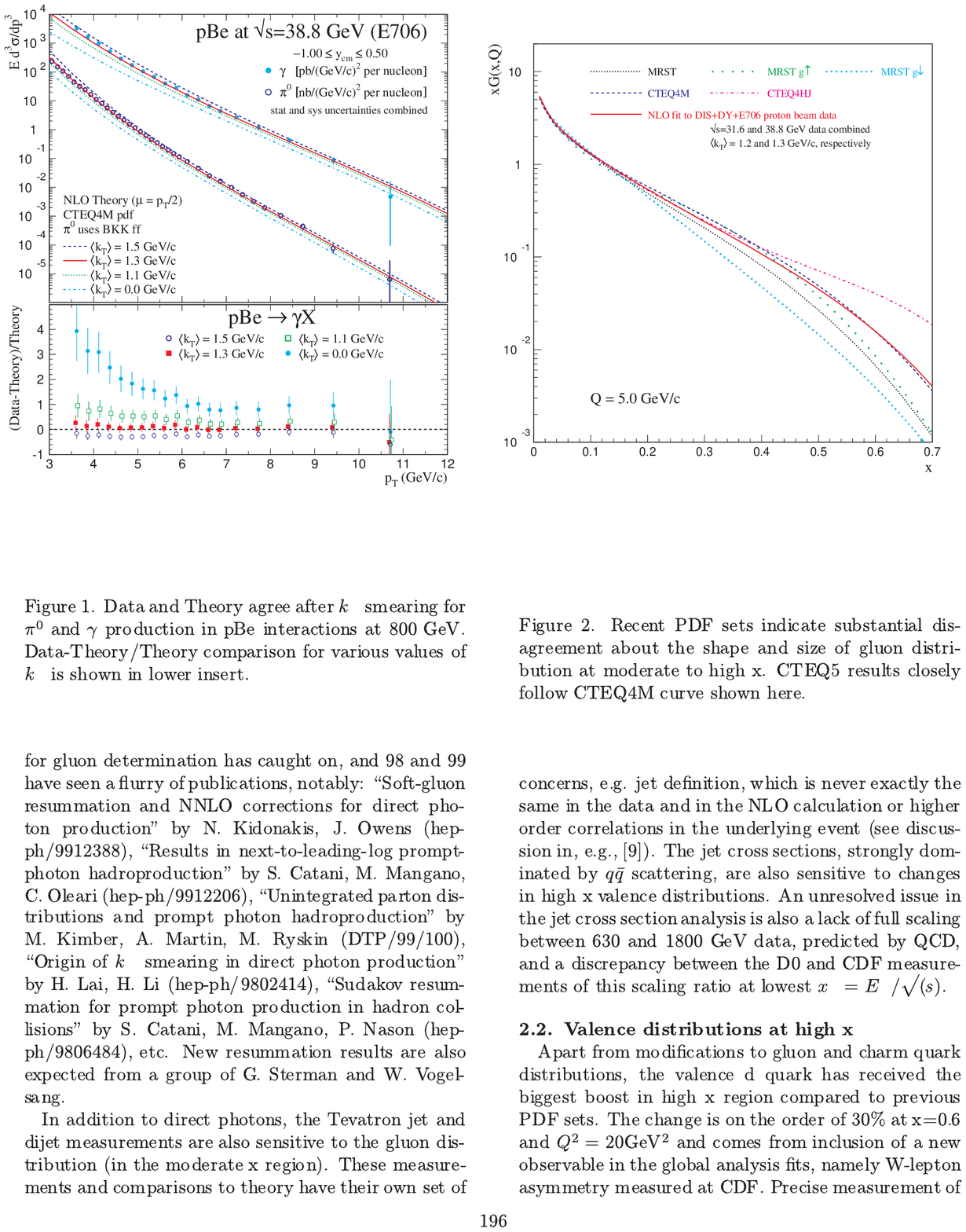,width=1\linewidth,clip}
\caption{\label{fig:ch2.qcd_fits}  Comparison of the photon and
direct-$\gamma$ measured by $E706$ experiment in $p+Be$ collisions
at $\sqs = 38.8$ GeV with pQCD predictions for different values of
$\kt$ smearing (left panel). Comparison of gluon structure
function for most resent PDF sets (right panel).~\cite{begel}}
\end{figure}

Fig.~\ref{fig:ch2.qcd_fits} shows the most striking agreement of
pQCD prediction and the results of $\pi^0$ and direct-$\gamma$
production in $p+Be$ collisions from $E706$ experiment. The
results show the existence of non-zero smearing of the transverse
$\kt \approx 1$ GeV of the parton (so called intrinsic $k_T$).
Experimental results helped to constrain the gluon distribution
function with very good accuracy. There is a big variety of PDF
functions developed by different theory groups. At the current
moment, the CTEQ group's PDF set~\cite{CTEQ5} seems to be the most
accurate in describing existing experimental data for the
structure of the proton.

Taking the assumption of the parton distribution, we can write the
total crossection for Heavy Flavor production ~\cite{whatK}

\begin{eqnarray}
\label{eq:ch2.total_charm} \sigma_{pp}(s,m_Q^2) = \sum_{i,j =
q,{\overline q},g} \int_{\frac{4m_Q^2}{s}}^{1} d\tau
\int_{\tau}^{1} \frac{dx_1}{x_1} \, f_i^p(x_1,\mu_F^2)
f_j^p\bigg(\frac{\tau}{x_1},\mu_F^2 \bigg) \, \sigma_{ij}(\tau
s,m_Q^2,\mu_R^2) \nonumber \\
\end{eqnarray}

where $\sigma_{ij}(\tau s,m_Q^2,\mu_R^2)$ is the partial partonic
crossection (Eq.~\ref{eq:ch2.partonic_crossection}), $\mu_F^2$ is
a momentum transfer scale (\textbf{factorization scale}) of the
PDF factorization. The usual assumption for pQCD calculations is
to expect the factorization and renormalization scales to be
similar $\mu_R^2=\mu_F^2=\mu^2$ since all PDFs uses this
assumption \textit{a priory} in calculations~\cite{whatK}.

\pagebreak
\section{\QQ\ fragmentation}

After the quark-antiquark pair is created, each of the produced
heavy quarks picks up a light quark from the proton in order to
create a color singlet object (\textbf{string}). The string is
fragmented into hadrons using a phenomenologically determined
\textbf{fragmentation function} $D(z)$ where $z$ is the momentum
fraction of the quark, carried by the hadron. $D(z)$ determines
the probability of producing hadron with given momentum.

By default, PYTHIA (a widely used pQCD Monte Carlo program) uses
the Lund fragmentation function~\cite{LUND} modified by
Bowler~\cite{bowler}:

\begin{equation}
D(z)_{LUND} \propto \frac{(1-z)^a}{z^{1+bm_Q^2}}
\exp{\left(\frac{-bm_T^2}{z}\right)} \label{eq:ch2.lund}
\end{equation}

where $m_T^2 = m_h^2+p_T^2$ is the transverse mass of the hadron.
The default parameter values in PYTHIA are $a=0.3$ and $b=0.58$
GeV$^{-2}$.

\pagebreak

A different parametrization of the fragmentation function was proposed
by Peterson~\cite{peterson}

\begin{equation}
D(z)_{Peterson} \propto \frac{1}{z(1-1/z-\epsilon /(1-z))^2}
\label{eq:ch2.peterson}
\end{equation}

The central value for $\epsilon$ in $D$-meson decay usually assumes to
be $\epsilon=0.06$~\cite{pet_eps,chrin} in case of LO theory fits
to the data, this parameter is much smaller if NLO fits are used
for parameter extraction
($\epsilon=0.01-0.02$)~\cite{two_lectures}.

Now we have everything to calculate the single differential
crossection for the produced Heavy Flavor hadrons. The
non-perturbative hadron production can be obtained using the
\textit{factorization theorem}:

\begin{equation}
\frac{d\sigma^H}{dp_T} = \int dp^Q_T\, dz\,
\frac{d\sigma^Q}{dp^Q_T}\,D(z)\,\delta(p_T-zp^Q_T)
\label{eq:ch2.factorization}
\end{equation}

where $\frac{d\sigma^Q}{dp^Q_T}$ is single differential
crossection for heavy quark.

Heavy Flavor hadroproduction is currently a subject to significant
uncertainty. The most accurate approach to deriving the
fragmentation function is to use the Mellin transforms of the
$D(z)$ and obtain the momenta of this transform from the
experimental data~\cite{saga,excess,pt_spectrum}. Mellin
transformation is a decomposition of the function on the powers of
$N$ according to the following formula~\cite{excess}:
\begin{equation}
 D_N \equiv \int D(z)z^N\frac{dz}{z}
 \label{eq:ch2.mellin}
\end{equation}

Assuming the quark $p_T$ distributions have power law behavior
$\frac{d\sigma}{d\hat{p}_T} = \frac{A}{\hat{p}_T^n}$ in the
neighborhood of some $p_T$, one can immediately find from
Eq.~\ref{eq:ch2.factorization} that

\begin{equation}
\frac{d\sigma^H}{dp_T} = \int d\hat{p}_T\, dz\,
\frac{A}{\hat{p}_T^n}\,D(z)\,\delta(p_T-z\hat{p}_T) =
\frac{A}{p_T^n}D_n \label{eq:ch2.factorization_decomp}
\end{equation}

\begin{figure}[t]
\centering
\epsfig{figure=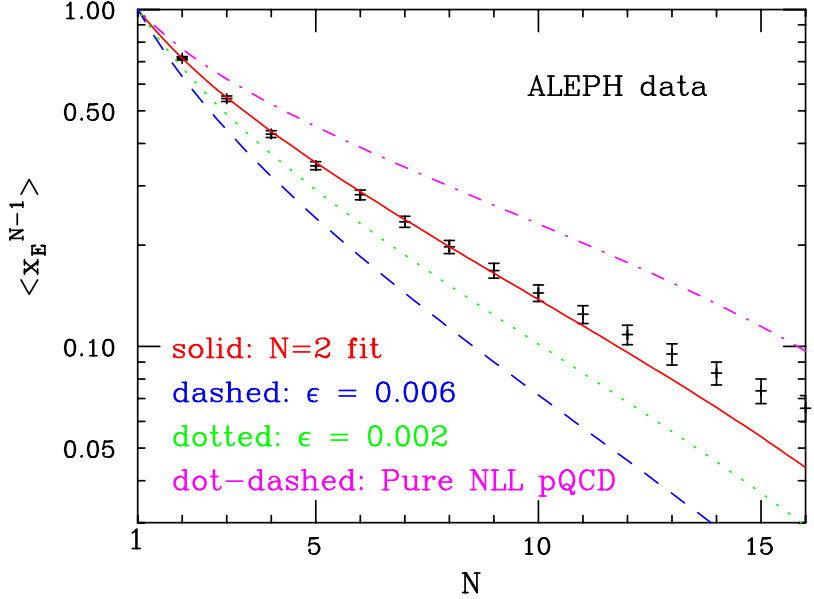,width=0.8\linewidth,clip,trim = 0in
0.1in 0.1in 0.1in} \caption{\label{fig:ch2.aleph} Momenta of
measured B meson fragmentation function compared with pQCD NLL
calculations with different assumptions for $D(z)$. The solid line
is one-parameter fit to the second momentum~\cite{excess}.}
\end{figure}

Thus, the hadronic crossection is given by the product of $n^{th}$
moment of the fragmentation function. Recent results from the
ALEPH collaboration on the bottom meson energy transfer indicates
that Peterson fragmentation form \textbf{does not} describe the
calculated momenta of the fragmentation function (see
Fig.~\ref{fig:ch2.aleph}). More studies of fragmentation function
form for Heavy Flavor mesons need to be performed to remove this
apparent uncertainty.

\section{Heavy Flavor meson decay}

\begin{figure}[h]
\centering
\epsfig{figure=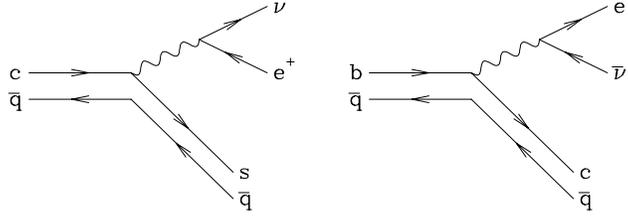,width=0.65\linewidth,clip,trim
= 0in 0.1in 0.1in 0.2in} \caption{\label{fig:ch2.sl_decay}
Semi-leptonic decay diagrams for Open Charm and Open Bottom meson
semi-leptonic decay.}
\end{figure}

Heavy flavor semi-leptonic decay diagrams for $D$ and $B$ mesons are
shown in Fig.~\ref{fig:ch2.sl_decay}. This weak decay channel
should be calculable using the standard technique, developed for
decay of free mesons (\textit{spectator model})~\cite{Ellis}.

The total semi-leptonic decay rate can be written as
\begin{equation}
 \Gamma_{sl}^Q=\frac {m_Q}{2^8\pi^3}\int dxdy\, \theta(x+y-x_m)\,
 \theta(x_m-x-y+xy)\times \overline{\sum}|M^Q|^2
 \label{eq:ch2.sl_rate}
\end{equation}

where $x = 2E_e/m_Q$ and $y=2E_\nu/m_Q$ are the rescaled energies
of charged and neutral lepton in the heavy quark rest frame.
$x_m=1-(m_q/m_Q)^2$ is kinematic limit to the energy transfer.

Matrix decay elements equal to:
\begin{equation}
 \overline{\sum}|M^c|^2 = 64 G_F^2|V_{cs}|^2\, c\cdot e^{+}\, s\cdot
 \nu \ \ \
 \overline{\sum}|M^b|^2 = 64 G_F^2|V_{cb}|^2\, b\cdot \nu\, c\cdot
 e^{+}
\end{equation}

where $c$, $b$, $s$, $e^+$, $\nu$ are corresponding 4-momenta of
the decay particles. $|V_cs|$, $|V_cb|$ are elements of
Cabibbo-Kobayashi-Maskawa (CKM) flavor mixing matrix. $G_F =
1.16637 \cdot 10^{-5}\ GeV^{-2}$ is Fermi constant of weak
interaction.

From Eq.~\ref{eq:ch2.sl_rate} we can now calculate the
differential semi-leptonic decay rate for the Heavy Flavor meson.

\begin{figure}[b]
\centering
\epsfig{figure=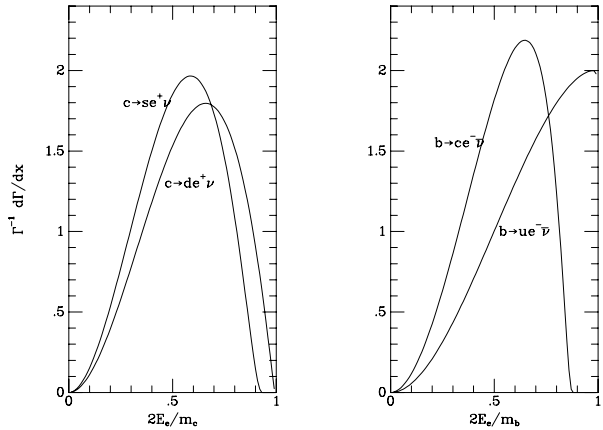,width=0.6\linewidth,clip}
\caption{\label{fig:ch2.rate} Semi-leptonic decay rate as a
function of $2E_e/M_Q$ for Open Charm (left panel) and Open Bottom
(right panel)~\cite{Ellis}.}
\end{figure}

\begin{eqnarray}
 \frac{d\Gamma_{sl}^c}{dx}&=& |V_{cs}|^2
 \Gamma_0(m_c)\left[\frac{12x(x_m-x)^2}{(1-x)}\right]\\
 \frac{d\Gamma_{sl}^b}{dx}&=& |V_{cb}|^2
 \Gamma_0(m_b)\left[\frac{2x^2(x_m-x)^2}{(1-x)^3}\right](6-6x+x\,x_m+2x^2-3x_m)\nonumber
 \label{eq:ch2.def_rate}
\end{eqnarray}
where $\Gamma_0(m_Q) = \frac{G_F^2\, m_Q^5}{192\,\pi^3}$.
Fig.~\ref{fig:ch2.rate} shows the resulting decay rate as a
function of relative electron energy for $c\rightarrow s$,
$c\rightarrow d$, $b\rightarrow c$, $b\rightarrow u$ semi-leptonic
decays.

Thus, pQCD theory provides us with the solid prescription to
calculate the Heavy Flavor production, hadronization and decay.

\chapter{PHENIX Experiment}\label{sec:ch3}
\section{PHENIX Detector}

PHENIX is one of the most advanced detector systems in High Energy
Nuclear Physics. It consists of 14 independent detector subsystems
collecting the information about the particles produced in Heavy
Ion collisions. The PHENIX tracking system provides accurate
measurements of charged particle tracks and particle
identification (electron, muon, hadron identification) over wide
range of momentum. Together with charged particle tracking, PHENIX
also performs a high precision measurements of photons via a large
area Electro-Magnetic Calorimeter. A sophisticated Level-1 and
Level-2 trigger system, enables a unique ability to address
specific $\bold{rare}$ physics events recording at high luminosity
($J/\psi$ leptonic decays, for example, by triggering on a pair of
highly energetic electron candidates). All those factors make
PHENIX both the most challenging and the most capable detector in
RHIC to study Heavy Flavor particle productions.

\begin{table} [h]
  \begin{center}
  \caption{Summary of the PHENIX Detector Subsystems~\cite{PHENIXCDR}.}
  \protect
  \label{tab:phenix_sub}
    \begin{tabular} {|l|c|c|p{2.1in}|} \hline
     Element & $\Delta\eta$ & $\Delta\phi$
                             & Purpose and Special Features\\ \hline
     Magnet: central (CM) & $\pm$0.35 &360$^{\circ}$
                                       & Up to 1.15 T$\cdot$m. \\
     $\;\;\;\;\;\;\;$$\;\;\;\;\;\;\,$ muon (MMS)
                & -1.1 to -2.2 &360$^{\circ}$ & 0.72 T$\cdot$m for
                                     $\eta=2$ \\
     $\;\;\;\;\;\;\;$$\;\;\;\;\;\;\,$ muon (MMN)
                 & 1.1 to 2.4 &360$^{\circ}$ & 0.72 T$\cdot$m for
                                     $\eta=2$  \\      \hline
     Silicon (MVD)  & $\pm$2.6 & 360$^{\circ}$
                        & $d^2N/d\eta d\phi$, precise vertex, \\
                        &&& reaction plane determination \\
     Beam-beam (BBC) &$\pm$(3.1 to 3.9)&360$^{\circ}$
                                 & Start timing, fast vertex. \\
   NTC &$\pm$(1 to 2) &320$^{\circ}$ & Extend coverage of BBC for p-p and p-A.
   \\
     ZDC & $\pm2$ mrad & 360$^{\circ}$ & Minimum bias trigger. \\ \hline
     Drift chambers (DC)&$\pm 0.35$&90$^{\circ}$$\times$2
                         & Good momentum and mass resolution, \\
      & & & $\Delta m/m=0.4$\% at $m = 1 {\rm GeV}$. \\
     Pad chambers (PC)  &$\pm 0.35$&90$^{\circ}$$\times$2
                               & Pattern recognition, tracking \\
               & & & for nonbend direction.\\
     TEC         &$\pm 0.35$&90$^{\circ}$
                               & Pattern recognition, $dE/dx$.\\ \hline
     RICH        &$\pm 0.35$&90$^{\circ}$$\times$2
                                   & Electron identification. \\
     ToF         &$\pm 0.35$&45$^{\circ}$
             &  Good hadron identification, $\sigma <$100 ps.  \\
     T0  & $\pm 0.35$ & 45$^{\circ}$ & Improve ToF timing for p-p and p-A.
     \\ \hline
     PbSc EMCal     &$\pm 0.35$ &90$^{\circ}$+45$^{\circ}$
                     & For both calorimeters, photon and electron detection.\\

     PbGl EMCal &$\pm 0.35$&45$^{\circ}$ & Good $e^{\pm}/\pi^{\pm}$ separation
             at $p>1$ GeV/c by EM shower and $p<0.35$ GeV/c by ToF.\\

             & & & $K^{\pm}/\pi^{\pm}$ separation up to 1 GeV/c by ToF.
             \\ \hline
     $\mu$ tracker: ($\mu$TS)& -1.15 to -2.25 &360$^{\circ}$
                              &  Tracking for muons. \\
$\;\;\;\;\;\;\;\;\;\;\;\;\;\;\;$ ($\mu$TN)& 1.15 to 2.44
&360$^{\circ}$
                        & Muon tracker north installed for year-3 \\
     $\mu$ identifier: ($\mu$IDS) & -1.15 to -2.25 &360$^{\circ}$
                        & Steel absorbers and Iarocci tubes for \\
$\;\;\;\;\;\;\;\;\;\;\;\;\;\;\;\;\;\;$ ($\mu$IDN) & 1.15 to 2.44 &
                  360$^{\circ}$ & muon/hadron separation. \\ \hline
    \end{tabular}
\end{center}
\end{table}

\begin{figure} [hvt]
\centering \epsfig{file=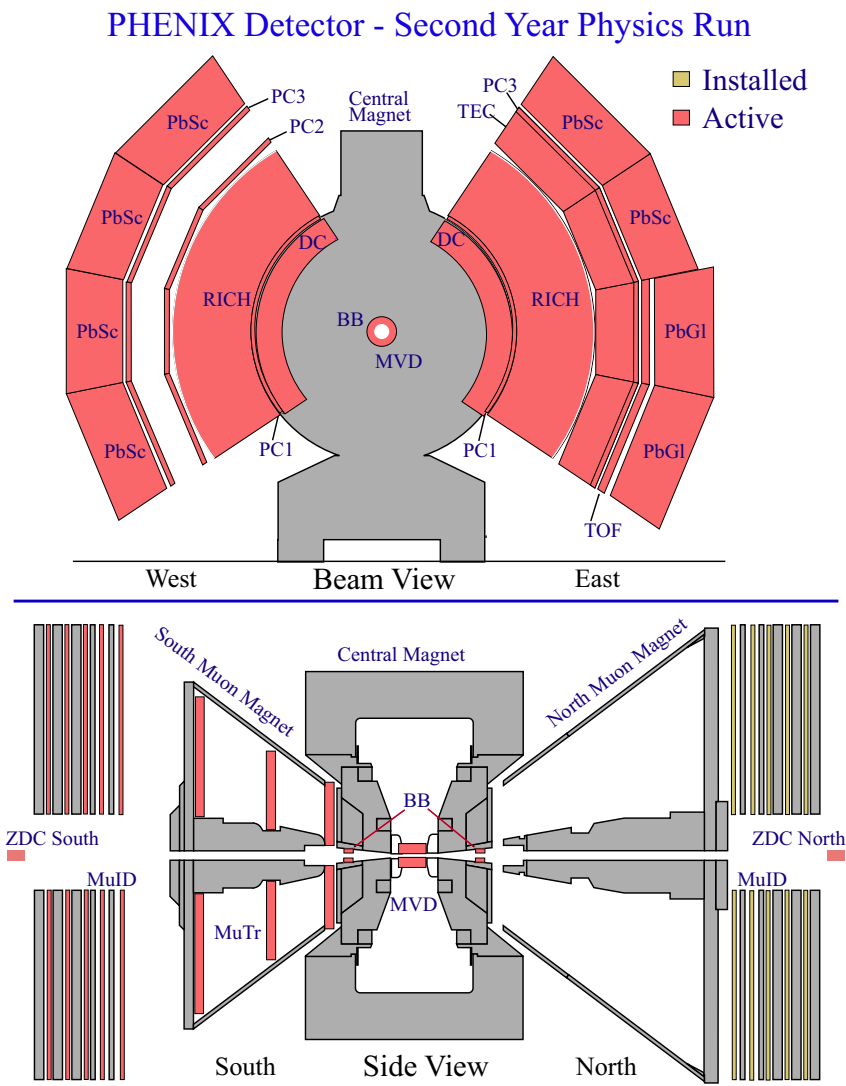,width=0.8\linewidth}
\caption{\label{fig:ch3.phenix} Layout for PHENIX Experiment. }
\end{figure}

The detectors in PHENIX are grouped into three major categories by
design and specific physics tasks they are intended to perform:
\begin{itemize}
    \item Global Trigger Detectors
    \item Central Arm Detectors
    \item Muon Arm Detectors
\end{itemize}
The PHENIX detector if shown in Figure~\ref{fig:ch3.phenix}.
It is symmetric around mid-rapidity with the interaction
point positioned in the center of the magnets, commonly referred to as
the Interaction Region (IR). Two Central Arm detectors (referred to as
East and West) are located around the interaction point and cover the
rapidity range $-0.35<\eta<0.35$. Each Central Arm covers $90^{\circ}$
in $\phi$ and populated with the particle detectors for precise
measurement of electrons, photons, and charged hadrons. Two Muon Arm
detector subsystems (referred to as South and North) measure
the muon and decay hadrons yield at forward and backward rapidity
region ($\eta \in(-2.25,-1.15)$ and $\eta \in(1.15,2.44)$). Global
Trigger detectors provide the Level-1 trigger information for the
collision by measuring the time and position of the interaction vertex
with high precision.

Detector parameters and performance,
rapidity and azimuthal angle ($\phi$) coverage, of each subsystem
are summarized in Table~\ref{tab:phenix_sub}.

We define a coordinate system relative to the beam axis. The origin is
located in the center of the IR with the $Z$-axis pointing along the
beam line from South to North. The $X$-axis is points horizontally
from East to West and the $Y$-axis points upwards, making a proper
right-handed coordinate system.

%

\section{Global Detectors}

Nuclear collisions are characterized by their impact parameter vector
(both magnitude and direction).  Both of these parameters can be
determined with reasonable collision on a event-by-event basis.
Nearly all physics measurements are then studied to determine their
variation with respect to these so-called ``global'' characteristics
of the event.

``Global'' detectors measure global characteristics of the
collisions and issue online trigger decision to read-out the
information. In PHENIX this task has been principally accomplished
by two subsystems: \linebreak the $\bold{Beam-Beam\ Counters}$
(BBC) and the $\bold{Zero\ Degree\ Calorimeters}$ (ZDC).  The BBC
consists of a set of two (South and North) fast trigger counters
providing information about $Z$ vertex position and collision time
with respect to the RHIC beam crossing clock ($t_{0\ BBC}$).  The
ZDCs are a set of two Tungsten calorimeters (South and North)
located far from the collision point and providing additional
information about the impact parameter of the collision. The third
global detector is the $\bold{Multiplicity\ Vertex\ Detector}$
(MVD), a silicon barrel detector designed to measure
${d^2N}/{d\eta\, d\phi}$ distribution for charged particles near
the collision point.  This detector is not yet functioning
reliably and was not used in my analysis.

\subsection{Beam-Beam Counter-BBC}\label{sec:ch3.BBC}

\begin{figure}[hbt]
\begin{tabular}{lr}
\begin{minipage}{0.5\linewidth}
\begin{flushleft}
\epsfig{figure=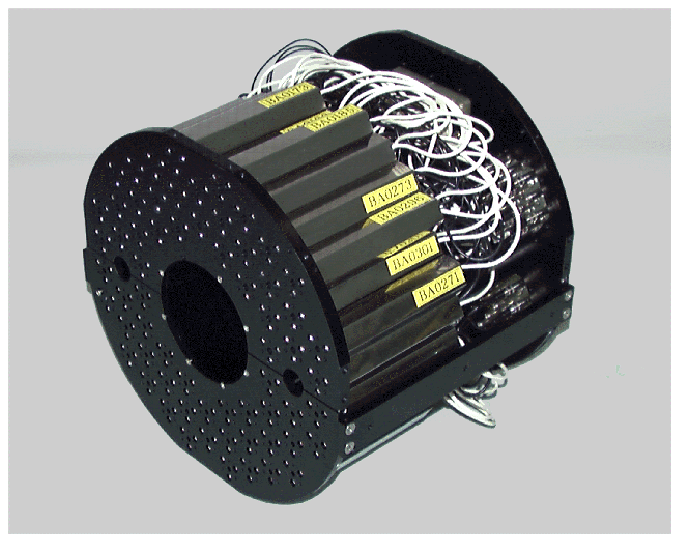,width=0.8\linewidth,clip}
 \caption{\label{fig:ch3.bbc_geo}
Picture of BBC barrel before the installation.}
\end{flushleft}
\end{minipage}
&
\begin{minipage}{0.5\linewidth}
\begin{flushright}
\epsfig{figure=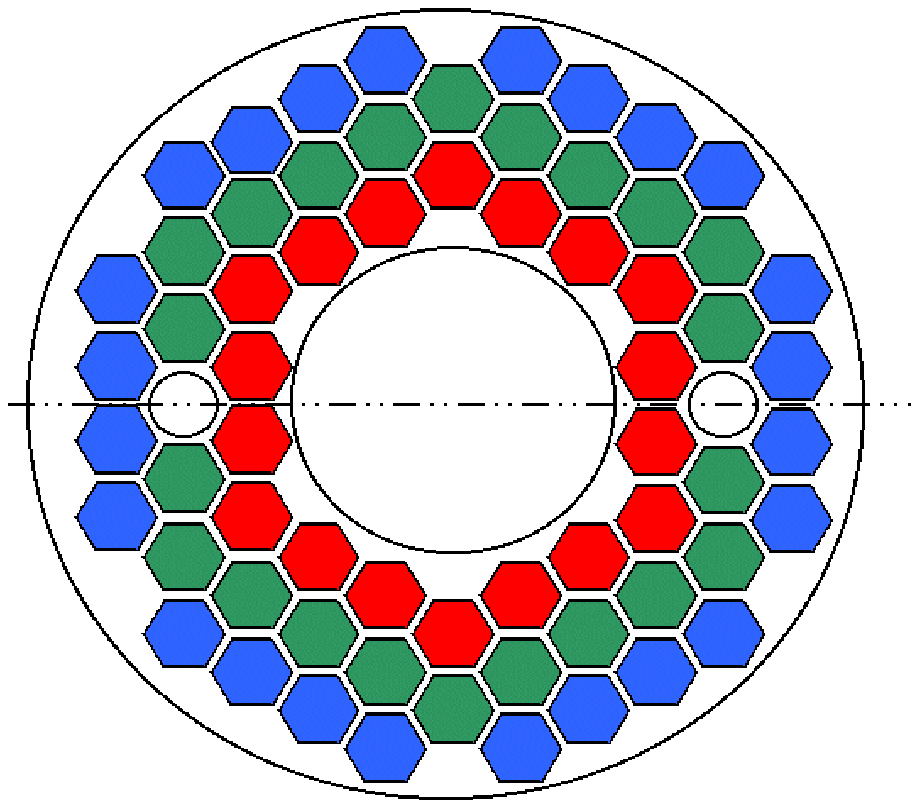,width=0.7\linewidth,clip}
 \caption{\label{fig:ch3.bbc_view}
Structural drawing of one BBC (beam view). Each box corresponds to
one PMT.}
\end{flushright}
\end{minipage}
\end{tabular}
\end{figure}

The BBC consists of two sets of quartz tube
Cerenkov arrays, which measure relativistic charged particles
produced in narrow cone around each beam axis
($3.0\leq\eta\leq3.9$, 2$\pi$ in $\phi$). Positioned at 1.4 m from
the PHENIX center point it has an outer radius of 30 cm and inner
radius of 5 cm (see Fig.~\ref{fig:ch3.bbc_geo}).

Each BBC counter consists of 64 photomultiplier tubes
(schematically shown in Fig.~\ref{fig:ch3.bbc_view}) equipped with
quartz Cerenkov radiators in front. The dynamic range of each tube
of the BBC allows to register 1-30 minimum ionizing particles
which makes it the main Minimum Bias trigger detector in PHENIX
for any collision species.

Each BBC PMT has intrinsic timing resolution of $\sigma_t = 50$
ps and provides high precision measurement of collision time and
vertex position. For each collision, the BBC measures time of the
collision with respect to the RHIC collider clock (synchronized
with beam bunches). This time is usually referred to as the
\textbf{BBC t-zero} and is one of the fundamental items of information
about the collision in that it sets the start time for all the subsystems
performing timing measurements.

$t_0^{BBC}$ is calculated as a half sum of average hit time over
individual BBC PMTs (denote as $t_N^{BBC}$, $t_S^{BBC}$ for North
and South detector). Obviously, vertex position can be calculated
as a half difference of those variable.
\begin{equation}
 t_0^{BBC} = (t_N^{BBC} + t_S^{BBC})/2 ;  \ \ \  \ \ \ \
 Z_{vtx}^{BBC} = (t_N^{BBC} - t_S^{BBC})/2c
 \label{eq:ch2.bbc_t0_z}
\end{equation}

The vertex resolution in $\pp$ collisions can be evaluated from the
assumption that the multiplicity in the BBC is small and we have one hit
in each BBC. Using Eq.~\ref{eq:ch2.bbc_t0_z}, we can estimate in
this case that $\sigma_Z = \sigma_t/\sqrt{2}c \approx 1.2$ cm. In
central $\AA$ collisions, accuracy of vertex measurement becomes
much better ($\sigma_Z \leq 0.3$ cm) due to the averaging effect
over all PMTs improves the timing resolution significantly.

The BBC timing is an essential input to Level-1 trigger providing
online information about the vertex of the collision. Level-1 trigger
electronics, generate trigger accept signals if the vertex lies within
$\pm 50$ cm from the PHENIX center point (to remove beam gas
interactions and interactions inside the magnet poles). This method
enables the experiment to obtain a clean sample of events.  Since the
BBC trigger is efficient for a wide variety of interaction processes, it
is referred to as a so-called ``Minimum Bias'' trigger.

\subsection{Zero Degree Calorimeter}

The Zero Degree Calorimeter is a small area hadron calorimeter
positioned $\approx 18$ m from the interaction point along the beam
axis. The main purpose of ZDC is measurement of the spectator neutron
rate in nucleus-nucleus collisions. Spectator protons and other
charged particles produced in the collision bend in the magnetic field
of RHIC dipole magnets and miss the ZDC acceptance. By measuring the
flux of neutrons in heavy ion collision in correlation with the
charged particles multiplicity in Beam-Beam counter, PHENIX obtains an
information about the impact parameter on the collision (typically
called the collision's ``centrality''). A single ZDC counter consists
of 3 modules each with a depth of two hadronic interaction lengths and
read out by a single PMT. The ZDC provides timing and amplitude
information similar to the BBC but with decidedly lower
resolution. The energy resolution at the one neutron peak is
approximately 21\%~\cite{zdcmickey,zdcnim}. In Au+Au collisions the
ZDC is an important part of Minimum Bias trigger, but in case of $\pp$
collisions, presented in the analysis, it has no particular use since
there are no spectator neutrons.

\begin{figure}[b]
\centering \epsfig{file=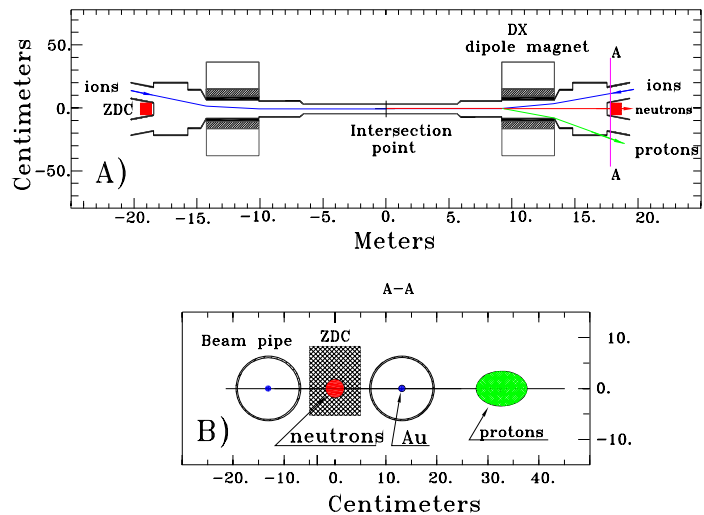,width=0.7\linewidth}
\caption{\label{fig:ch3.zdc} Schematic view of ZDC detector. A)
Top view of interaction region. B) Projected proton and neutron
deflection area at ZDC placement plane.}
\end{figure}

Fig.~\ref{fig:ch3.zdc} shows schematic view of ZDC detector. One
can also see the projected trajectory for $Au$ ions, spectator
protons and neutrons at the ZDC placement plane $A-A$.

\pagebreak

\section{Central Arm Detectors}

The Central Arm detectors provide unambiguous measurements of charged
particle track information in mid-rapidity range ($|\eta|<0.35$).  The
track's momentum is determined by measuring its deflection angle in
the magnetic field of the \textbf{Central Magnet} by the multi-wire
\textbf{Drift Chamber} (DCH).  A set of three \textbf{Pad Chambers}
(PC1, PC2, PC3) helps to reconstruct the $Z$ information of the track
and remove the background tracks. The \textbf{Ring Image Cherenkov}
detector (RICH) is essential for electron identification and provides
good $e/\pi$ separation for $p_T< 4.8$ GeV/c. Finally, the
\textbf{Electromagnetic Calorimeter} (EMC) measures the energy,
deposited by the charged particle or photon. The EMC also provides
significant hadron/electron separation, crucial for electron analysis.

The central arm subsystems are the most essential component of my electron
analysis and the final results rely on the stable
performance of all of the described detectors.

\subsection{Central Magnet}

\begin{figure}[hbt]
\begin{tabular}{lr}
\begin{minipage}{0.5\linewidth}
\begin{center}
\epsfig{figure=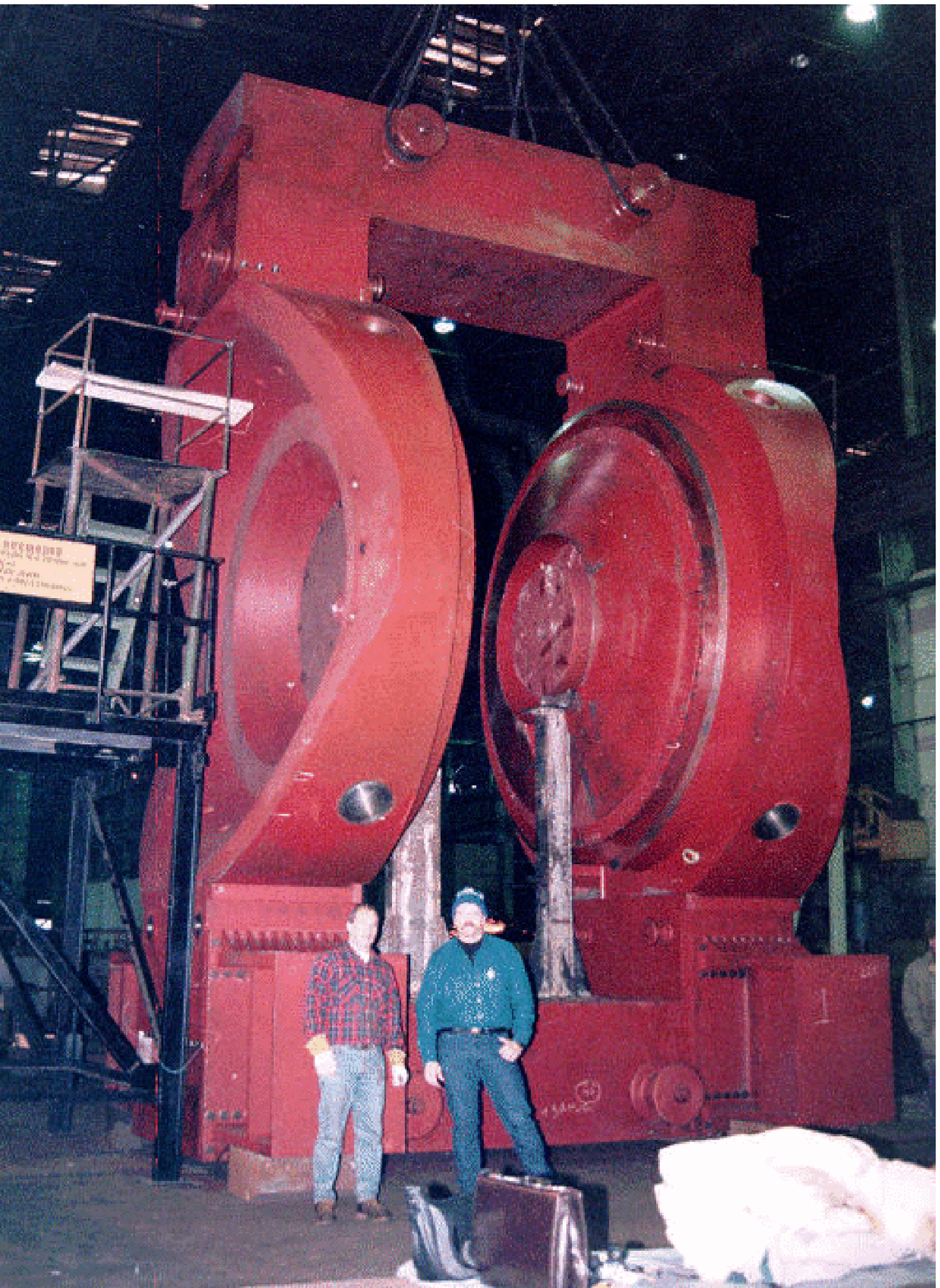,width=0.7\linewidth,clip}
 \caption{\label{fig:ch3.cm_view}
Central Magnet during assembly (Izhorskyi plant, St. Petersburg,
Russia).}
\end{center}
\end{minipage}
&
\begin{minipage}{0.5\linewidth}
\begin{center}
\epsfig{figure=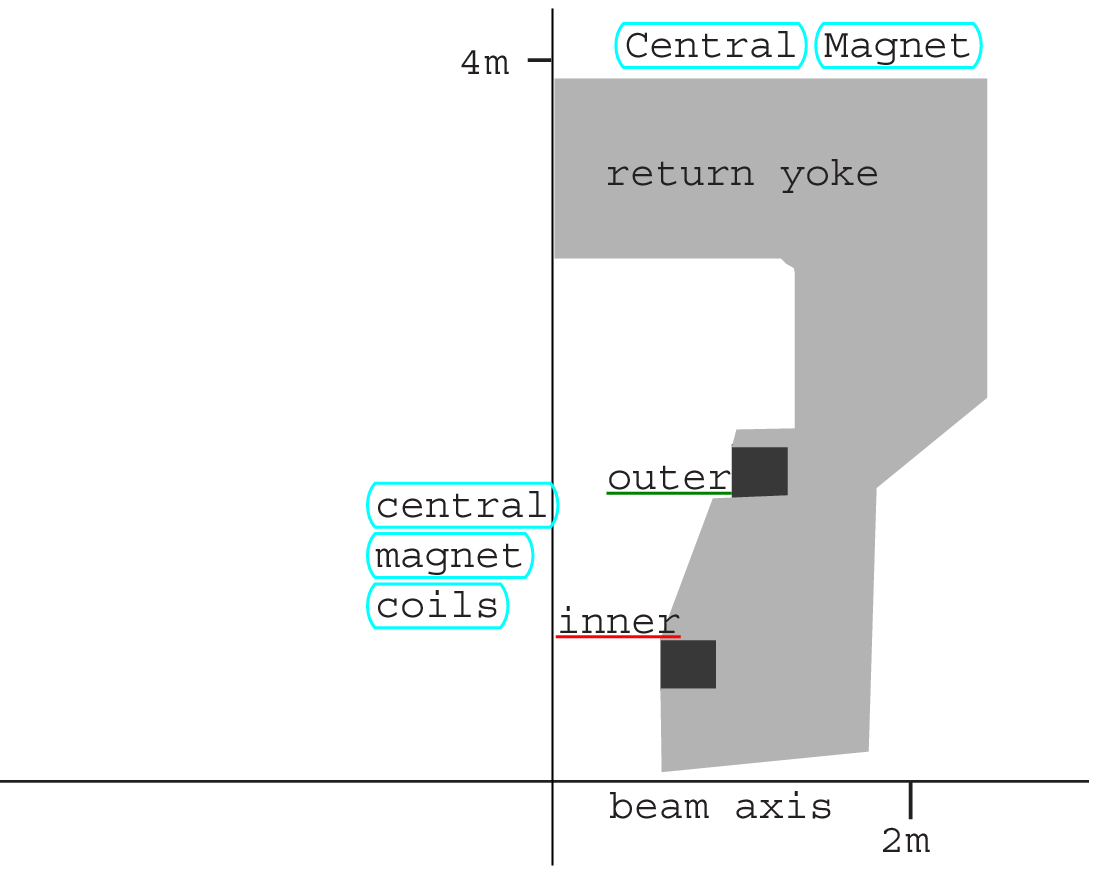,width=0.8\linewidth,clip}
 \caption{\label{fig:ch3.cm_cross}
Crossection view of Central Magnet coils and yolk.}
\end{center}
\end{minipage}
\end{tabular}
\centering
\epsfig{file=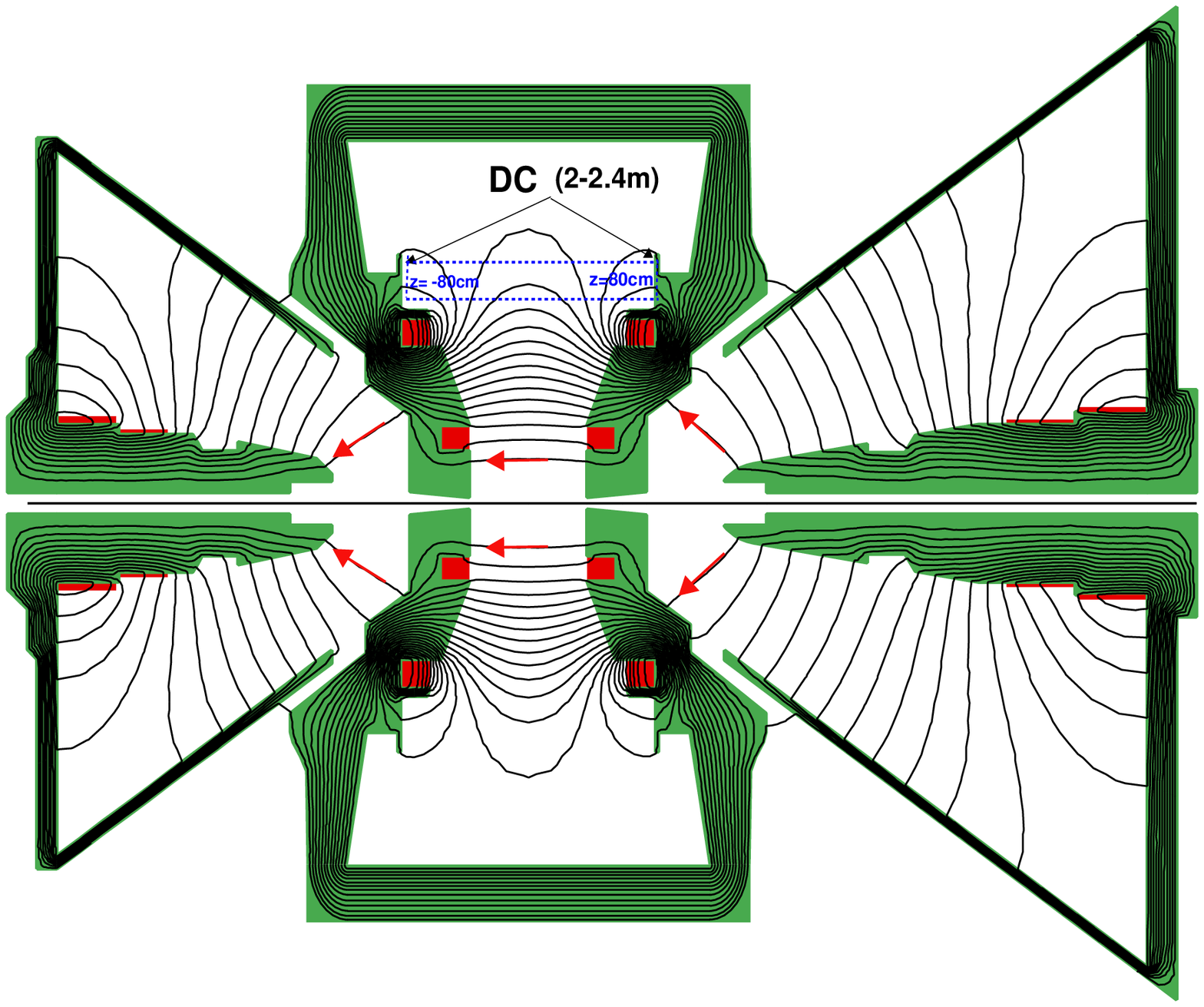,width=0.8\linewidth,clip}
\caption{\label{fig:ch3.mag_lines} Magnetic field lines in PHENIX
Magnet System. Drift Chamber location is shown by dashed box. }
\end{figure}

Although the Central Magnet (CM) is not a detector subsystem,
central arm tracking relies on its stable operation. The Central
Magnet consists of two coils (outer and inner) embedded into the
massive steel yolk that generate an axially symmetric magnetic
field in the region close to the interaction point. A picture of
the CM during production is shown in Fig.~\ref{fig:ch3.cm_view}
and the crossection schematic drawing is shown in
Fig.~\ref{fig:ch3.cm_cross}.

The main requirements to the Central Magnet listed below:
\begin{itemize}
    \item Provide an smoothly varying magnetic field that can be mapped
    with 0.2\% precision.
    \item Two coil operation, enabling creation of zero field at
    $R=0$ region in case of opposite coil polarization.
    \item Minimal material in the PHENIX Central Area detector
    acceptance in order to minimize photon conversion background.
    \item Magnetic Field strength should be significantly lower in the region of
    tracking detectors ($R >200$ cm). This allows to assume a
    straight track model for tracking.
    \item Movable configuration of the magnets in order to
    simplify detector assembling and commissioning.
\end{itemize}

\pagebreak

\begin{figure}[h]
\centering \epsfig{file=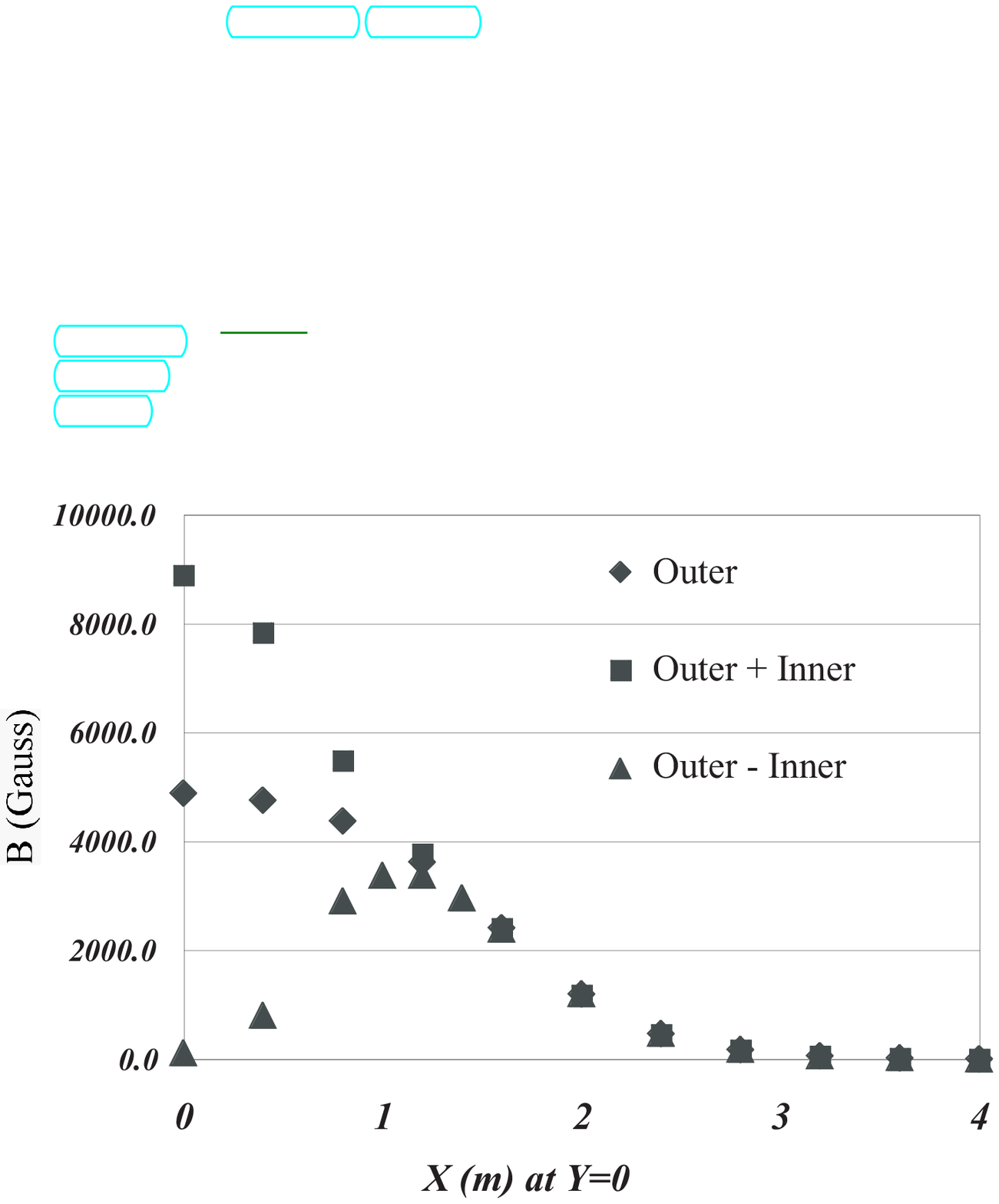,width=1\linewidth,clip}
\caption{\label{fig:ch3.mag_str} Magnetic field strength as
function of radius for three different configurations of the
magnetic coil polarizations.}
\end{figure}

All the tasks listed were successfully accomplished, the magnetic
field line contours in PHENIX Magnet system are shown in
Fig.~\ref{fig:ch3.mag_lines}. Magnetic field strength as a
function of radial distance $R$ on $Z=0$ plane for different
polarization of ``outer'' and ``inner'' magnetic coil is shown in
Fig.~\ref{fig:ch3.mag_str}. During Run02 PHENIX running period,
used for this analysis, only the ``outer'' coil was energized
which lead us to effective field integral for the charged track
$\int B dx = 0.78\ [T\cdot m]$. An additional ``inner'' coil used
in later runs helps to create an even stronger magnetic field in
the acceptance in order to improve the momentum resolution for
high $p_T$ tracks.

The field reaches 0.096~T (0.048~T) at the DCH inner and outer
radius at $Z=0$ and reach even smaller values at $|Z|=80$ cm which
allows us to assume straight track in the DCH drift volume.

\pagebreak

\subsection{Drift Chamber}\label{sec:DC}

PHENIX Drift Chamber system contains a set of two ``jet''-type
multiwire detectors and is the main tracking device in PHENIX. It is
placed at a radius of $R\in (202;246)$ cm and consists of two
identical arms each of which cover $90^\circ$. The DCH performs the
following tasks:

\begin{itemize}
    \item Accurate measurement of charged particle tracks in
$r$-$\phi$ plane for determination of their transverse momentum
$p_T$.
    \item Measure $Z$ and $\theta$ (inclination angle of the
    track with respect to Z axis) of the charged particle tracks together with PC1 and
    BBC.
    \item Provides input information for global tracking in
    PHENIX.
\end{itemize}

During the construction of the Drift Chamber the following design
requirements have been applied~\cite{PHENIXCDR}:

\begin{itemize}
\item Single wire resolution better than 150 $\mu m$ in $r$-$\phi$
direction \item Single track reconstruction efficiency better than
99\% \item Two track resolution better than 1.5 $mm$. \item
Spacial resolution in the $z$ direction better than 2 $mm$.
\end{itemize}

\begin{figure}
\begin{tabular}{lr}
\begin{minipage}{0.5\linewidth}
\centering \epsfig{figure=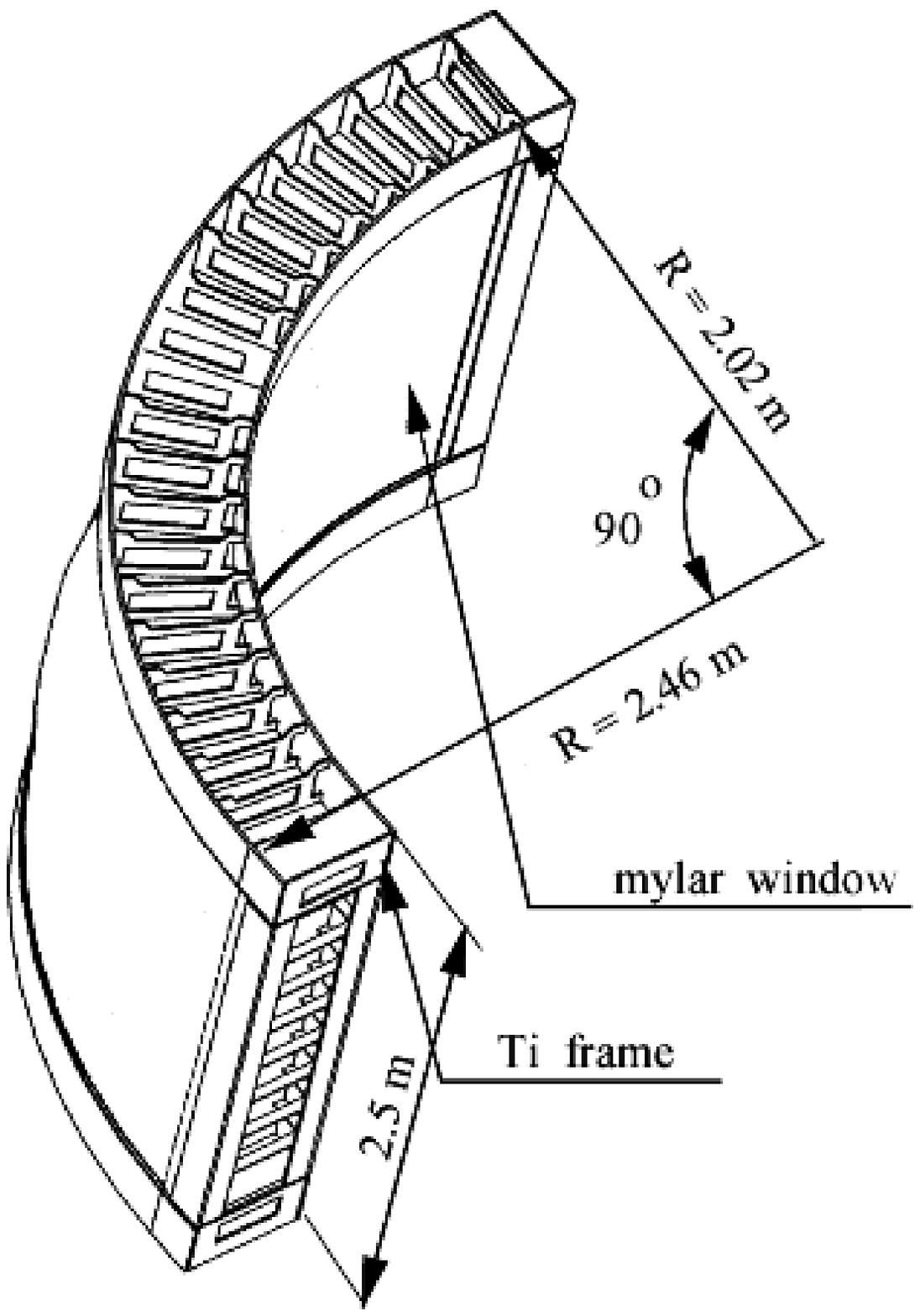,width=1\linewidth,clip}
 \caption{\label{fig:ch3.dc_arm}
Schematic drawing of one Drift Chamber Arm.}
\end{minipage}
&
\begin{minipage}{0.5\linewidth}
\centering \epsfig{figure=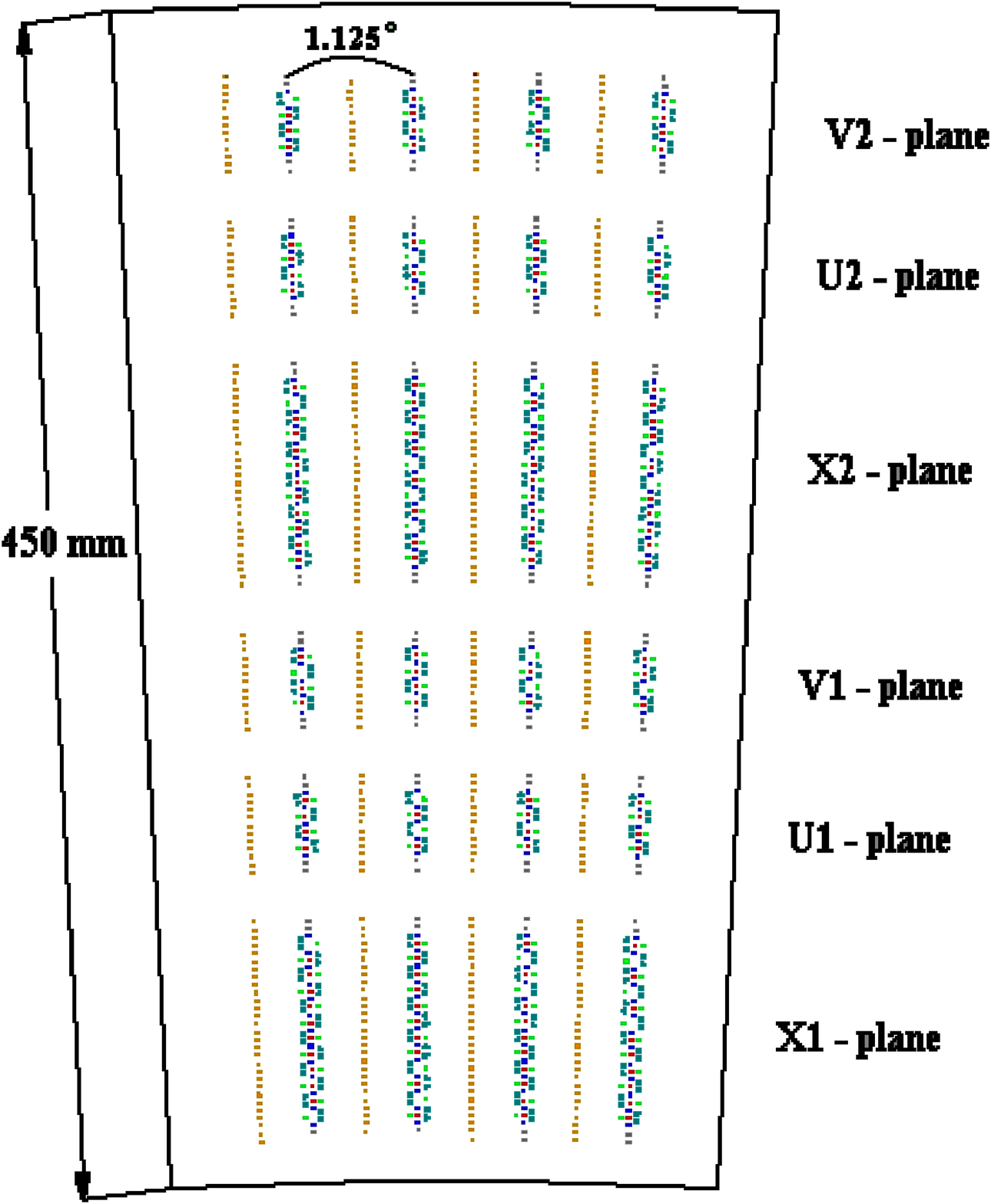,width=1\linewidth,clip}
 \caption{\label{fig:ch3.keystone_wire}
Wire structure of DCH Keystone.}
\end{minipage}
\end{tabular}
\centering \epsfig{figure=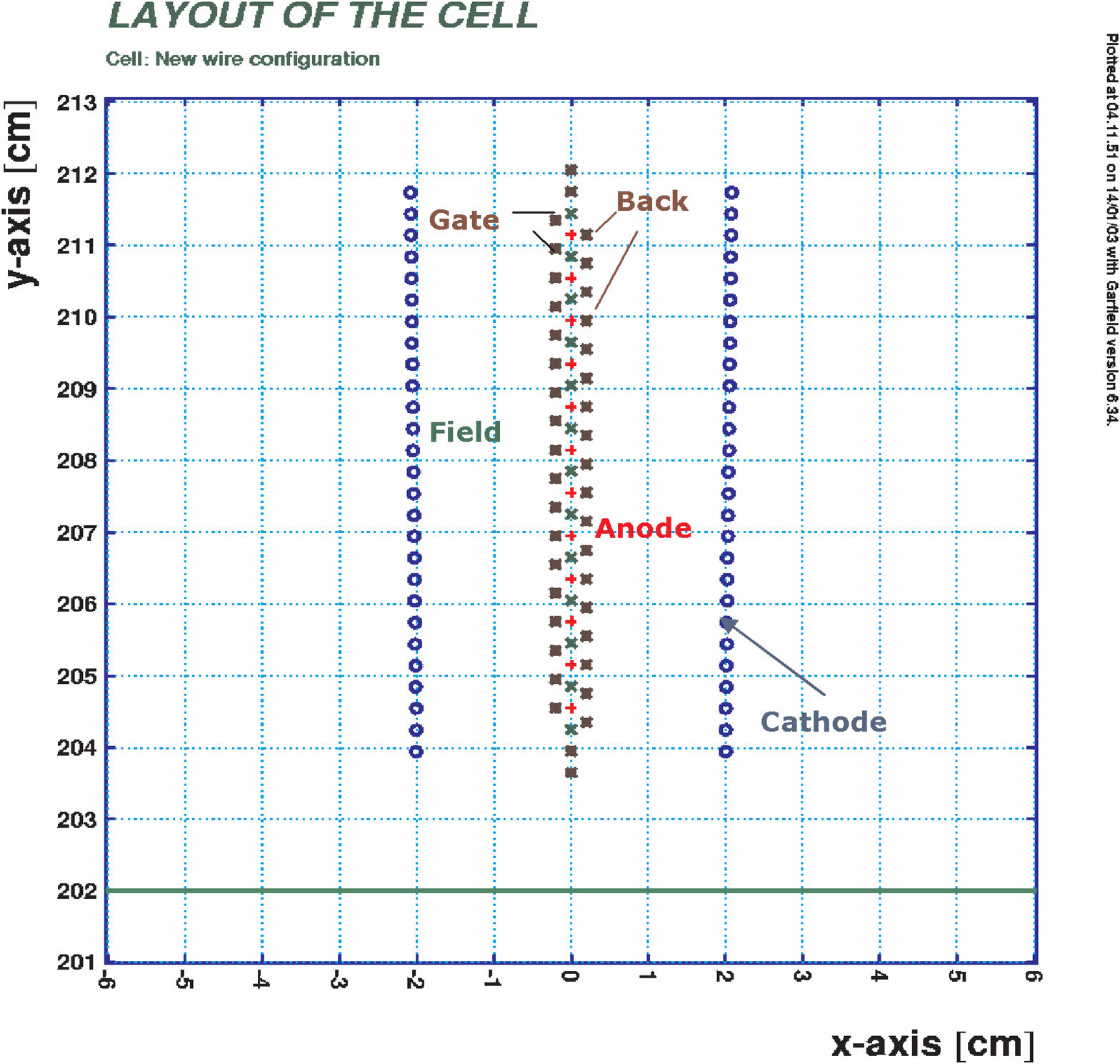,width=0.7\linewidth,clip}
 \caption{\label{fig:ch3.X1_cell}
Layout of the wire structure of X1 cell.}
\end{figure}

In order to reach the specifications, the following design concept
was implemented. A cylindrically shaped $Ti$ frame was built as
the support for wire nets with inner and outer radii of 202 and
246 cm and a length of 180 cm. A schematic view of one of the DCH
arms is shown in Fig.~\ref{fig:ch3.dc_arm}. The active volume of
the DCH is filled with Argon($50\%$)/Ethane($50\%$) gas mixture at
STP. Charged particle tracking is done by a set of wire nets
placed inside the gas volume of the chamber. There are total of 80
identical wire structures called nets around the azimuthal angle.
Each net covers $1.125^\circ$ in azimuth and is designed to
measure the position of the track within its coverage by measuring
the \textbf{drift time} ($t_{dr}$) of the charge clusters, ionized
by the incoming charged particle in the vicinity of the sense
(anode) wire.

The wire nets are subdivided into 6 separate sections along the
radius, named in the following order (from inside to outside radius)
X1, U1, V1, X2, U2, V2. Each X wire net consists of 12 anode wires and
measures the track trajectory in $r-\phi$ plane.  Each U, V net has 4
sense wires and designed to reconstruct $Z$ information for the
track. Groups of 4 wire cells share the same electronics set and high
voltage supply.  Such a grouping is called a \textbf{keystone}.
Fig.~\ref{fig:ch3.keystone_wire} shows the net configuration within
one keystone.

\newpage

One cell consists of an anode wire net, surrounded by a pair of
cathode nets.  The region of $\approx 2$ cm between cathode and
anode net is called drift region or charge collection region.
Fig.~\ref{fig:ch3.X1_cell} shows the wire structure of X1 wire
cell.

The wire nets have a complicated wire and high voltage configuration
in order to create a specially tailored electric field in the drift
region. In total 5 different values of high voltage are applied to
different wires in order to create a narrow and isochronous alley in
the drift region that supplies charge to the anode wire as shown in
Fig.~\ref{fig:ch3.X1_cell}.  The 5 voltages are named for the drift
characteristic they control:

\begin{itemize}
    \item ``Anode'' (sense) wires - Read-out the charge, induced by drifting
    ion current.
    \item ``Cathode'' wires - Create uniform electric field in the
    drift region \\
    ($V_C\approx -4100$ V).
    \item ``Back'' wires - Block charge drift from one side of the
    anode wire to solve left-right ambiguity ($V_B\approx -850$ V).
    \item ``Gate'' wires - Create a localized isochronous charge collection region and
    increase the field strength close to the anode wire ($V_G\approx -2000$ V).
    \item ``Field'' wires - Separate individual anode wires drift regions and
    create a strong electric field around the anode wire ($V_F\approx -2000$
    V).
\end{itemize}

\begin{figure}[t]
\centering \epsfig{figure=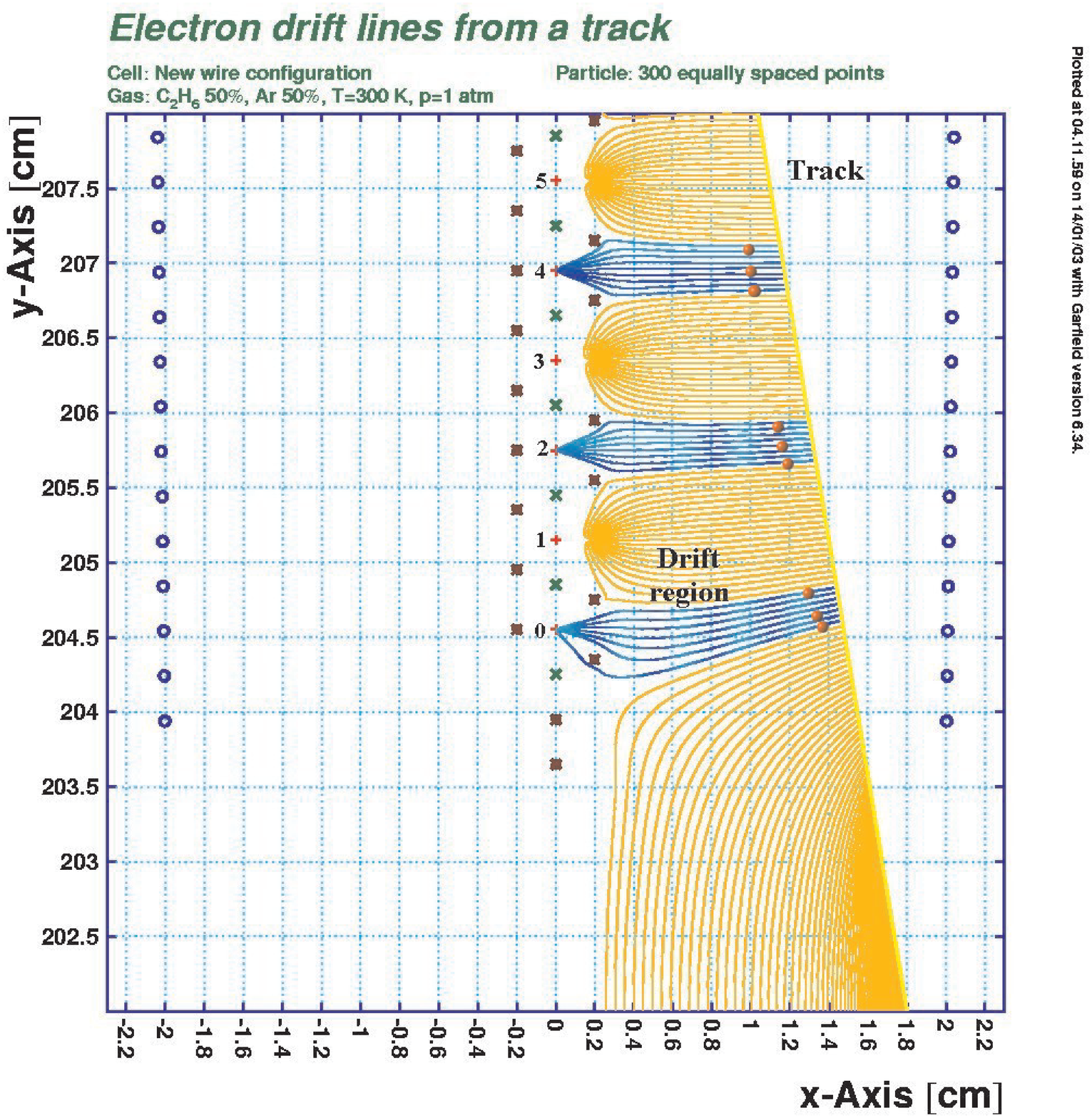,width=0.6\linewidth,clip}
 \caption{\label{fig:ch3.X1_drift}
GARFIELD simulation of electric field lines inside X1 cell of the
DCH. Marked region display the charge collection zone of each
anode wire. Circles represents charge clusters drifting towards
the anode wire.}
\end{figure}

As a result, the electric field, created within one cell generates a
well-localized charge collection region directed to one side of the
net for odd anode wires and to the opposite side for the even
wires. Fig.~\ref{fig:ch3.X1_drift} illustrates GARFIELD~\cite{garf}
simulation of the regions which allow the ionized charge (dots) to
drift to the anode wires.

\begin{figure}
\begin{tabular}{lr}
\begin{minipage}{0.5\linewidth}
\centering \epsfig{figure=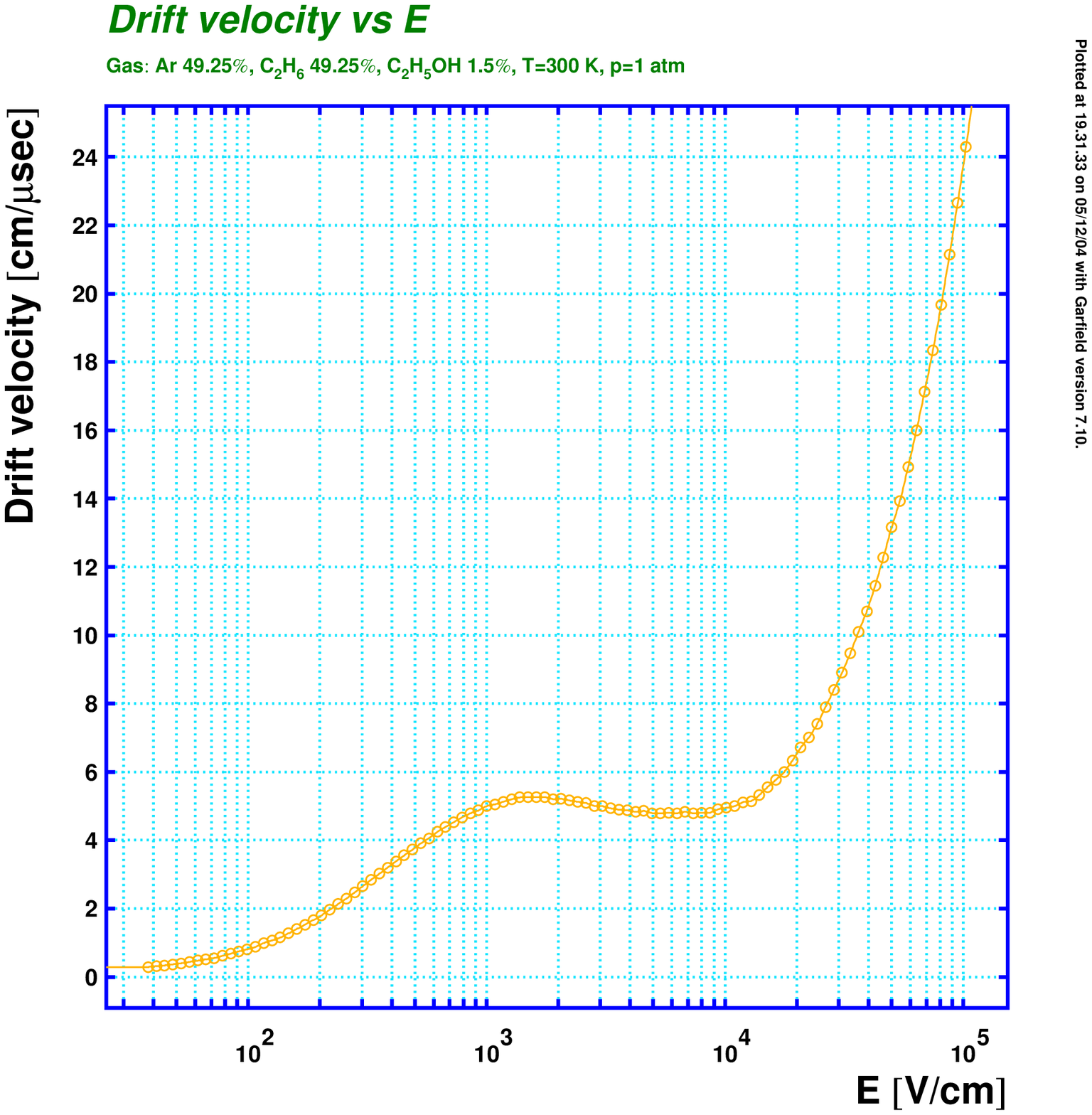,width=0.9\linewidth,clip}
 \caption{\label{fig:ch3.dv_E}
Drift velocity as a function of electric field.}
\end{minipage}
&
\begin{minipage}{0.5\linewidth}
\centering\epsfig{figure=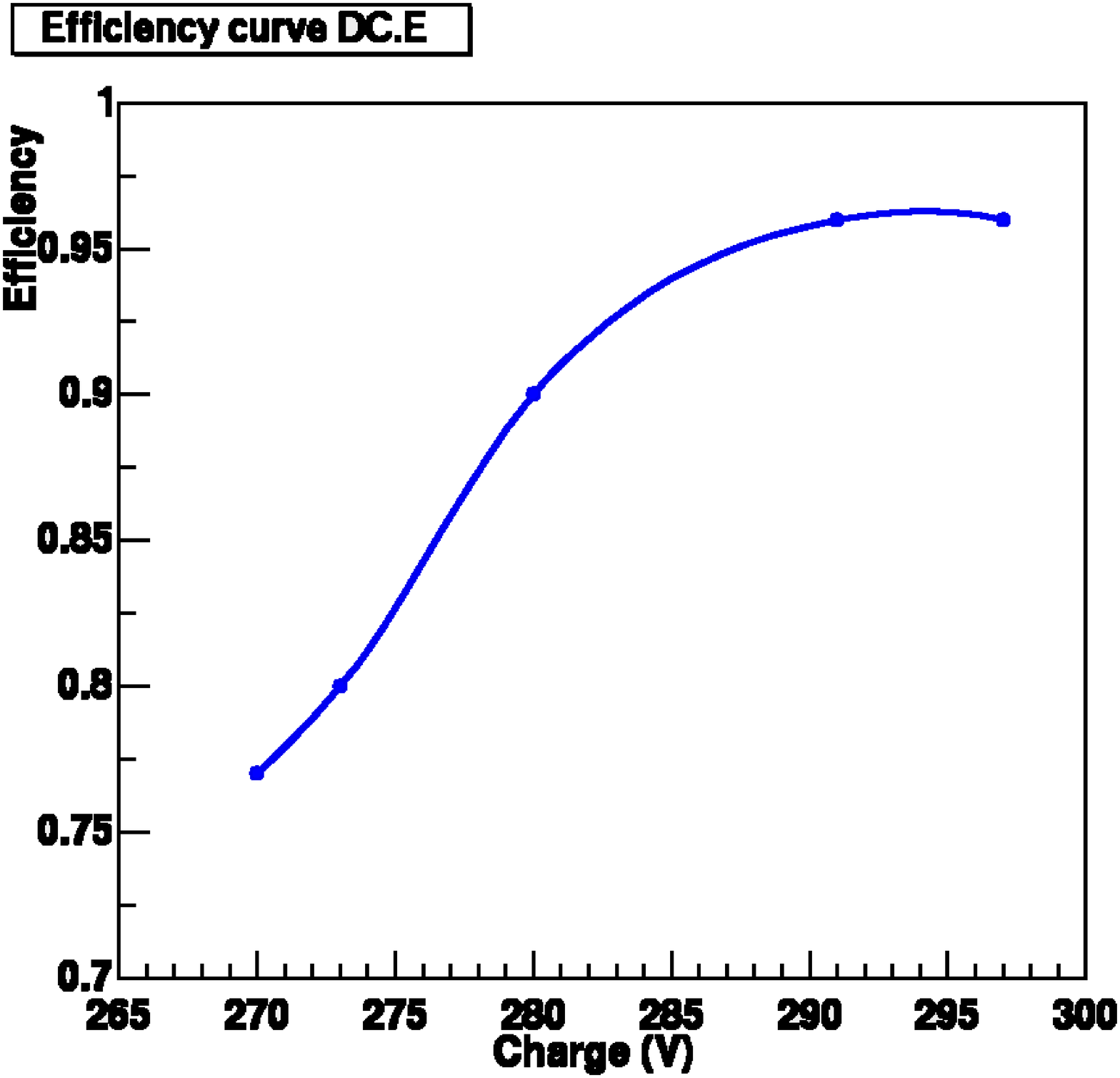,width=1\linewidth,clip}
 \caption{\label{fig:ch3.sw_eff} Single wire efficiency as a
 function of charge per unit length $Q$.}
\end{minipage}
\end{tabular}
\end{figure}

The \textbf{drift velocity} ($v_{dr}$) within the drift region has
a weak dependence on the electric field by the choice of gas mixture
and typical field strength. The drift velocity as a function of
electric field is shown in Fig.~\ref{fig:ch3.dv_E}. The electric
field is on the order of $E\approx 0.8 -1$ kV/cm in the area between
the Gate and Cathode and is significantly higher between the Gate and Anode
wire. The average drift velocity in the drift region is on
the order of $\langle v_{dr} \rangle \approx 50\ \mu m/ns$.

The position of a hit within the cell can be calculated as using
$x-t$ relation:
\begin{equation}
x = v_{dr}\cdot(t_{dr}-t_0) \label{eq:ch3.x_t}
\end{equation}

where $t_0$ is an important reference constant, corresponding to
creation of the charge in the area of the anode wire.

Detection of the ionization signal would be impossible using room
temperature electronics without \textbf{gas amplification}. Electrons
travelling in the strong electric field can obtain enough energy
between the collisions, to knock-out a secondary electron from a gas
molecule, this secondary electron can then knock out another one and
so on causing avalanche type multiplication of the charge.  The
threshold electric field is usually on the order of $E_{thr} \approx
10~kV/cm$ and is reached very close to the anode wire. The electric
field in the vicinity of the anode wire can be expressed as $E =
\frac{Q}{r}$ where $Q$ is a charge per unit length in
$[V]=\frac{[C]}{2\pi\epsilon_0}$. Using GARFIELD we can simulate the
distribution of the charges on each wire for given values of the
potentials and wire geometry. From the test runs measurements (see
Fig.\ref{fig:ch3.sw_eff}), it was found that $Q = 290$ V produce a
single wire efficiency of 90 \%. The standard voltage configuration
was selected so that the charge on all planes exceed this limit.

\begin{figure}[t]
\centering
\epsfig{figure=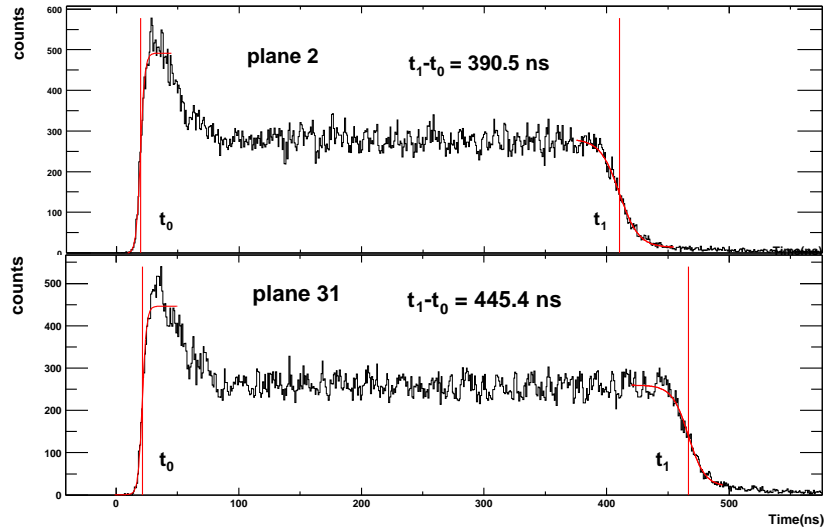,width=0.8\linewidth,clip}
 \caption{\label{fig:ch3.time}
Time distribution for two DCH planes. Left and right edges fitted
with error function~\cite{jjiathesis}.}
\end{figure}

The DCH Front End electronics digitize the leading and trailing
edge of the charge signal using the so-called ``ASD-8'' chip
(Analog - Shaper - Discriminator)~\cite{PHENIXNIM}. The thresholds
for each input channel are set via an eternal DC voltage
established by a DAC this is downloaded through an ARCNET network.
The typical running configuration uses a $q_{thr} = 6$ fC
threshold.  The time of the leading edge is being digitized with a
granularity of $\frac{1}{128}^{th}$ of the RHIC clock period or
$\approx 0.8$ ns.  , The trailing is digitized with twice coarser
binning allowing the pulse width of each trigger to be measured.
This measurement is used to reject especially narrow pulses as
noise. The ``Time Memory Chip'' (TMC) functions by storing the
continual running history of the leading and trailing edges for
the previous 6 $\mu$s in a circular memory buffer.  Triggers force
the chip to store one memory frame (an range from the past of user
selected delay and depth) into one of 5 static memories.  This
allows the TMC chip to buffer hits from triggers that occur during
the readout of previous data.  This nearly eliminates deadtime
from the PHENIX data collection system.  The value of the
``offset'' (delay between the real data and the trigger arrival)
is selected so that full drift time range could be read-out.
Fig.~\ref{fig:ch3.time} shows the typical time distribution shape
from the DCH. The left edge of the timing distribution corresponds
to the particles, depositing charge close to the anode
wire\footnote{In the region between Gate and Anode wire we do not
have Back side cancellation mechanism, this cause apparent
double-counting of the signal and left-right ambiguity.}. The
right edge of the timing distribution corresponds to the drift
time of the electrons from the area of the Cathode wire. By
measuring the half height time of the edges of the timing
distribution we can make a rough estimation for $t_0$ and $v_{dr}$
using Eq.~\ref{eq:ch3.x_t}.

\subsubsection{DCH fine tuning}

It quickly became clear that in order to reach the design
resolution, fine tuning of the calibration parameters ($t_0$ and
$v_{dr}$) need to be performed for each wire. This method was
called \textbf{internal alignment} of the DCH and included:

\begin{itemize}
    \item Slewing correction - removal of $t_0$ dependance of
    the width of the signal.
    \item $v_{dr}$ channel-by-channel alignment.
    \item $t_0$ channel-by-channel alignment.
\end{itemize}

The most appropriate way of performing those type of corrections
was based upon \textbf{residual} distributions. Residual by
definition was denoted as $\Delta t = t_0 -(t_1+t_2)/2$ where
$t_0$ is the time of the hit on the trial wire, $t_1$ and $t_2$
are corresponding times for the neighboring wires hits. For
convenience (as those calculations were performed on-line) the
time unit for all time variables is going to be TMC time-bin ($1$
tb $\approx 0.8$ ns). Using the straight track assumption, it is
clear that in optimal case $\Delta t$ should be independent of all
parameters and have a mean of zero. The side-standing wires (wires
at the edge of the cell) have no neighbors and we can not
calculate the residuals for them, they play a role of reference
wires.

\begin{figure}[h]
\epsfig{figure=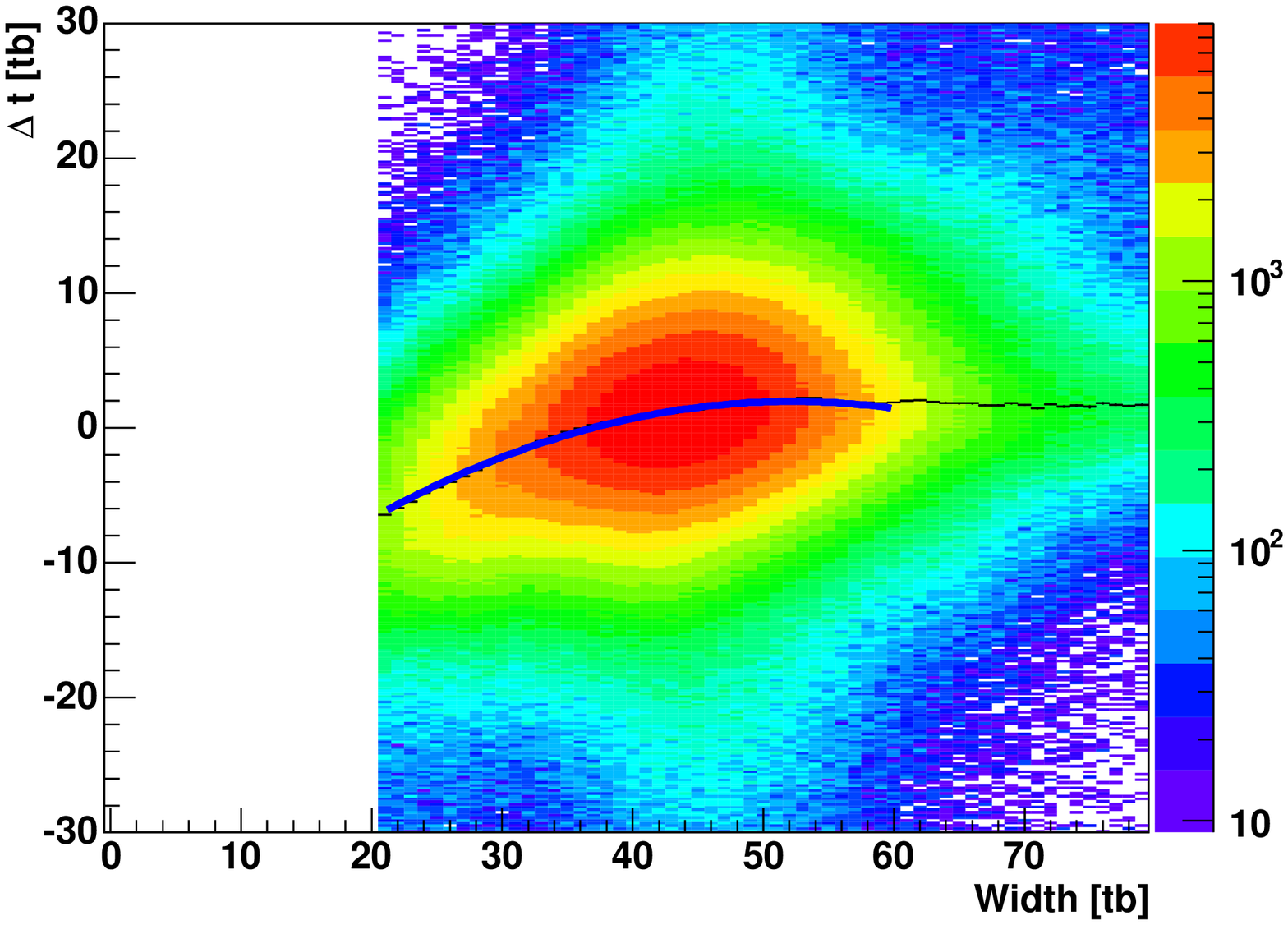,width=0.47\linewidth,clip}
\epsfig{figure=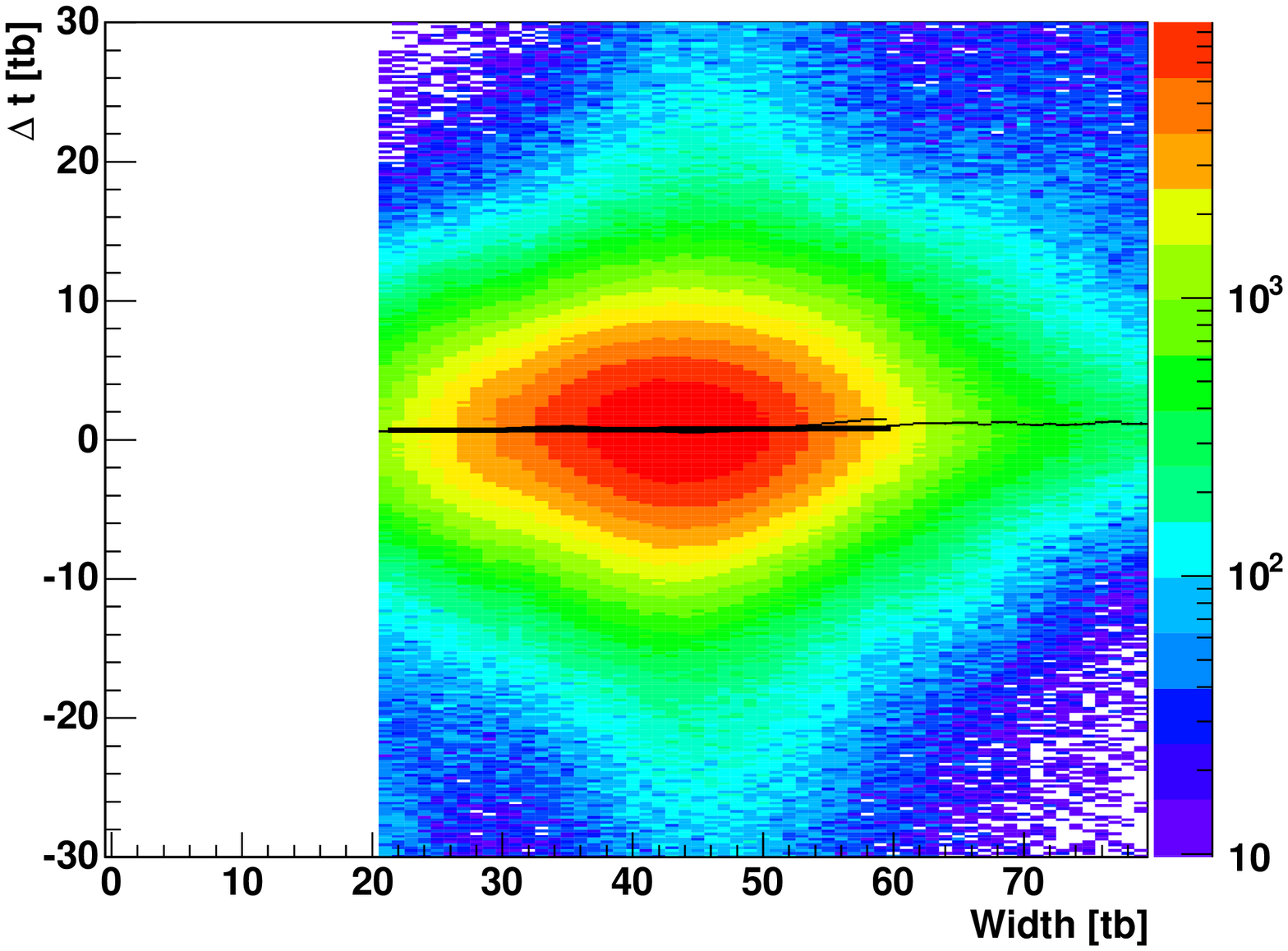,width=0.45\linewidth,clip}
\caption{\label{fig:ch3.slewing} Residual distribution as a
function of hit width before (left) and after (right) the slewing
corrections.}
%
\centering\epsfig{figure=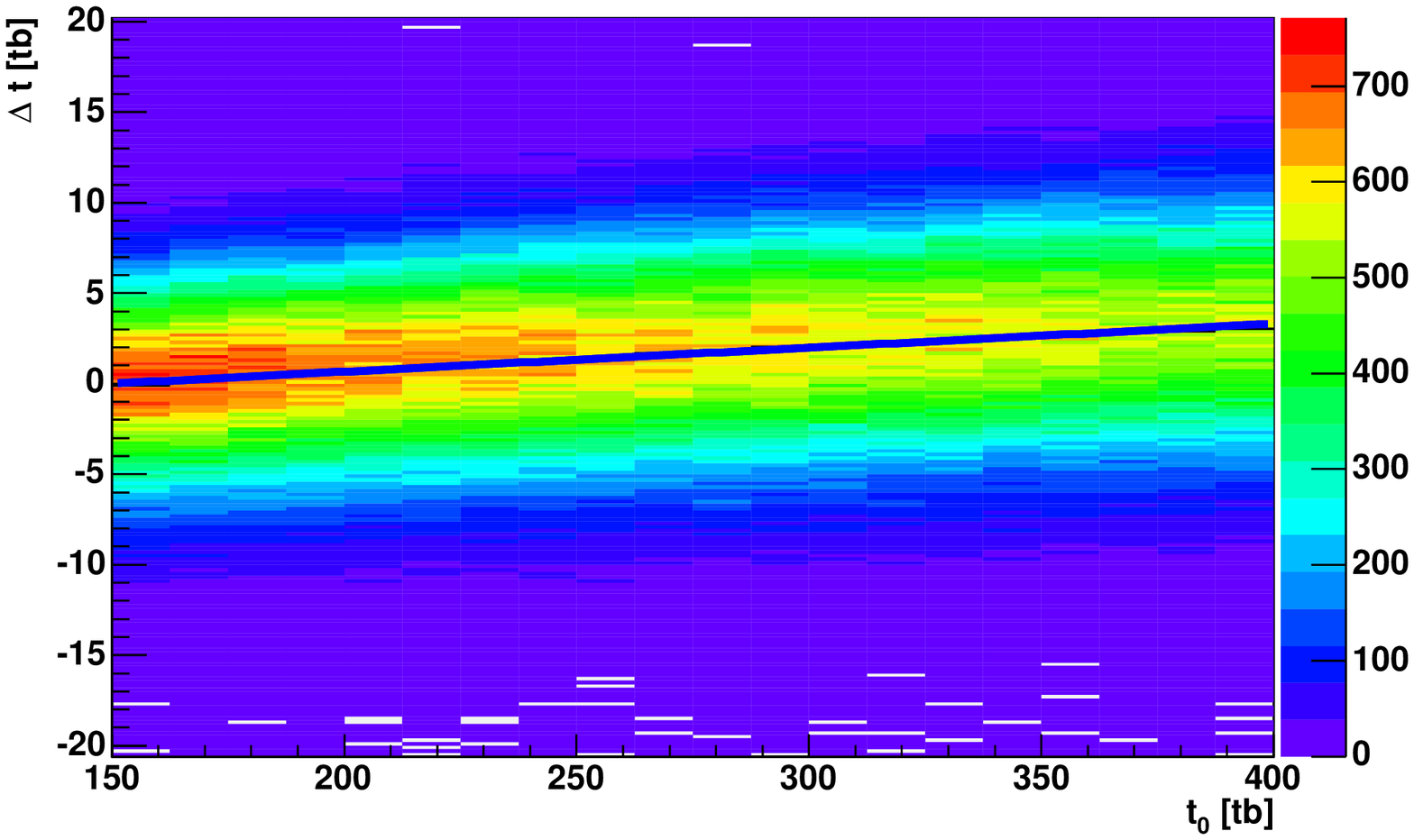,width=0.6\linewidth,clip}
\caption{\label{fig:ch3.dv_corr} Residual as a function of time
fitted with linear function.}
\begin{tabular}{lr}
\begin{minipage}{0.5\linewidth}\epsfig{figure=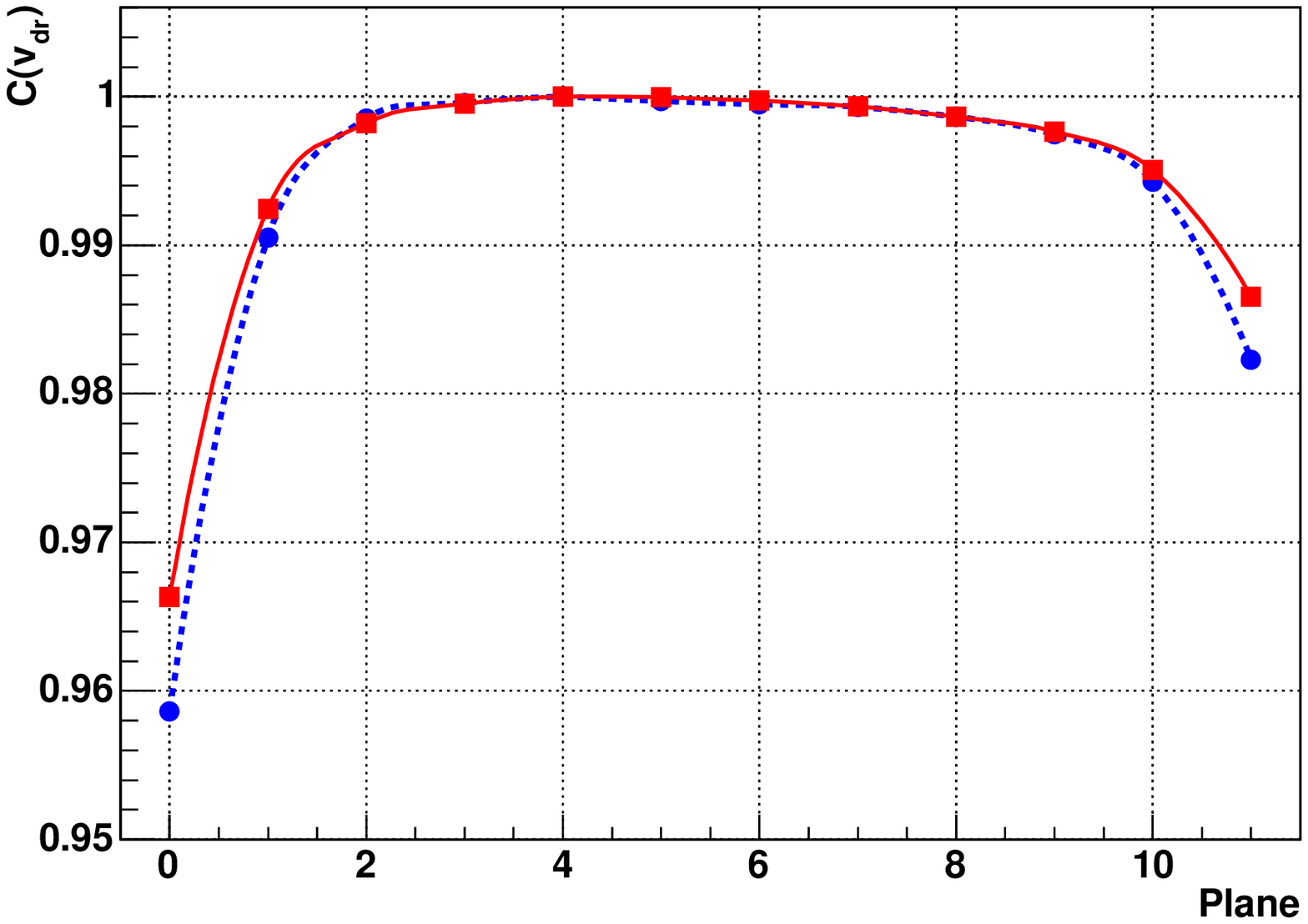,width=0.8\linewidth,clip}
\caption{\label{fig:ch3.dv_comp} Comparison of $v_{dr}$ for East
X1 wires obtained from the data residual slope (circles) and
GARFIELD simulation (squares).}
\end{minipage}
&
\begin{minipage}{0.5\linewidth} \centering
\epsfig{figure=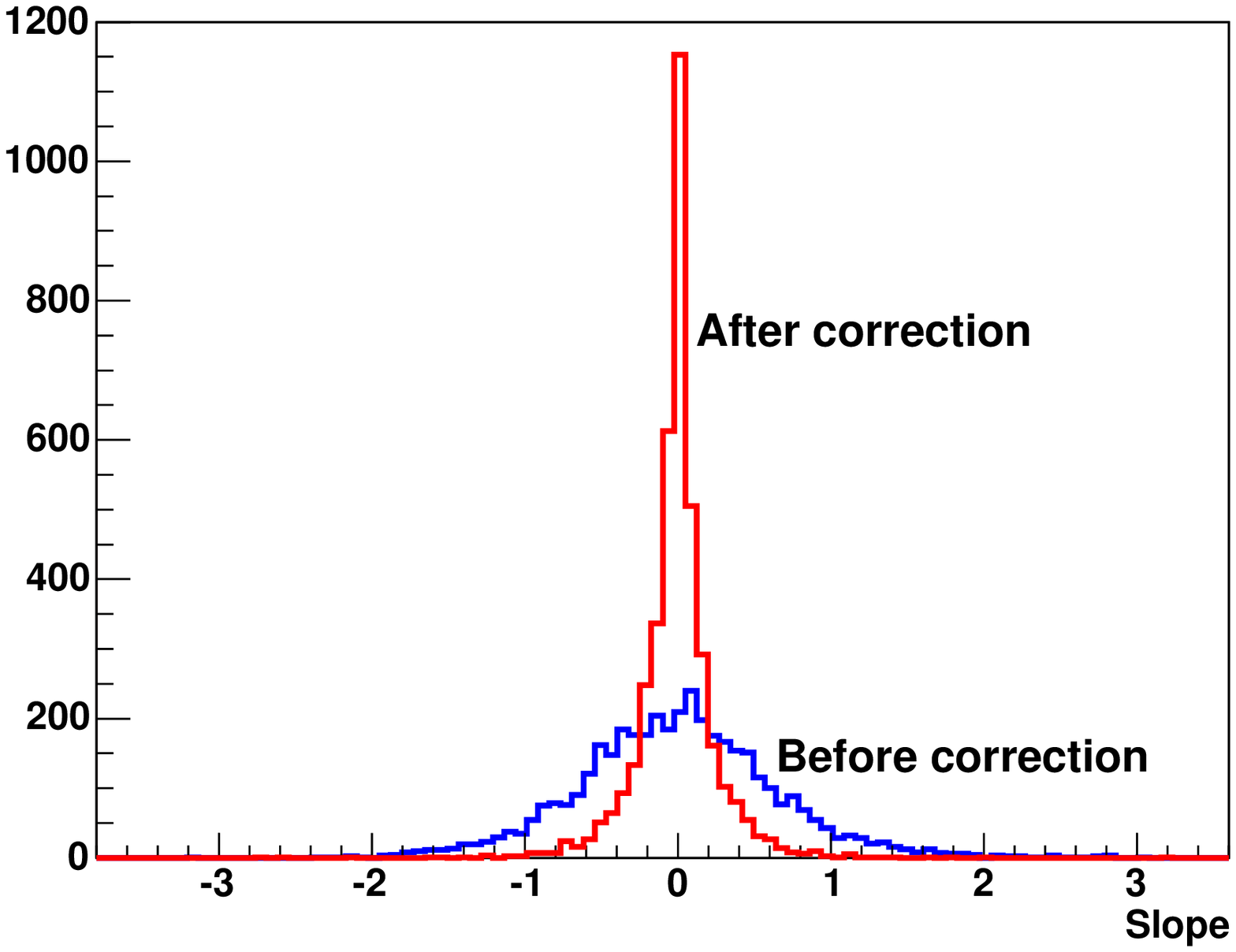,width=0.8\linewidth,clip}
 \caption{\label{fig:ch3.slopes_all}
Distribution of the residual distribution slopes (see
Fig.~\ref{fig:ch3.dv_corr}) for all the DCH wires before and after
drift velocity corrections.}
\end{minipage}
\end{tabular}
\end{figure}
\pagebreak

We first look at the dependence of the residual as a function of
the hit width (see Fig.~\ref{fig:ch3.slewing}). After including
the fitted value of the mean shift as a correction to t-zero $t_0
= t_0 - dt^{slew}(w)$ and performing at least three iterations we
remove the dependence of residual on a hit width.

The Drift velocity can be fine-adjusted using the field-off data. In
this case, the majority of the tracks (not coming from decays or
scatters) should go radially from the vertex with no deflection angle
$\alpha \approx 0^{\circ}$. If we look at the distribution of the
residual for particular wire as a function of time (shown in
Fig.~\ref{fig:ch3.dv_corr}), the linear slope of the distribution that
can \textbf{only} be caused by the difference of the drift velocities
between the wire and its neighboring wires. It is possible to solve
the corresponding system of linear equations for the corrections to
the $v_{dr}$ variations with respect to the reference drift
velocity. The solution require two constrains which can be selected
arbitrarily.  The best choice for choosing the constraints is by
comparing the results of the corrections to the variation of $x-t$
relation slope from GARFIELD simulation. Fig.~\ref{fig:ch3.dv_comp}
shows the comparison of drift velocity profile obtained from data with
the one produced from GARFIELD code. One can see nice agreement
between expectation and the measurement. This method was performed on
the plane-by-plane basics, the second order wire-by-wire corrections
have been performed as a perturbation to the plane-by-plane
corrections. Comparison of the individual wire slopes after
wire-by-wire fine tuning of the drift velocity is shown in
Fig.~\ref{fig:ch3.slopes_all}. The final set of $v_{dr}$ correction
coefficients was recorded as a multiplier for the global drift
velocity, measured from the edges of the timing distribution.

Fine adjustments to the calibration parameters were also performed on
the $t_0$. Similar to the drift velocities, $t_0$s have strong
dependence on the plane (probably due to the combined effect of the
$x-t$ relation non-linearity and the propagation time of the signal on
on the ASD board).  It also has a strong variation on the wire-by-wire
basis due to possible geometrical displacement (driven by
electrostatic sag) of individual wires and possibly different gas
gain. The wire-by-wire variation of $t_0$ was applied as an additional
correction on top on plane-by-plane correction.

\begin{figure}[h]
\begin{tabular}{lr}
\begin{minipage}{0.5\linewidth}\epsfig{figure=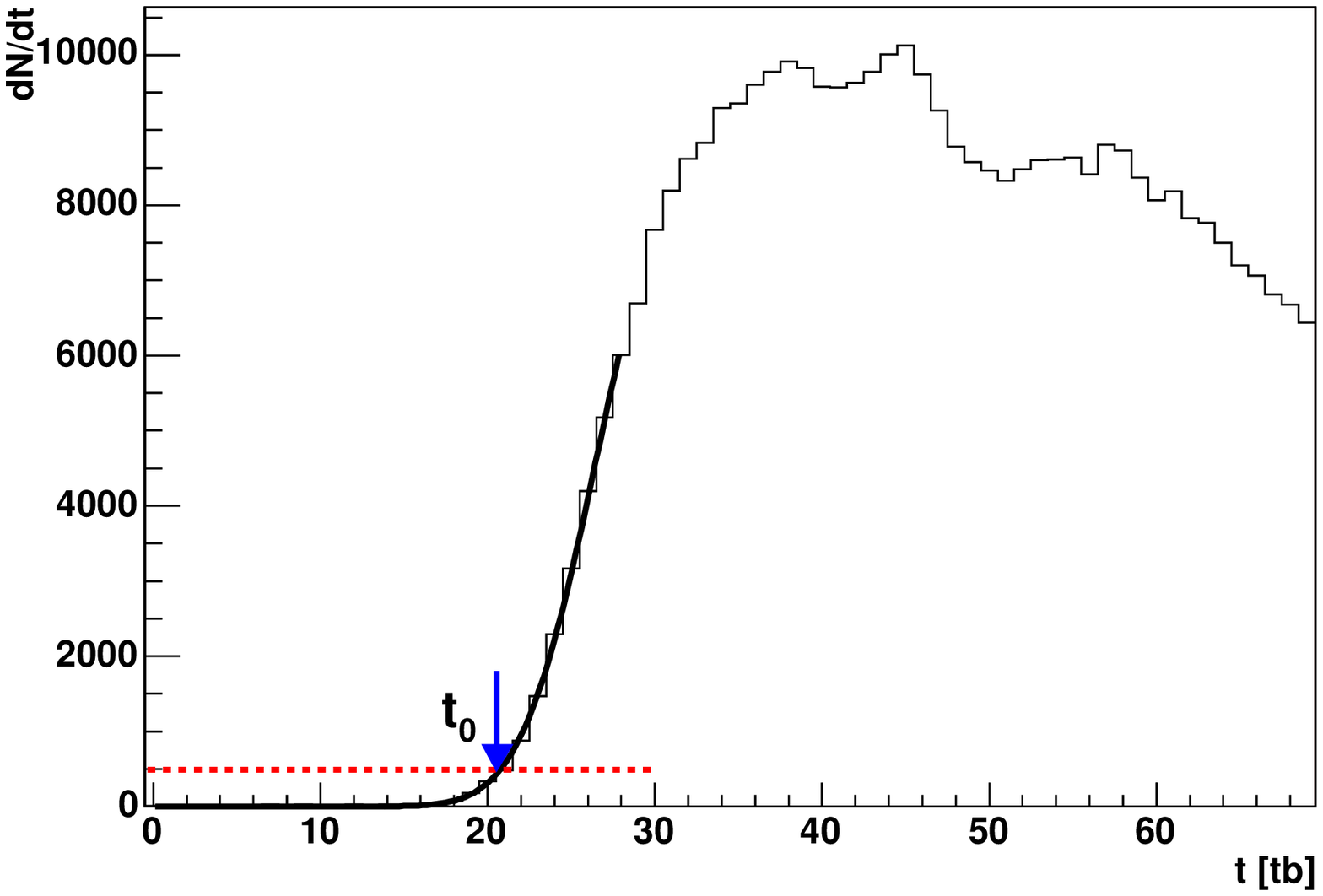,width=0.8\linewidth,clip}
\caption{\label{fig:ch3.t0_fit} Determination of $t_0$ from the
slope of timing distribution at constant height for one DCH plane.
Slope is fitted by Error function.}
\end{minipage}
&
\begin{minipage}{0.5\linewidth} \centering
\epsfig{figure=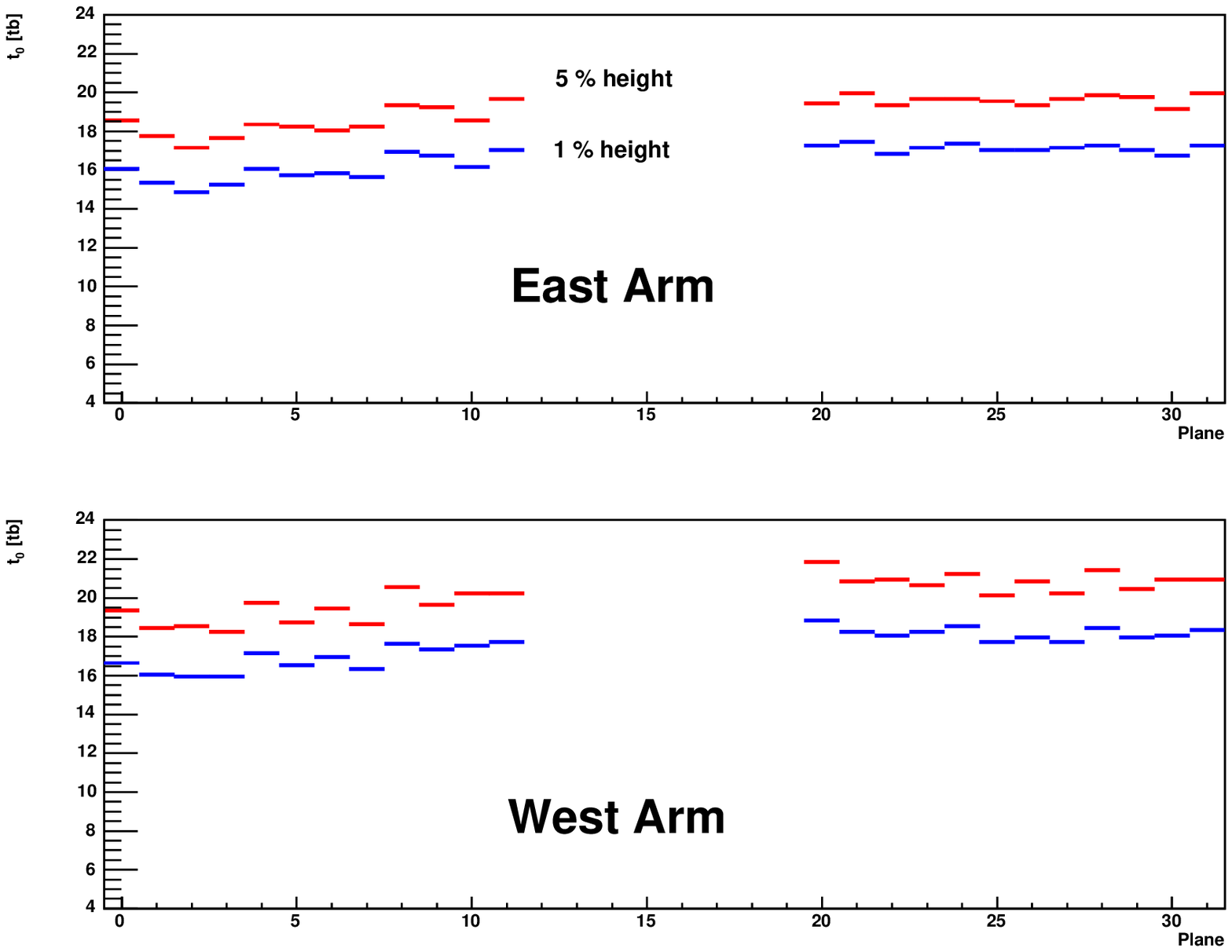,width=0.8\linewidth,clip}
 \caption{\label{fig:ch3.t0_corr}
Comparison of plane-by-plane $t_0$ corrections for different DCH
arms and different height cut.}
\end{minipage}
\end{tabular}
\end{figure}

In order to estimate the plane-by-plane variation of $t_0$,
the rising edge of the timing distribution was fitted at the
constant level (1\% - 5\% of the maximum) as shown in
Fig.~\ref{fig:ch3.t0_fit}. The results of the fit do not depend
on the fit level (except for constant shift of the resulting time) and
behavior is similar for both DCH arms (see
Fig.~\ref{fig:ch3.t0_corr}).

 The local wire-by-wire $t_0$ fine tuning was done plane by plane
using the side-standing wires as a references and moving the
middle wires $t_0$ in order to zero the mean of the residual. The
problem is easily reducible to a set of linear equations for
$dt_{0\ i}$ as a function of $\Delta t_{0\ i}$.

\begin{figure}[h]
\centering \epsfig{figure=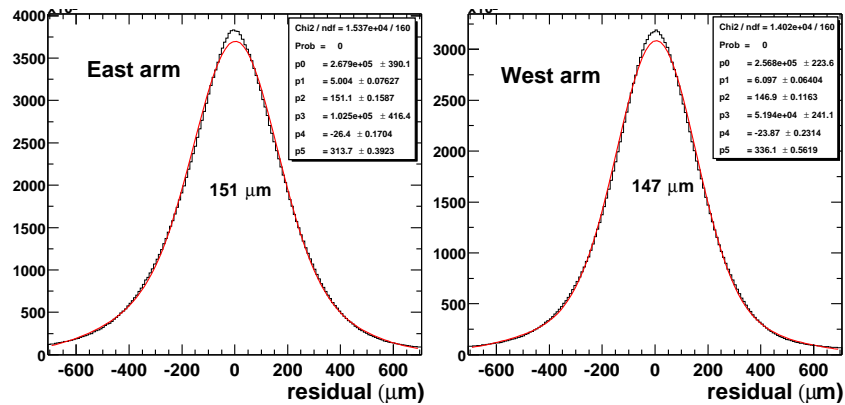,width=0.8\linewidth,clip}
 \caption{\label{fig:ch3.resid}
Residual distribution of the hit to the track for East (left
panel) and West (right panel) DCH arm fitted with
double-gaussian~\cite{jjiathesis}.}
\end{figure}

After all the corrections are applied, the residuals distribution of
the hit to the track reaches the design values of $\approx 150 \mu$m
which is shown in Fig.~\ref{fig:ch3.resid} for both DCH arms. If we
only look at the tracks far from the anode and cathode, this value can
go down to $\approx 100 \mu$m - which is probably the physical
limitation due to the cluster arrival time smearing and the gas
diffusion coefficient. Needless to say that this value only indicates
the relative accuracy of the tracking within one cell, data-based
geometrical alignment of the nets within the DCH and matching to the
outer detectors and vertex need to be done in order to perform
\textbf{global alignment} of the Drift Chamber~\cite{jjiathesis}.

\subsubsection{Tracking algorithm}

Tracking is the most critical step of the particle identification.
Not only must the tracking operate with high single track
reconstruction efficiency, but also it needs to be reliable in the
high multiplicity environment of central $\Au$ collisions and have a
low ``ghost'' track rate.  \footnote{Ghost tracks are pattern
recognition solutions that did not actually come from real
tracks.} The rate of the ``ghost'' tracks can be later reduced by
association to other detectors (PC2, PC3, EMC).

Much of the tuning to optomize the tracking performance was performed
by me and I will summarize the key features of the method in this
chapter.

\begin{figure}[h]
\centering
\epsfig{figure=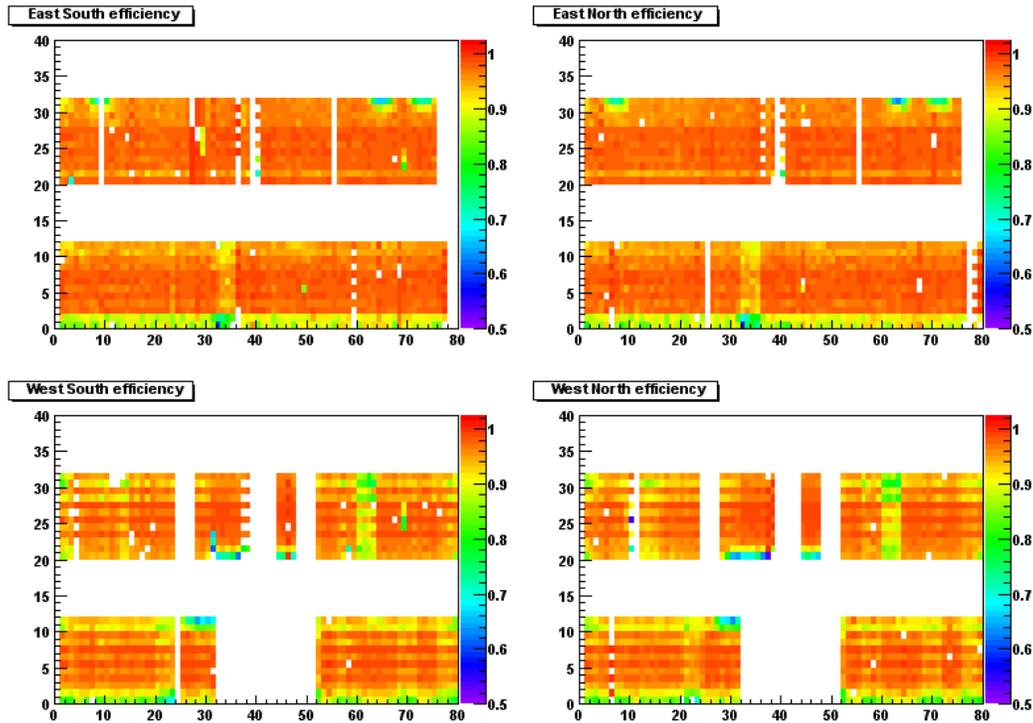,width=1\linewidth,clip}
 \caption{\label{fig:ch3.eff_map}
Single wire efficiency map for $\pp$ Run02.}
\end{figure}

The ideal track (meaning 100 \% single wire efficiency) should leave
at least 6 hits in X1 section and 6 hits in the X2 section of the
DCH. In reality, inefficiencies cause the track to lose hits with a
certain probability. The single wire efficiency can be calculated for
each wire using the ratio of tracks, having a hit on this wire within
some wide association window to the total number of tracks passing
within that wire's active area. In order to remove the complications
of the cathode or anode crossing tracks, only the region confined
close to the center of the cell is used for this calculation. The
final efficiency map for $\pp$ Run02 is shown in
Fig.~\ref{fig:ch3.eff_map}. One can see that the single wire
efficiency is on the order of 95\% for the middle wires and on the
order of 90\% for the side-standing wires (especially closest to mylar
window of the chamber).  This effect is well understood and was
predicted from the Garfield simulations.  It results from ``edge
effects'' that reduce the field and hence the gain on the
side-standing wires.

\begin{figure}[h]
\centering
\epsfig{figure=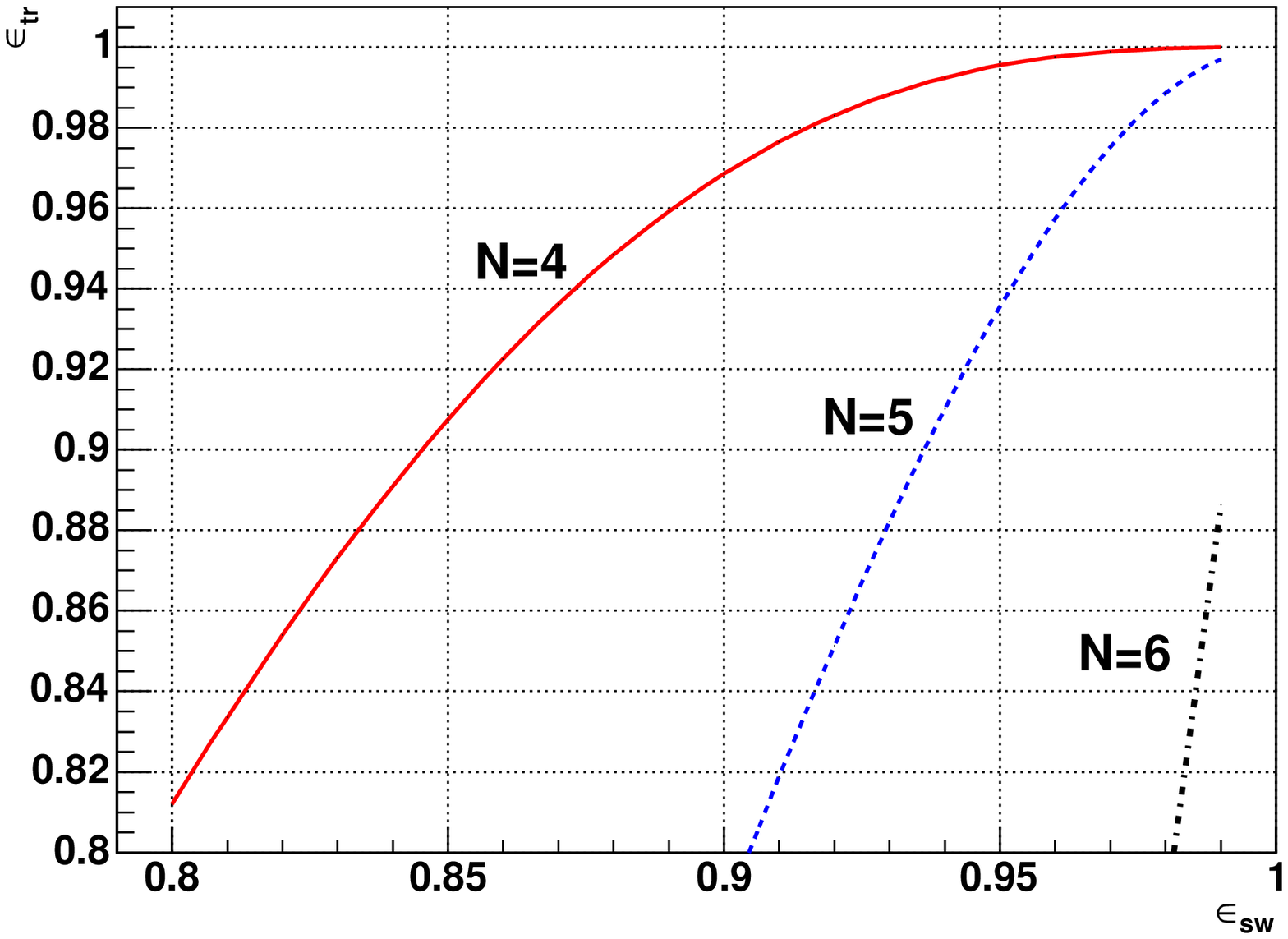,width=0.75\linewidth,clip}
 \caption{\label{fig:ch3.tracking_eff}
Tracking efficiency as a function of single wire efficiency for
different requirement on the number of hits in X1 and X2 DCH
plane.}
\end{figure}

Now we can estimate the tracking efficiency assuming that we
consider a track to have at least $N$ hits in X1 and $N$ hits in
X2 for a given (constant) level of single wire efficiency. This
can be exactly calculated using probability theory, the results of
calculations for $ N=4,\ 5,\ 6$ are shown in the
Fig.~\ref{fig:ch3.tracking_eff}. This plot show that having 95\%
single wire efficiency we already at the region of $> 99\%$
tracking efficiency if we chose $N=4$ which was chosen as a
minimum number of hits in X1 or X2 layer for the track.

\begin{figure}[h]
\centering
\epsfig{figure=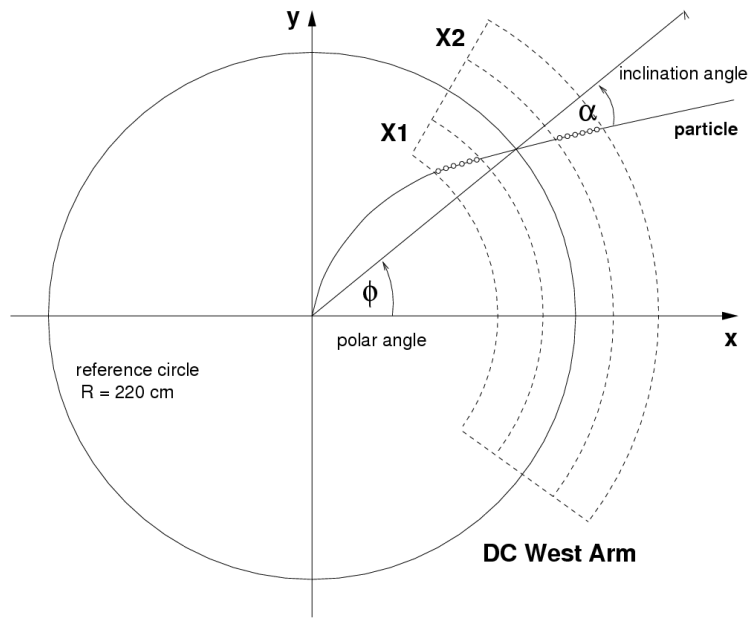,width=0.75\linewidth,clip}
 \caption{\label{fig:ch3.alpha_phi}
Single wire efficiency map for $\pp$ Run02.}
\end{figure}

The tracking in PHENIX is based on the assumption of the track having
a straight line trajectory inside the Drift Chamber volume. First, the
track is reconstructed in $X-Y$ plane projection, determining $\alpha$
and $\phi$ angles of the track, defined as indicated in
Fig.~\ref{fig:ch3.alpha_phi}.

First stage of the trackfinding utilizes the \textbf{``combinatorial
hough transform''} algorithm. Basic idea of the ``combinatorial hough
transform'' is very simple: we are looking for a straight tracks,
having X1 and X2 hits.  Every straight line is described by two
parameters (y=mx+b), both of which can be determined by any set of two
points lying along the line.  If we make all possible combinations
between each X1 and each X2 hit (laying within reasonable vicinity
range from each other) we can calculate and histogram the line
parameters.  The line parameters (m,b) have infinite bounds and are
impractical to histogram, so we instead calculate, for each pair, the
local angles $\alpha_p$ and $\phi_p$ (same notation as on
Fig.~\ref{fig:ch3.alpha_phi}) to fill a 2-D histogram of $\alpha_p$
vs. $\phi_p$. As the result - all the pairs of hits belonging to one
track will create a localized peak on this histogram. The histogram
can be replaced by 2-D array, which is called ``hough array'' and the
local maxima in this array provide a ``guess'' value for $\alpha$ and
$\phi$ of a track. In order to remove the possible bin-edge effects, a
threshold is applied to the 3x3 bin sum around the local maximum of the
``hough array'' elements.  This way we are able to clearly separate
the background random combinations from the real tracks. to improve
the initial guess parameters, we use weighted average of the
neighboring array bins to improve the initial guess for $\alpha$ and
$\phi$.

The threshold of the hough array is chosen $N_{thr}=15$, low enough to
allow ``4 hits in X1 + 4 hits in X2'' tracks to survive the cut. The
binning of the hough array is also very critical issue: the coarser
the binning, the more the probability to have all hits from one track
localized to one bin.  However if the bins are too coarse the accuracy
for the determination of the initial guess parameters in
decreased. Making the bins finer also has the drawback that hits from one
track start to smear around the bins of ``hough array'' and the track may
fall below the cut threshold. The bin size of the hough array was
optimized by studying the reconstruction probability of one chosen
track in high multiplicity $\Au$ events (worst case scenario). It does
not matter at this stage whether we have a lot of background tracks,
it is more important not to lose any.

\begin{figure}[h]
\centering \epsfig{figure=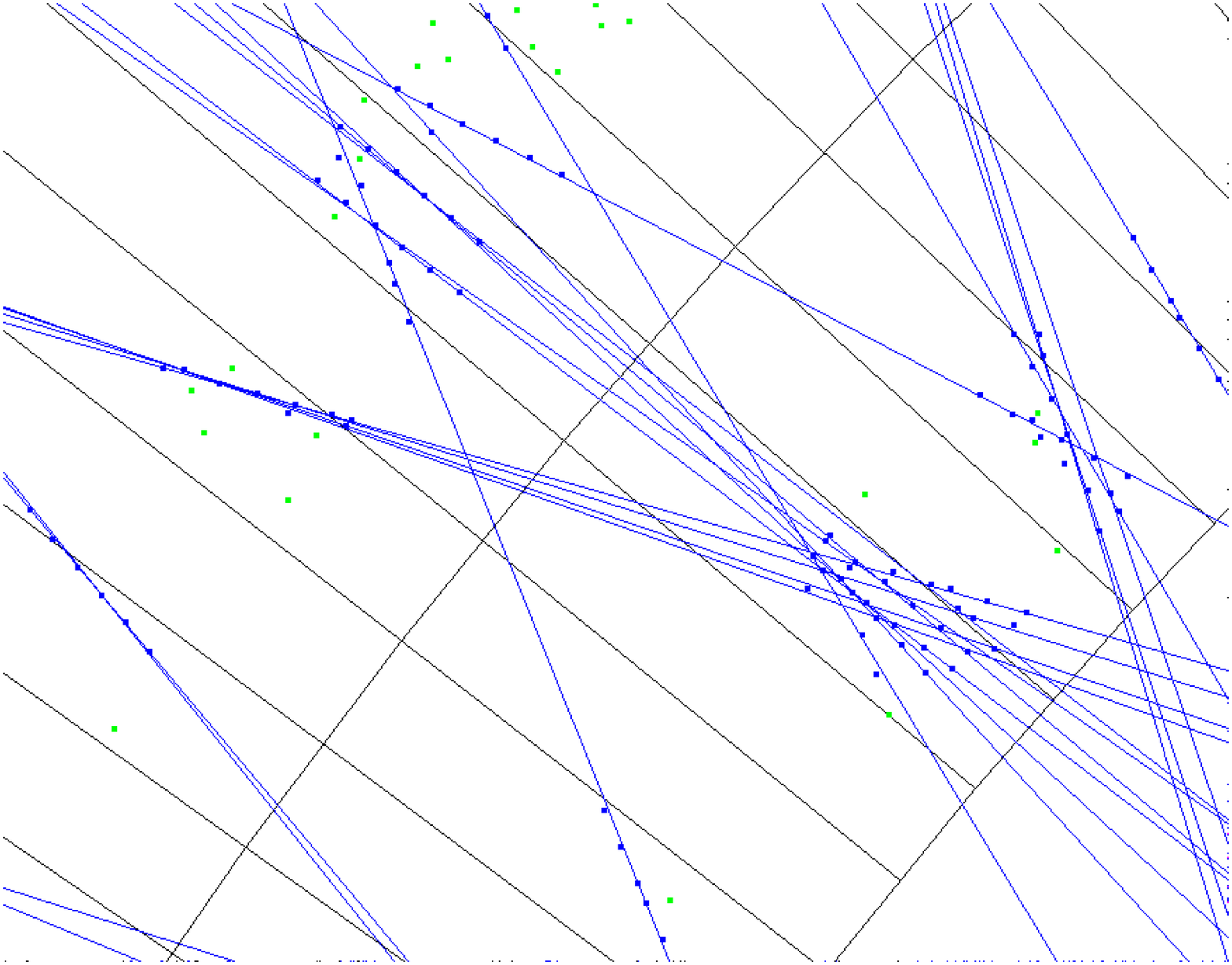,width=0.6\linewidth,clip}
\caption{\label{fig:ch3.1st_stage} Event display snapshot with
track candidates after the initial hit fitting.} \centering
\epsfig{figure=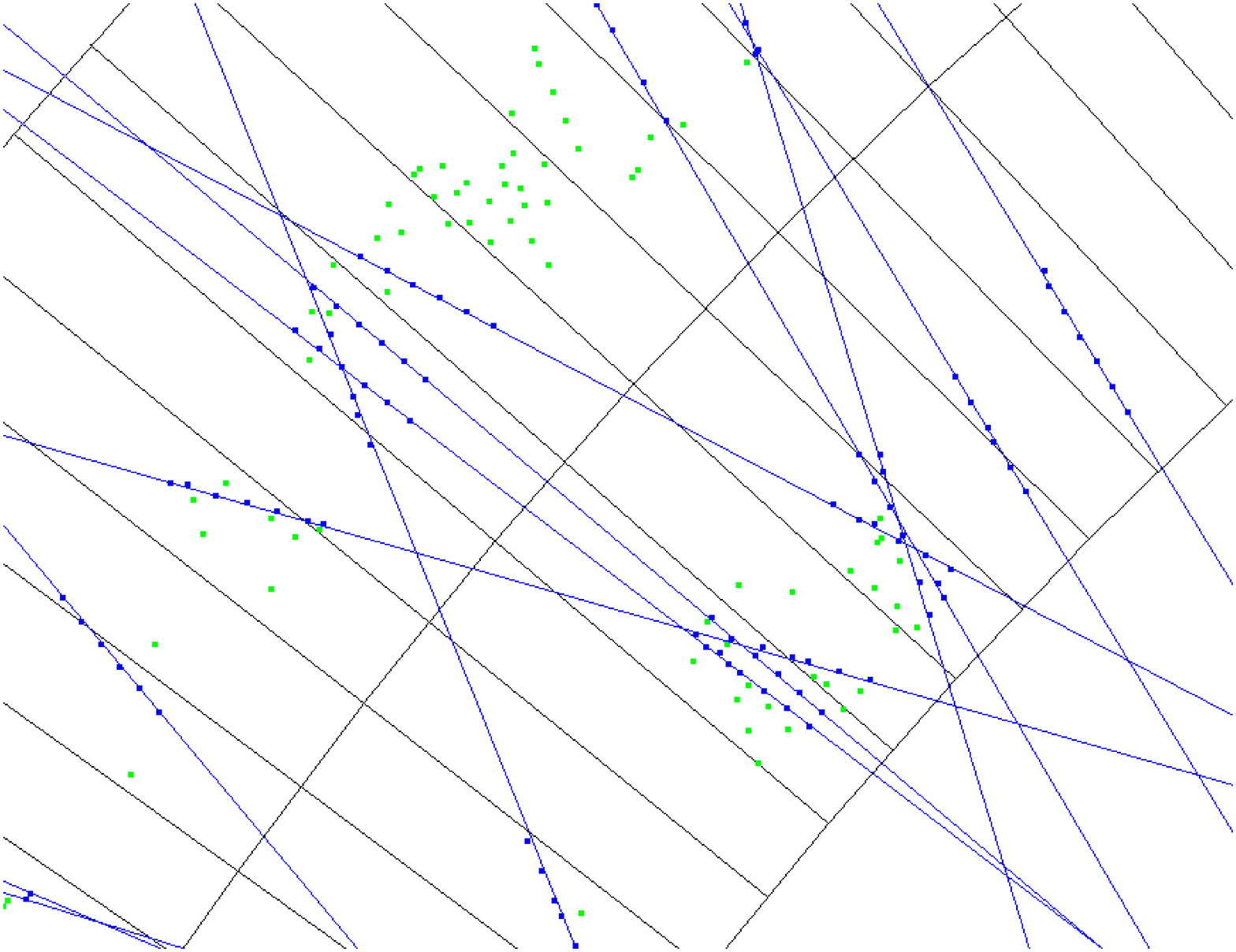,width=0.6\linewidth,clip}
 \caption{\label{fig:ch3.2nd_stage} Event display snapshot with tracks on the
 final stage of trackfinding. Same event as for Fig.~\ref{fig:ch3.1st_stage}}
\end{figure}

The next step of the trackfinding - is a ``gentle'' removal of the
background:
\begin{itemize}
    \item Association of the hits to the track with ($one\ hit
    \leftrightarrow many\ track$) correspondence. The association
    algorithm calculates the closest approach of the hit to the
    projected track guess and associates it if the hit is at least 4 mm
    from the track guess.
    \item Robust fitting of the track - iterative linear fitting using
    a weighting the hits, deweighting hits in accordance to their
    deviation from the mean of the previous iteration. This helps to
    remove randomly associated hits that are far off from the
    projected track. 5 fitting iterations performed, gradually
    rejecting mis-associated hits. Fig.~\ref{fig:ch3.1st_stage} shows
    the track candidates (improved accuracy as compared to track
    guesses) after the first fitting stage.
    \item The next step is sequential removal of the excess track candidates. 3
    stages of ``gentle'' removal perform the following procedure:
    \begin{enumerate}
        \item Associate hits to the track with $one\ hit
    \leftrightarrow one\ track$ correspondence. In this step each
    hit will be associated \textbf{only} to the closest track.
    Association window is 4 mm for 1st and 2nd stage and 2 mm for
    the final association stage.
        \item Removal of the tracks that have less his than a
        threshold $n_{thr}$ associated. $n_{thr} = 4$ for the 1st
        stage, $n_{thr} = 6$ for the 2nd stage and $n_{thr} =
        8$ for the final stage. This procedure
        \textbf{gradually} removes the tracks that have to few
        hits associated, returning their mis-associated hits to
        the real tracks.
    \end{enumerate}
    \item Final fitting of the remaining tracks to the hits.
    Fig.~\ref{fig:ch3.2nd_stage} shows the tracks that were filtered
    to the output. Most all extraneous tracks are removed by the
    method which proved to be extremely robust and efficient enabling
    as to reach $\approx 75 \%$ tracking efficiency in the most
    central full energy AuAu collisions. In low multiplicity
    environment of $\pp$ collision the tracking efficiency exceedss
    $98 \%$.
\end{itemize}

\pagebreak
\subsection{Pad Chambers}\label{sec:PC}

The PHENIX Pad Chambers are unique background rejection devices using
a novel pixelation scheme. Their primary purposes are
following~\cite{PHENIXCDR}:
\begin{itemize}
\item Measurement of non-projective three dimensional spatial
points, which are used for both momentum determination ($p_z$) and
pattern recognition.
\item Rejection of decays and photon conversion
background at high $\pt$ by tight matching requirements to the
tracks measured by the DC.
\item Distinguishing electrons from other
particles by accurate pointing of charged track to the RICH and
EMC.
\item Charged particle veto in front of EMC.
\item Providing seed for tracks in charged high $\pt$ Level-2 triggers
and electron Level-2 triggers.
\end{itemize}

The space points provided by PC3 and EMC allow us to more accurately
determine the track's actual trajectory through the RICH, an essential
improvement to the electron identification. All $Z$ information for
the track is obtained from PC1 high precision $Z$
measurement\footnote{The DCH can provide information about $Z$ of the
track using UV layers, but, the accuracy of this measurements is less
precise than that provided by PC1 alone.}

During the PC design of the following requirements were
applied~\cite{sashathesis}

\begin{itemize}
\item
Very high efficiency ($>99$\%) and low occupancy(few \% in most
central $\Au$ collisions).
\item
Good spatial resolution.
\item
Low mass, in order to minimize secondary particle production and
multiple scattering.
\end{itemize}

The PCs are multi-wire proportional chambers placed at radial
positions of 2.5m, 4m, and 5m. Each detector contains a single plane
of wires inside a gas volume bounded by two cathode planes. One
cathode is segmented into an array of interlaced ``pixel-pads''. Each
track fires three pixelpads.  The coincidence reduces the false hit
rate to be entirely negligible and localizes the track 3X better than
a standard pixel chamber with the same number of channels.

\begin{figure}[ht]
\begin{center}
\epsfig{file=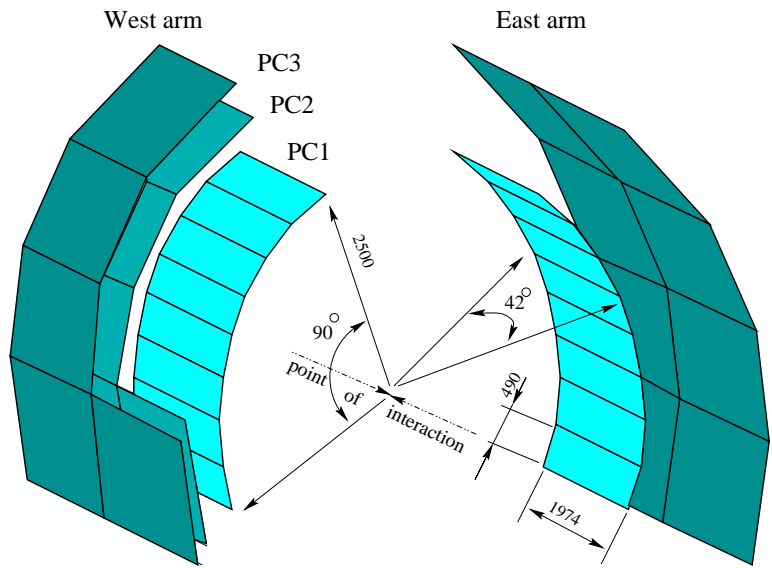,width=4in} \caption{\label{fig:ch2.pc1}
Schematic view of PHENIX Pad Chamber set. Several sectors of PC2
and PC3 in the west arm are removed for clarity of the
picture.~\cite{sashathesis}.}
\end{center}
\end{figure}

A schematic view of the PC subsystem is shown in Fig.~\ref{fig:ch2.pc1}.
The important PC specifications achieved in RUN-2 are listed in
Table~\ref{tab:ch2.pc1}. The pad size for PC1 is 0.84 $cm\times$
0.845 $cm$ to achieve less than $8$\% occupancy in most central
$\Au$ collisions. This gives a position resolution of 1.7 $mm$
along $z$ and 2.5 $mm$ in $r-\phi$. The pad size for PC2 and PC3
is chosen such that they have similar angular resolution compared
to PC1.

\begin{table}[ht]
\caption{\label{tab:ch2.pc1}Performance of Pad Chambers in
RUN-2~\cite{sashathesis}.}
\begin{tabular}{|c|c|c|c|}\hline
Parameters&PC1&PC2&PC3\\\hline Pad Size ($r$-$\phi\times
z)[cm^2]$&$0.84\times0.845$&$1.355\times1.425$&$1.6\times1.67$\\\hline
Single hit resolution &&&\\
($r$-$\phi, z$) in [$mm$]& (2.5,1.7) &(3.9,3.1)&(4.6,3.6)\\\hline
Double hit resolution &&&\\
($r$-$\phi$, $z$) [$cm$] &(2.9,2.4)&(4.6,4.0)&(5.3,5.0)\\\hline
Radiation Length [\%]&1.2&2.4&2.4\\\hline Efficiency
&$>$99\%&$>$99\%&$>$99\%\\\hline
\end{tabular}
\end{table}

\subsection{Ring Imaging Cerenkov Detectors}\label{sec:RICH}

The Ring Image Cherenkov (RICH) detector is the key component of
PHENIX leptonic program. Not only does it have a nearly perfect
rejection of pions over electrons up to $p_T \approx\ 5$ GeV/c
(~$1\times 10^{-4}$ error rate), but it also provides the Level-1
electron trigger decision that enables us to collect rare electron and
dielectron events. The main functions of RICH are:
\begin{itemize}
\item Identification of electrons below $\pt<4.8$ GeV/$c$.
\item Enable charged pion identification at $\pt>4.8$
GeV/$c$~\cite{jjiathesis}.
\item Provide a fast Level-1 trigger decision. In combination with
EMC tile trigger, helps us \textbf{significantly enrich} the electron
sample in high luminosity $\pp$ collisions. Unfortunately, electron
rate in $Au$ collisions is too high to make the electron trigger
effective.
\end{itemize}

\begin{figure}
\centering
\epsfig{file=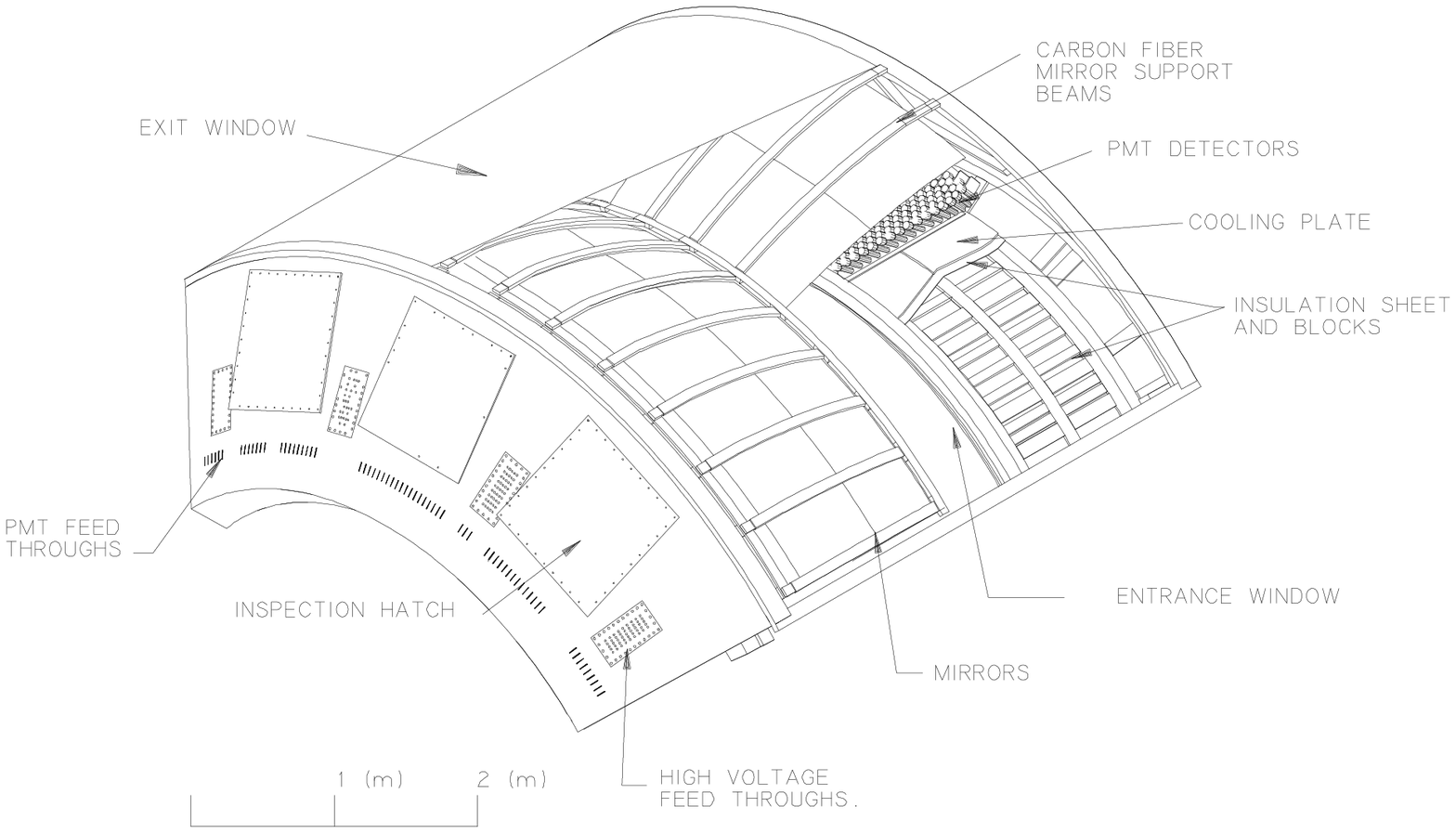,width=5in}\caption{\label{fig:rich0} A
cutaway view of the RICH detector~\cite{richnim}.}
\centering
\epsfig{file=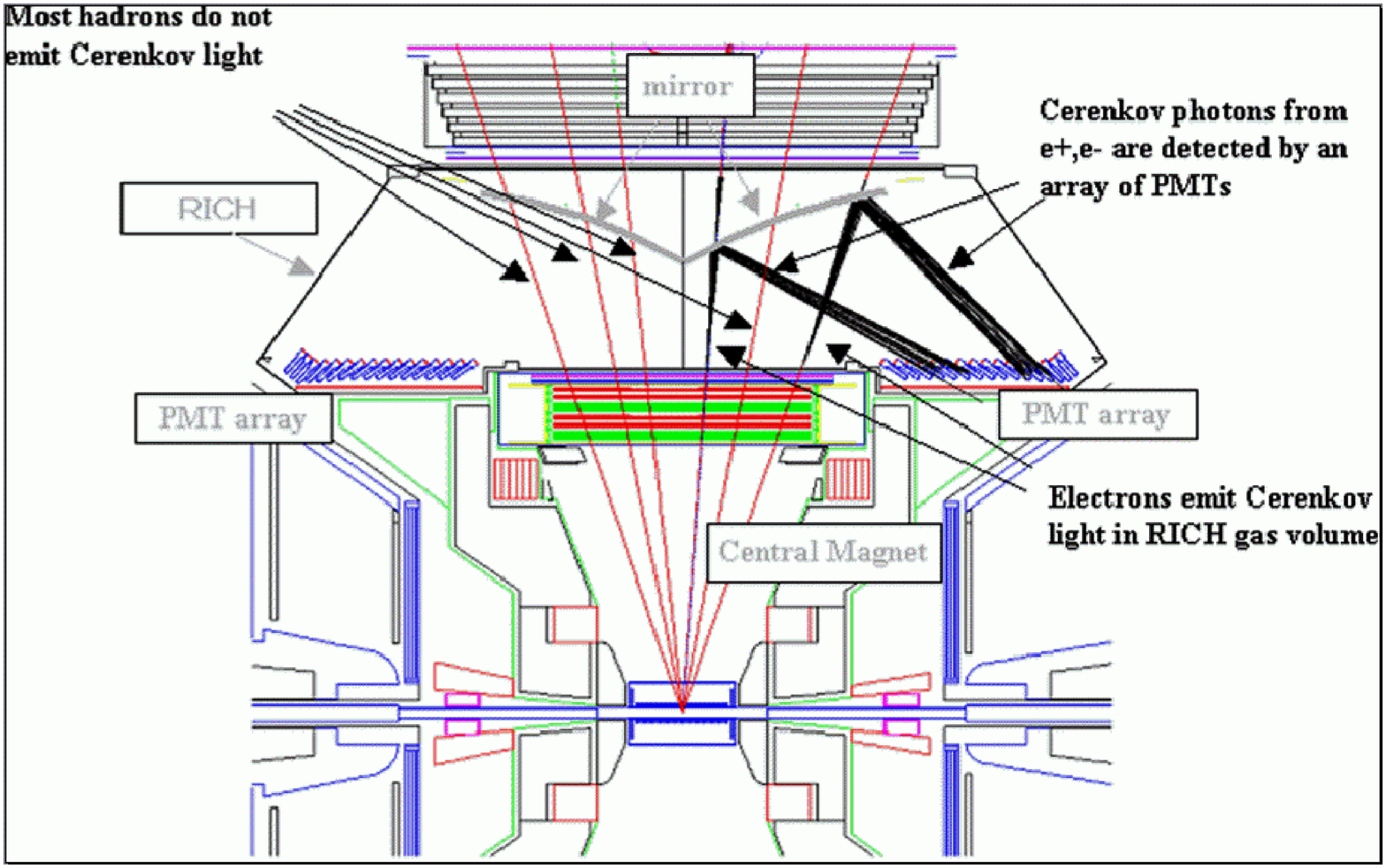,width=1\linewidth,clip}
\caption{\label{fig:rich1} Top view of the RICH and its optics.
The tracks of the electrons and the emitted Cherenkov light cone
are shown. Courtesy of Takashi Hachiya.}
\end{figure}

A schematic view of RICH detector is shown in Fig.~\ref{fig:rich0}.
Each RICH detector has a volume of 40 $m^3$. The spherical mirrors
focus Cerenkov light onto two arrays of photomultiplier tubes
(PMT), each located on one side of the RICH entrance window. In
order to achieve the design requirements, RICH performance has to
satisfy the following specifications~\cite{PHENIXCDR}:
\begin{itemize}
\item
$e/\pi$ separation at the $10^4$ level for single tracks.
\item
The Photo Multiplier Tube (PMT) should have high single photon
efficiency ($>99\%$). It should also have good timing resolution (~300
ps) to reduce noise and contamination from albebo electrons generated
during emc showers.
\item
Minimal radiation length to reduce conversions inside RICH.
\end{itemize}

The entrance of each PMT features a ``Winston cone'' of 50 $mm$
diameter.  The cone funnles light to the tube and increases the active
area by reflecting light into the sensitive area of the tube that
otherwise would have been missed.  Each tube also has a magnetic
shield that allows it to operate in magnetic field up to 100
Gauss. The radiator gas length seen by electron is 87 cm at $\eta=0$
and 150 cm at $\eta=0.35$, the average path length through radiator
gas is 120 cm.

The RICH active volume is filled with $CO_2$ gas at a pressure
slightly above atmospheric. The gas has a Cerenkov threshold of
$\gamma_{thr}=35$ which is about 17 MeV/c for an electron and 4.8
GeV/c for pion. The RICH can also be used to identify pions at
$\pt>4.8$ GeV/$c$~\cite{jjiathesis} Fig.~\ref{fig:rich1}
illustrates the principle of electron detection in the RICH. The
Cerenkov photons generated by $e^+, e^-$ and high momentum hadrons
are reflected by spherical mirrors placed within the radiator
volume. The photons are focused onto PMTs placed just behind the
PHENIX central magnet. The pole tips of the magnet thus serve as
hadron shields for the PMTs.

Cerenkov photons are emitted at an angle of $\theta=cos^{-1}(1/(\beta
n))$. These photons are focused as a ring of photons onto the PMT
array,

The total number of photo electrons for a charged particle above
the Cerenkov threshold can be written as~\cite{PDG}
\begin{equation}
N_{npe} = L\frac{\alpha^2 z^2}{r_e m_e c^2
}\int\epsilon_{c}\epsilon_{d} sin^2\theta dE \label{eq:ch3.npe}
\end{equation}

where $\frac{\alpha^2 z^2}{r_e m_e c^2 } = 370 cm^{-1} eV^{-1}$ ,
$L$ is path length of particles in the gas volume, $\epsilon_{c}$
is the PMT Cerenkov light collecting efficiency and $\epsilon_{d}$
is the quantum efficiency of the PMT.
\begin{eqnarray}
N_{npe} &=& N^0_{npe}Lsin^2\theta\qquad, \mbox{where}\nonumber\\
N^0_{npe}&=&\frac{\alpha^2 z^2}{r_e m_e c^2}\langle \epsilon_c
\rangle \langle \epsilon_d \rangle. \label{eq:rich2}
\end{eqnarray}
which quantifies the RICH electron detection performance. This
number takes into account acceptance and quantum efficiency of the
PMT and the property of the gas. In RUN-2, it is measured to be
116 $cm^{-1}$ for $CO_2$ gas.

\begin{figure}
\centering
\epsfig{file=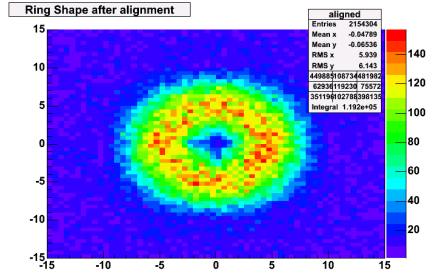,width=0.4\linewidth,clip}
\caption{\label{fig:rich2} Profile of the ring around the
projected charged track intersection point. Courtesy of
T.~Kajihara.}
\end{figure}

Each PMT has a diameter of about 2.5 $cm$, while the ring of
photons reflected onto the PMT array has a radius of 11.8 $cm$.
Fig.~\ref{fig:rich2} shows the association are of the RICH ring
with respect the the incident charged track. The RICH ring can be
clearly seen as expected from the ring diameter for $CO_2$ gas. To
reduce false hit rate, the number of PMTs for a given charged
track are counted within 3.4-12.8$cm$ from the projection - this
value is called $n0$.

\subsection{Electromagnetic Calorimeters}\label{sec:emcal}

The Electromagnetic Calorimeter (EMC) in the PHENIX is used to
measure the spatial position and energy of electrons and photons
produced in Heavy Ion collisions. EMC also provides the means to
trigger on rare events (high $p_T$ electrons and photons). The
hadrons with kinetic energy more then 200 MeV will not deposit
significant energy in the calorimeter as the design and the
thickness is deliberately chosen to be ``uncompensated''. The
detector covers the full Central Arm acceptance of $-0.35 < \eta
<0.35$ and has two arms 90$^\circ$ in azimuth each.

The main specifications to the EMC design are listed
below~\cite{PHENIXCDR}:
\begin{itemize}
    \item Good energy and position resolution for electromagnetic
    showers.
    \item Sub-nanosecond time resolution.
    \item Comparatively low cost.
\end{itemize}

\begin{figure*}[h]
\centering
\epsfig{figure=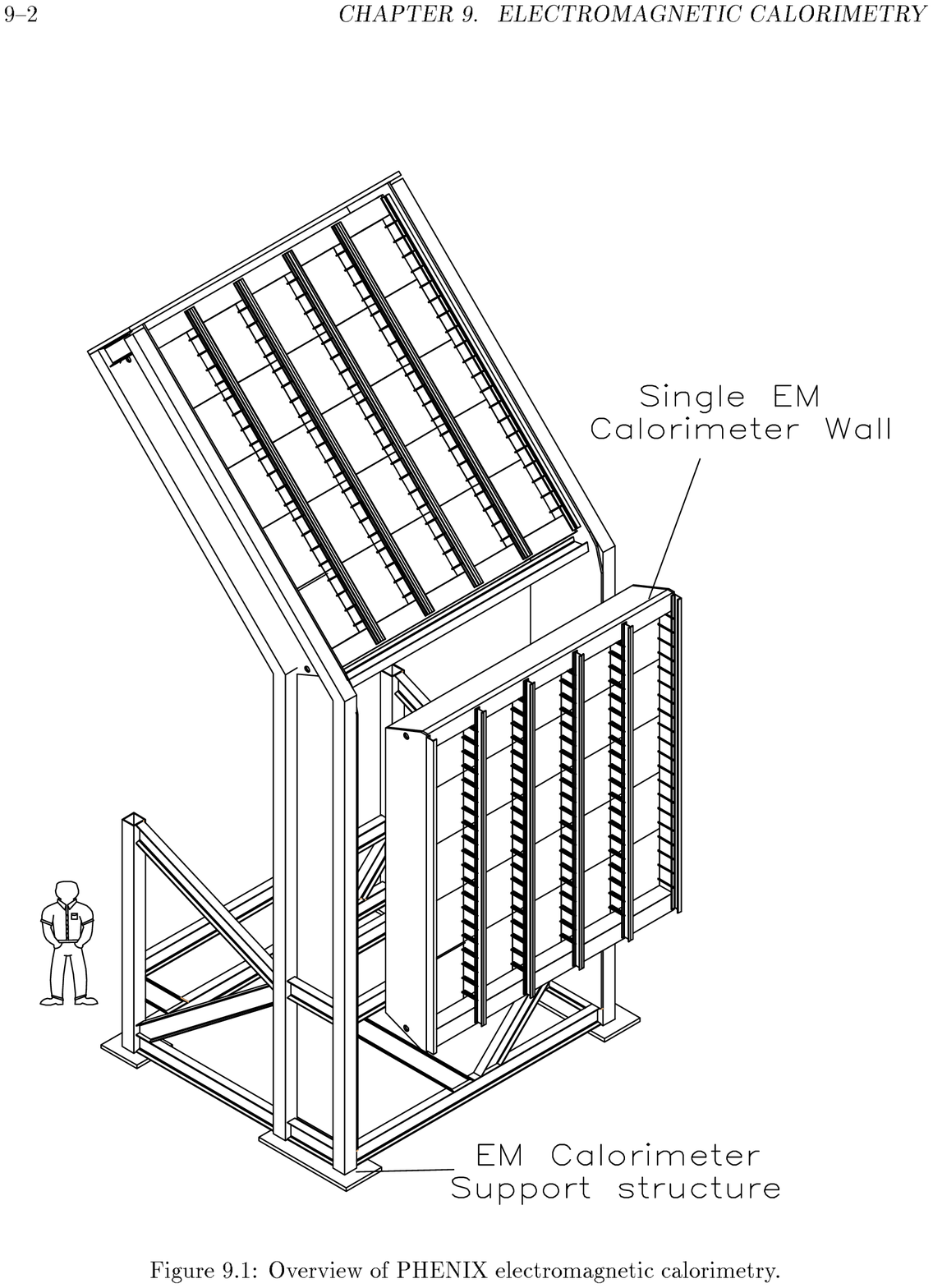,width=0.5\linewidth,clip}
 \caption{\label{fig:ch3.emc_sect}
Structural design of EMC sector~\cite{PHENIXCDR}.} \centering
\epsfig{figure=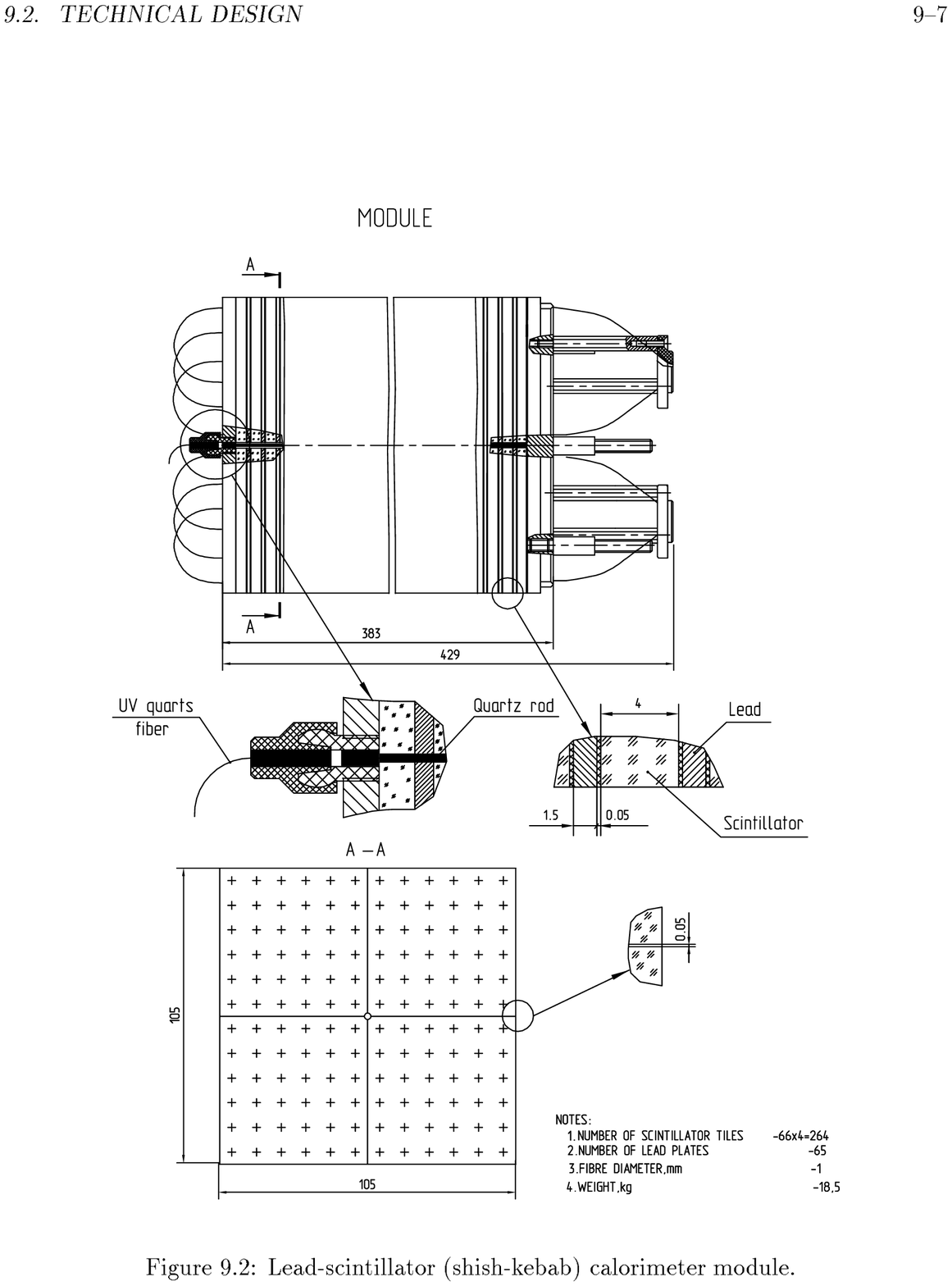,width=0.5\linewidth,clip}
 \caption{\label{fig:ch3.pbsc_tower}
Design of PbSc EMC tower~\cite{PHENIXCDR}.}
\end{figure*}

To accomplish those goals, the basic design of EMC was selected to
consist of 8 large sectors covering in total 60 $m^2$ of the
PHENIX acceptance, 6 of the EMC sectors (E2,E3,W0-3)
\textbf{``Plumbum-Scintillator''} modules (PbSc) and 2 sectors of
\textbf{``Plumbum-Glass''} modules (PbGl). The PbGl represents the
greatest cost savings since the device was recycled from a
previous experiment.  The schematic view of the EMC sector is
shown in Fig.~\ref{fig:ch3.emc_sect}. PbGl sectors of EMC were
previously used by WA98 experiment and were re-installed in
PHENIX. The PbSc sectors was built specifically for PHENIX in 1992
and designed as a set of lead-scintillator sandwich with readout
by wavelength shifting (WLS) fibers penetrating the entire length
of the detector cell (usually referred to as ``tower''). The
dimensions of one PbSc tower are 5.25x5.25x37.0 $cm^3$ and the
effective depth of the EMC corresponds to 18 radiation length. The
depth is chosen to obtain the optimal $e/\pi$ separation via $E/p$
matching. For this analysis we use only PbSc sectors of EMC. Each
tower measure the deposit energy of the electromagnetic shower in
localized place on the detector surface. This enables us to look
for the \textbf{``clusters''} of energy, localized to a particular
block of towers and measure the total deposited energy under
assumption of electromagnetic shower. The electron and photon
leave all its energy in the EMC and localized to 2x2 towers with
85\% probability, making the EMC an unique electron ID device in
PHENIX.

\newpage

\pagebreak
 The schematic design of one PbSc tower is shown in
Fig.~\ref{fig:ch3.pbsc_tower} there are total of 18240 individual
towers in PbSc EMC sectors grouped into \linebreak 25
``supermodules'' each containing 64 towers. The module consists of
a stack of alternating layers of 1.5 mm thick layers of lead,
white reflective paper and 4.0 mm thick polystyren-based
scintillator tiles. Each stack has a drilled array of holes for
the WLS fibers with 500 nm emission peak. The light is collected
from fibers by a conventional PMT tube at the base of the tower.
This design (so called ``shish-kebab'' type) prove to provide a
perfect light collection and perfect time uniformity of the
signal. The dynamic range of the PMT was chosen to perform the
energy measurements starting from 0.1 up to 50 GeV with good
linearity.

\begin{figure}[h]
\centering
\epsfig{figure=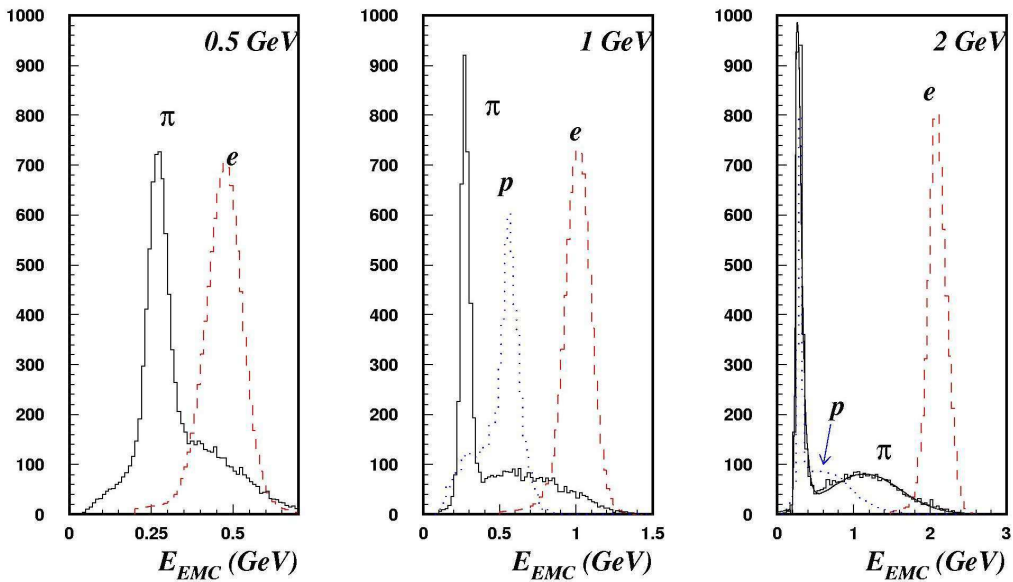,width=1\linewidth,clip}
 \caption{\label{fig:ch3.pbsc_energy}
Energy deposited in PbSc EMC by the pion, proton and electron of E
= 0.5, 1.0 and 2.0 GeV.}
\end{figure}

The energy resolution of the PbSc EMC was measured on the test
beam to be on the order of $\frac{\sigma(E)}{E} = 1.2 \oplus
\frac{8}{\sqrt{E}} \%$ and spatial resolution on the order of
$\sigma(x) = \frac{10}{\sqrt{E(GeV)}}$ mm.
Fig.~\ref{fig:ch3.pbsc_energy} shows the energy deposited in the
EMC by pion, proton and electron with different particle energy.
One can clearly see very good separation of the electrons and
hadrons (especially at $E>1.0$ GeV) which is the advantage of the
electron identification capabilities of the EMC.

\chapter{Data Analysis}\label{sec:ch4}

This chapter describes the analysis procedure used to extract and
isolate the electron sample from heavy flavor semi-leptonic
decays. Section~\ref{sec:ch4.QA} describes the quality control, run and
event selection used in the analysis. Electron identification
cuts and their optimization is presented in Section
\ref{sec:ch4.eIDCuts}. Section~\ref{sec:ch4.Inclusive} explains the
inclusive electron invariant crossection calculation. The estimation
of the electron component from "photon" related decays of light mesons
("photonic" electron component) through the EXODUS Cocktail is
summarized in Section~\ref{sec:ch4.Cocktail}. The final results for
"non-photonic" electron crossection subtraction are presented in
Section~\ref{sec:ch4.Subtraction}. In order to crosscheck the results
of "non-photonic" electron component measurement, the independent
"Converter subtraction" analysis was performed which is described in
Section~\ref{sec:ch4.Converter}.  Section~\ref{sec:ch4.Systematics}
presents the Systematic Error estimations for inclusive electron
crossection, Cocktail prediction and the subtracted "non-photonic"
electron crossection.

\section{Quality Assurance and run selection}\label{sec:ch4.QA}

The accurate run selection is essential for high precision
measurement of Open Charm decay electron component. The
contribution of "photonic" electron background is on the order of
80\% at low $p_{T}$ of total electron signal and even a small
variation of total electron yield can cause a significant variation
in the background-subtracted result.

The other complication for the electron analysis is that we need
to be certain that we have uniform distribution of material in the
acceptance. Any additional piece of equipment in the acceptance
can cause a significant increase to creation of conversion
electrons. Thus we need to apply an elaborate acceptance cuts in
order to make a conversion rate uniform in the acceptance.

\subsection{Acceptance cuts}\label{sec:ch4.acc_cuts}

The acceptance for the electrons in PHENIX is best represented in
terms of the track inclination angle $\alpha$ and the azimuthal angle
$\phi$. The transverse momentum of the particle is inversely
proportional to $\alpha$ and can be approximated to the first
order as $p_{T} \approx \frac{0.086}{\alpha}$ GeV/c.

\begin{figure}[vht]
\begin{center}
\epsfig{figure=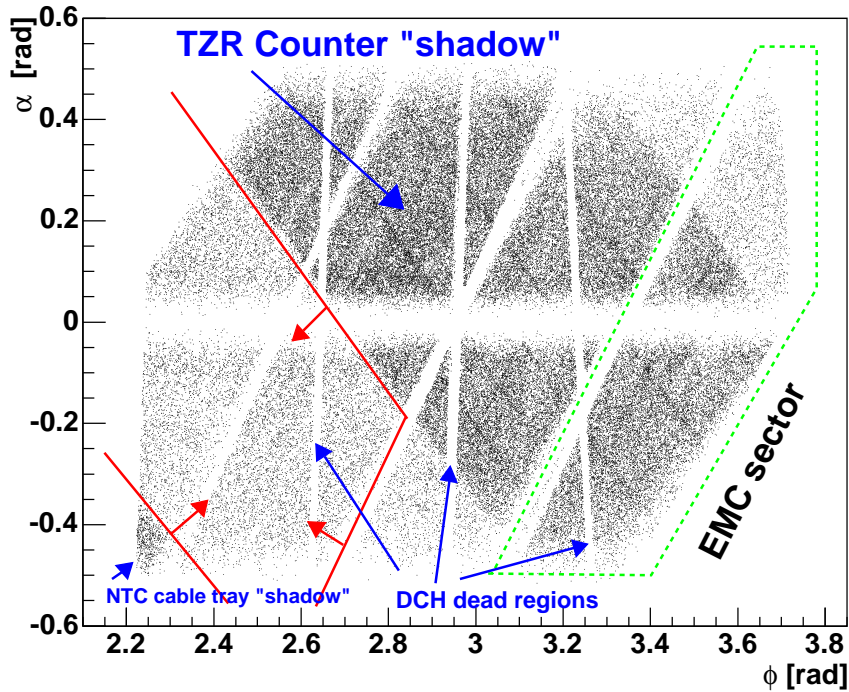,width=5.in}
\caption{\label{fig:ch4.alpha_phi} Density of the electrons in
$\alpha$ vs. $\phi$ space for East Arm. The area indicated by red
arrows is the acceptance region used in the analysis. The various
acceptance holes and additional photon conversion "shadows"
indicated on the plot.}
\end{center}
\end{figure}

The Drift Chamber performance in the East arm was much more stable
then that of the West arm and for this analysis we decided to use
only the East arm acceptance. Applying very loose electron ID cuts
($ n0 > 1$, $|d_{EMC}| < 5$ ) we can plot the density of the
electron candidates in $\alpha$ vs. $\phi$ space.
Fig~\ref{fig:ch4.alpha_phi} shows the electron acceptance of the
East arm. One can see that big portion of acceptance is "shadowed"
by conversions from Time Zero counter (TZR). This detector was
installed into the PHENIX acceptance about 60 cm from the
interaction point in order to improve the measurement of "start"
time for the Time-of-Flight detector. Unfortunately due to very
large radiation length of TZR counter ($X_{TZR} \approx 5.0 \%$!!)
it creates a very large rate of conversions far from vertex that
creates a huge conversion background in the region of its shadow.
\footnote{Since the TZR detector debacle, any new detector placed
in the PHENIX aperture has been required to submit a ``detector
impact statement'' prior to its inclusion in our apparatus.}  The
region effected by the TZR counter is removed by the fiducial cut
shown in the acceptance plot. Stripes on this figure depict
acceptance holes for various PHENIX detectors. There is also a
small portion of acceptance (circled on the plot) affected by
conversions from cable tray of the ``New Trigger Counter'' NTC
detector, it is also removed by fiducial cut\footnote{The NTC also
failed to pass its ``detector impact study'' and is removed.
Neither the NTC nor the TZR were ever used to produce a physics
result in PHENIX.}. The small acceptance region at $\phi
> 3.2$ rad that is not affected by TZR "shadow" does not contain
high momentum electrons which are of particular interest for Open
Charm electron measurements. That is why analysis is using only
EMC PbSc sectors E2, E3.

Non-uniform conversion rate acceptance cuts are listed below:
\begin{itemize}
\item $|\phi+ 0.85\cdot\alpha| < 2.68;$         TZR counter cut

\item $|\phi+ 0.85\cdot\alpha| > 1.95;$         NTC cable tray
shadow

\item E2,E3; EMC sector cut
\end{itemize}
\pagebreak

 The additional holes in the acceptance of the detector
were studied starting from those closest to the interaction point
({\it i.e.} the DCH). Due to the bending of the track in the
magnetic field, the azimuthal angle $\phi'$ at which track
intercepts each detector component of the PHENIX will be shifted
with respect to DCH $\phi$, which is calculated at the "reference
radius" $R_{ref} = 220$ cm. The shift is proportional to $\alpha$
and is negative for interception with radius $R > R_{ref}$ and is
positive otherwise.  $\phi_{X1} = \phi +0.06\cdot\alpha$
corresponds to track angle in X1 DCH plane, $\phi_{X2} = \phi
-0.04\cdot\alpha$ corresponds to track intersection of X2 DCH
plane. By plotting track density in $\phi_{X1}$, $\phi_{X2}$
coordinates we can clearly identify the dead DCH regions.

\begin{figure}[ht]
\begin{center}
\epsfig{figure=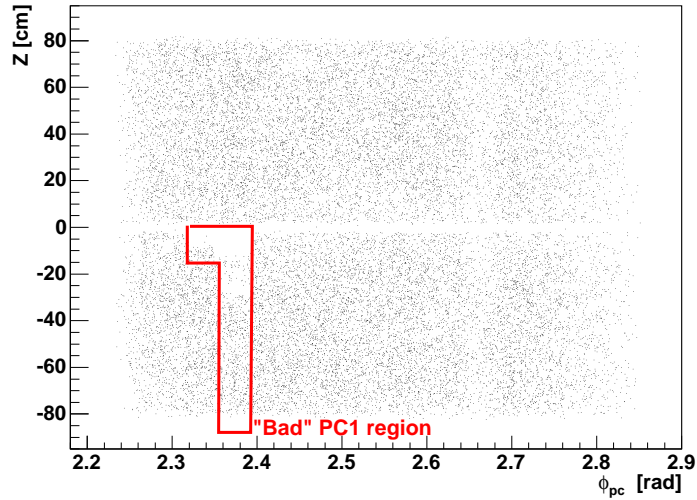,width=4.in}
\caption{\label{fig:ch4.zed_phipc} Density of the electrons in $Z$
vs. $\phi_{PC}$ space.}
\end{center}
\end{figure}

The same analysis can be performed for the Pad Chamber (PC1) dead
regions. $\phi_{PC}$ can be approximated as $\phi_{PC} = \phi
-0.13\cdot\alpha$. Fig.~\ref{fig:ch4.zed_phipc} shows the track
density in $\phi_{PC}$, $Z$ space. There is a PC1 region of
unstable performance that was cut-out.

Final list of tracking fiducial cuts is summarized below:

\begin{itemize}
\item $Not(|\phi+ 0.06\cdot\alpha -2.562 | < 0.005 \;\& \;(Z<0))$
DCH dead region

\item $Not(|\phi- 0.13\cdot\alpha -2.365 | < 0.025 \;\& \;(Z<0))$
PC1 dead region

\item $Not(|\phi- 0.13\cdot\alpha -2.325 | < 0.025\;\& \;(Z<0)\;
\&\; (Z> -15))$ PC1 dead region
\end{itemize}

\begin{figure}[vht]
\begin{center}
\epsfig{figure=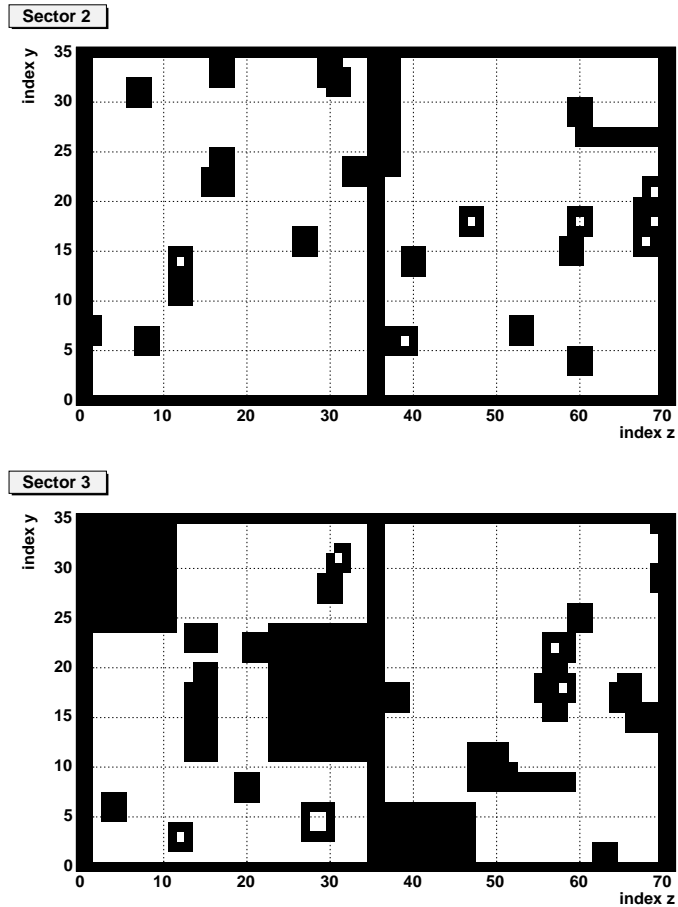,width=4.in}
\caption{\label{fig:ch4.dead_emc} Dead/noisy EMC tower map for E2,
E3 EMC sectors.}
\end{center}
\end{figure}

EMC dead area was calculated by photon density measurements in
each EMC tower on a run-by-run basis~\cite{ana143,pp_pi0}. The
final dead/noise map used in the analysis included all the towers
that had a dead/noise flag set at least in one run. 3x3 tower
region around the "bad" tower was fiducially removed in order to
have more precise energy measurement in the vicinity of the "bad"
tower. Fig.~\ref{fig:ch4.dead_emc} shows the map of the dead
towers in EMC that were removed from the analysis.\\

\subsection{Event selection}

The ``Minimum Bias'' event trigger in Run02 P+P was based on the
coincidence of at least one hit each in the North and South
Beam-Beam Counters (BBC). In order to keep up with high luminosity
and to keep the constant bandwidth for Data Acquisition System, a
trigger $\it{prescale}$ logic was implemented. The live trigger
rate was artificially reduced by only storing each one event out
of $R_{scale}$. $R_{scale}$ is called trigger prescale factor
which depended on RHIC store luminosity and could be set to four
possible values of 10, 20, 40, 80. The Minimum Bias trigger was
always prescaled. \pagebreak

 In order to increase the rate of
events containing energetic electron, \linebreak PHENIX uses
special Level-1 electronics trigger called the ERT (EMC-RICH
trigger). The basic principle of the trigger lies in the online
summing of the energy signals in the EMC over a 2x2 tower region
called a tile. If the signal from particular tile exceeds the
tunable threshold value the specific bit is set in the data
stream. To avoid edge effects, the tiles are overlapped and with
2x2 summing the number of tiles equals the number of towers.  The
other bit is set once a RICH tile (4x5 tubes, overlapping) have a
signal exceeding threshold. A spatial match between emc and rich
tiles is an indication that the high momentum electron may have
been detected in the particular region of detector.\footnote{low
momentum particles have displaced RICH and EMC tiles and fail the
trigger.} The trigger electronics issues the Local Level 1 (LL1)
trigger decision for the PHENIX Global Level 1 system (GL1). The
energy threshold of the ERT trigger can be adjusted by the
threshold settings and was set to have a 50\% registration
probability for 800 MeV electron. The efficiency of the ERT
trigger is discussed in more details in
Section~\ref{sec:ch4.Inclusive}. The same electronics can be used
to fire on high energy photons by skipping the coincidence of the
RICH bit. This trigger called "Gamma1" was successfully used for
high momentum $\pi^0$ measurements~\cite{pp_pi0,ana224}. The ERT
trigger has a significant rejection power (the rate of the
triggered events compared to Minimum Bias rate) $R_{ERT} \approx
40-50$, does not limit the DAQ bandwidth, and requires no prescale
factor. The proper normalization of the ERT trigger data should be
done to the total number of $\it{live}$ Minimum Bias triggers
corresponding to the particular Run. It is crucial to $\bold{not}$
use the number of ERT trigger events for normalization as any
noisy channel can artificially increase the rate of this trigger
and a strong bias would be applied to the results.

\begin{figure}[vht]
\begin{center}
\epsfig{figure=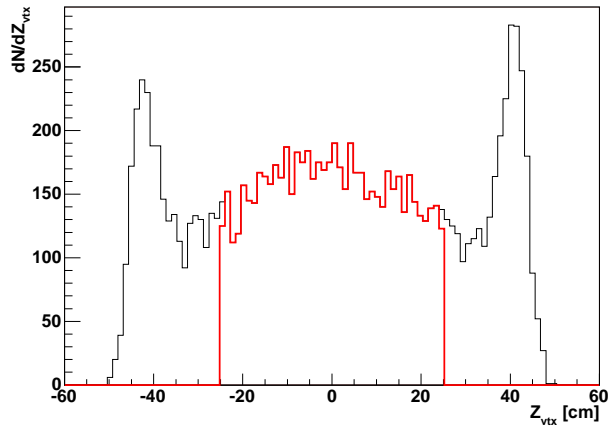,width=3.5in}
\caption{\label{fig:ch4.bbcz} BBC $Z_{vtx}$ distribution for
electron candidates ($n0>3$). Bold line shows the vertex region
used in the analysis.}
\end{center}
\end{figure}

The collision vertex is measured by the Beam-Beam Counters (see
Section ~\ref{sec:ch3.BBC}). The vertex resolution in $\pp$ Run02
was $\delta_{Z_{vtx}} = 1.2$ cm. Due to the specific geometry of
PHENIX, the tracks originating from collisions that are far from
the center of the detector ($Z_{vtx}=0$ cm) have a higher
probability to interact with the material of the Central Magnet
thus creating additional conversion electron background. The
collision vertex distribution for the particles that are primarily
electrons (tight eID cut $n0 >3$) is shown in
Fig.~\ref{fig:ch4.bbcz}. One can see that the vertex distribution
has an almost Gaussian shape with some additional structure for
high $Z_{vtx}$. In order to minimize the conversion background, we
use tight vertex cuts for this analysis and only look at the
events with $|Z_{vtx}| < 25$ cm.

\subsection{Run selection and event counting}

    To filter the bad runs we looked at the $\phi$  distribution
of all charged tracks ($p_{T} > 0.4$ GeV/c) with standard
electron identification cuts (except for RICH $n0>1$ cut,see
Section~\ref{sec:ch4.eIDCuts}) and all acceptance cuts. The
$\frac{dN}{d\phi}$ distribution for each run was normalized to the
number of recorded Minimum Bias events $N_{MB}$. Then the ratio of
the $\frac{1}{N_{MB}}\frac{dN}{d\phi}$ distribution for given run
to the same distribution for the chosen "reference" run (run
having significant statistics and stable acceptance) was fitted
with a constant $R$. Any significant deviation of the fit
parameter $R$ from one is an indication of additional dead area in
the trial run. The criteria for the selection of the run was
chosen to be:
\begin{itemize}
\item $R > 0.94$ \item $\chi^2_{\nu} < 2.0$
\end{itemize}

\begin{figure}[]
\begin{center}
\epsfig{figure=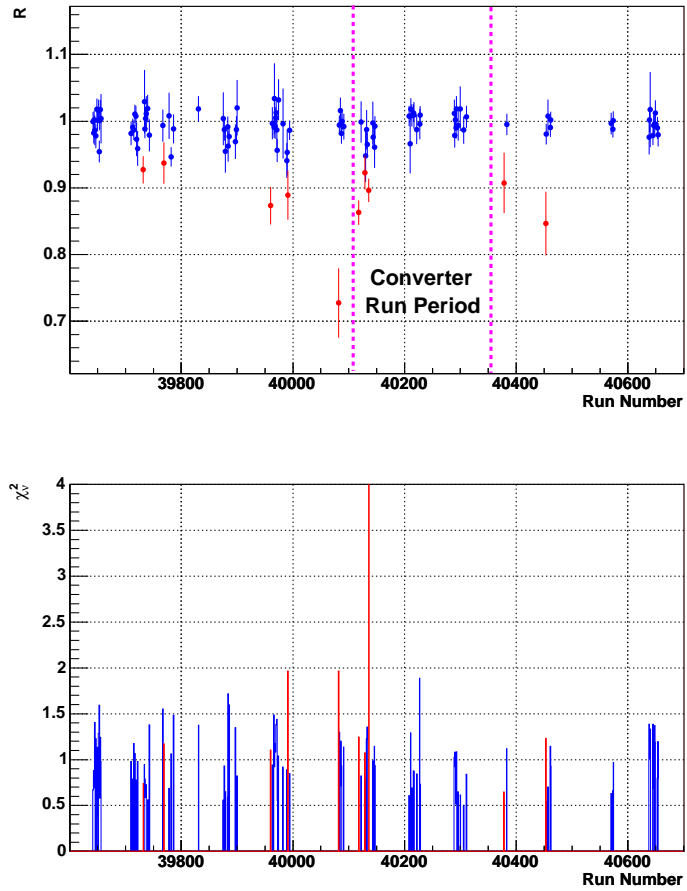,width=4.1in}
\caption{\label{fig:ch4.run_qa} Run-by-run variation of charged
particle yield.}
\end{center}
\end{figure}

The run by run variation of $R$ and $\chi^2_{\nu}$ presented on
Fig.~\ref{fig:ch4.run_qa} (runs shown as red are considered to be
"bad"). During Run02 we had an period with additional "photon
converter" installed inside the PHENIX acceptance. "Converter"
subtraction method explained in details in
Section~\ref{sec:ch4.Converter}. Converter run period should be
treated separately from Non-Converter run period. The total event
statistics of the Converter and Non-Converter run periods summarized in
Table~\ref{tab:runqatable}.

\begin{table}

\caption{Statistics summary for Non-Converter and Converter run
period.}
\begin{center}
\begin{tabular}{|c|r|r|}

\hline    &  $N_{MB}$ &$N_{MB\ live}$\\
\hline
Non-Converter "Total"&   15 931 737   &  475 849 920\\
Non-Converter "Bad"&    540 061    & 10 683 600\\
Non-Converter "Good"& $\bold{15\ 391\ 676}$    &  $\bold{465\ 166\ 320}$\\
\hline
Converter "Total"&   4 851 787 & 264 284 240\\
Converter "Bad"&     361 847& 28 947 760\\
Converter "Good"& $\bold{4\ 489\ 940}$&$\bold{235\ 336\ 480}$\\
\hline

\end{tabular}
\end{center}
\label{tab:runqatable}
\end{table}

\newpage
\section{Electron identification cuts}\label{sec:ch4.eIDCuts}

Electron identification is one of the most critical parts of the
analysis and required a precise tuning of the eID parameters.
PHENIX is able to identify the electrons using the following
parameters:
\begin{itemize}
\item Number of RICH photomultipliers that have hit within the
projected track ring - $n0$.
\item EMC matching - distance between
track projection and EMC cluster in $\phi$ and $Z$ coordinates. We
denote $d_{EMC}$ as $d_{EMC} = \sqrt{d\phi_{EMC}^2+dZ_{EMC}^2}$.
\item Ratio of EMC deposited energy to particle momentum - $E/p$.
\end{itemize}

Those variables have been used for all the current PHENIX electron
results~\cite{jpsi,ppg011,ppg035}. All eID parameters were
adjusted both for Data and Monte Carlo simulation to be identical
and uniform (i.e. matching parameters have the mean value of zero
and width of $1\sigma$ independent of momentum and uniform
throughout the detector acceptance).

\subsection{n0 cut optimization}

The rejection power of separate eID cuts was tested by studying the
$E/p$ distribution of electron candidates before and after the cut.

\begin{figure}[t]
\begin{center}
\epsfig{figure=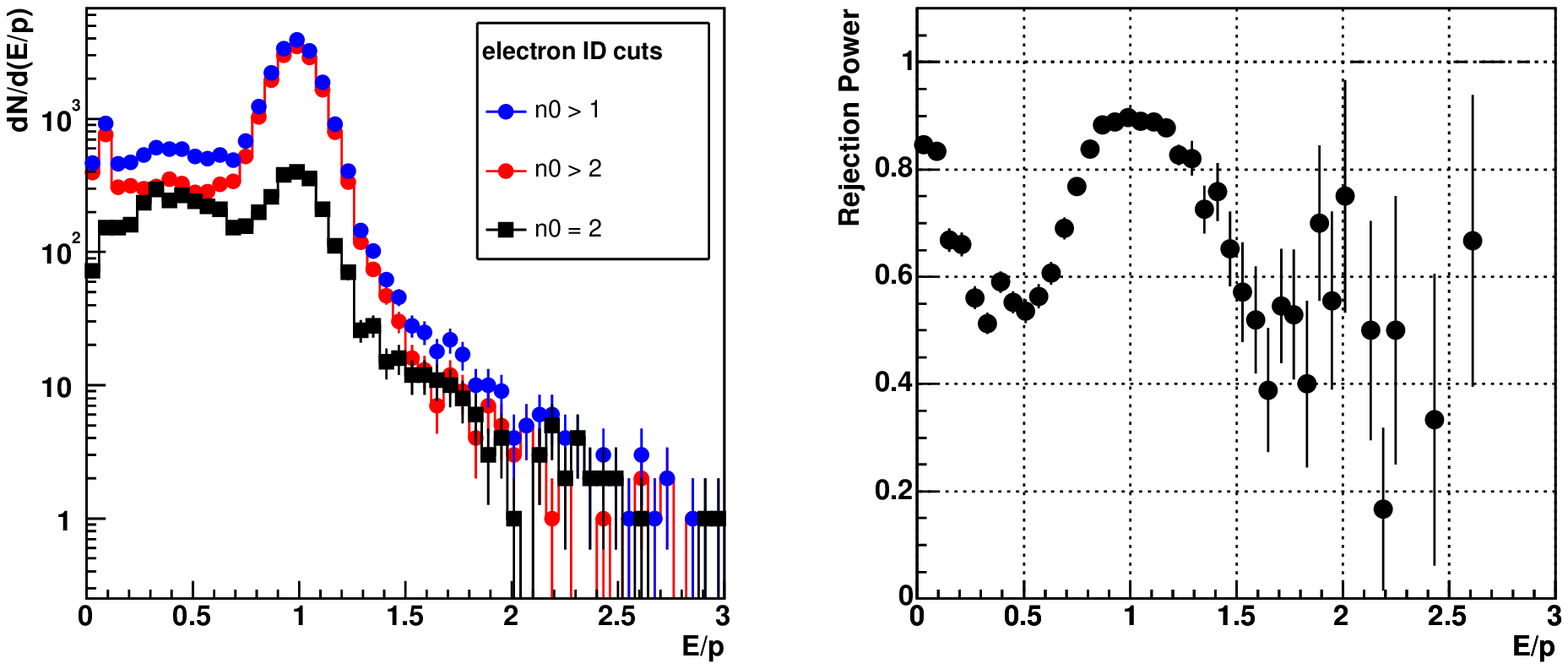,width=0.95\linewidth}
\caption{\label{fig:ch4.n0_cut} a) $E/p$ distribution for the
electron candidates for different n0 cuts b) Rejection power of
eID cut $n0>2$ in comparison with $n0>1$ cut.}

\epsfig{figure=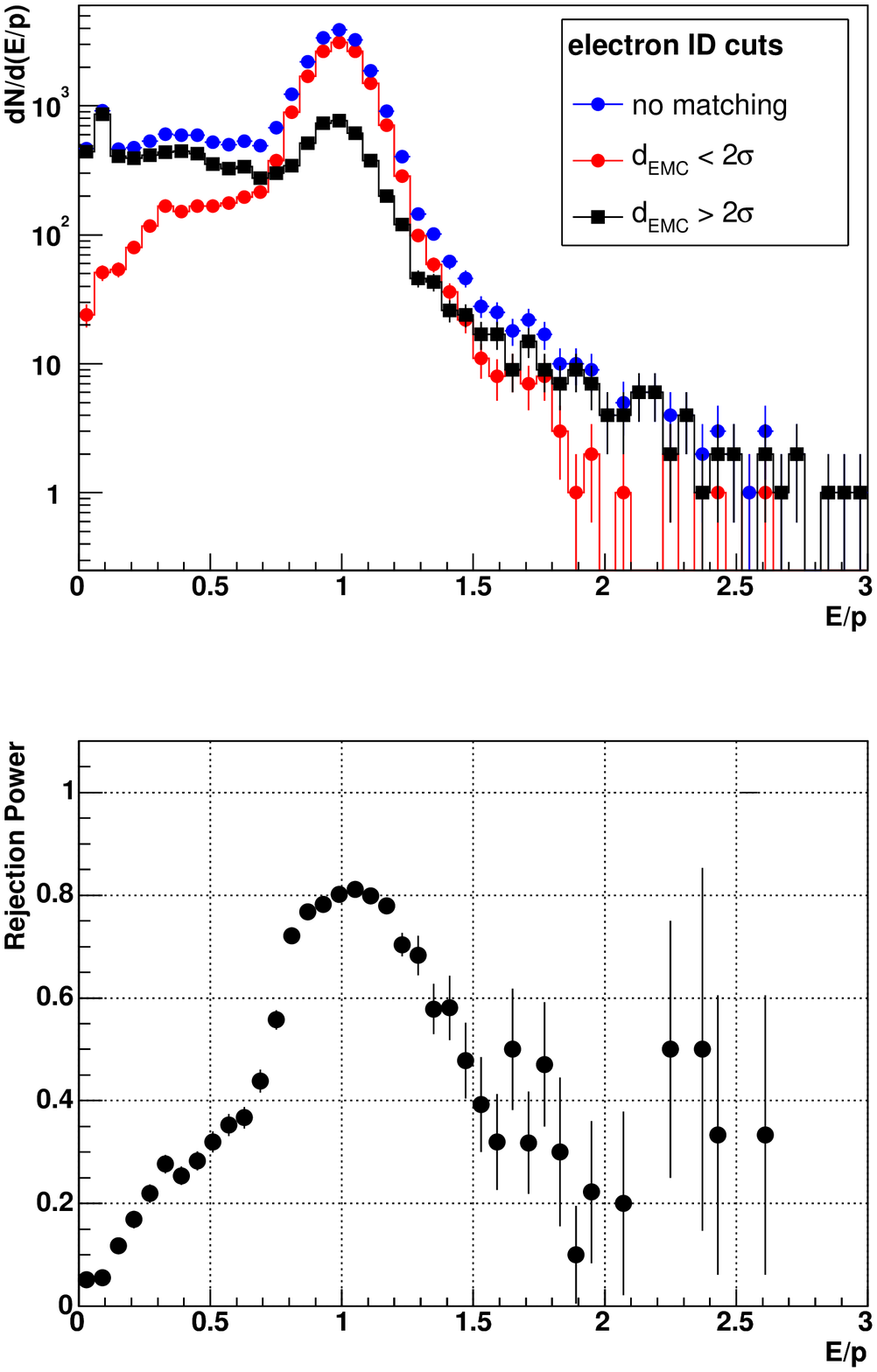,width=0.3\linewidth,clip,trim =
0.2in 0in 0.7in 0in}
\epsfig{figure=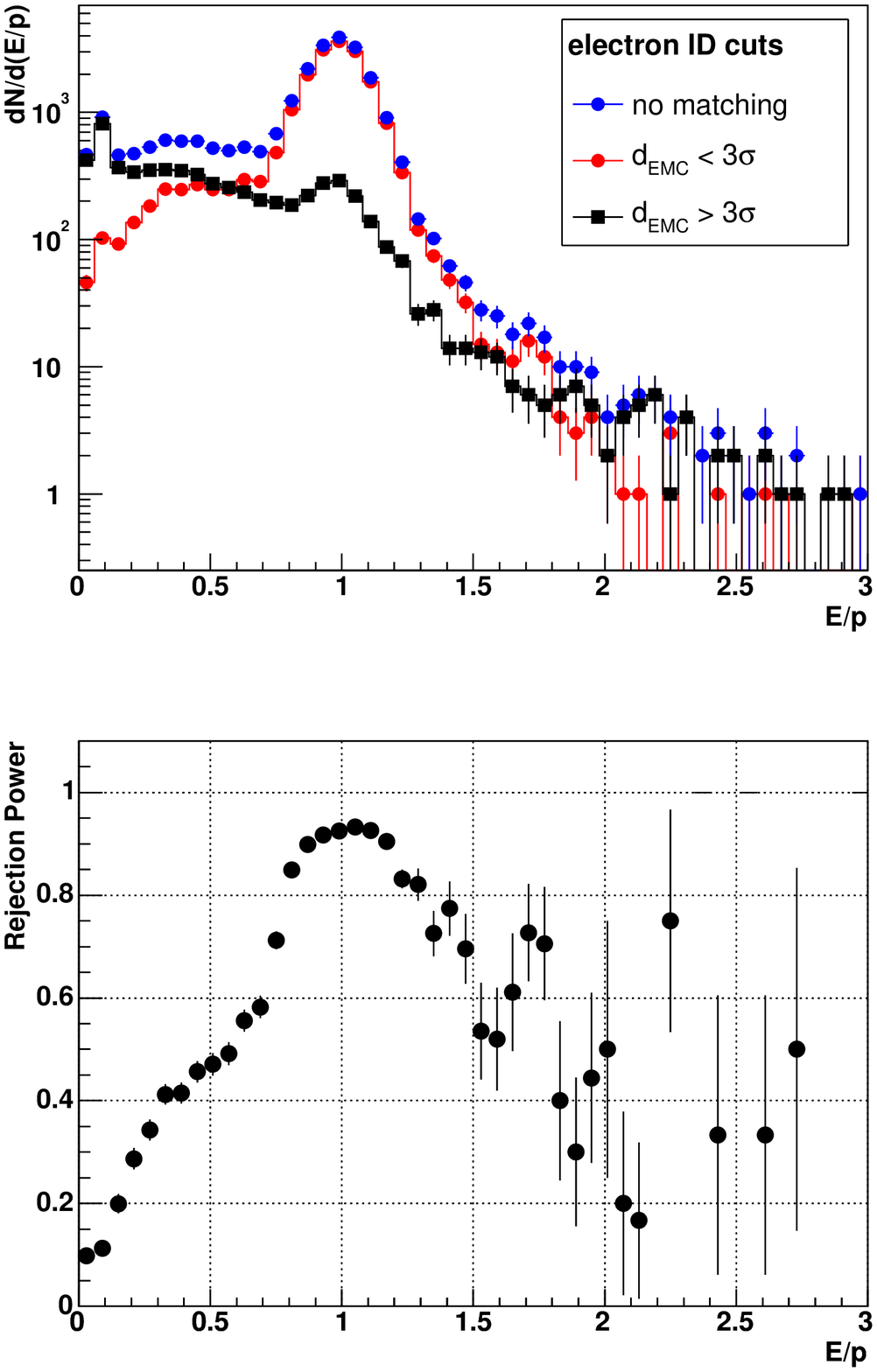,width=0.3\linewidth,clip,trim =
0.2in 0in 0.7in 0in}
\epsfig{figure=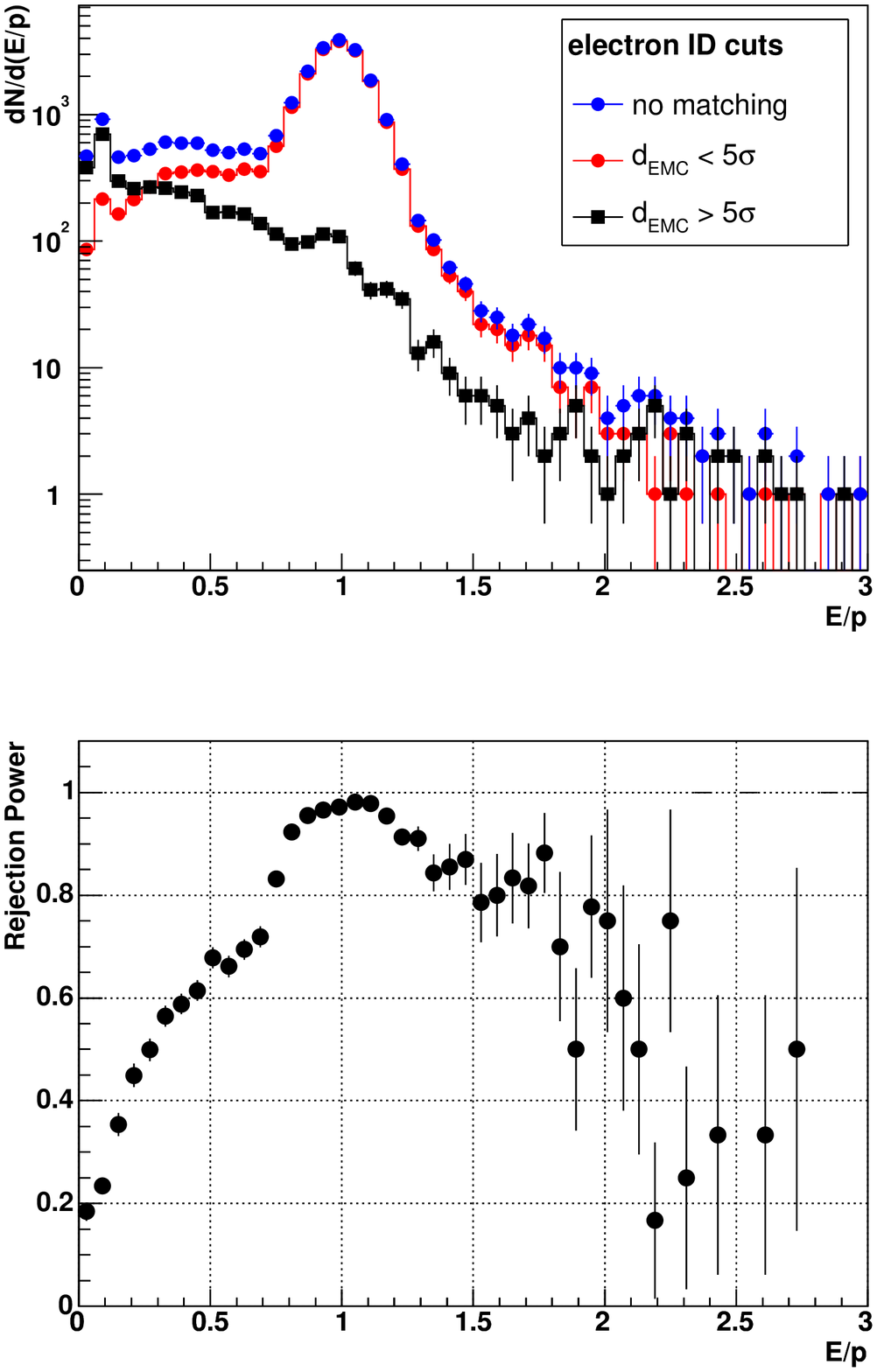,width=0.3\linewidth,clip,trim =
0.2in 0in 0.7in 0in} \caption{\label{fig:ch4.demc_cut} $E/p$
distribution for the electron candidates and rejection power for
different $d_{EMC}$ cuts a) $d_{EMC} < 2\sigma$ b) $d_{EMC} <
3\sigma$ c) $d_{EMC} < 5\sigma$.}
\end{center}
\end{figure}

An initial assumption is chosen for the $n0$ cut.  This cut can not be
set as low as one phototube since the random association background
due to electronics noise would be too high. Thus $n0>1$ was assumed to
be lowest possible $n0$ cut. The next step is to study what happens as
the cut is tightened.  Fig.~\ref{fig:ch4.n0_cut} shows the effect of
$n0>2$ cut on the initial electron candidate sample ($n0>1$) for
$p_{T} > 0.4$. The bottom left inlet shows the rejection power of the
cut which shows the remaining portion of the particles after the
cut. One can see a distinctive peak in $E/p$ distribution which
defines "real" electrons. The tail at $E/p < 0.5$ consists principally
of random charged hadron tracks associated with real RICH hits and
off-vertex conversion electrons with mis-measured momentum. In order
to estimate the effectiveness of the eID cut we check how much signal
it removes in the $E/p$ electron peak region and how strongly it
cuts away the low $E/p$ background.

From the plot we can conclude that we lose $\approx 10\%$ of the
electron signal and $\approx 50 - 60 \%$ of the background. This
loss of the electron efficiency is too large to afford  and the
rejection is not significantly high so we decided to use $n0>1$
cut for the analysis.

\subsection{EMC matching cut optimization}

The similar studies have been done on the "adjusted" track matching
$d_{EMC} = \sqrt{d\phi_{EMC}^2+dZ_{EMC}^2}$ parameter. $d_{EMC}$
variable cut has been tested for values of $d_{EMC}<2$, $d_{EMC}<3$,
and $d_{EMC}<5$. The resulting rejection power of is shown in
Fig.~\ref{fig:ch4.demc_cut}.

It is clear that $d_{EMC} < 3$ already cuts a big portion of low
$E/p$ electron candidates leaving the peak statistics almost
intact. The efficiency of a \linebreak $d_{EMC} < 2$ cut at the
peak is $\approx 80 \%$ which is a significant statistics loss.
Therefore, $d_{EMC} < 3$ was chosen as an optimum for the
analysis.

\subsection{$E/p$ cut parametrization}

We expect electrons to generate an electromagnetic shower in the EMC
and therefore register an energy equal to their momentum.  The energy
over momentum distribution not only enables us to identify electrons
by also allows us to measure both the energy and momentum resolution.
The resolution of E/p can be directly derived from $\frac{\sigma
(p)}{p}$ and $\frac{\sigma (E)}{E}$ and can be written as:
\begin{eqnarray}
    \frac{\sigma(p)}{p} &=& \sqrt{{\sigma}_{MS}^{2} + ({\sigma}_{DCH}\cdot
    p)^{2}}\nonumber\\
    \frac{\sigma(E)}{E} &=& \sqrt{{\sigma}_{C}^{2} +
(\frac{{\sigma}_{EMC}}{\sqrt{E}})^{2}}
 \label{eq:ch4.resol}
\end{eqnarray}
\\ where $\sigma_{MS}$ is term due to the multiple scattering,
$(\sigma_{DCH} \cdot p)$ is DCH angular resolution, $\sigma_{C}$ is a
constant term of EMC energy resolution,
$\frac{\sigma_{EMC}}{\sqrt{E}}$ - is an EMC energy resolution
depending upon fluctuations in the number of particles produced in
the EM shower.

The fluctuations are independent and so

\begin{equation}
    \sigma(\frac{E}{p}) \approx \sqrt{\frac{\sigma(E)^{2}}{p^{2}} + \frac{E^{2} \cdot
    \sigma(p)^{2}}{p^{4}}} \approx \sqrt{\sigma_{C}^{2} + \sigma_{MS}^{2}
    + \frac{\sigma_{EMC}^{2}}{p_{T}} + (\sigma_{DCH}\cdot
    p)^{2}}
 \label{eq:ch4.ep_resol}
\end{equation}
\\
Eq.~\ref{eq:ch4.ep_resol} was obtained assuming $E \approx p
\approx p_{T}$. From this equation one can see that at low $p_{T}$
the main contributor to $E/p$ resolution is energy resolution
which is been overcome by momentum resolution term at high $p_{T}$
and starts grow linearly.

\begin{figure}[ht]
\begin{center}
\epsfig{figure=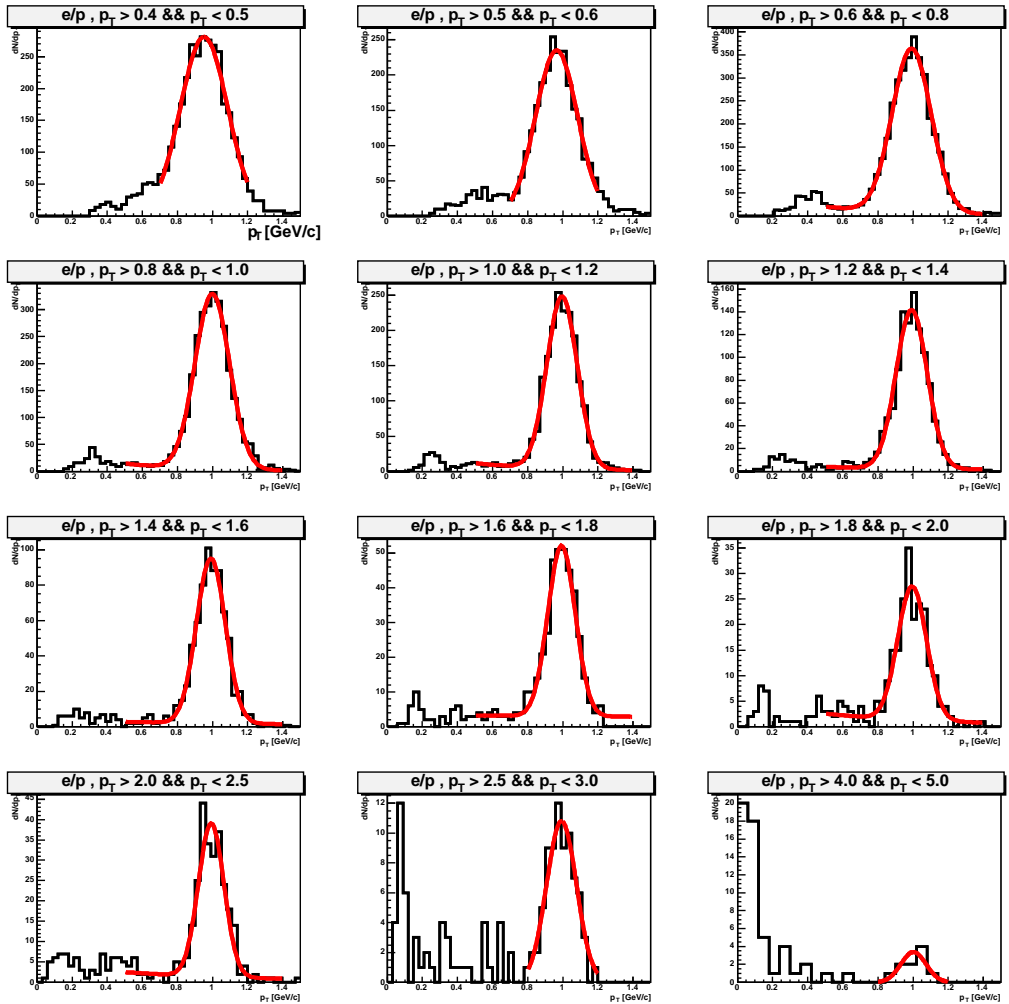,width=1.0\linewidth}
\caption{\label{fig:ch4.fit_ep} Fits to $E/p$ distribution of
electron candidates for different $p_T$ bins.}
\end{center}
\end{figure}

The mean and sigma of $E/p$ distribution for electron candidates
was obtained as a function of $p_T$ by fitting each slice with
Gaussian + exponential background (or Gaussian wherever background
is negligible or non-exponential). Fig.~\ref{fig:ch4.fit_ep} shows
the fit results for different $p_T$ bins starting from 0.4 GeV/c
up to 5 GeV/c. \pagebreak

\begin{figure}[]
\begin{center}
\epsfig{figure=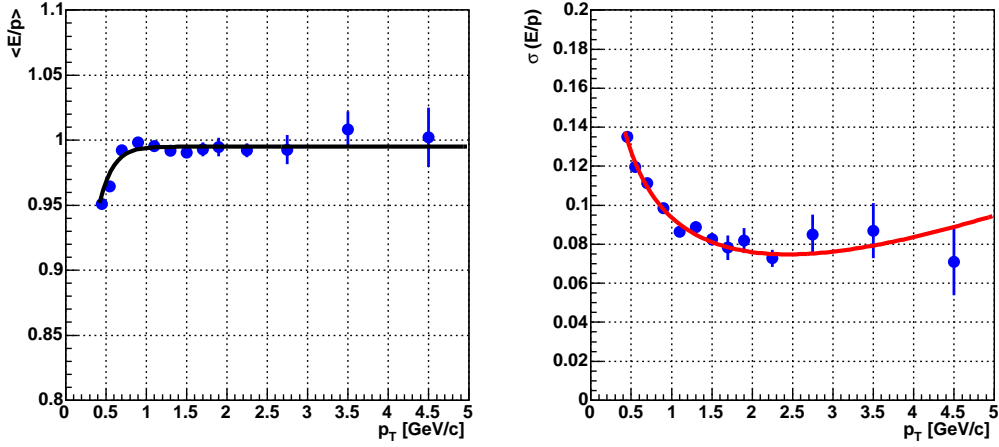,width=1\linewidth}
\caption{\label{fig:ch4.mean_sigma_ep} a) Mean and b) Sigma of
$E/p$ Gaussian fit to electron signal as a function of $p_T$
(Sigma is fitted with Eq.~\ref{eq:ch4.ep_resol} function).}
\end{center}
\end{figure}

Fig.~\ref{fig:ch4.mean_sigma_ep} shows the mean and sigma
distribution for the Gaussian component of the fit as a function
of electron transverse momentum. One can see that we have a very
clean electron sample with the background contribution slowly
drifting to the lower $E/p$ values going to higher momentum.

From Fig.~\ref{fig:ch4.fit_ep} one can see that our energy and
momentum measurements are in good agreement. The apparent fall of
the mean $E/p$ is possibly related with the fact that at low
momentum the inclination angle of the track becomes significant
and EMC cluster starts to spread spatially and we start measure
only a fraction of its total deposited energy\footnote{The EMC
cluster algorithm is tuned for photons that land at near-normal
incidence.}.

The $E/p$ resolution is fitted with Eq.~\ref{eq:ch4.ep_resol} and
the following energy and momentum resolution were derived:

\begin{itemize}
\item $\sigma_{C} \oplus  \sigma_{MS} = (3.82 \pm 0.86)\%$

\item $\sigma_{EMC} = (8.47 \pm 0.28)\%$

\item $\sigma_{DCH} = (1.48 \pm 0.41)\%$
\end{itemize}

This results are in good agreement with PHENIX measurements of energy
and momentum resolution~\cite{pp_pi0,ana172} via other independent
techniques.

As the background level in p+p collisions is very low and
significantly suppressed by EMC matching cuts, we decided to use a
loose ($\pm 3\sigma$) $E/p$ cut for the analysis
\\ \\
    $|\frac{E/p- <E/p>}{\sigma(E/p)}| < 3$
\\ \\
Thus, the cut width follows the fitted sigma as a function of the
track's momentum.

\newpage
\section{Inclusive electron invariant crossection}\label{sec:ch4.Inclusive}

This section of the Thesis explains the procedure of single
differential crossection calculation for single electrons starting
from the "raw" $\frac{N_{e}}{\Delta p_T}$ distribution, correction of
the "raw" electron distribution to full azimuthal \& one unit in
rapidity, estimation of background level, combining ERT and MB
data sample, trigger bias correction of the final crossection and
bin width related corrections. The final expression for MB and ERT
inclusive crossection can be written the following way

\begin{eqnarray}
    E\frac{d\sigma}{dp_{T}^3}_{MB} &=&
    \frac{1}{N_{MB}}\cdot\frac{1}{2\pi}\cdot\frac{1}{2}\cdot\frac{1}{p_{T}}\cdot
    \frac{N_{e\ MB}}{\Delta p_{T}}\cdot\frac{1}{\Delta y}\cdot\frac{\sigma_{pp\ tot}\cdot\epsilon_{BBC}}
    {\epsilon_{reco}(p_{T})\cdot\epsilon_{bias}(p_{T})}\nonumber\\
    E\frac{d\sigma}{dp_{T}^3}_{ERT} &=&
    \frac{1}{N_{MB\ live}}\cdot\frac{1}{2\pi}\cdot\frac{1}{2}\cdot\frac{1}{p_{T}}\cdot
    \frac{N_{e\ ERT}}{\Delta p_{T}}\cdot\frac{1}{\Delta y}\cdot\frac{\sigma_{pp\ tot}\cdot\epsilon_{BBC}}
    {\epsilon_{reco}(p_{T})\cdot\epsilon_{bias}(p_{T})\cdot\epsilon_{ERT}(p_{T})}\nonumber\\
 \label{eq:ch4.inv_cross}
\end{eqnarray}

where
\\
\begin{tabular}{ll}
\\
$N_{MB}$ &- number of scaled minimum bias events in MB
sample (Table~\ref{tab:runqatable})\\
$N_{MB\ live}$ &- number of live minimum
bias events in ERT sample (Table~\ref{tab:runqatable})\\
$\frac{N_{e}}{\Delta p_{T}}$ &- "raw" electron count in
$p_{T}$ bin\\
$\Delta y $&- rapidity range ($\pm 0.5$ units in rapidity)\\
$\epsilon_{reco}(p_{T})$ &- reconstruction and acceptance
efficiency (correction function)\\
$\epsilon_{bias}(p_{T})$ &- BBC trigger bias\\
$\sigma_{pp\ tot}$ &- total p+p inelastic crossection~\cite{ana148} $(42.2\pm1.9)$ mb\\
$\epsilon_{BBC}$ &- BBC efficiency for Minimum Bias~\cite{ana148} $(0.516\pm0.031)$\\
$\epsilon_{ERT}(p_{T})$ &- ERT trigger efficiency\\
\\
\end{tabular}

\subsection{"Raw" electron yield}

A standard procedure for any spectroscopic measurement is to start
with "raw" signal counting. First of all we selected an
appropriate $p_{T}$ binning which was chosen to match the bin
boundaries of previous Au+Au single electron measurements
~\cite{ppg035}.  Those bins that had significant statistics compared
to Au+Au were split into two. The final choice for the binning is
listed below:
\begin{itemize}
\item $\{ 0.4,\ 0.5,\ 0.6,\ 0.8,\ 1.0,\ 1.2,\ 1.4,\ 1.6,\ 2.0, \
2.5,\ 3.0,\ 4.0,\ 5.0 \} $
\end{itemize}
\pagebreak

\begin{figure}[ht]
\centering
\epsfig{figure=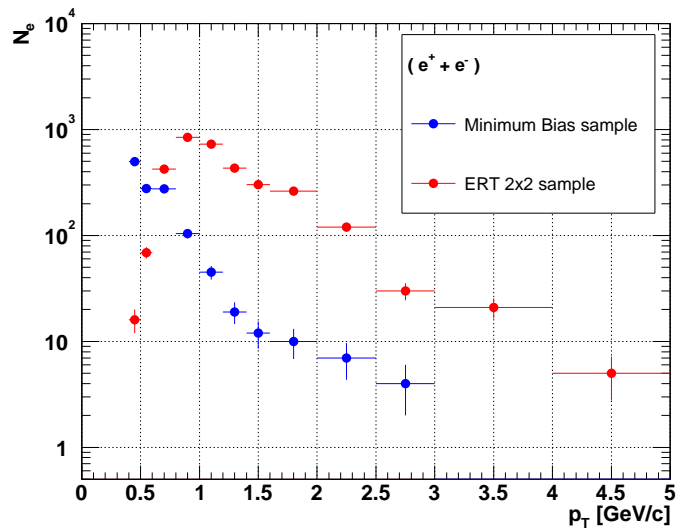,width=0.7\linewidth}
\caption{\label{fig:ch4.raw_electrons} "Raw" electron count in
$p_T$ bin for Minimum Bias and ERT trigger sample.}
\end{figure}

\begin{table}[h]
\centering \caption{"Raw" electron count in $p_T$ bin for Minimum
Bias and ERT trigger sample.}
\begin{tabular}{|c|r|r|r|r|}
\hline $p_T$ bin [GeV/c] &  $N_{e\ MB}$ & $\delta N_{e\ MB}$
&$N_{e\ ERT}$ & $\delta N_{e\ ERT}$\\
\hline
0.4-0.5& 498& 22.31& 16&  4.00\\
0.5-0.6& 278& 16.67& 69&  8.30\\
0.6-0.8& 276& 16.61& 424& 20.59\\
0.8-1.0& 104& 10.20&  846& 29.09\\
1.0-1.2& 45& 6.71&  728& 26.98\\
1.2-1.4& 19&  4.36&  433& 20.81\\
1.4-1.6& 12& 3.46&  301& 17.35\\
1.6-2.0& 10&  3.16& 263& 16.22\\
2.0-2.5& 7& 2.65& 120& 10.95\\
2.5-3.0& 4&   2.00&   30&  5.48\\
3.0-4.0& 0&   0&   21& 4.58 \\
4.0-5.0& 0&   0& 5&   2.24\\
\hline
\end{tabular}
\label{tab:raw_table}
\end{table}

Fig.~\ref{fig:ch4.raw_electrons} shows the electron statistics per
bin for the ERT and Minimum Bias trigger data samples.
Table~\ref{tab:raw_table} summarizes "raw" electron counting
results. One can see that statistics in three highest $p_{T}$ bins
is quite low which is a limiting factor for Run02 single electron
analysis.

\subsection{ERT trigger efficiency}

The ERT trigger for Run02 was calculated for the $J/\psi$ analysis in
$\pp$ Run02~\cite{jpsi,ana139} using single photons.  The photon
analysis uses only the EMC bit of the ERT trigger.  However, the RICH
trigger part the trigger does not introduce a significant efficiency
loss. We can not use single electrons for the trigger efficiency
measurement due to low statistics. The ERT efficiency calculation is
trivial and described below.

\begin{itemize}
\item Find a single photon cluster of energy $E$ in EMC with tight
identification cuts from a Minimum Bias event.

\item Check whether the ERT EMC trigger bit was set for this event
and whether $\bold {this\ particular\ photon}$ fired the trigger.

\item The ratio of ERT registered yield $dN_{r}/dE$  to the total
yield $dN_{t}/dE$ will give us ERT trigger efficiency
$\epsilon_{ERT}(E)$

\end{itemize}

Taking into account the fact that the electron momentum resolution
at low $p_{T}$ is much better then energy resolution we use the
trigger efficiency as a function of particle momentum instead of
energy $\epsilon_{ERT}(p)$.

\begin{figure}[]
\centering \epsfig{figure=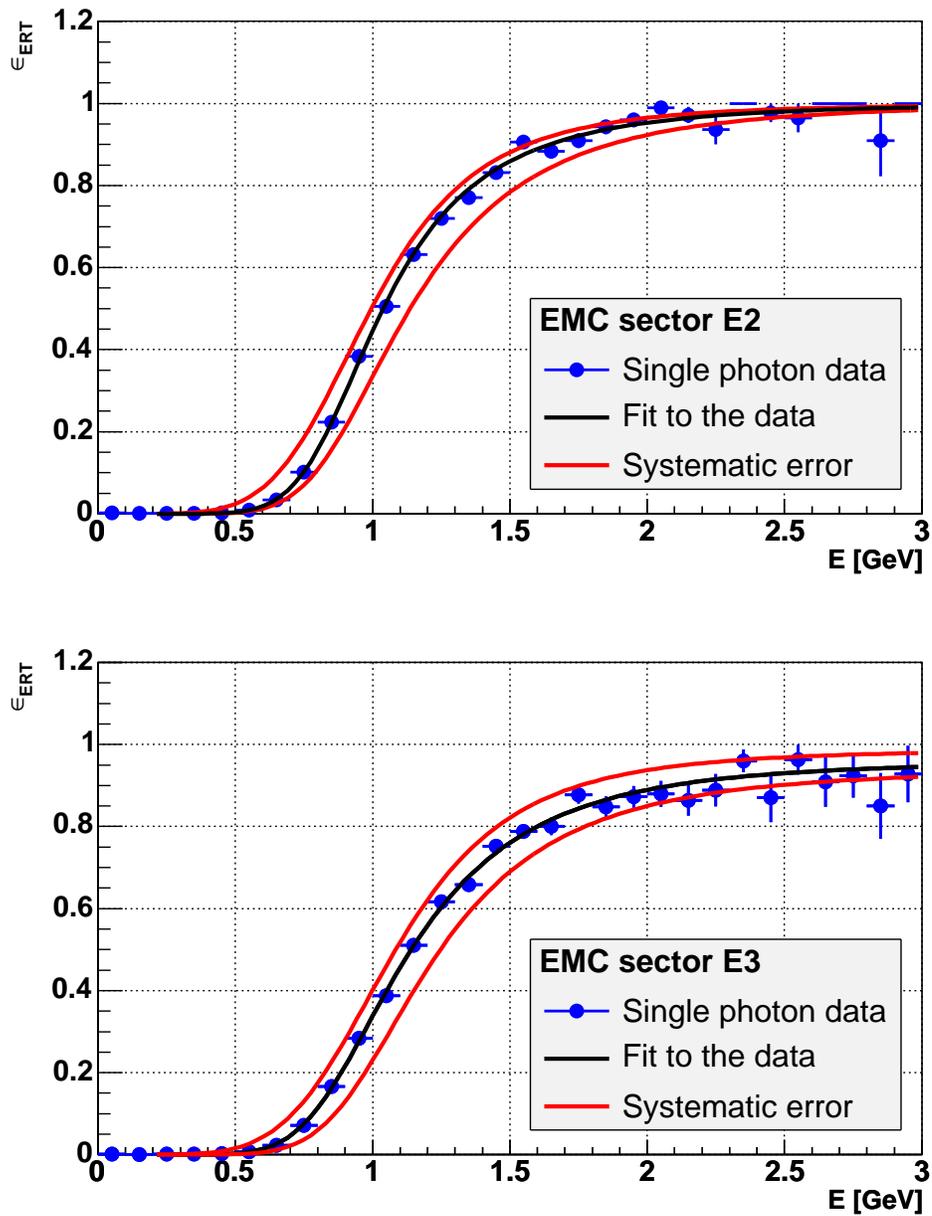,width=1\linewidth}
\caption{\label{fig:ch4.ert_eff} ERT trigger efficiency for E2, E3
EMC sector. The systematic error band to the efficiency from
trigger simulation.}
\end{figure}

The measured ERT trigger efficiency for the E2, E3 EMC sectors is
shown in Fig.~\ref{fig:ch4.ert_eff}. The systematic error shows the
maximum extent error of efficiency variation obtained by a 10\%
variation of the number of dead/noisy towers in the trigger
simulation. The trigger efficiency may be underestimated for this
analysis because we remove additional "bad" EMC towers (see
Fig.~\ref{fig:ch4.dead_emc}) as compared to the $J\Psi$ analysis, but
this difference is easily covered by the systematic error. Both
the trigger efficiency and hi-lo limits of the systematic error were
fitted with arbitrary functions, presented in
Table~\ref{tab:ert_eff_table}

\begin{table}[ht]
\centering \caption{Fit results for ERT trigger efficiency and
hi-lo systematic error band for E2,E3 EMC sectors.}
\begin{tabular}{|c|c|c|c|}
\hline  &  ERT efficiency & Systematic error (hi limit)
&Systematic error (lo limit)\\
\hline
&&&\\
E2 & $\frac{1}{1+0.771/p^{4}+0.473/p^{9}}$
&$\frac{1}{1+0.451/p^{4}+0.525/p^{6}}$ &
$\frac{1}{1+1.307/p^{4}+0.685/p^{9}}$\\
\hline
&&&\\
E3 & $\frac{0.96}{1+1.252/p^{4}+0.588/p^{9}}$
&$\frac{0.99}{1+0.681/p^{4}+0.787/p^{6}}$ &
$\frac{0.94}{1+1.649/p^{4}+1.435/p^{9}}$\\
\hline
\end{tabular}
\label{tab:ert_eff_table}
\end{table}

\newpage
\subsection{Hadronic background}\label{sec:ch4.hadr}

\begin{figure}[h]
\begin{tabular}{lr}
\begin{minipage}{0.5\linewidth}
\begin{flushleft}
\epsfig{figure=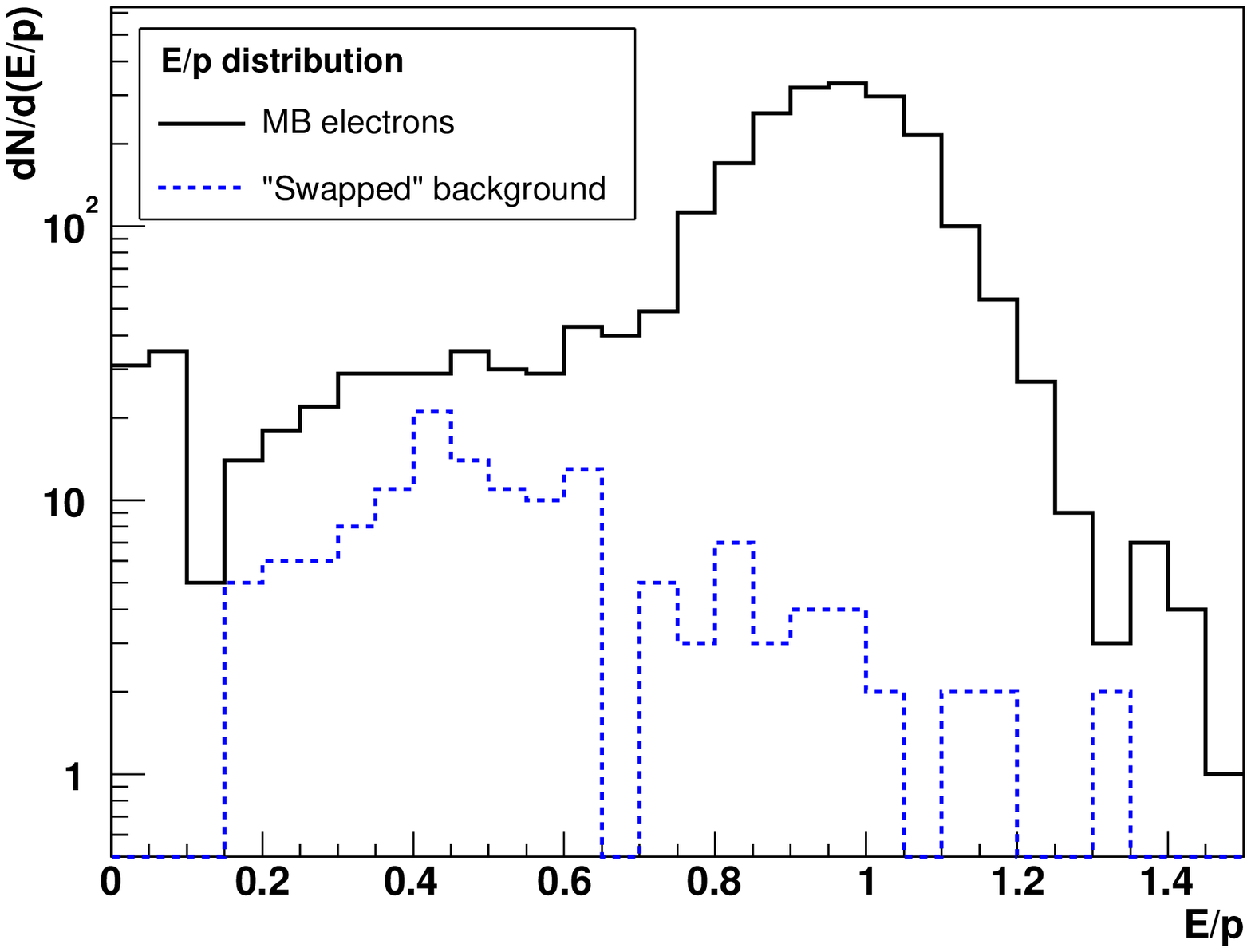,width=1\linewidth}
\caption{\label{fig:ch4.ep_sn0} $E/p$ distribution for electron
candidates (solid curve) and random association tracks $sn0
>1$ (dashed curve) for Minimum Bias events $p_T
> 0.4$ GeV/c.}
\end{flushleft}
\end{minipage}
&
\begin{minipage}{0.5\linewidth}
\begin{flushright}
\epsfig{figure=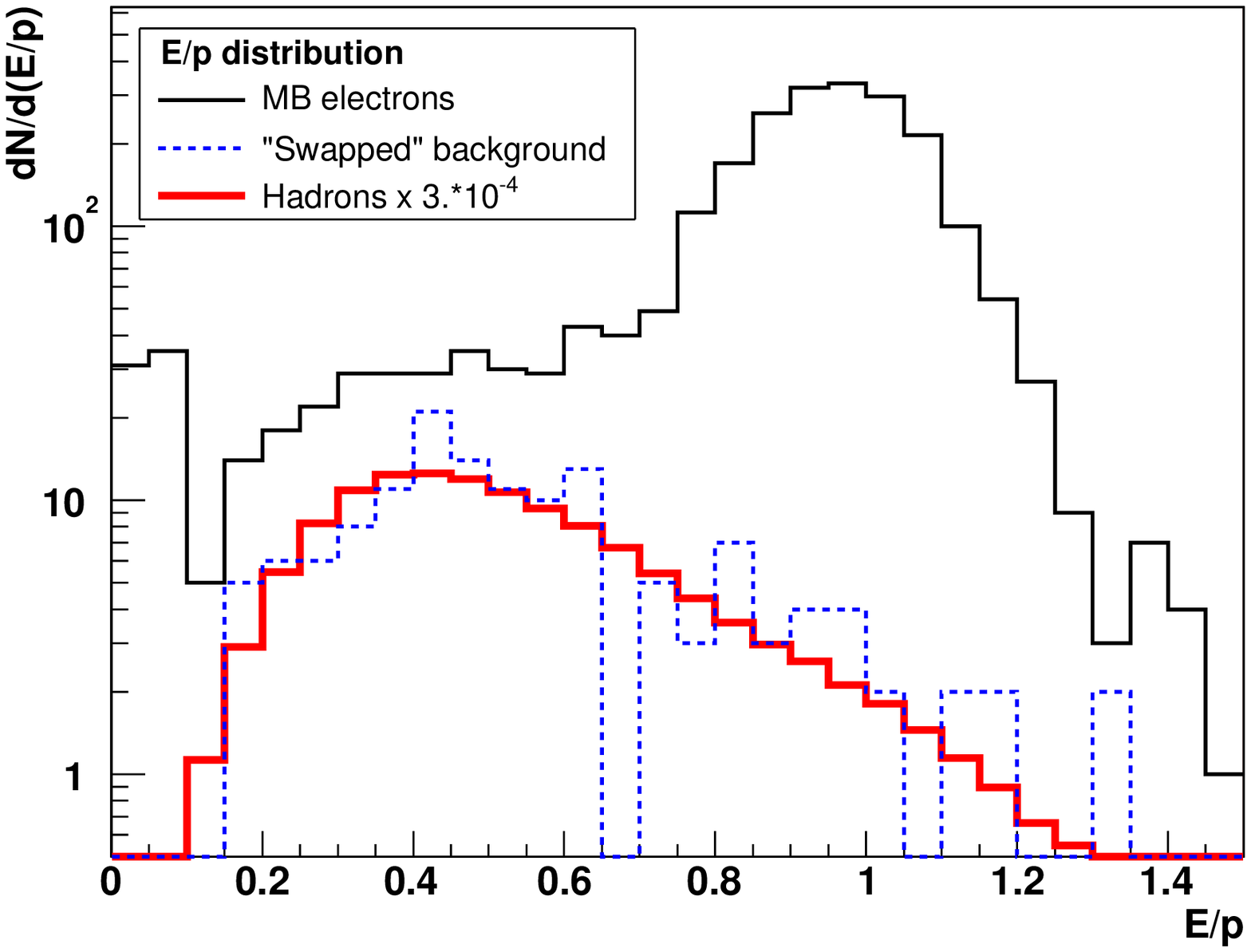,width=1\linewidth}
\caption{\label{fig:ch4.ep_hadrons} $E/p$ distribution for
electron candidates (solid curve), random association tracks
(dashed curve) and charged hadron tracks scaled by
$\epsilon_{rand} = 3\cdot10^{-4}$ (thick solid curve) for Minimum
Bias events $p_T
> 0.4$ GeV/c.}
\end{flushright}
\end{minipage}
\end{tabular}
\centering
\epsfig{figure=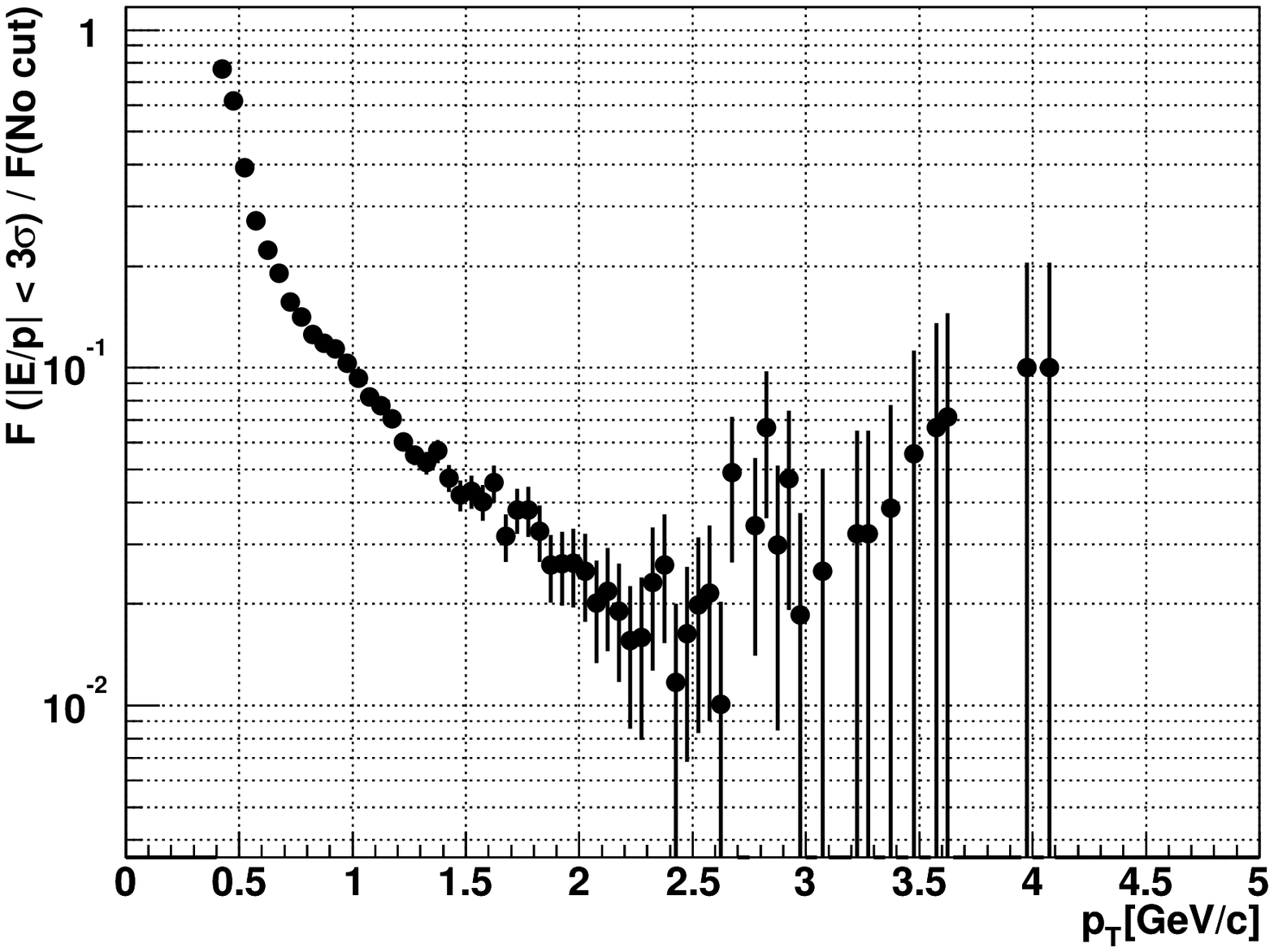,width=0.8\linewidth,clip}
\caption{\label{fig:ch4.ep_rejection} Rejection power of $|E/p|
<3\sigma$ for charged hadrons.}
\end{figure}

The major source of background for this analysis is a random
coincidence of a charged track with a RICH cluster (thereby
falsely identifying the track as an electron). This was studied
previously in great detail for the $\Au$ single electron
measurements~\cite{ppg011,ppg035} where this contribution is much
more significant. In $\pp$ collisions the multiplicity is low,
thus the chance of random coincidence is significantly reduced
(see Fig.~\ref{fig:ch4.fit_ep} for background level estimation).
The standard technique that is used in PHENIX offline software is
so called "$\it {flip\ and\ slide}$" method which is based on
creation of a fake ("swapped") charged tracks by exchanging the
North and South hits in all detectors except the drift chamber.
This way we create an unbiased random track that then is being
associated with outer PHENIX detector. The number of RICH
phototube that are associated with the "swapped" charged track
denotes as $sn0$. The distribution of $E/p$ for the $n0 > 1$ and
$sn0 > 1$ is shown in Fig.~\ref{fig:ch4.ep_sn0} for Minimum Bias
data sample. Unfortunately, we can not use ERT trigger sample for
those studies because the electron content of ERT events is
strongly biased by trigger efficiency. One can see that the
statistics of the purely random association of Minimum Bias sample
is very small and alternative way to estimate the background
contribution must to be found.

We want to make an assumption at this point that the random
association rate should not depend on the inclination angle of the
track and, thus, it is not a function of the particles momentum.
We would then find a constant probability $\epsilon_{rand}$ that
the charged track is associated with a RICH ring. We can plot
the $E/p$ distribution for $\bold {charged\ tracks}$ and normalize
it to the $sn0 >1$ $E/p$ distribution at low $p_{T}$. The
normalization constant will be $\epsilon_{rand}$ by construction.
Fig.~\ref{fig:ch4.ep_hadrons} shows the $E/p$ distribution for
$n0>1$ , $sn0 >1$ , and charged hadrons scaled by $\epsilon_{rand}
= (3\pm1.5(sys))\cdot10^{-4}$ . In order to account for a
qualitative comparison in the normalization, we put a large (50\%)
systematic error on this value.

This probability does not include the effect of the $\pm 3 \sigma\
E/p$ cut rejection. It adds an additional suppression of the
random hadron association component by a factor $\approx 10-100$.
Fig.~\ref{fig:ch4.ep_rejection} shows the probability that
randomly associated charged track "survives" the $E/p$ cut.

\newpage

\subsection{$\delta$-electron background}

$\delta$-electrons (also called "knock-on" electrons or $\delta$-rays)
refer to energetic electrons, that were knocked from the atomic shell
of some atom in the detector volume. Depending on the construction of
the detector and particle identification principles "knock-on"
electrons may create a significant background.

In the case of PHENIX we need to take into account the rate of
electrons emitted in RICH gas volume. The difference with previous
effect is that delta electron, emitted with a reasonably small
angle with respect to the initial hadron can create a hit in RICH
and be misidentified as an electron. Thus, we need to estimate the
probability for a hadron to emit delta electron that may fire RICH
detector. To estimate the yield of $\delta$-rays in RICH volume
\linebreak (100 cm of $CO_2$) we used the formula
(~\ref{eq:ch4.delta_rate},
~\ref{eq:ch4.tmax})~\cite{PDG}\footnote{In calculations below we
assume the validity of Rutherford crossection and \\spin-0
incident pion}.

\begin{equation}
    \frac{d^2 N}{dTdx} =
    \frac{1}{2}Kz^2\frac{Z}{A}\frac{1}{\beta^2}\frac{(1-\beta^2T/T_{max})}{T^2}\\
 \label{eq:ch4.delta_rate}
\end{equation}

\begin{equation}
    T_{max} =
    \frac{2m_{e}c^{2}\beta^2}{1+2\gamma m_{e}/M+(m_{e}/M)^2}\\
 \label{eq:ch4.tmax}
\end{equation}

where
\\
\begin{tabular}{ll}
\\
$K$&- $4\pi N_{A}r_{e}^{2}m_{e}c^2 = 0.307075\ [MeV\,cm^2]$\\
$Z$ &- atomic number of absorber\\
$A$ &- atomic mass of absorber $[g\, mol^{-1}]$\\
$z$ &- charge of the incident particle\\
$T$&- kinetic energy of the electron\\
$\gamma$&- $\gamma$ the incident particle\\
$\beta$&- $\beta$ the incident particle\\
$M$&- mass of the incident particle\\
$T_{max}$&- maximal possible kinetic energy of the
$\delta$-electron
(Eq.~\ref{eq:ch4.tmax})\\
\\
\end{tabular}

\begin{figure}[ht]
\centering
\epsfig{figure=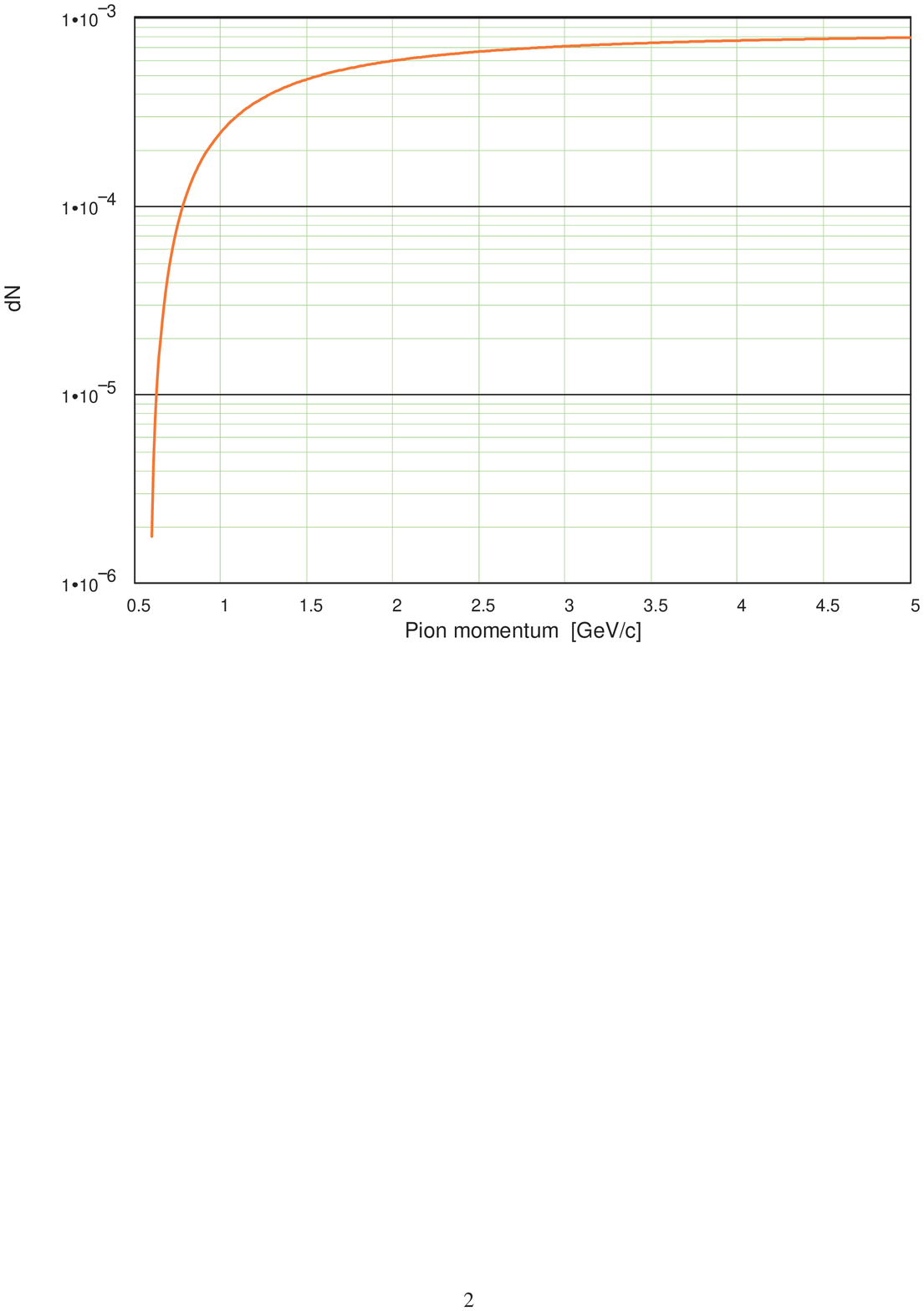,width=0.7\linewidth,clip}
\caption{\label{fig:ch4.Init_delta} Total $\delta$-electron rate
as a function of incident pion momentum.}
\end{figure}

RICH threshold for the electron is  $\gamma_{thr} = 35$. Thus,
minimal energy of electron that can "fire" RICH is $E_{min} =
\gamma_{thr} \cdot m_e c^2 \approx 17.9$ MeV. Integrating
Eq.~\ref{eq:ch4.delta_rate} on T from $E_{min}$ to $T_{max}$ we
obtain the total yield of $\delta$-electrons as the function of
pion momentum (Fig.~\ref{fig:ch4.Init_delta}).

The angle of the "knock-on" electron with respect to the incident
pion can be calculated by the following formula~\cite{PDG}:

\begin{equation}
    \cos(\theta) = (T_{e}/p_{e})(p_{max}/T_{max})\\
 \label{eq:ch4.cos_theta}
\end{equation}

where
\\
\begin{tabular}{ll}
$T_{e},p_{e}$&- kinetic energy and momentum of the electron\\
$T_{max},p_{max}$&- maximum available kinetic energy and momentum of the electron\\
\\
\end{tabular}

\begin{figure}[ht]
\centering
\epsfig{figure=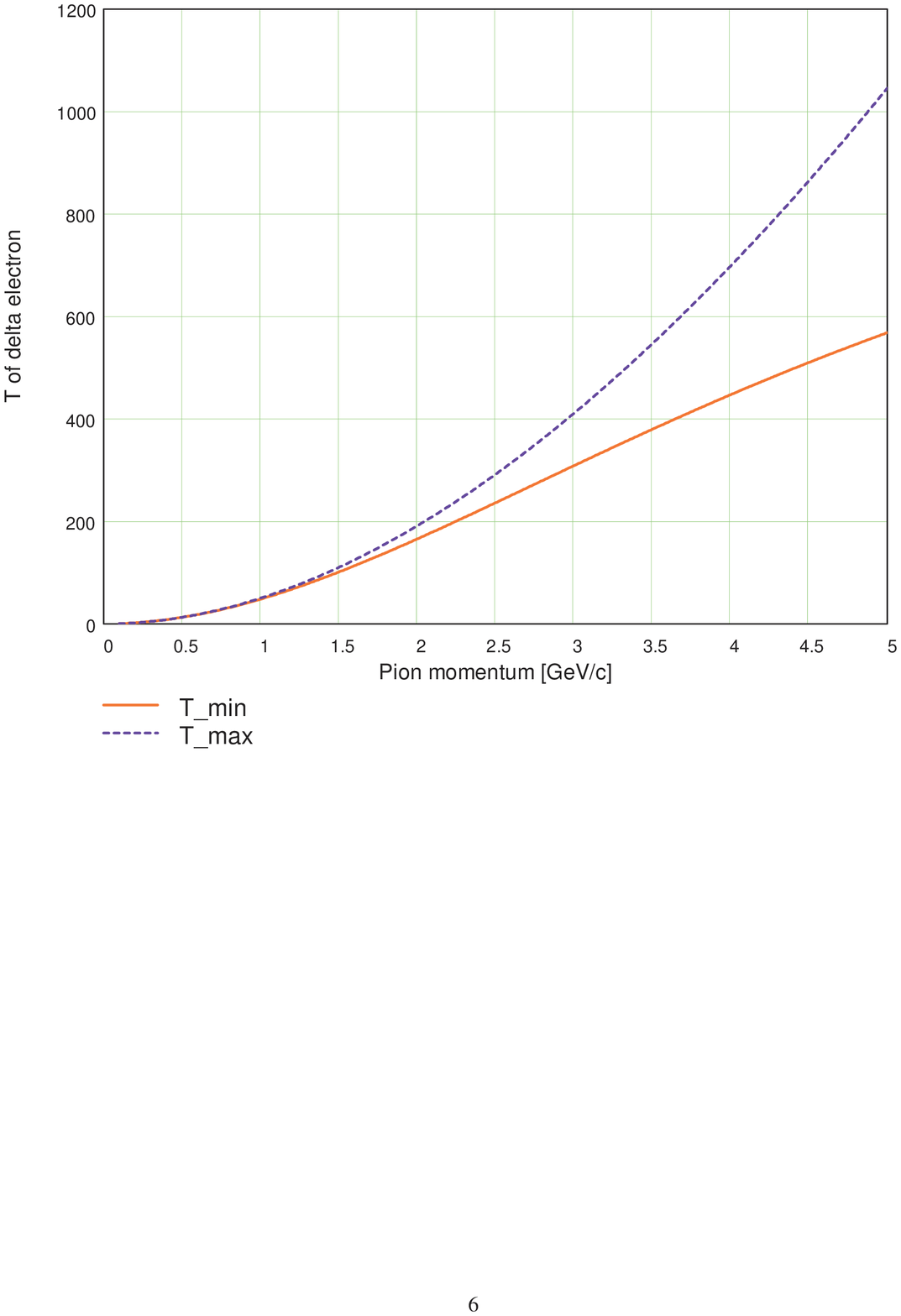,width=0.78\linewidth,clip}
\caption{\label{fig:ch4.T_max_min} The range of $\delta$-electron
kinetic energy that can be reconstructed by RICH, $T_{min}$
(solid) and $T_{max}$ (dashed).} \vspace*{-0.05in}
\epsfig{figure=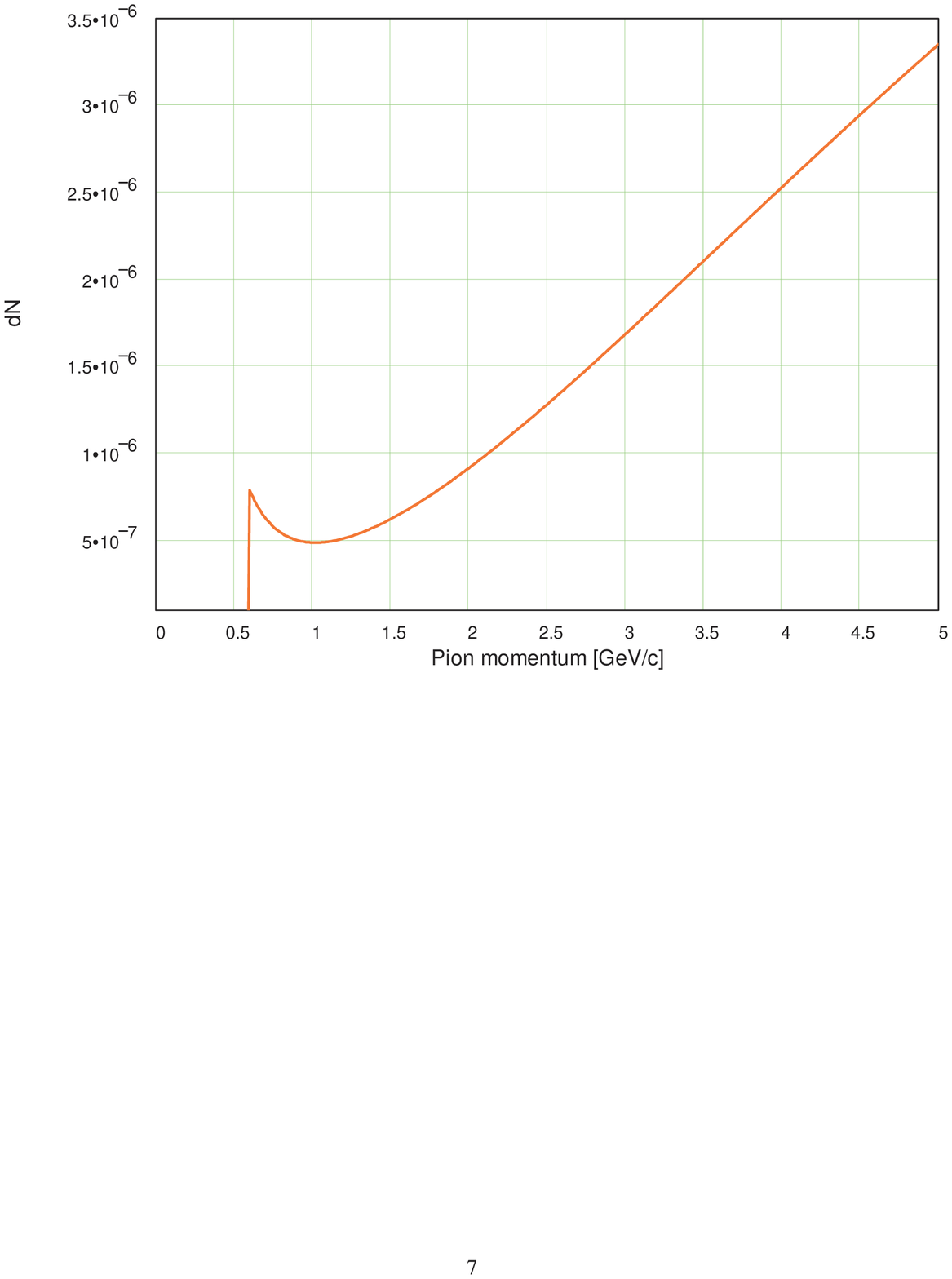,width=0.78\linewidth,clip}
\caption{\label{fig:ch4.rate_delta_final} Rate of
$\delta$-electrons reconstructible by RICH as a function of
incident pion momentum.}
\end{figure}

Due to the RICH's geometrical acceptance, only when the $\delta$-ray is
produced within $\cos(\theta_{max}) > 1/n$ with respect to the
pion direction will its Cerenkov radiation overlap in the RICH
"ring" constructed around the pion projection point. The
refraction index for $CO_{2}$ gas $n = 1 + 410\cdot10^{-6}$ which
gives the value for the maximum angle $\theta_{max} = 28.63$ mrad.
From Eq.~\ref{eq:ch4.cos_theta} one can calculate the minimal
kinetic energy $T_{min}$ of $\delta$-electron which is deflected
to an angle $\theta = \theta_{max}$. Fig.~\ref{fig:ch4.T_max_min}
shows the range of kinetic energies for electrons that are emitted
in $\theta<\theta_{max}$ cone around the direction of incident
pion.

Now we can obtain the yield of $\delta$-electrons that can be
reconstructed in the RICH by integrating Eq.~\ref{eq:ch4.delta_rate}
from $T_{min}$ to $T_{max}$. $\delta$-electrons rate as a function
of incident pion momentum is shown in
Fig.~\ref{fig:ch4.rate_delta_final}. One can see that we have a
probability of $\epsilon_{\delta}\approx 10^{-6}$ for a pion to
create such an electron (the rate is even smaller for incident
particle of higher mass). The yield is significantly lower then
previously calculated random association background of
$3\cdot10^{-4}$. Thus, $\delta$-electron contribution can be
neglected in PHENIX electron analysis.
\newpage

\subsection{Acceptance Correction function}\label{sec:ch4.cf}

In order to properly normalize the results of the measurements, we
need to calculate the factor which describes the difference
between an ideal $4\pi$ detector and PHENIX. This factor (see
Eq.~\ref{eq:ch4.inv_cross}) is called the correction function
$\epsilon_{reco}(p_{T})$ and takes into account the limited
acceptance and track reconstruction efficiency.

The standard way to obtain the correction function is through the
full simulation of particle detection probability assuming an ideally
distributed input particle density. In the case of single electron
analysis we "throw" single electrons, generated by the
EXODUS~\cite{EXODUS} event generator, with the following input
parameters
\begin{itemize}
\item Uniform azimuthal angle distribution $0<\phi<2\pi$
\item
Uniform vertex $Z_{vtx}$ distribution\footnote{There is no strong
dependence of reconstruction efficiency on $Z_{vtx}$. This fact allows
us to use a "flat" vertex distribution instead of realistic "Gaussian"
distribution (shown in Fig.~\ref{fig:ch4.bbcz})} $|Z_{vtx}|<25$
cm
\item Uniform rapidity distribution $-0.6 <y< 0.6$ units
\item Uniform $p_{T}$ distribution $0.0 <p_{T}< 5.0$ GeV/c
\end{itemize}

The total statistics of our simulation sample was $3.98\cdot10^6$
single particles ($2.00\cdot10^6$ positrons and $1.98\cdot10^6$
electrons).

The particles pass through the full detector simulation chain
called PISA ($\it{PHENIX\ Integrated\ Simulation\
Application}$~\cite{PISA}). PISA is a GEANT-3 based simulation
code that has been successfully used since 1992 to simulate
realistic particle propagation and detector response. The particle
is "swimmed" through the tabulated Magnetic Field and GEANT-3
simulates the interaction of the primary particle with the
material inside the PHENIX aperture. Both primary and secondary
particles create a $\bold{hit}$ every time they enter the active
area of the detector. This "Monte Carlo hit" information
is stored in the $\it{"PISA\ output\ file"}$.

The next step in simulation is applying a realistic detector
response to the MC hits. This procedure includes a smearing of hit
position and timing with appropriate resolution, digitization of
the timing and analog information, hit merging, applying
time-of-flight effects, and reproduction of registration efficiency
\& dead map for each detector subsystem, e.t.c.

The final step of the simulation is the $\it{reconstruction}$ of the
simulated data using standard PHENIX offline code which is used
for real data analysis.

As an output we have a collection of reconstructed tracks. The offline
software ($\it{evaluation\ package}$) maintains the relationship
between MC track and reconstructed track. A "main contributor" scheme
is used in that the MC track that provided most hits to any
reconstructed track is then considered to be the $\bold{"main\
contributor"}$ to that track and is considered as the source of that
reconstructed track. This helps us to obtain a direct correspondence
between the reconstructed track parameters and the input track parameters.

\begin{figure}[h]
\centering
\begin{tabular}{ll}
\epsfig{figure=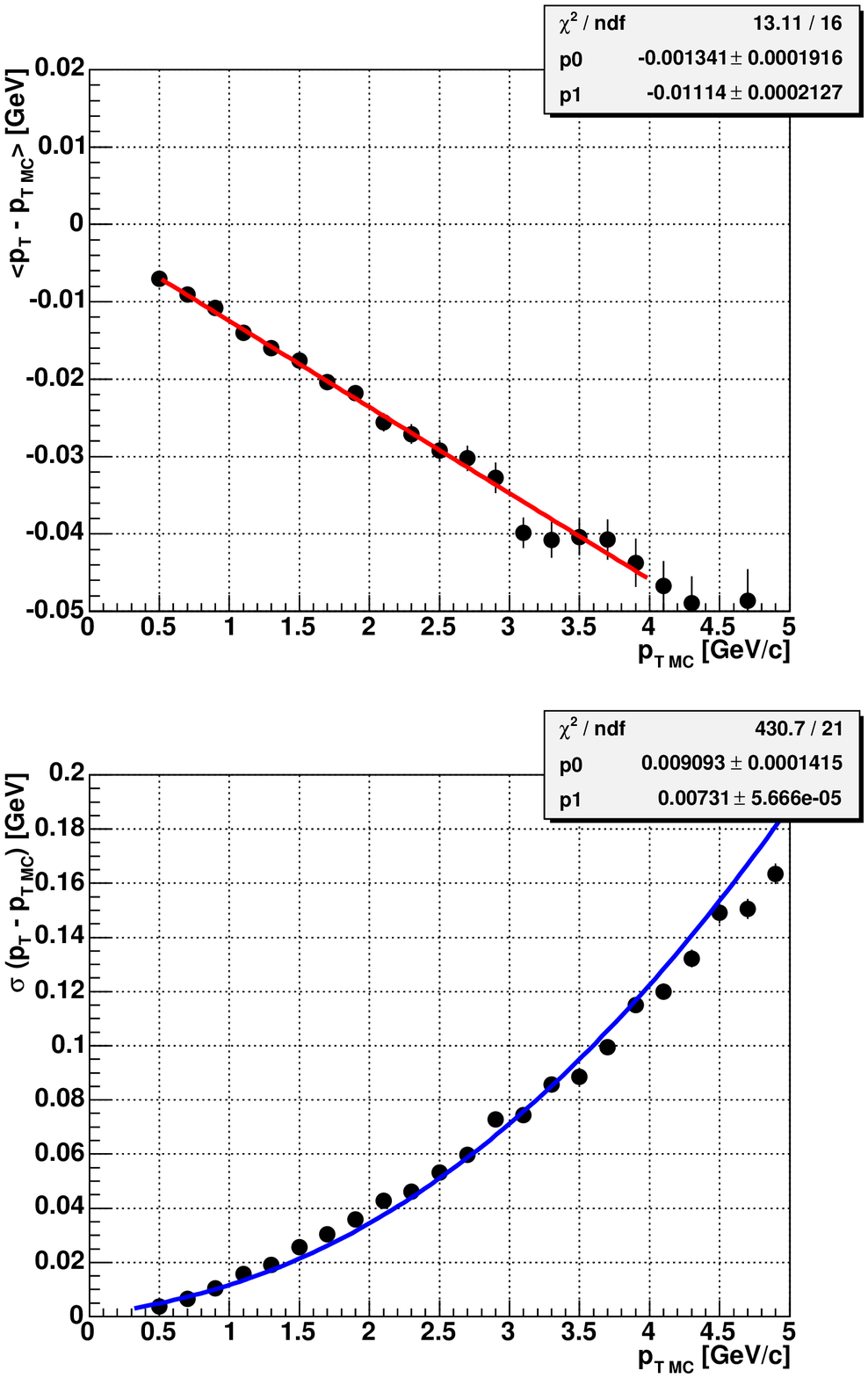,width=0.42\linewidth,clip}
\epsfig{figure=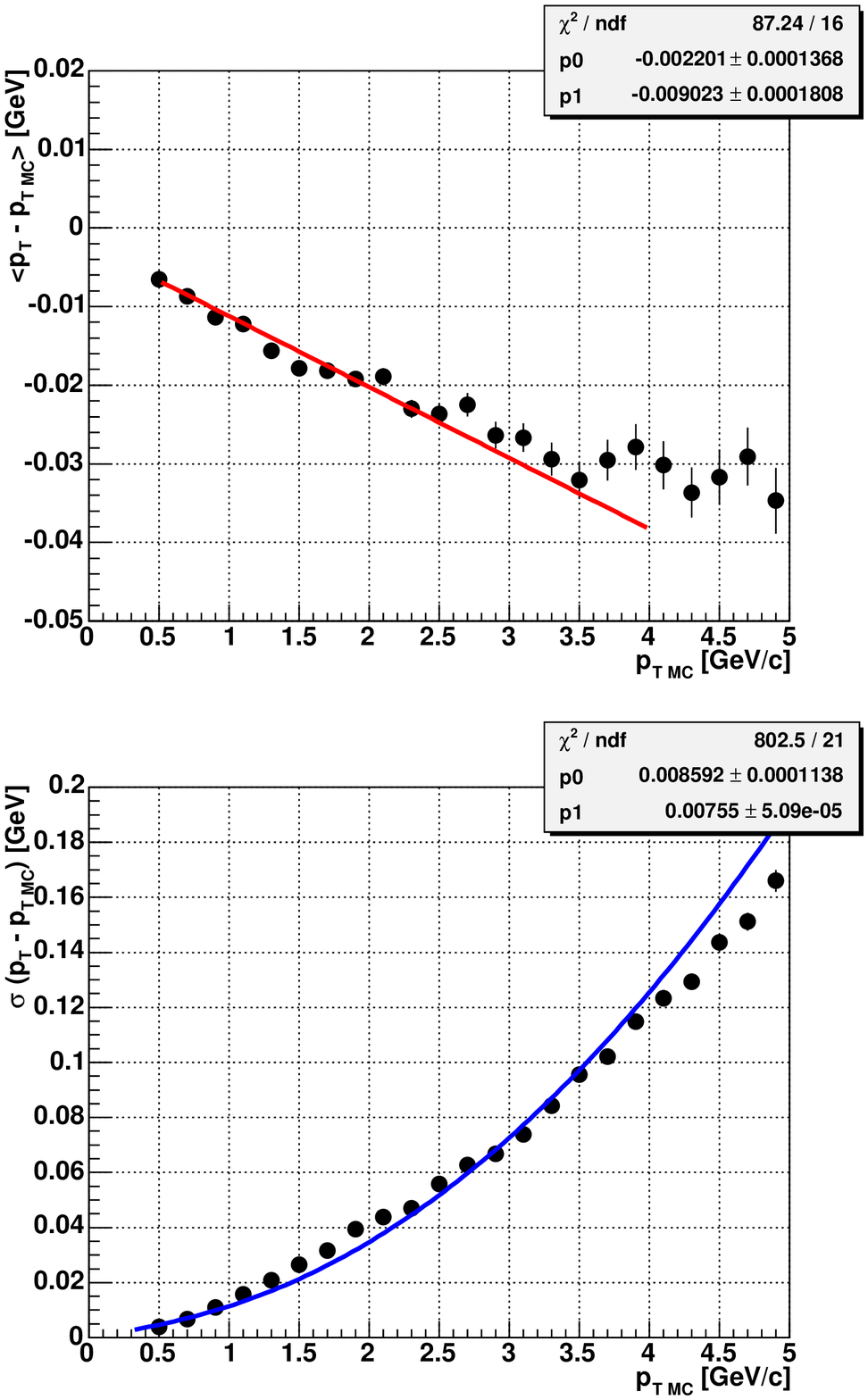,width=0.42\linewidth,clip}
\end{tabular}
\caption{\label{fig:ch4.mom_res_sim} Mean and sigma of difference
between reconstructed and ideal $p_{T}$ in Simulation for
electrons (left) and positrons (right). Linear fit is shown for
mean distribution, sigma is fitted with Eq.~\ref{eq:ch4.resol}
functional form.}
\end{figure}

The first thing that needs to be checked in simulation is the momentum
and energy resolution.  Fig.~\ref{fig:ch4.mom_res_sim} shows the mean
and $\sigma$ of the difference between reconstructed $p_{T}$ and
initial $p_{T\ MC}$ transversal momentum as a function of $p_{T\ MC}$
for electrons and positrons. One can see that there is a linear
dependence of the $\delta(p_{T})$ which needs to be removed from the
simulation (we can not justify that this effect should exist in
simulation and need to remove it and later treat it as a systematic
error).

The momentum resolution can be estimated from the fit to
$\sigma(p_{T}-p_{T\ MC})$ using functional form for the momentum
resolution (Eq.~\ref{eq:ch4.resol}). The Drift Chamber momentum
resolution term in simulation is factor of two smaller then in the
data:

\begin{itemize}
\item $\sigma_{MS} = (0.87 \pm 0.05)\%$

\item $\sigma_{DCH} = (0.74 \pm 0.02)\%$
\end{itemize}

this means that we need to artificially worsen the momentum
resolution in simulation and study what effect it may cause on the
correction function (see Section~\ref{sec:ch4.Systematics}).

\begin{figure}[]
\centering
\epsfig{figure=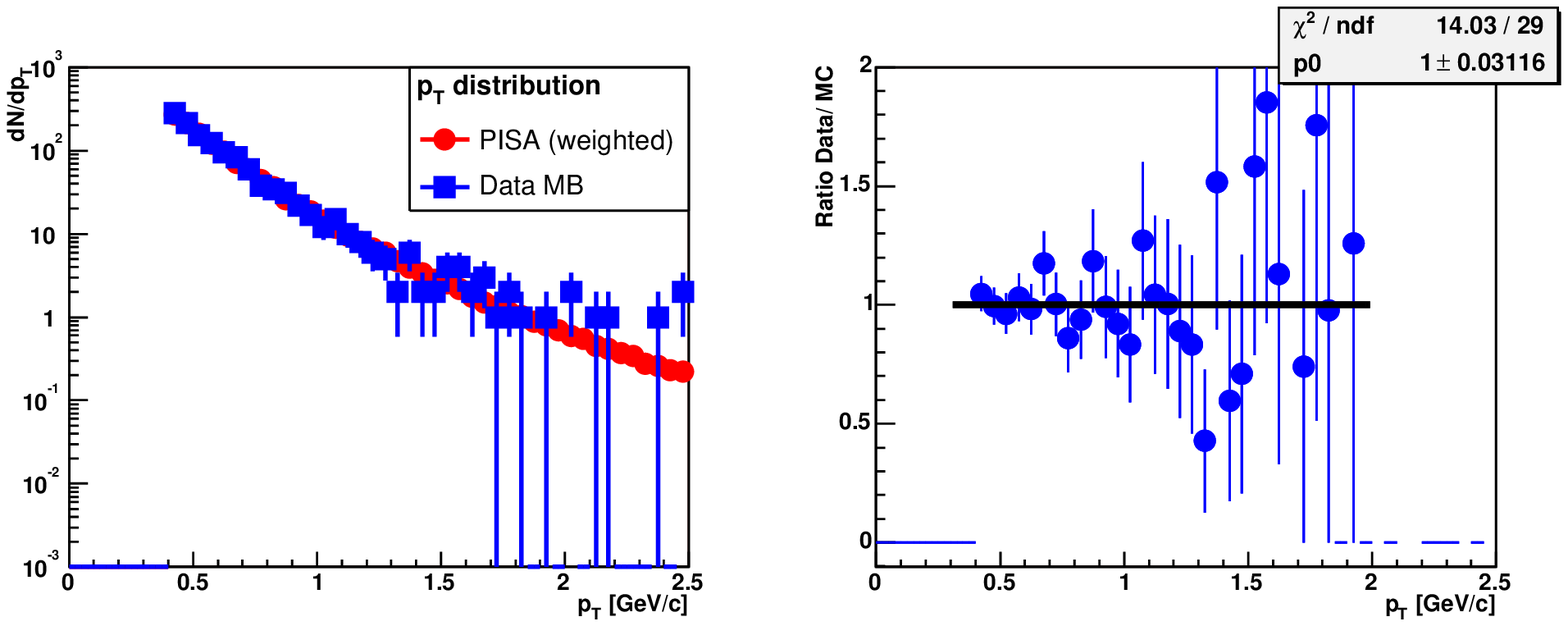,width=0.9\linewidth,clip}
\caption{\label{fig:ch4.dNdpt_sim} Comparison of
$\frac{dN_{e}}{dp_{T}}$ distribution for weighted PISA simulation
(circles) and MB data (squares). Ratio of $\frac{dN_{e}}{dp_{T}}$
in MB data to simulation (right).}
\epsfig{figure=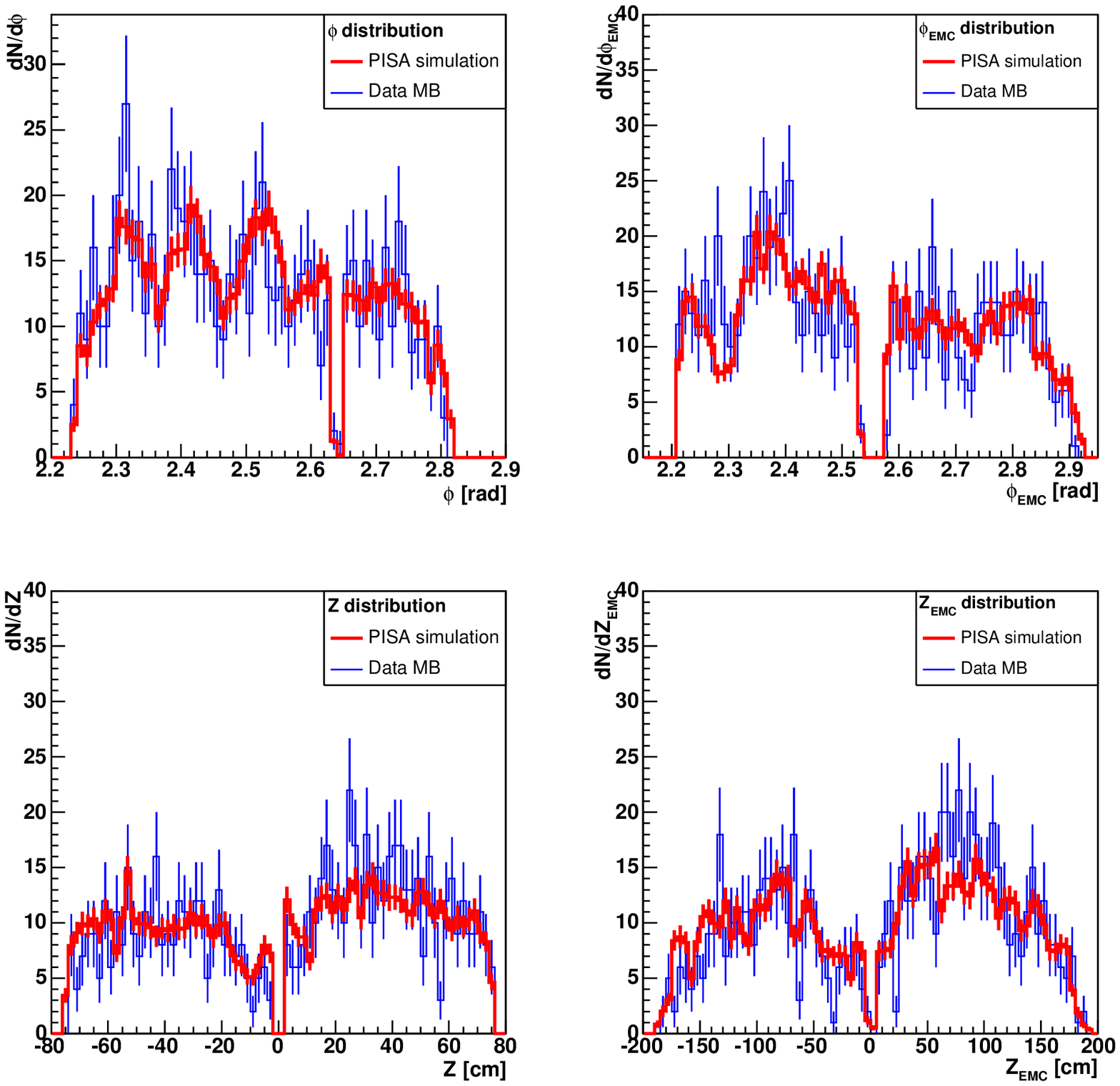,width=0.9\linewidth,clip}
\caption{\label{fig:ch4.comp_acc_mb} Comparison of acceptance in
$\phi$, $\phi_{EMC}$, $Z$, $Z_{EMC}$ for MB data (thin line) and
weighted PISA simulation (thick line) for $0.5 <p_{T} <2.0$
GeV/c.}
\end{figure}

The momentum distribution of input particles is uniform and different
from the $\frac{dN_{e}}{dp_{T}}$ of real data. In order to take into
account this difference, a weighting factor of
$(\frac{dN_{e}}{dp_{T}})_{Data}$ dependent upon $p_{T\ MC}$ is
applied to each Monte Carlo variable. This weighting
procedure "artificially" adjusts the shape of the input MC momentum
distribution to match the final Data momentum shape. The shape of
the final inclusive electron distribution can be taken from
Fig.~\ref{fig:ch4.final_inclusive} and we can assume the weighting
function to be:
\begin{equation}
    w(p_{T\ MC})=\left(\frac{dN_{e}}{dp_{T}}\right)_{Data} = \frac{p_{T\ MC}}{(p_{T\ MC}+0.406)^{7.249}}
 \label{eq:ch4.weight_mc}
\end{equation}
The absolute scale is taken arbitrary in this formula as we are only
interested in the shape of the input spectrum.
Fig.~\ref{fig:ch4.dNdpt_sim} shows the comparison of the reconstructed
tracks in Simulation weighted with $w(p_{T\ MC})$ and reconstructed
Minimum Bias (not final!)  One can see that the shape of reconstructed
tracks in the data agrees very well with Monte Carlo.

\begin{figure}[]
\centering
\epsfig{figure=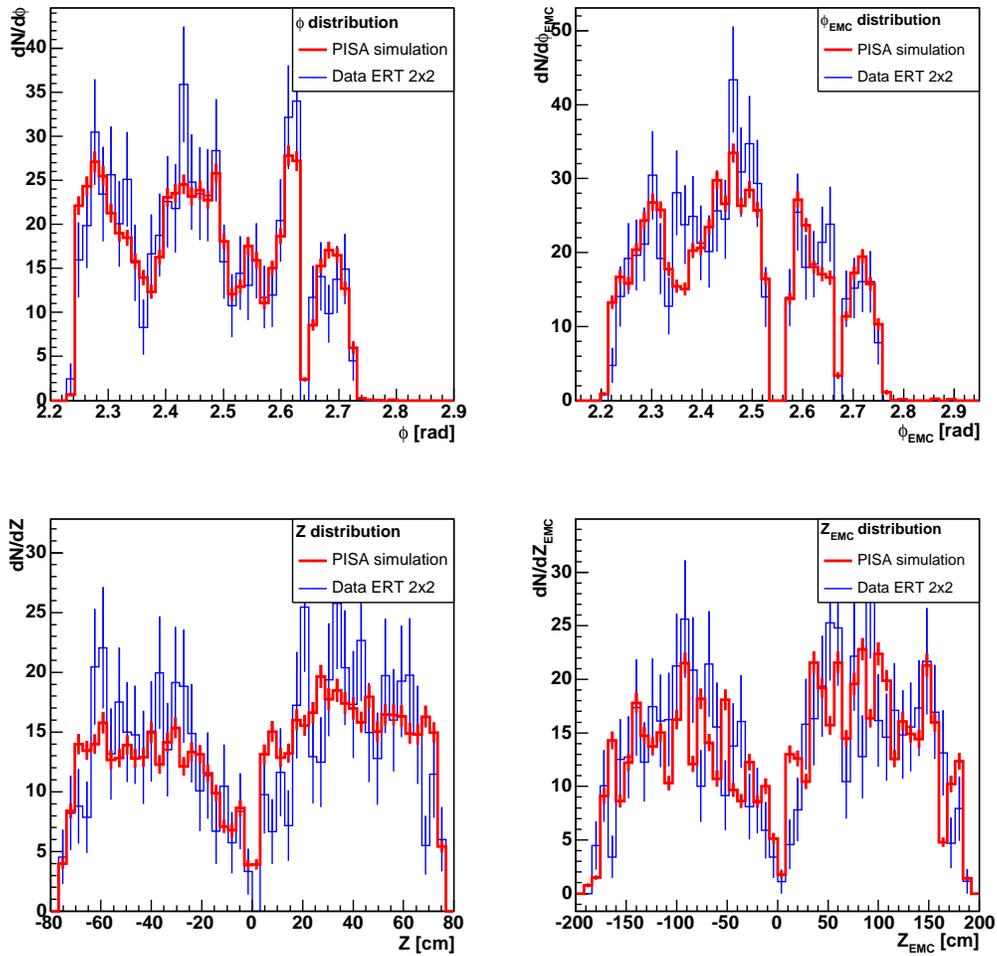,width=1\linewidth,clip}
\caption{\label{fig:ch4.comp_acc_ert} Comparison of acceptance in
$\phi$, $\phi_{EMC}$, $Z$, $Z_{EMC}$ for ERT data (thin line) and
weighted PISA simulation (thick line) for $1.5 <p_{T} <5.0$
GeV/c.}
\end{figure}

Matching and eID cut parameters of the Simulations were adjusted
in the same way as was done for the data. A comparison of the
acceptance in the simulation and real data after applying the full
eID cuts is shown in Fig.~\ref{fig:ch4.comp_acc_mb} for MB data
sample ($0.5 <p_{T} <2.0$ GeV/c) and in
Fig.~\ref{fig:ch4.comp_acc_ert} for ERT data sample scaled by ERT
trigger efficiency ($1.5 <p_{T} <5.0$ GeV/c).  The acceptance
agrees well in all projections ($\phi$, $\phi_{EMC}$, $Z$,
$Z_{EMC}$).

\newpage

Now we have everything to calculate the correction function for
simulation. By definition of the correction function:

\begin{equation}
    \epsilon_{reco}(p_{T\ Reco})=\frac{\frac{dN}{dp_{T}}_{Reco}\cdot w(p_{T\
    Output})}{\frac{dN}{dp_{T}}_{Input}\cdot w(p_{T\
    Input})}
 \label{eq:ch4.weight_mc_2}
\end{equation}

where the ratio means the ratio of the histograms (for a given
$p_{T}$ bin) filled with $\frac{dN}{dp_{T}}_{Output}\cdot w(p_{T\
Output})$ and $\frac{dN}{dp_{T}}_{Reco}\cdot w(p_{T\ Input})$
correspondingly. $p_{T\ Input}$ denotes the transverse momentum of
the input EXODUS particle, $p_{T\ Output}$ is the transverse
momentum of the "main contributor" PISA track associated to the
reconstructed track with $p_{T\ Reco}$. This method of correction
function calculation treats the weighting of the input and output
distributions correctly.

A rapidity cut of $|y| <0.5$ is applied to the input tracks in order
to normalize the correction function to one unit of rapidity.

\begin{figure}[hb]
\centering
\epsfig{figure=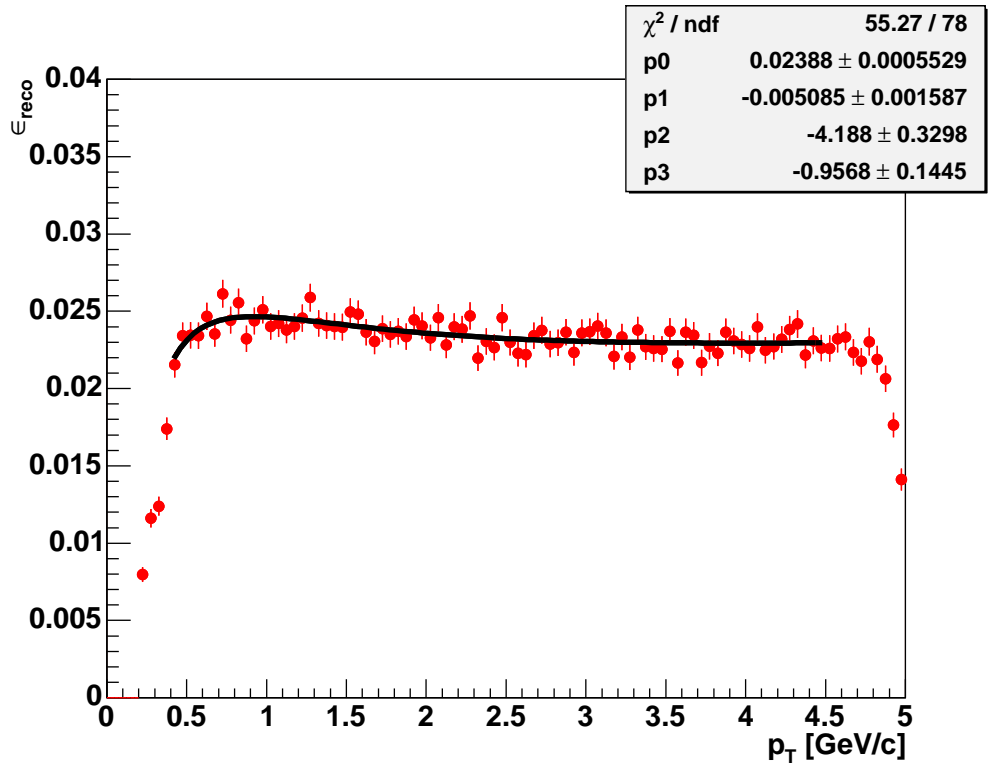,width=0.95\linewidth,clip}
\caption{\label{fig:ch4.corr_function} Correction function
$\epsilon_{reco}(p_{T})$ for $e^{+}+e^{-}$ (full electron ID
cuts).}
\end{figure}

The correction function for the $e^{+}+e^{-}$ simulation is shown
on Fig.~\ref{fig:ch4.corr_function}. The points are fitted in a
range $0.4 <p_T<4.5$ GeV/c with a functional form that well-describes the
shape.

The correction function indicates that we register $\approx 2.5\%$
of the simulated particles in our acceptance almost independent on
$p_{T}$. The apparent drop of $\epsilon_{reco}(p_{T})$ at
$p_T>4.8$ GeV/c is non-physical and is caused by the high momentum cut-off
$p_{T\ Input} < 5.0$ GeV/c in the simulated particle sample.

\begin{figure}[hb]
\centering
\epsfig{figure=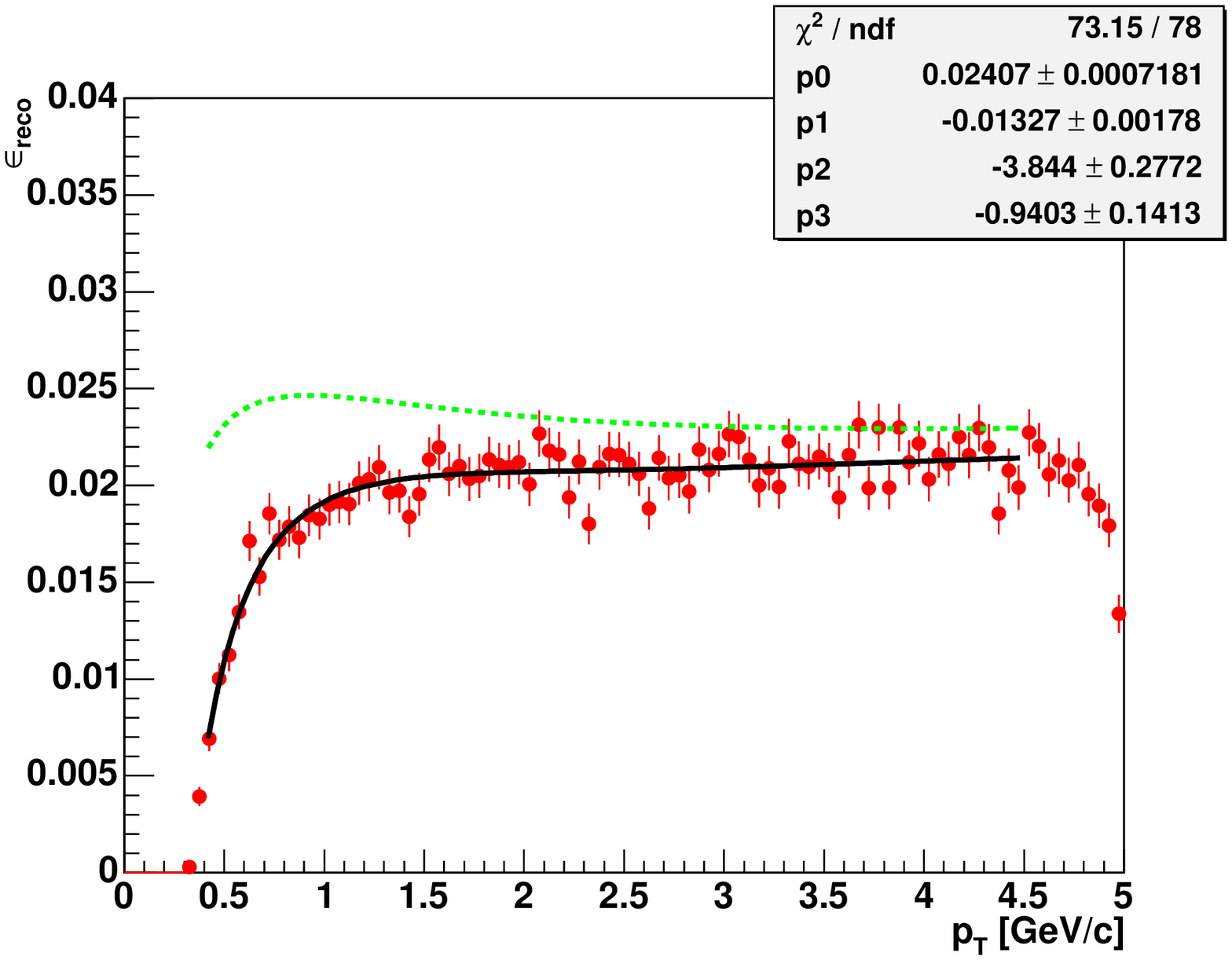,width=0.55\linewidth,clip}
\caption{\label{fig:ch4.corr_function_electrons} Correction
function $\epsilon_{reco}(p_{T})$ for $e^{-}$. Total correction
function (dashed curve) shown for comparison.}
\epsfig{figure=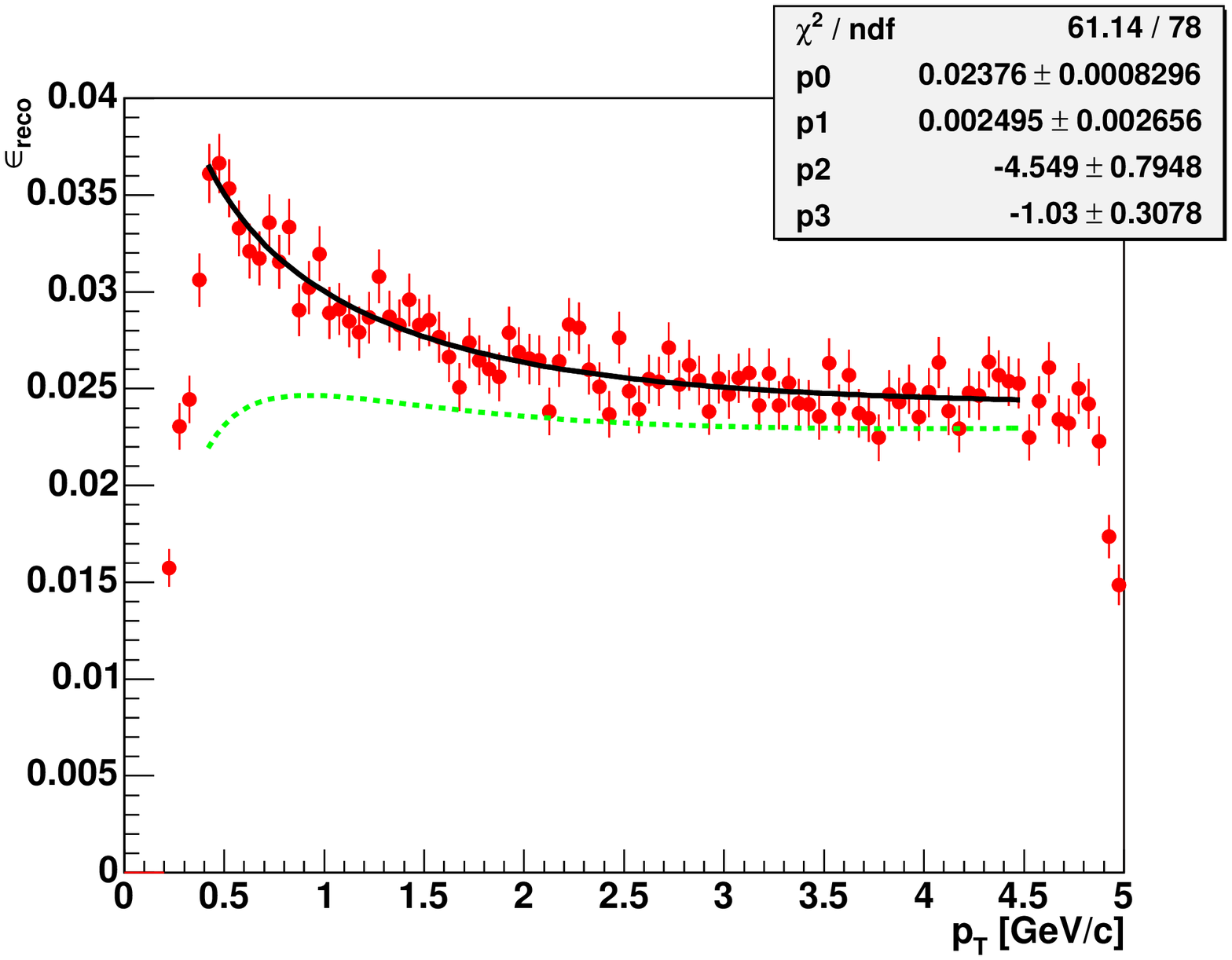,width=0.55\linewidth,clip}
\caption{\label{fig:ch4.corr_function_positrons} Correction
function $\epsilon_{reco}(p_{T})$ for $e^{+}$. Total correction
function (dashed curve) shown for comparison.}
\end{figure}

Correction functions for electrons and positrons separately are
shown in
Fig.~\ref{fig:ch4.corr_function_electrons}~\ref{fig:ch4.corr_function_positrons}.
The shape of the correction functions for different charges is
\linebreak different because of highly asymmetric acceptance of
the TZR counter "shadow" cut (see Fig.~\ref{fig:ch4.alpha_phi}).

\newpage

\subsection{BBC Trigger Bias}\label{sec:ch4.trig_bias}

The remaining unknown in Eq.~\ref{eq:ch4.inv_cross} is the $\bold{BBC\
trigger\ bias}$ $\epsilon_{bias}(p_{T})$. ``BBC trigger bias'' is
PHENIX-specific term referring to the probability at which the BBC
counter issues a Level 1 trigger decision for an event containing
specific particle of interest. The overall BBC efficiency describes
the fraction of the total $p+p$ crossection registered by the BBC and
was measured by Vernier scan to be $\epsilon_{BBC} =(0.516\pm0.031)$
~\cite{ana148}. It is obvious that events with a hard parton
scattering are more likely to be registered because the track
multiplicity in the BBC is higher for these events.  As an example,
soft partonic scattering or worse still single- or double-diffractive
scattering produce far fewer tracks in the BBC and are more likely to
fail in generating a trigger. This means that of all events that
contain a hard scattering process, the fraction recorded will be
higher than the ``inclusive'' BBC trigger cross section.  The fact
that the trigger cross section depends upon the physics process is
what we term ``Bias''.

$\epsilon_{bias}(p_{T})$ was calculated ~\cite{pp_pi0} for $\pi^0$
production using the following technique:

\begin{itemize}
\item An unbiased sample of events was selected to be ERT 4x4 trigger
with no BBC requirements.

\item $\pi^0$ was reconstructed through the $\pi^0 \rightarrow \gamma
+ \gamma$ decay in the EMC using the formula $M_{\gamma \gamma} = 4\cdot
E_{1}\cdot E_{2}\cdot sin^{2}(\theta_{\gamma \gamma}/2)$.

\item The BBC trigger bias was calculated as a ratio of events with
BBC vertex information reconstructed to the total number of ERT
events.

\end{itemize}

\begin{figure}[ht]

\epsfig{figure=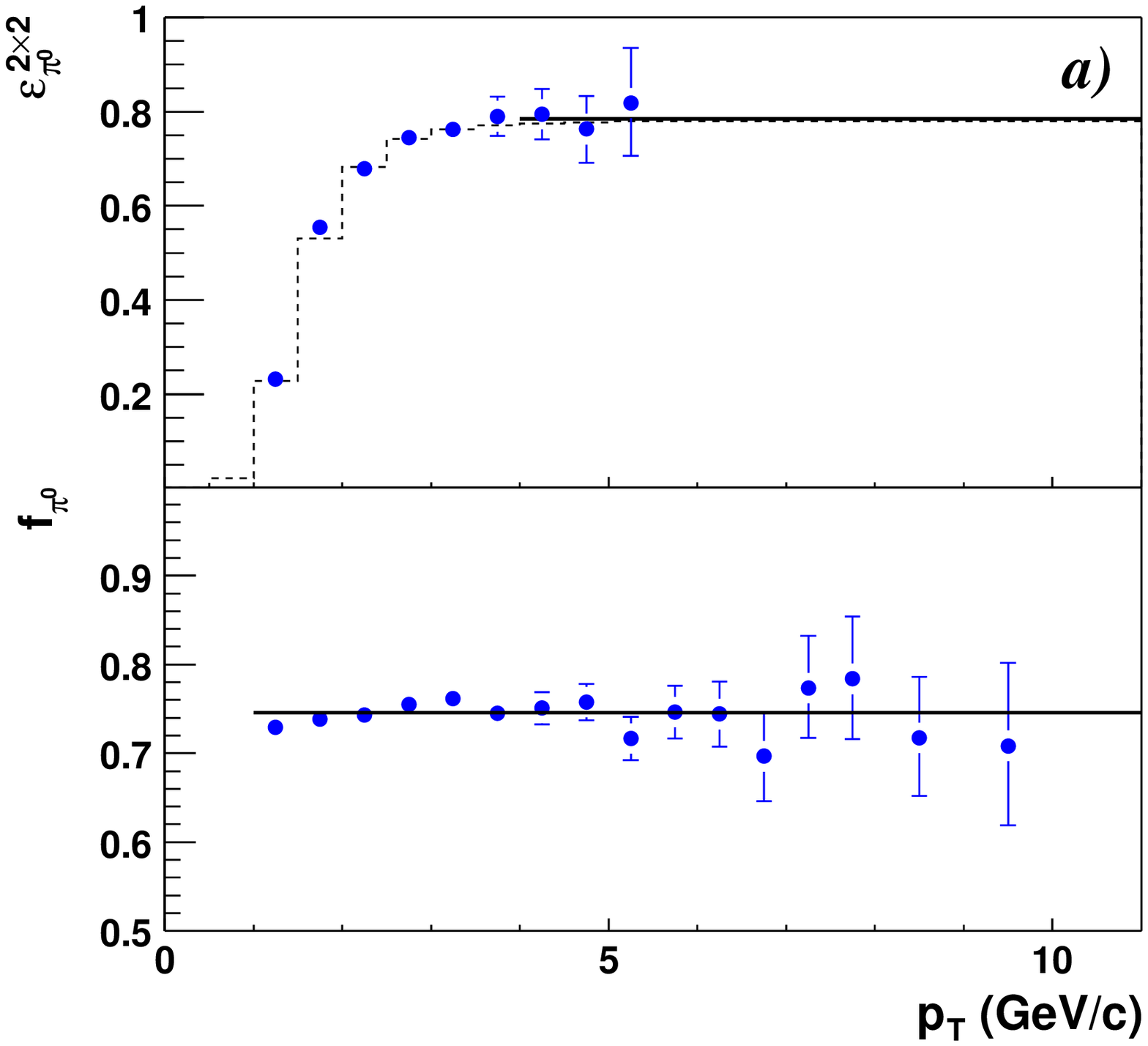,width=1\linewidth,clip}
\caption{\label{fig:ch4.pi0_trig_bias} BBC trigger bias for
neutral pions as a function of $\pi^0\ p_{T}$ with the constant
fit to the data~\cite{pp_pi0}.}

\centering
\epsfig{figure=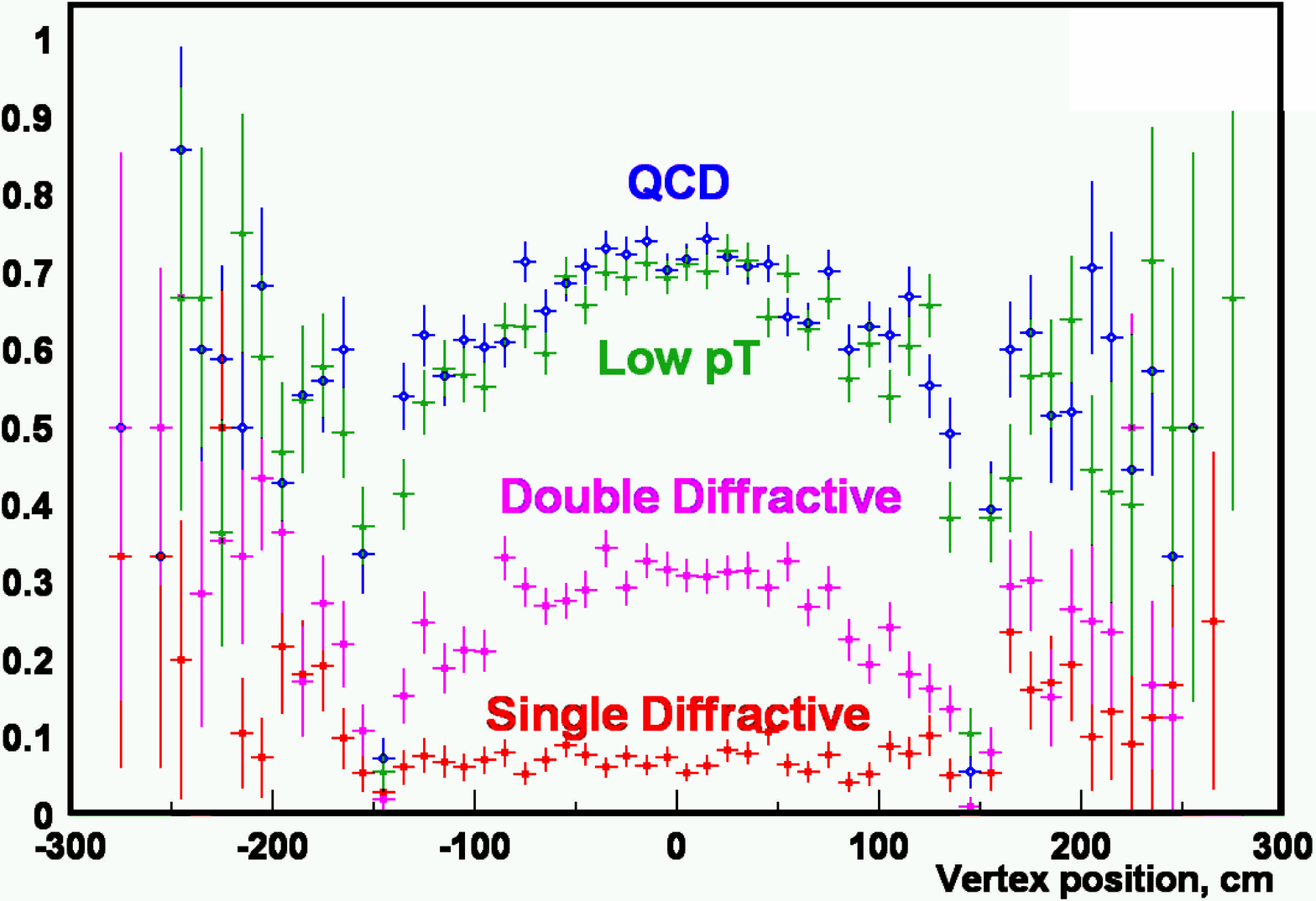,width=1\linewidth,clip}
\caption{\label{fig:ch4.pythia_trig_bias} PYTHIA+PISA calculations
for BBC efficiency as function of $Z_{vtx}$ for different physical
processes. }
\end{figure}

Fig.~\ref{fig:ch4.pi0_trig_bias} shows BBC trigger bias as a
function of neutral pion $p_{T}$. One can see that the results
agree with the value of $\epsilon_{bias} = (0.75\pm0.02)$
independent of $p_{T}$~\cite{pp_pi0} and, as expected,
significantly higher than the inclusive BBC efficiency,
$\epsilon_{BBC} =(0.516\pm0.031)$.

This measured value of the constant BBC trigger bias is in good
agreement with PYTHIA calculations of the BBC efficiency for hard
pQCD partonic scattering processes~\cite{ana148}.
Fig.~\ref{fig:ch4.pythia_trig_bias} shows the PYTHIA simulation
results for the BBC trigger efficiency as a function of the
collision vertex, $Z_{vtx}$, for different physical processes. One
can see that the expected efficiency for pQCD hard processes is
$\approx$ 0.75 independent on the vertex position. Open Charm
production should use this value of trigger bias since any
collision process that can generate the charm quark's mass energy
will certainly be a hard process. \pagebreak

\pagebreak

 At low $p_{T}$ we have a large contribution of soft
parton scattering and in this case the best way to estimate
trigger bias is by looking at charged hadrons yield using similar
approach as for $\pi^0$. This analysis was performed
in~\cite{ana148} using purely random $\bold{Clock\ trigger}$.
"Clock trigger" (or "Forced accept trigger") is a special trigger
mechanism which force a random event ($f_{Clock} = 1$ Hz) to be
stored at a certain bunch crossing independent on Level 1 trigger
decision. This feature makes the "Clock trigger" completely
unbiased.

\begin{figure}[hb]
\centering
\epsfig{figure=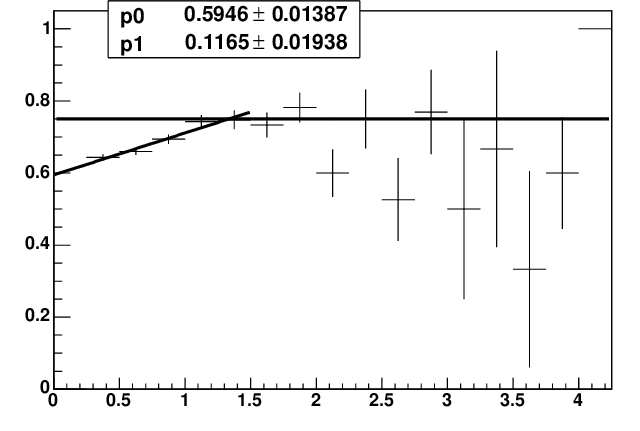,width=0.7\linewidth,clip}
\caption{\label{fig:ch4.charged_trig_bias} BBC trigger bias for
charged hadrons as a function of $p_{T}$ with fits to the data in
two intervals~\cite{ana148}.}
\end{figure}

The DCH uses time to measure position.  The appropriate time is the
time of the DCH hit as compared to the collision time (as measured by
the BBC).  Thus, the drift chamber requires the BBC to perform its
tracking and cannot be used to determine the unbiased efficiency.  To
solve this problem ~\cite{ana148}, the charged hadron analysis group
uses tracks, constructed using only the PC2, PC3, EMC cluster position
information. This method works quite well for our low multiplicity
$p+p$ collisions.  Moreover, the $Z_{vtx}$ position is approximately
evaluated by a PC hit $Z$ extrapolation to the vertex point. By
making a rough vertex cut, and taking the ratio of "Clock trigger"
events having a valid BBC vertex to the total number of "Clock trigger"
events, they estimated the BBC bias correction for charged hadrons as a
function of $p_{T}$. Their final result is shown in
Fig.~\ref{fig:ch4.charged_trig_bias}. At $p_{T}>1.3$ GeV/c the BBC trigger
bias becomes independent of $p_{T}$ and agrees well with the $\pi^0$ results. At
lower $p_{T}$ we observe a linear behavior that can be approximated by
$\epsilon_{bias}(p_{T}) = 0.59 +0.12\cdot p_{T}$.

\begin{figure}[h]
\centering
\epsfig{figure=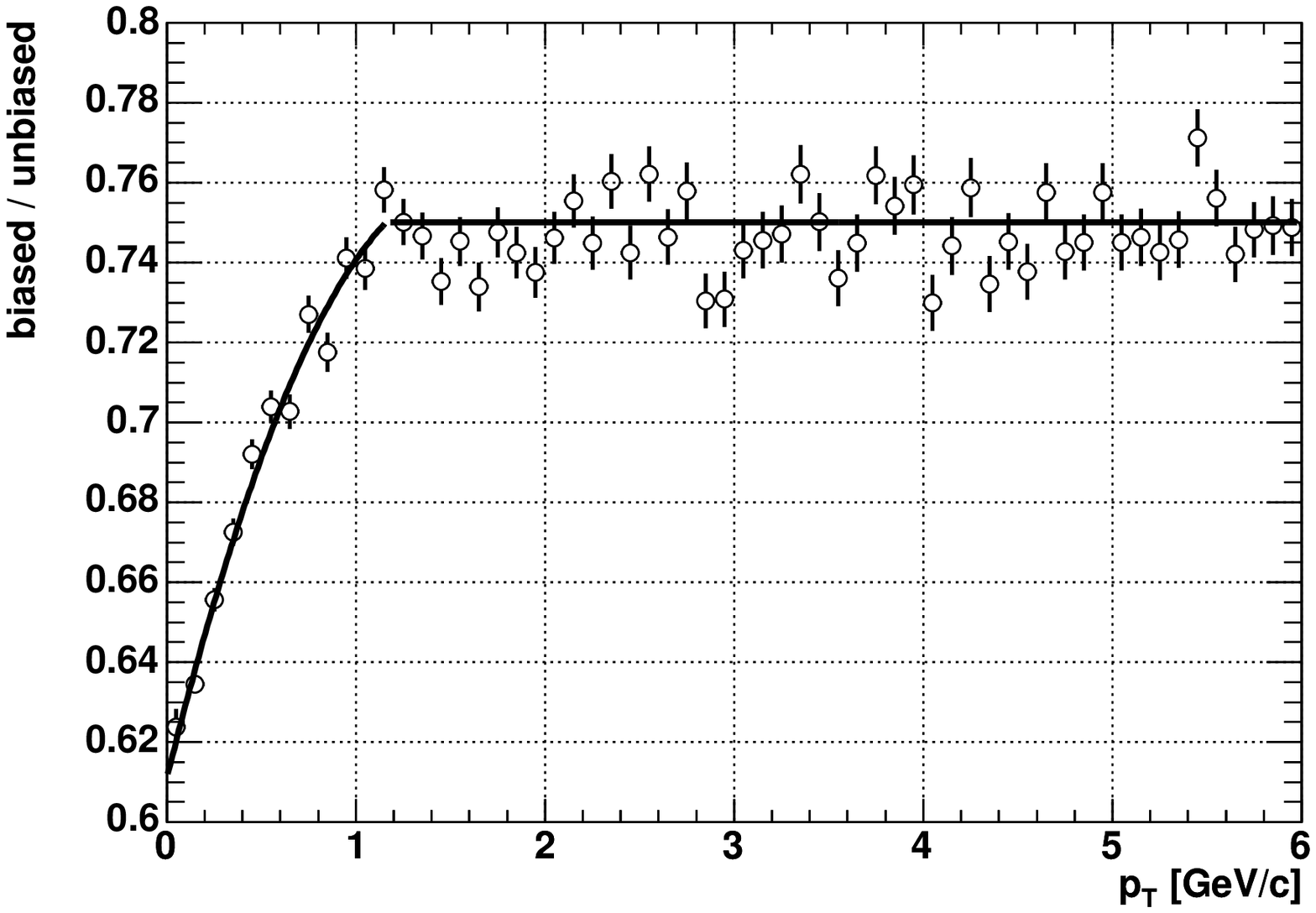,width=0.7\linewidth,clip}
\caption{\label{fig:ch4.photonic_trig_bias} BBC trigger bias for
EXODUS Cocktail electrons as a function of $p_{T}$ with fits to
the data in two $p_{T}$ intervals.}
\end{figure}

\pagebreak For the "Photonic" electron case, we must look at the
Dalitz decay of light vector mesons that have a given probability
to fire the BBC trigger depending on their momentum. In order to
evaluate the trigger bias for the decay electrons we need to use
an indirect method of trigger bias estimation because the electron
momentum is about a factor of two lower than its parent meson's
momentum. To make a translation of hadronic trigger bias to the
one for the electrons we use the EXODUS~\cite{EXODUS} decay
machine (see Section~\ref{sec:ch4.Cocktail}). Two versions of the
electron Cocktail were simulated, one without trigger bias on the
input pion crossection, and one with the trigger bias estimation
from Fig.~\ref{fig:ch4.charged_trig_bias}. The ratio of these two
Cocktail predictions gives an estimate for the "Photonic" electron
BBC trigger bias. Fig.~\ref{fig:ch4.photonic_trig_bias} shows the
final "Photonic" electron BBC trigger bias as a function of
electron transverse momentum. The final trigger bias results are
summarized in Table.\ref{tab:trig_bias}.

\begin{table}[h]
\caption{BBC trigger bias fit results for different particles. All
values have 3\% systematic uncertainties~\cite{ana148}} \centering
\begin{tabular}[b]{|c|c|c|c|}
\hline Particle & Low $p_T$ &
High $p_T$  & Cut-off [GeV/c]\\
\hline Charged hadrons & $0.59 + 0.12\cdot p_{T}$ & $0.75$
&1.33\\
\hline "Photonic" electrons & $0.61 + 0.19p_{T} - 0.061 p_{T}^2$
&0.75&1.16\\
\hline "Non-photonic" electrons &\multicolumn{2}{c|}{$0.75$}\\
\cline{1-3}
\end{tabular}
\label{tab:trig_bias}
\end{table}

\subsection{Combining statistics}

Now we have everything necessary to calculate the inclusive electron
crossection using the formula from Eq.~\ref{eq:ch4.inv_cross}. Since
the electron spectrum at low $p_{T}$ is dominated by electrons from
"Photonic" sources, we use the "Photonic" trigger bias to
normalize the crossection. This is a valid assumption for the first
iteration, as at $p_{T} < 1.0$ GeV/c the "photonic" contribution is about
80\% of the total electron signal.

\begin{figure}[h]
\centering
\epsfig{figure=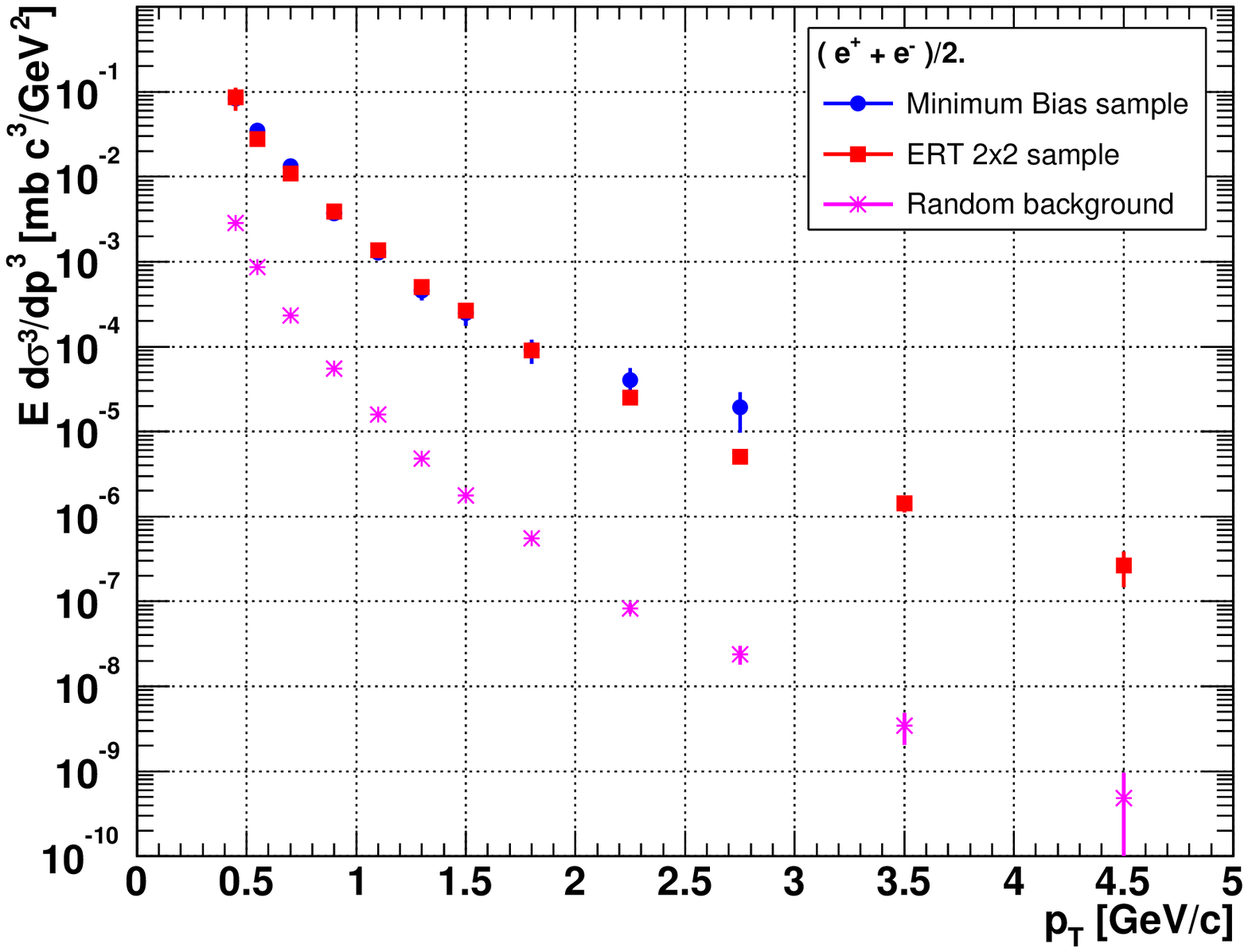,width=1\linewidth,clip}
\caption{\label{fig:ch4.initial_inclusive} Initial inclusive
electron invariant crossection for Minimum Bias (circles) , ERT
trigger (squares). Random Hadronic background level (asterisk) is
$\bold{not}$ subtracted from the data. }
\end{figure}

The inclusive electron crossection for the Minimum Bias and the ERT trigger
sample is shown in Fig.~\ref{fig:ch4.initial_inclusive}.  The hadronic
background contribution is not subtracted from the spectra but
is shown in the plot for comparison. The same trigger bias correction
as for the electrons is applied to the hadronic background
contribution. Only statistical errors are shown in this plot. The data
points are tabulated in Table~\ref{tab:init_incl}.

\begin{table}[ht]
\caption{ Initial inclusive electron invariant crossection for
Minimum Bias, ERT trigger. Random Hadronic background level is
$\bold{not}$ subtracted from the data. Crossection and statistical
errors are in units of $\cunit$.}
\begin{center}
\begin{tabular}[b]{|c|c|c|c|c|c|c|}
\hline $p_{T}$ bin & MB &MB & ERT &ERT &Random&Random\\
$[GeV/c]$ & cros-&stat.&cros-&stat.&cros-&stat.\\
 & section&error&section&error&section&error\\
\hline
0.4-0.5 &8.29e-02 &3.73e-03 &8.61e-02 &2.63e-02 &2.85e-03 &1.20e-05 \\
0.5-0.6 &3.51e-02 &2.11e-03 &2.79e-02 &4.05e-03 &8.66e-04 &5.76e-06 \\
0.6-0.8 &1.33e-02 &8.05e-04 &1.09e-02 &6.67e-04 &2.31e-04 &1.84e-06 \\
0.8-1.0 &3.69e-03 &3.63e-04 &3.86e-03 &1.43e-04 &5.47e-05 &7.66e-07 \\
1.0-1.2 &1.27e-03 &1.90e-04 &1.37e-03 &5.18e-05 &1.58e-05 &3.68e-07 \\
1.2-1.4 &4.53e-04 &1.04e-04 &5.02e-04 &2.43e-05 &4.80e-06 &1.86e-07 \\
1.4-1.6 &2.47e-04 &7.14e-05 &2.64e-04 &1.53e-05 &1.75e-06 &1.05e-07 \\
1.6-2.0 &9.14e-05 &2.89e-05 &9.02e-05 &5.58e-06 &5.55e-07 &3.89e-08 \\
2.0-2.5 &4.04e-05 &1.53e-05 &2.52e-05 &2.31e-06 &8.29e-08 &1.21e-08 \\
2.5-3.0 &1.9e-05 &9.68e-06 &5.08e-06 &9.28e-07 &2.39e-08 &5.81e-09 \\
3.0-4.0 &0.00e+00 &0.00e+00 &1.42e-06 &3.11e-07 &3.45e-09 &1.41e-09 \\
4.0-5.0 &0.00e+00 &0.00e+00 &2.65e-07 &1.19e-07 &4.80e-10 &4.80e-10 \\
\hline \end{tabular} \label{tab:init_incl} \end{center}
\end{table}

The standard way of combining two statistically independent data sets
is to use the weighted average of both data sets. We assumption two
independent measurements, of a normally distributed value $n$, $n_{1}$
and $n_{2}$, with corresponding statistical errors, $\sigma_{1}$ and
$\sigma_{2}$.  In this case the unbiased estimator and the absolute
uncertainty of the value can be calculated as:
\begin{eqnarray}
\langle n \rangle &=&(\frac{n_1}{\sigma_1^2}
+\frac{n_2}{\sigma_2^{2}})/(\frac{1}{\sigma_1^2}
+\frac{1}{\sigma_2^2})\nonumber \\
\sigma_n^2 &=& 1/(\frac{1}{\sigma_{1}^{2}}
+\frac{1}{\sigma_{2}^{2}}) \label{eq:ch4.weight_avg}
\end{eqnarray}

\begin{figure}[h]
\centering
\epsfig{figure=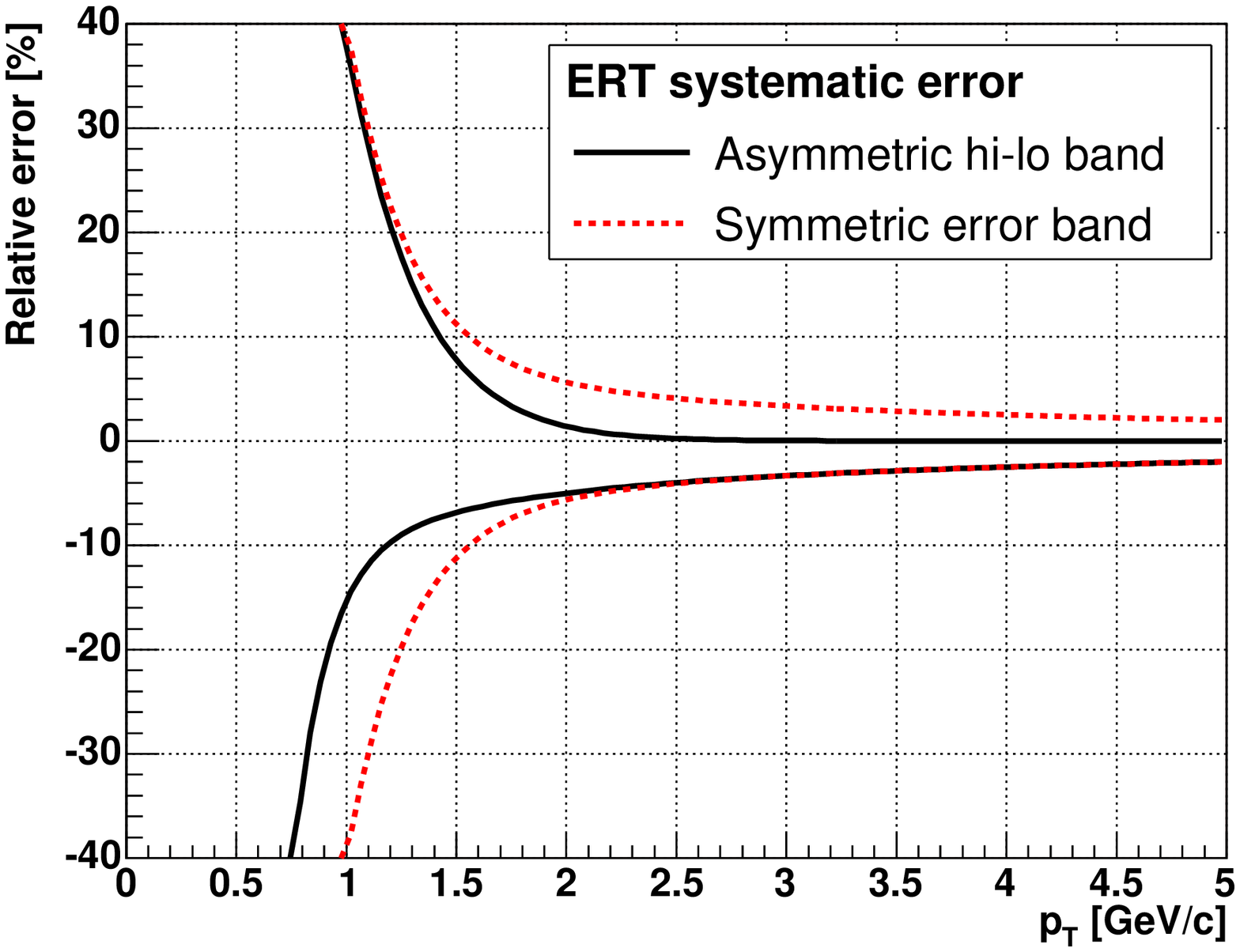,width=0.6\linewidth,clip}
\caption{\label{fig:ch4.ERT_variation} Systematic error band on
the ERT inclusive electron crossection due to uncertainty of ERT
trigger efficiency (Table~\ref{tab:ert_eff_table}). Asymmetric
error band (solid) uses the exact variations. Symmetric error
 band (dashed) is used in the analysis.}
\epsfig{figure=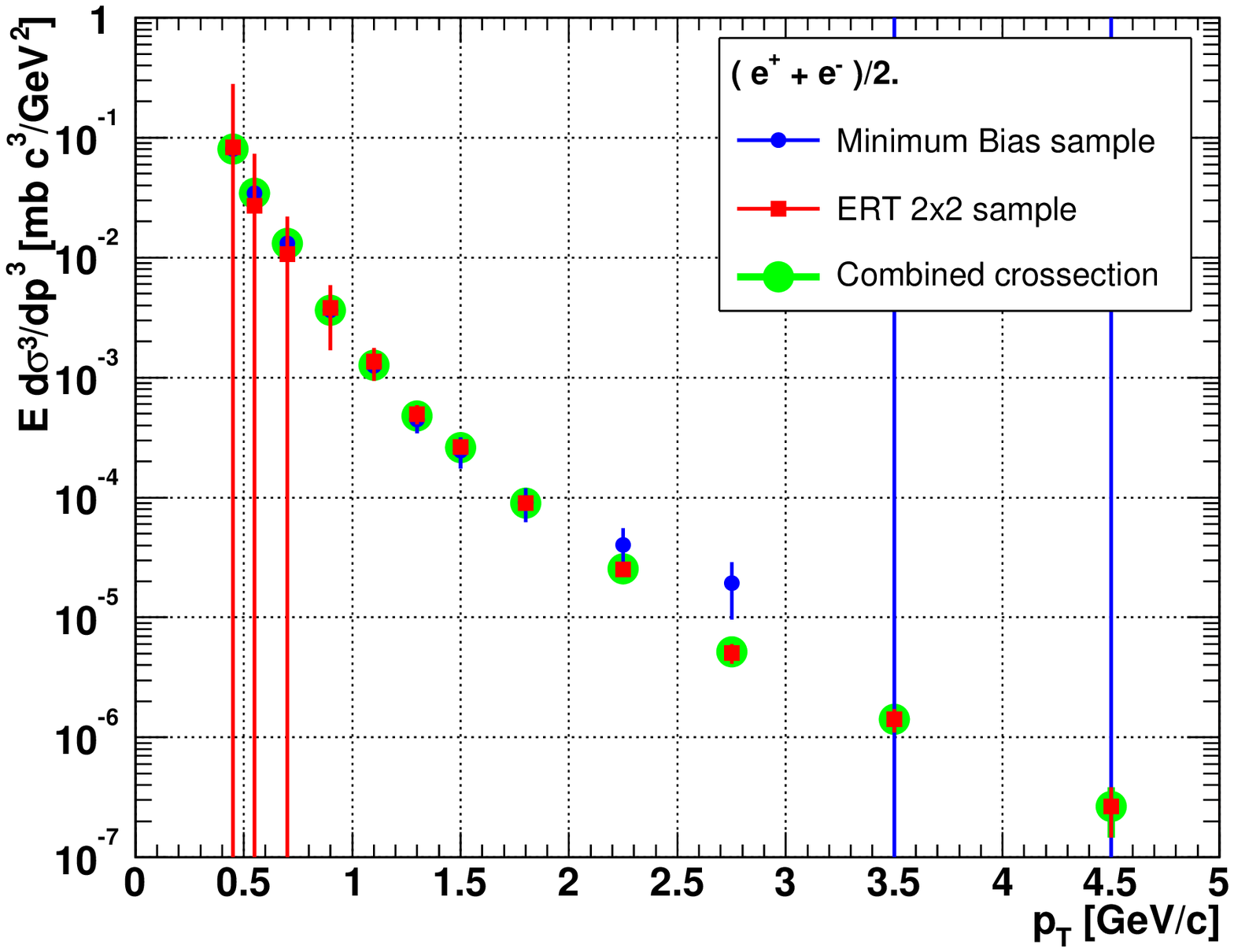,width=0.8\linewidth,clip,trim
= 0in 0in 0in 0.5in} \caption{\label{fig:ch4.comb_incl} Inclusive
electron invariant crossection for Minimum Bias (circle) and ERT
data sample (squares) (ERT statistical errors includes the
systematic error due to ERT trigger efficiency). Combined by
Eq.~\ref{eq:ch4.weight_avg} inclusive electron invariant
crossection (large circle). Hadronic background is subtracted from
both data samples (see Table~\ref{tab:init_incl}). }
\end{figure}

In the case of the ERT trigger data, the absolute error of the result
will strongly depend on the systematic error of the ERT trigger.
Thus, in order to to use Eq.~\ref{eq:ch4.weight_avg} we need to
$\bold{combine}$ the statistical error of ERT data with the systematic
error. Using the systematic error band previously quoted in
Table~\ref{tab:ert_eff_table}, we can obtain the variation level of
the ERT inclusive electron spectrum shown in
Fig.~\ref{fig:ch4.ERT_variation}. To simplify the error propagation, a
symmetric error band, engulfing the hi-lo variations, was assumed for the
total systematic error to the ERT inclusive sample. One can see that
the systematic error is significant at low $p_{T}$ and falls to
$\approx 5\%$ at $p_{T} > 2.0$ GeV/c.

\newpage
\pagebreak

 The statistical error of the ERT inclusive electron
crossection (Table~\ref{tab:init_incl}) is combined with the
systematic error as a squared sum. The final error on the MB-only
and ERT-only inclusive electron invariant crossection together
with the weighted average of both data sets is shown in
Fig.~\ref{fig:ch4.comb_incl} and summarized in
Table~\ref{tab:comb_incl}. The hadronic background is already
$\bold{subtracted}$ at this point. Those MB sample $p_{T}$ bins
that have zero statistics are assumed to have infinite statistical
error and do not contribute to the combined crossection.

\begin{figure}[h]
\centering
\epsfig{figure=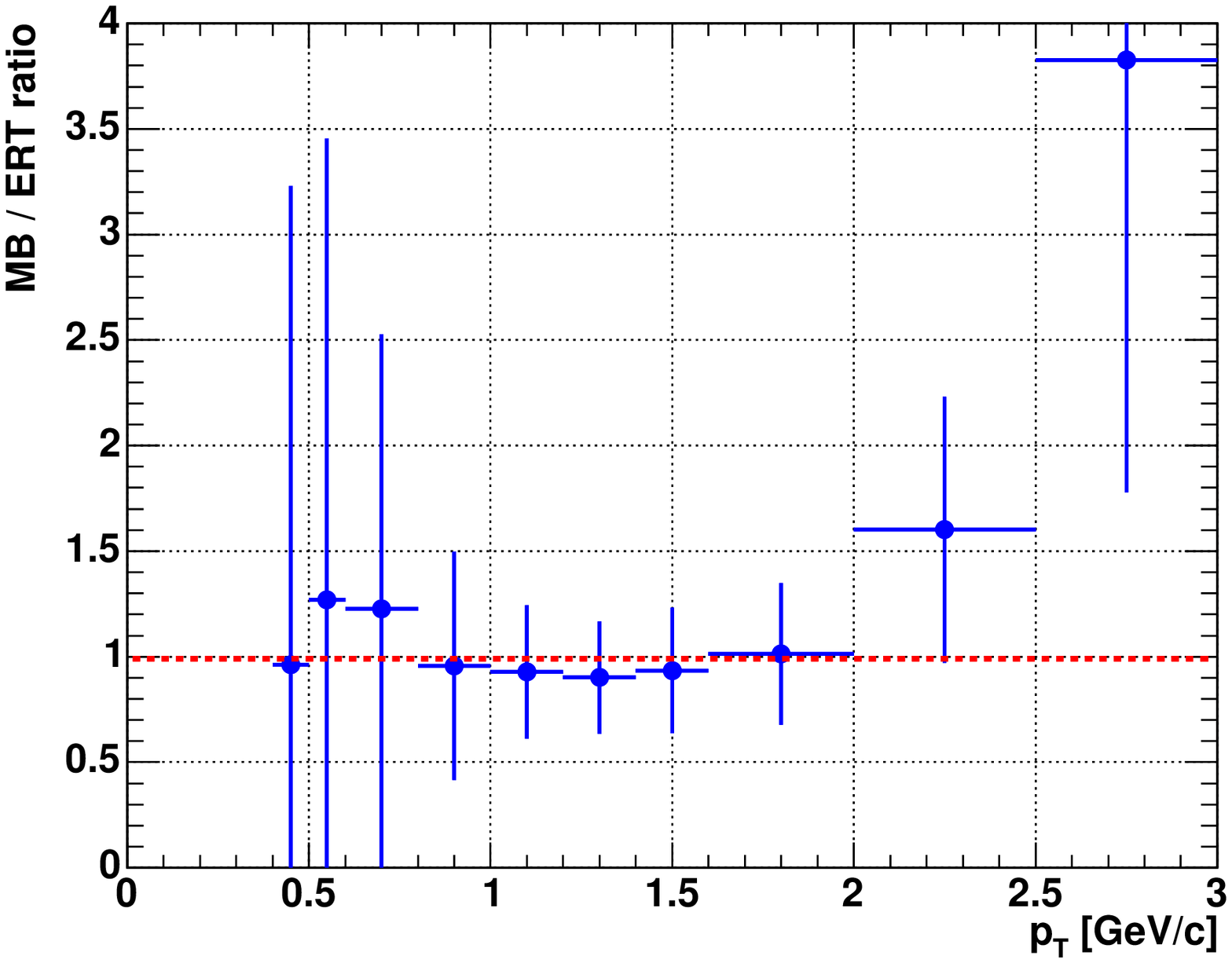,width=0.6\linewidth,trim = 0in
0in 0in 0.5in} \caption{\label{fig:ch4.ratio_MB_ERT} Ratio of MB
electron inclusive crossection to ERT crossection as a function of
$p_{T}$. }
\end{figure}

The ratio of MB and ERT inclusive electron crossection is shown in
Fig.~\ref{fig:ch4.ratio_MB_ERT}. The two independent data sets are
seen to be in good agreement with each other. There is an indication
that ERT signal is ~5\% higher then the MB in $p_{T}$ region from 1.0
to 2.0 GeV/c which is possibly due to the underestimation of ERT
trigger efficiency in this region as discussed previously. Please note
that the systematic error on the ERT trigger efficiency more than
covers this systematic difference. The apparent discrepancy of the MB
crossection at $p_{T} > 2.5 $ GeV/c is due prinarily to low statistics
in those bins (see Table~\ref{tab:raw_table}).

\newpage
\begin{table}[ht]
\caption{ Inclusive electron invariant crossection for MB and ERT
data sample (ERT statistical errors includes the systematic error
due to ERT trigger efficiency). Combined inclusive electron
crossection with statistical error. Hadronic background is
subtracted from both data samples (see Table~\ref{tab:init_incl}).
Crossection and statistical errors are in units of $\cunit$.}
\begin{center}
\begin{tabular}[b]{|c|c|c|c|c|c|c|}
\hline $p_{T}$ bin & MB &MB& ERT &ERT&Comb.&Comb.\\
$[GeV/c]$ & cros-&stat.&cros-&stat.&cros-&stat.\\
 &section&error&section&error&section&error\\
\hline
0.4-0.5 &8.01e-02 &3.73e-03 &8.32e-02 &1.96e-01 &8.01e-02 &3.73e-03 \\
0.5-0.6 &3.43e-02 &2.11e-03 &2.70e-02 &4.65e-02 &3.43e-02 &2.11e-03 \\
0.6-0.8 &1.31e-02 &8.05e-04 &1.06e-02 &1.12e-02 &1.31e-02 &8.02e-04 \\
0.8-1.0 &3.64e-03 &3.63e-04 &3.80e-03 &2.11e-03 &3.64e-03 &3.58e-04 \\
1.0-1.2 &1.26e-03 &1.90e-04 &1.35e-03 &4.12e-04 &1.27e-03 &1.73e-04 \\
1.2-1.4 &4.49e-04 &1.04e-04 &4.98e-04 &9.09e-05 &4.76e-04 &6.85e-05 \\
1.4-1.6 &2.45e-04 &7.14e-05 &2.63e-04 &3.33e-05 &2.59e-04 &3.02e-05 \\
1.6-2.0 &9.08e-05 &2.89e-05 &8.96e-05 &8.40e-06 &8.97e-05 &8.06e-06 \\
2.0-2.5 &4.03e-05 &1.53e-05 &2.52e-05 &2.59e-06 &2.56e-05 &2.56e-06 \\
2.5-3.0 &1.93e-05 &9.68e-06 &5.05e-06 &9.47e-07 &5.19e-06 &9.42e-07 \\
3.0-4.0 &0.00e+00 &1.00e+15 &1.42e-06 &3.13e-07 &1.42e-06 &3.13e-07 \\
4.0-5.0 &0.00e+00 &1.00e+15 &2.64e-07 &1.19e-07 &2.64e-07 &1.19e-07 \\
\hline \end{tabular} \label{tab:comb_incl} \end{center}
\end{table}

\pagebreak
\subsection{Bin width corrections}

The width of the bins in the current analysis are significant.  Thus,
the assumption that the mean $p_{T}$ of the electrons within a bin
equals the $p_T$ at the center of the bin is not correct for our
steeply falling distribution. We need to correct for this shift of the
mean $p_{T}$ within the histogram bin. The official PHENIX bin width
correction method ~\cite{ana073} has been successfully implemented
for all major spectrometric PHENIX results including this one.

The details of the iterative bin width correction method are described
below:

\begin{itemize}
\item{$\bold{First\ iteration}$: Make an assumption about the
shape of the spectrum to be corrected. In our case both the inclusive
and subtracted crossections are well fit with modified power law
function
\begin{equation}
f(p_T)=\frac{A}{(p_0+p_T)^n} \label{eq:ch4.hagedorn}
\end{equation}
where $A,\ p0,\ n$ are fit function parameters }
\item{Fit data
with the function~\ref{eq:ch4.hagedorn} to obtain initial guess
parameters.}
\item{Calculate the mean $p_{T}$ for given $p_T$ bin
$[a;b]$ by solving the integral equation for $p_{T\ mean}$
\begin{equation}
f(p_{T\ mean})=\frac{\int_{a}^{b} f(p) dp}{b-a}
\label{eq:ch4.mean_pt}
\end{equation}
} \item{Move data points from bin center to $p_{T\ mean}$}
\item{$\bold{Next\ iteration}$:Repeat the same steps using the
data points corrected on previous iteration }
\end{itemize}

This process converges after 5-10 iterations, producing fit parameters
that are nearly identical (to the computer's finite precision) to
those from the previous iteration. Upon convergence of the iterative
process, we move the data points to be in the center of the bin by
recalculating the bin height and the corresponding statistical and
systematic errors at the center of the bin \footnote{ The process of
moving point along the ordinate axis while keeping it in the bin
center does not converge if repeated multiple time and should be used
only once at final stage. The vertical shift is done in order to
simplify the comparison with different experimental data
sets such as AuAu.}. \nopagebreak
\begin{equation}
y' = y\cdot \frac{f(\frac{b-a}{2})}{f(p_{T\ mean})},\ \ \delta y'
= \delta y\cdot \frac{f(\frac{b-a}{2})}{f(p_{T\ mean})}
\label{eq:ch4.y_shift}
\end{equation}

\begin{figure}[h]
\centering
\epsfig{figure=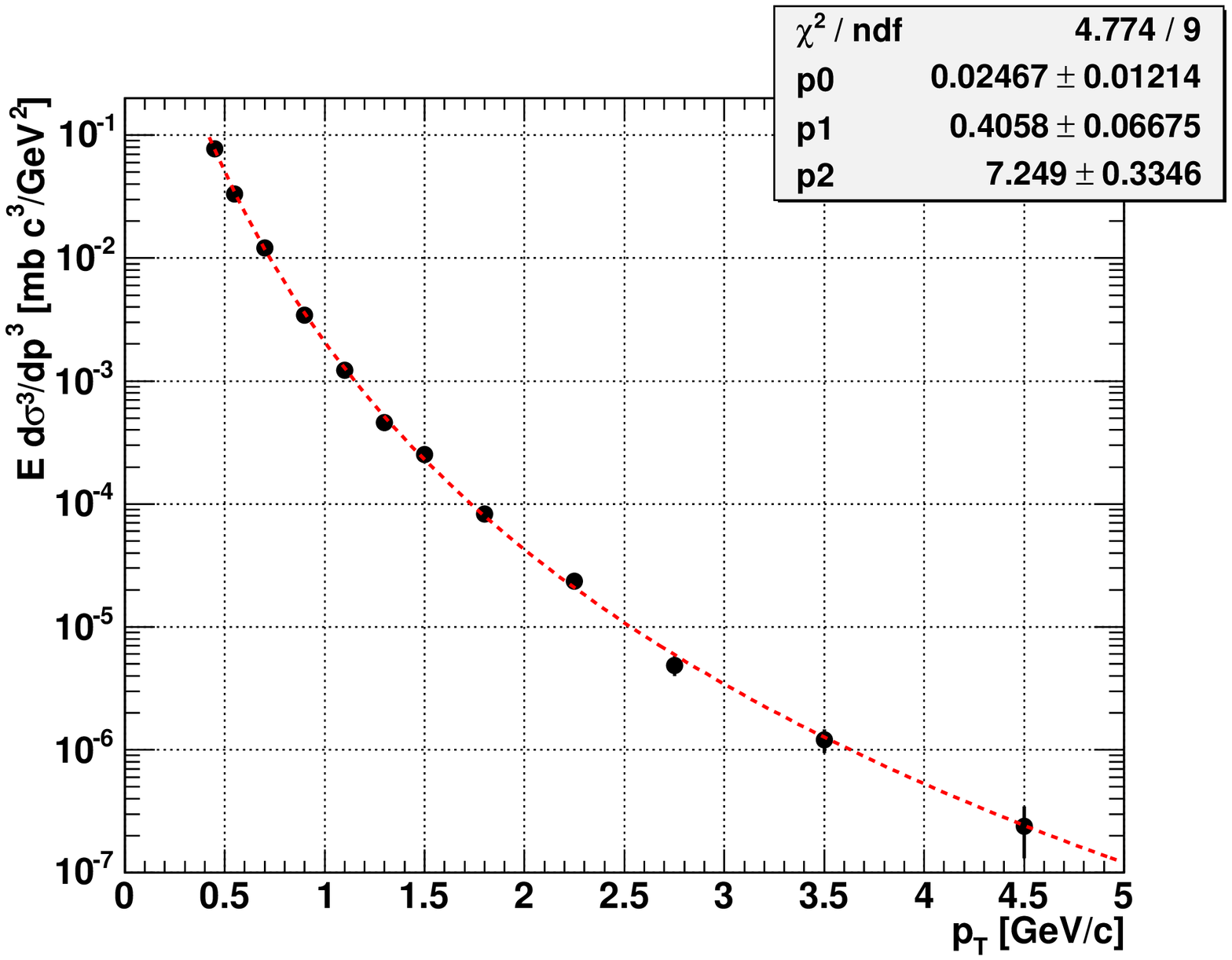,width=0.6\linewidth,clip}
\caption{\label{fig:ch4.final_inclusive} Final bin-width corrected
inclusive electron invariant crossection fitted with modified
power law function $\frac{p0}{(p1+p_{T})^{p3}}$.}
\end{figure}

After applying the described method to the combined electron
spectrum, we obtain our $\bold{final\ inclusive\ electron\ invariant\
crossection}$ as shown in Fig.~\ref{fig:ch4.final_inclusive} and
summarized in Table~\ref{tab:final_incl}. The final fit result by
modified power law function is shown in the figure.

\begin{table}[h]
\caption{ Final inclusive electron invariant crossection before
and after the bin width correction.}
\centering
\begin{tabular}[b]{|c||c|c||c|}
\hline  Uncorrected &Final& Final& Correction\\
 crossection&crossection&stat. error&factor\\ \hline
8.010e-02 &7.742e-02 &3.606e-03 &0.967\\
3.426e-02 &3.334e-02 &2.054e-03 &0.973\\
1.307e-02 &1.204e-02 &7.398e-04 &0.922\\
3.642e-03 &3.435e-03 &3.374e-04 &0.943\\
1.274e-03 &1.219e-03 &1.650e-04 &0.957\\
4.765e-04 &4.605e-04 &6.619e-05 &0.966\\
2.595e-04 &2.525e-04 &2.935e-05 &0.973\\
8.971e-05 &8.266e-05 &7.433e-06 &0.921\\
2.557e-05 &2.342e-05 &2.342e-06 &0.916\\
5.187e-06 &4.873e-06 &8.854e-07 &0.939\\
1.417e-06 &1.204e-06 &2.664e-07 &0.850\\
2.645e-07 &2.386e-07 &1.072e-07 &0.902\\
\hline \end{tabular} \label{tab:final_incl}
\end{table}
\pagebreak

\section{Cocktail Estimation of the "Photonic" Electron Component}\label{sec:ch4.Cocktail}

Our principal goal is not to learn the inclusive electron
spectrum, but rather the component of that spectrum that comes
from heavy flavor \linebreak decays.  Photonic electrons such as
those resulting from the decays of light vector mesons are a
physical background to our signal that must be estimated and
subtracted. To estimate the background electron rate from these
known photonic sources we again used EXODUS~\cite{EXODUS}
$\bold{Cocktail}$ generator. The EXODUS-based cocktail simulation
uses the following basic steps:
\begin{itemize}
\item {Simulate the Dalitz decays of light vector mesons ($\pi^0$,
$\eta$, $\eta'$, $\omega$) using the standard Dalitz decay
formalism~\cite{Kroll_Wada}.}
\item {Establish a realistic input distribution for the mesons using
the measured $\pi^0$ crossection. The abundance of higher mass vector
mesons is fixed by assuming $m_T$ scaling of the meson ratios at high
$p_{T}$~\cite{ppg011}.}
\item {Normalize conversion electrons rates using the ratio of conversion electrons to
Dalitz, as obtained from full the PISA simulation.}
\end{itemize}

\subsection{Cocktail input}

\subsubsection{Neutral pions}

\begin{figure}[ht]
\centering
\epsfig{figure=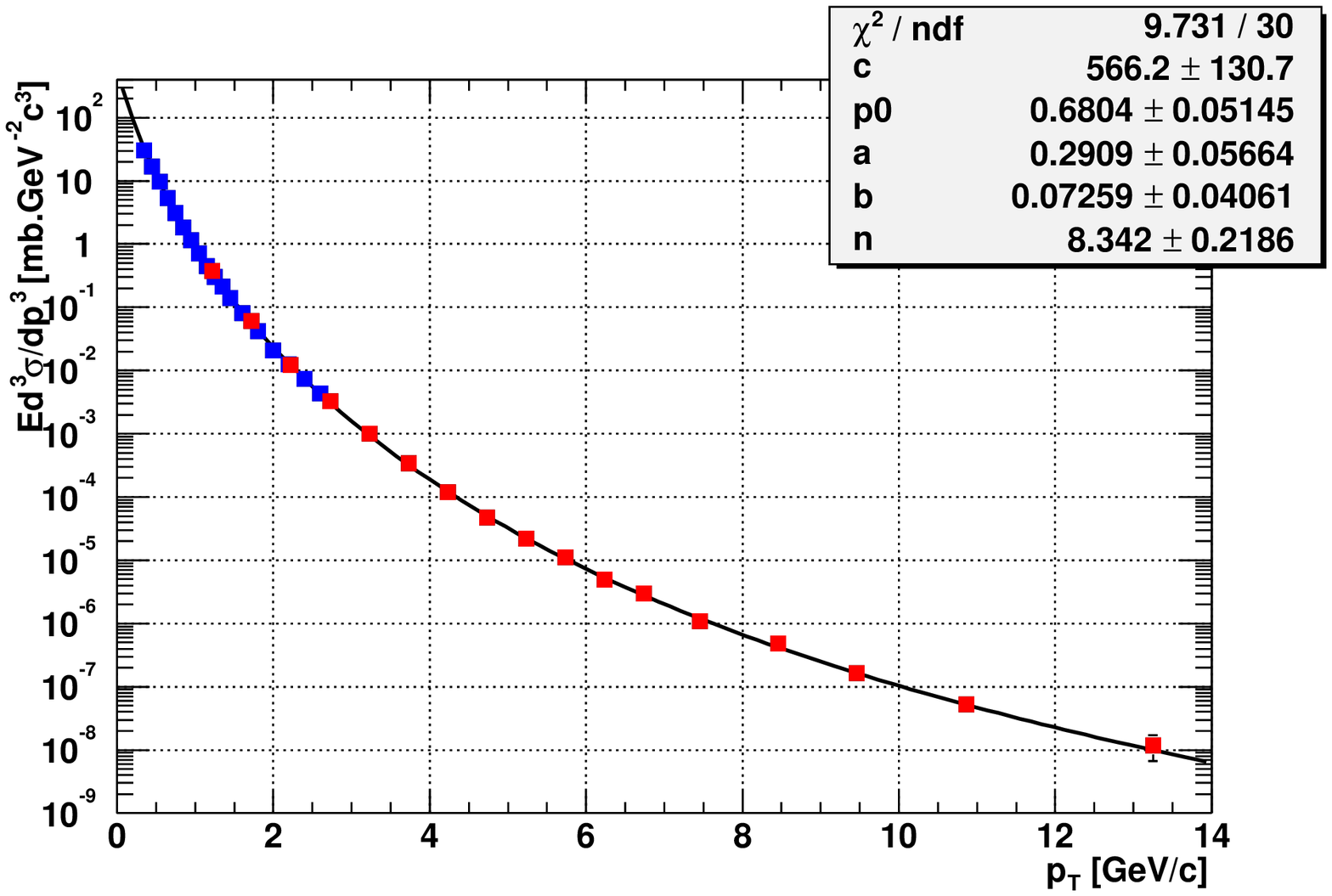,width=0.8\linewidth,clip}
\caption{\label{fig:ch4.pion_fit} Invariant crossection of charge
averaged pions $(\pi^{+} +\pi^{-})/2$ measured in Run03
 $p+p$ collisions~\cite{ppg029} (blue) and Run02 $p+p$
$\pi^0$ crossection~\cite{pp_pi0} (red) fitted with a modified
power law function (Eq.~\ref{eq:ch4.pion_fit_form}).} \centering
\epsfig{figure=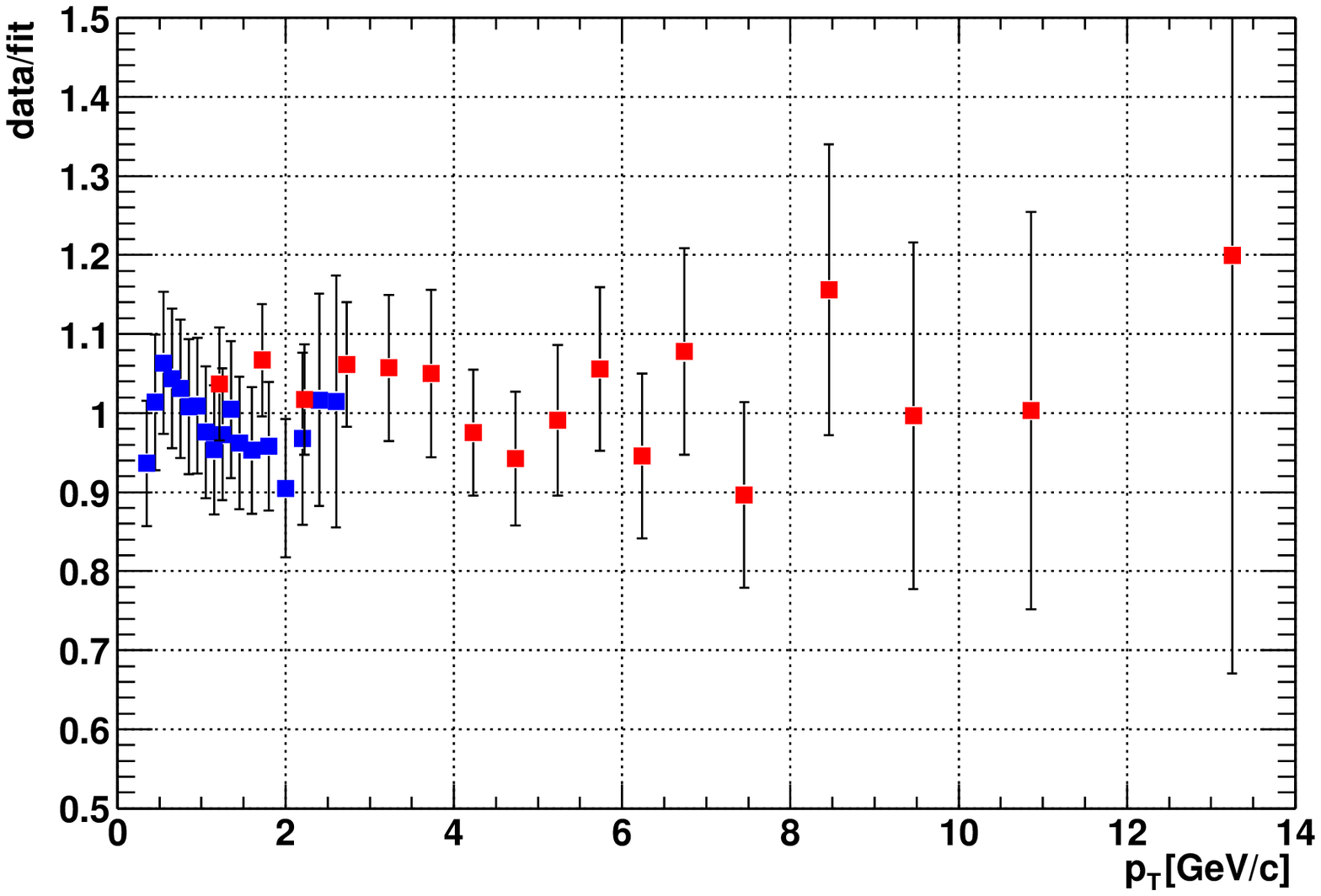,width=0.7\linewidth,clip}
\caption{\label{fig:ch4.ratio_pion_fit} Ratio of data to fit from
Fig.~\ref{fig:ch4.pion_fit}.}
\end{figure}

The $\pi^0$ Dalitz decay ($\pi^0 \rightarrow \gamma e^{+}e^{-}$)
is the main source of electrons in PHENIX. As input we use all
available PHENIX measurements of the pion crossection. At high
$p_{T} (> 1.0$ GeV/c) we use the published $\pi^0$
results~\cite{pp_pi0}. At low $p_T$ we can use the charged pion
measurements from Run03 $p+p$ at $\sqs = 200$ GeV that are
currently being prepared for publication~\cite{ppg029}. \linebreak
In combining these two data sets, we assume that the $\pi^0$
spectrum is the same as the average charged pion spectrum.
Fig.~\ref{fig:ch4.pion_fit} shows the result of the combined pion
invariant crossection fit with a modified power law fit
function\footnote{The quadratic term is not necessary to achieve a
superb fit of the pion spectra in $\pp$ collisions, but it turns
out to be important in case of $Au+Au$ collisions at the same
energy. Therefore, we include it for consistency with different
colliding systems.}  (Eq.~\ref{eq:ch4.pion_fit_form}). The ratio
of data to fit is shown in Fig.~\ref{fig:ch4.ratio_pion_fit}:

\begin{equation}
E \frac{d\sigma}{dp_T^3} = \frac{c}{(e^{(-a\cdot p_T - b\cdot
p_T^2)} + p_T/p_0)^n} \label{eq:ch4.pion_fit_form}
\end{equation}

Fit parameters listed below:\\
$c= 566 \pm 130$\\
$p_0= 0.68 \pm 0.05$\\
$a= 0.29 \pm 0.06$\\
$b= 0.073 \pm 0.041$\\
$n= 8.34 \pm 0.22$ \pagebreak

The parameterization chosen here is not meant to be of any physical
relevance. Our only goal is to have an excellent parameterization of
the pion spectrum as input to the EXODUS simulations. Only the
statistical errors are shown in Fig.~\ref{fig:ch4.pion_fit}
~\ref{fig:ch4.ratio_pion_fit}.  The systematic error on the initial
pion spectrum will be accounted for as an uncertainty applied to the
final Cocktail result. (see
Section~\ref{sec:ch4.Cocktail_Systematics}).

\subsubsection{Other light mesons}

Other light mesons contributing to the inclusive electron crossection
via their decays are the ($\pi^0$, $\eta$, $\eta'$, $\omega$, $\rho$)
mesons. Of all these, only the $\eta$ meson contributes a sizable
fraction of the inclusive decay electrons, particularly at high
$p_{T}$. The cocktail input for these other light mesons is prepared
in the established way:

\begin{itemize}
\item The rapidity distributions are assumed to be flat around mid
rapidity.
\item Shapes of the transverse momentum distributions are obtained via
$m_T$ scaling from the pions. The pion'ss modified power law fit
parameterization from the previous section
(Eq.~\ref{eq:ch4.pion_fit_form}) is used as input for other mesons
where transverse momentum $p_{T}$ is replaced by $\sqrt{p_{T}^2 +
m_{lm}^2 - m_{\pi}^2}$.  Here $m_{lm}$ denotes the mass of the light
meson.
\item The relative normalization of the light meson yield is fixed by
forcing the ratio $\frac{meson}{\pi^0}$ at high $p_{T}$ to the known
values measured at other $\sqs$. These ratios and corresponding
systematic errors are summarized in Table.~\ref{tab:part_ratios}
~\cite{ana101}.  We conservatively assign 30\% as the $1\sigma$
systematic error on all meson ratios except the $\eta$, for which the
systematic uncertainty is smaller since it was recently measured by
PHENIX ~\cite{ana333,ana337}.
\end{itemize}

\begin{table}[b]
\caption{ Ratios of light mesons to $\pi^0$ at high $p_T$.}
\begin{center}
\begin{tabular}[b]{|c|c|c|}
\hline  &Ratio& Sys. Error\\ \hline
$\eta /\pi^0$& 0.45 & 0.10 \\
$\rho /\pi^0$& 1.00 & 0.30 \\
$\omega /\pi^0$& 1.00 & 0.30 \\
$\eta' /\pi^0$& 0.25 & 0.08 \\
$\phi /\pi^0$& 0.40 & 0.12 \\
\hline
\end{tabular} \label{tab:part_ratios} \end{center}
\end{table}

\begin{figure}[h]
\centering
\epsfig{figure=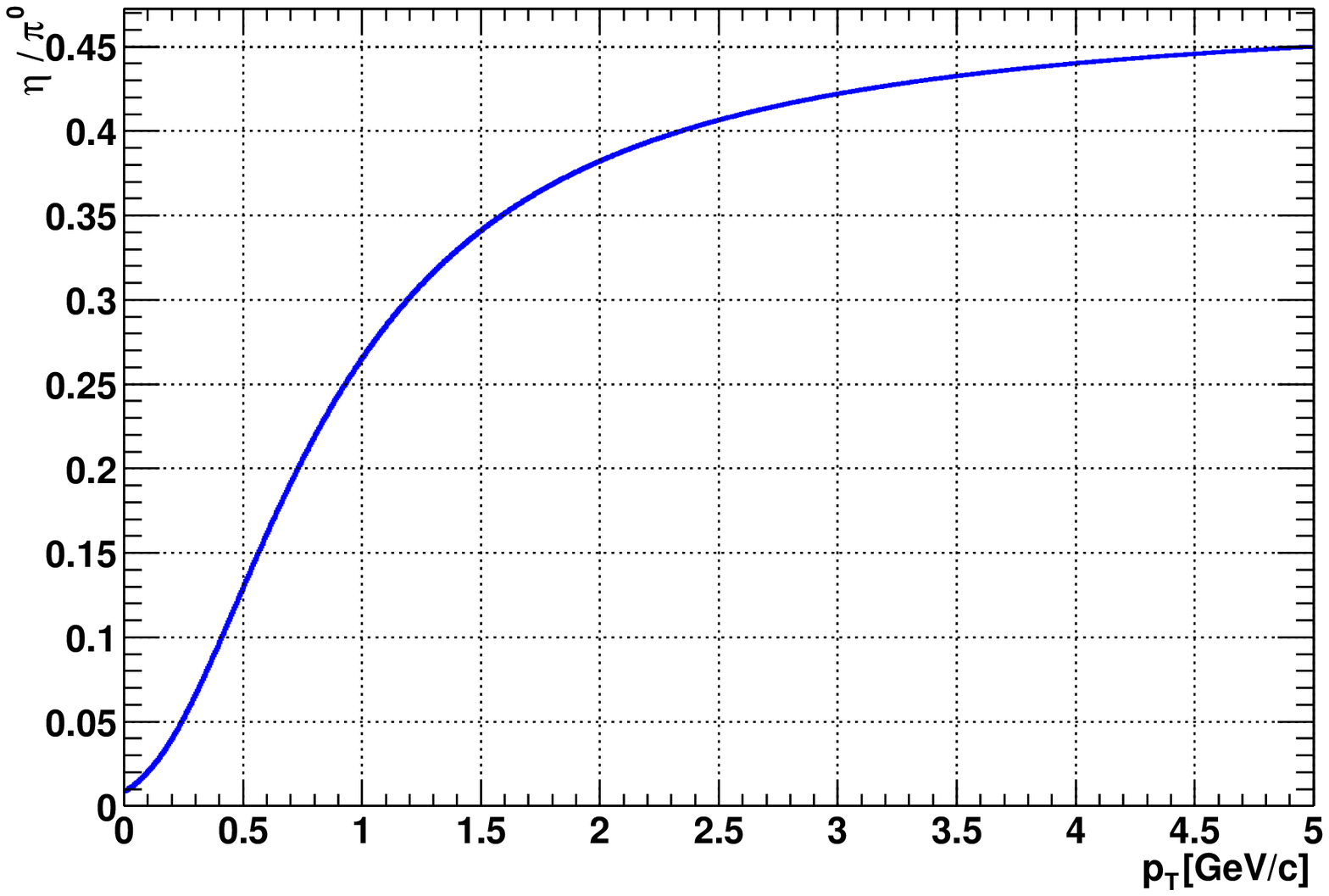,width=0.6\linewidth,clip}
\caption{\label{fig:ch4.eta_pion} Ratio of $\eta$ to $\pi^0$ as
function of $p_{T}$ in Cocktail.}
\end{figure}

Although all higher mass mesons beyond the $\eta$ are essentially
irrelevant (due to small yield), it is important to explicitly
demonstrate that the chosen parameterization is reasonable for
$\eta$-meson. Fig.~\ref{fig:ch4.eta_pion} shows the ratio of eta/pion in the
Cocktail using $m_T$ scaling assumption and a constant particle
ratio at high $p_T$ of 0.45.

\begin{figure}[h]
\centering
\epsfig{figure=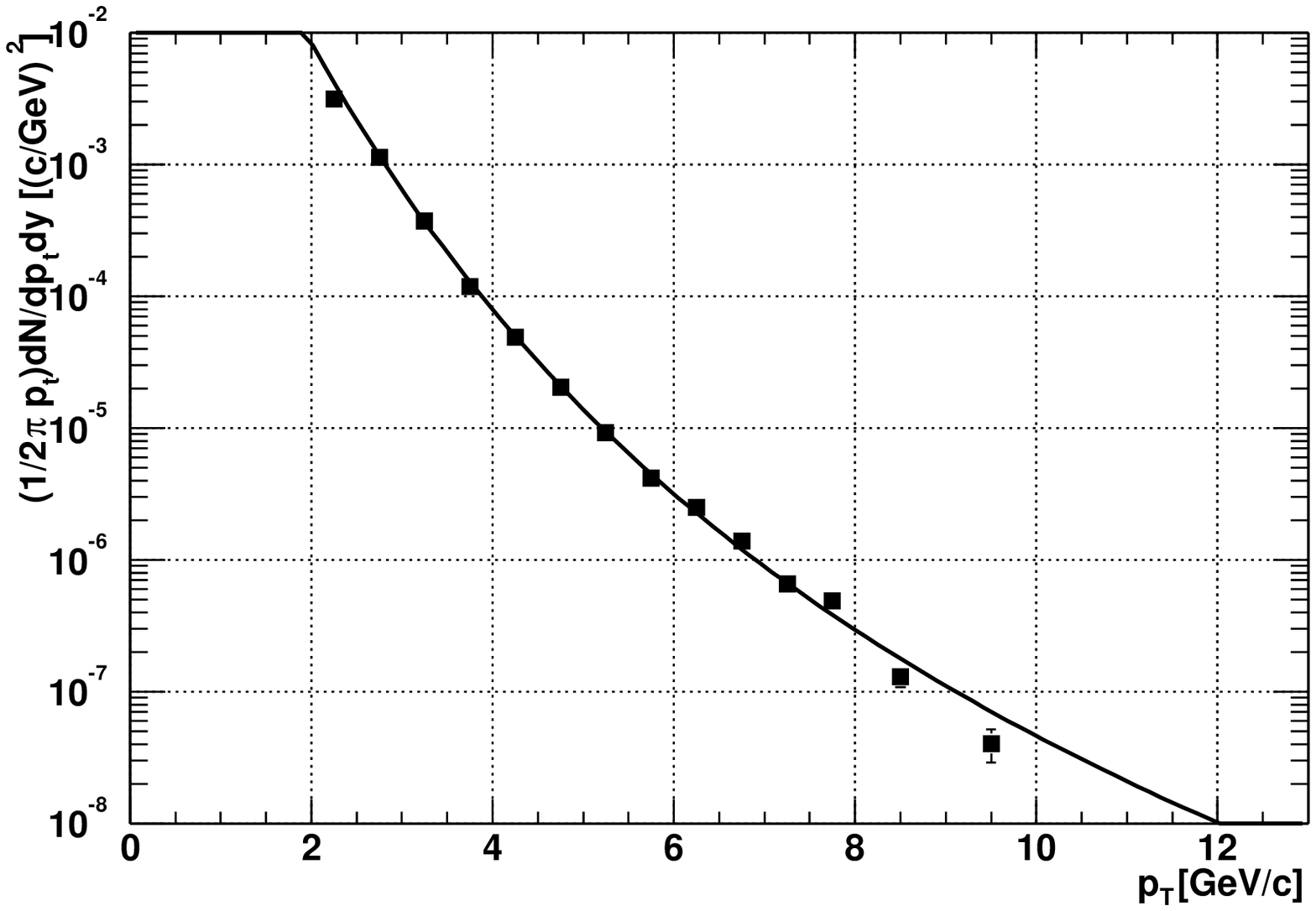,width=0.45\linewidth,clip}
\epsfig{figure=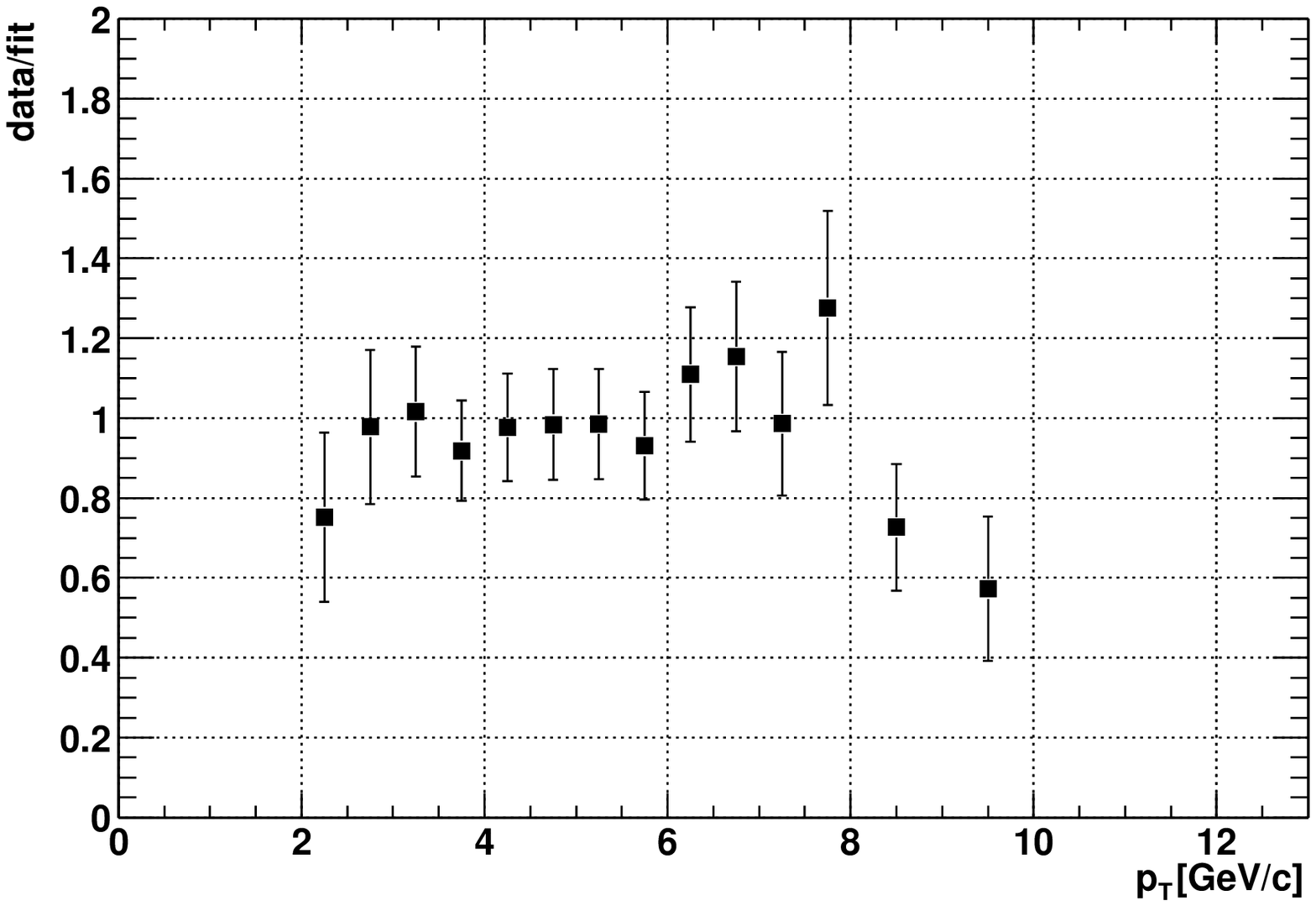,width=0.45\linewidth,clip}
\caption{\label{fig:ch4.eta_crosscheck} (Left panel) $\eta$-meson
invariant crossection (points) compared with the parameterization
used in the current cocktail (statistical errors only). (Right
panel) Ratio of data to Cocktail parameterization is shown in the
lower panel (error bars are the quadratic sum of statistical and
systematic errors).}
\end{figure}

 To check whether $m_{T}$ scaling actually holds in PHENIX, we compare
the Cocktail for $\eta$ meson invariant crossection with Run03
$p+p$ $\eta$ measurements~\cite{ana333}. The results of this
comparison are shown in Fig.~\ref{fig:ch4.eta_crosscheck}.  In the
measured $p_{T}$ range, the agreement between Cocktail and data is
perfect, thereby validating our initial assumptions.

\pagebreak

\subsubsection{Photon conversions}

The estimate of the level of photon conversion is the principal source
of systematic error in the current analysis. The treatment of
conversion level depends on how accurately we represent the material
in the PHENIX acceptance in our simulation. Very detailed studies were
done for the $Au+Au$ single electron analysis~\cite{ppg035,ana305} by
matching the conversion rate in Data and Simulation. The conclusion
from those studies was that the Data and the representation of the
material in the PISA simulation were consistent within the $\approx
8\%$ systematic error.

A full PISA simulation of $63\cdot 10^6$ generated $\pi^0$ decays
was used to estimate the ratio of $\pi^0$ Conversion to $\pi^0$
Dalitz electron ratio. The decay mode of each registered electron
could be determined by the displacement of the Monte Carlo decay
vertex in the $XY$ plane $d_e$. During track reconstruction, all
the particles are assumed to have been emitted from the
interaction point.  Thus the momentum of off-vertex conversions
electrons is slightly mis-measured. The amount of momentum shift
increase linearly with $d_e$, eventually causing electrons to fail
the eID cuts (mainly $d_{EMC}< 3$ cut).

\begin{figure}[t]
\centering
\epsfig{figure=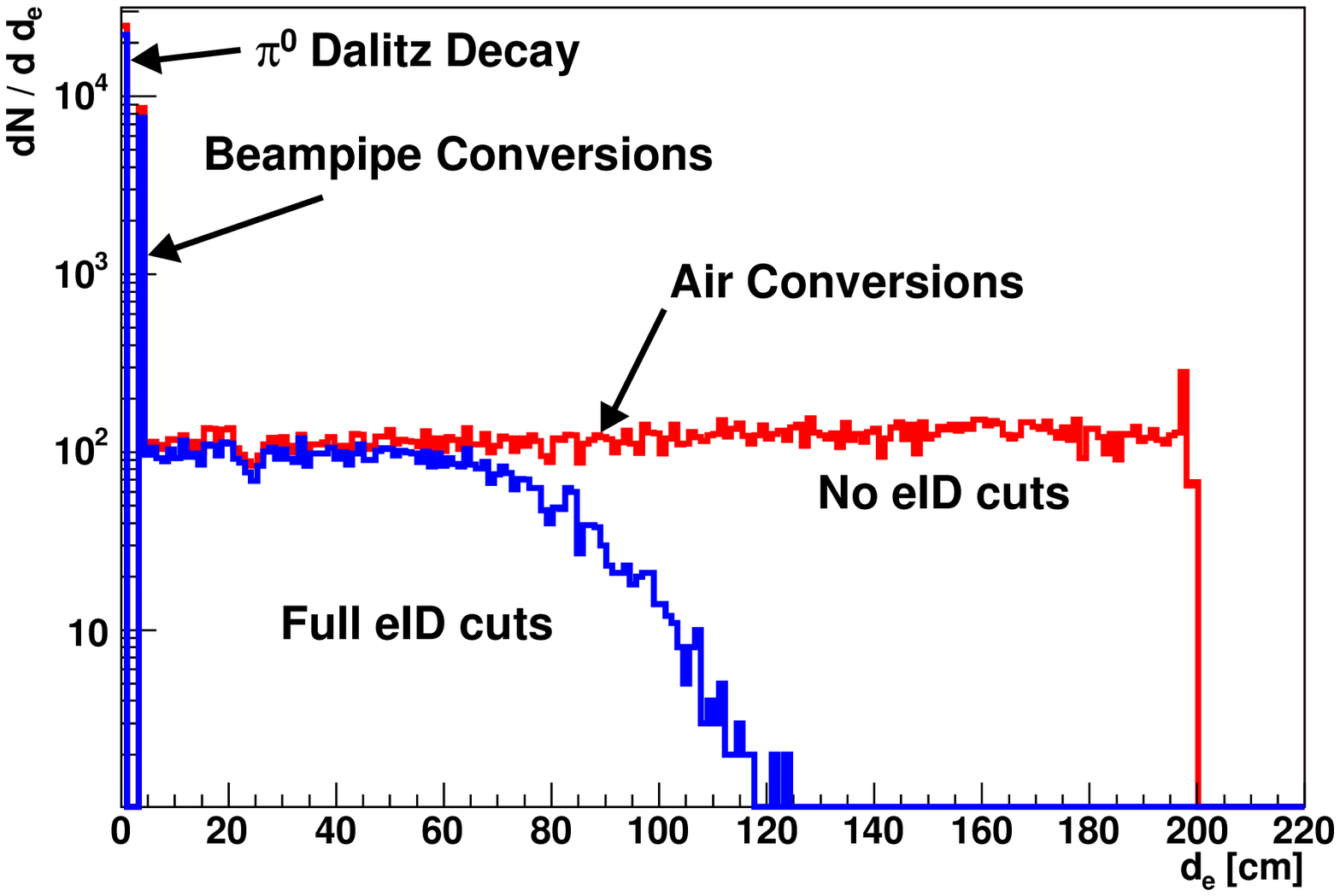,width=0.75\linewidth,clip}
\caption{\label{fig:ch4.vertex_disp} Electron rate as a function
of vertex displacement $d_e$ for $pi^0$ decay electrons with (blue
curve) and without (red curve) the electron ID cut. Contributions
from Dalitz, Beam pipe and air conversions indicated by arrows.}
\vspace*{0.2in}
\epsfig{figure=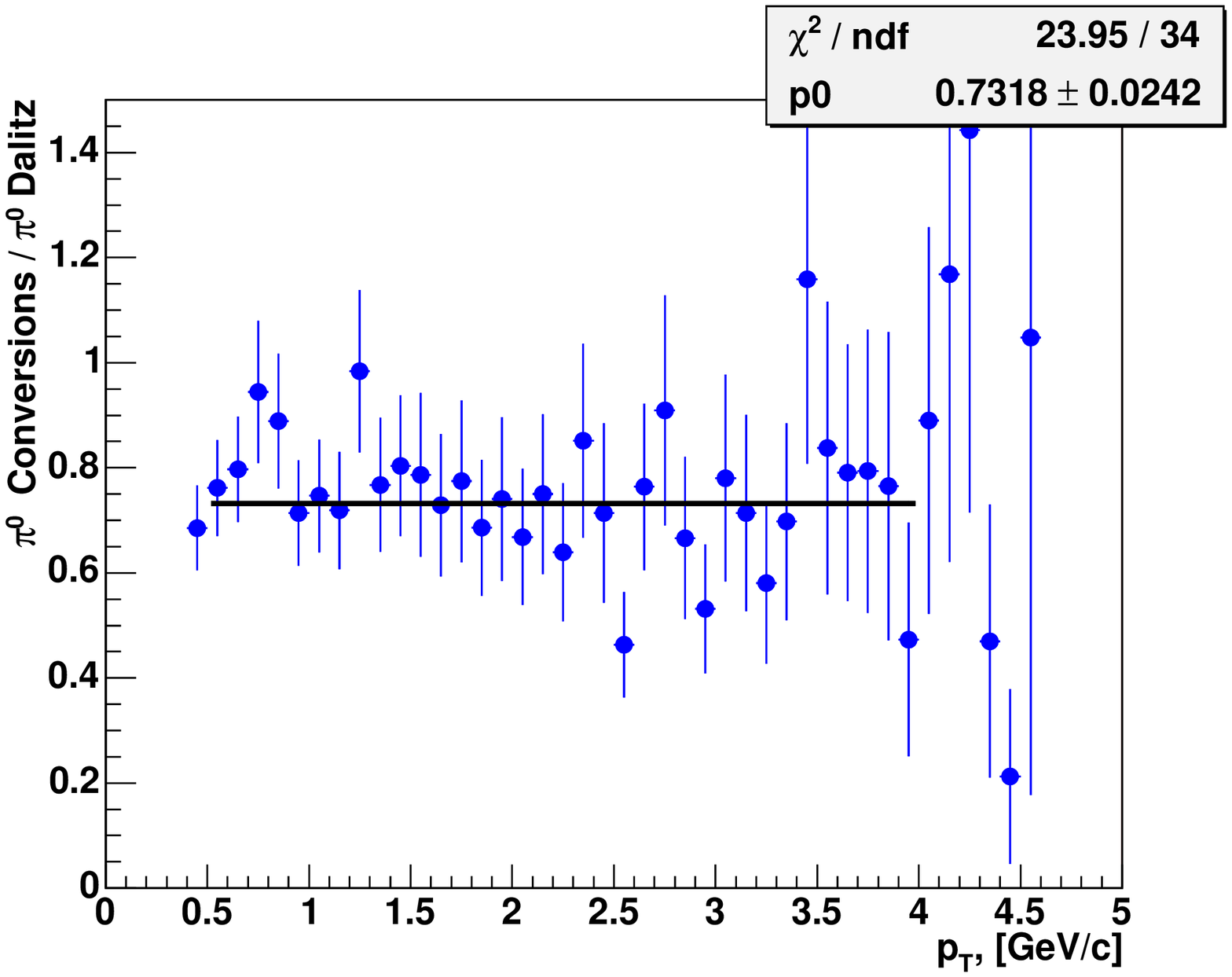,width=0.75\linewidth,clip}
\caption{\label{fig:ch4.conv_dalitz} Ratio of $\pi^0$ conversion
electrons to $\pi^0$ Dalitz electrons as a function of electron
$p_T$ fitted with a constant.}
\end{figure}

The main contributors to the overall conversion rate in PHENIX
are: (1) beryllium $\bold{beam\ pipe}$ - radiation length $X_{bp}
\approx 0.3 \%$~\cite{Tsai}, (2) 200 cm of $\bold{air}$ -
radiation length $X_{air} \approx 0.7 \%$~\cite{Tsai}. Beam pipe
conversions happen very close to vertex $r_{bp} \approx 4$ cm and
do not cause any significant reconstruction efficiency loss.
Conversely, air conversions far from vertex will be rejected
significantly by the electron ID cuts.
Fig.~\ref{fig:ch4.vertex_disp} shows the contributions of
different leptonic modes of $\pi^0$ decay from the full PISA
simulation as a function of decay vertex displacement with and
without eID cuts.  One can see that after $d_e > 60$ cm we have a
significant drop in the electron registration efficiency due to
the eID cuts. This means that we only reconstruct conversion
electrons in \linebreak $\approx 60 - 70$ cm of air. This reduces
the effective radiation length of air to $X_{air}^{eff} \approx
0.28 \%$.

The ratio of conversion electrons to Dalitz can be estimated from
PISA simulation and is shown in Fig.~\ref{fig:ch4.conv_dalitz}.
Realistic assumptions for the input pion distribution are applied
to the simulation input. As one can see this ratio is $p_T$
independent.  A constant fit gives a value of $\frac{\pi^0
Conv}{\pi^0 Dalitz} = 0.73 \pm 0.02(stat)$. This value is in a
good agreement with the previously calculated radiation length in
the acceptance. At high $p_T$, a Dalitz "effective radiation
length"~\cite{PHENIXCDR,Tsai} should be equal to $X_{Dalitz}
\approx 0.6\% \cdot \frac{9}{7} = 0.77 \%$. Then the ratio of
Conversion to Dalitz radiation lengths gives:
$\frac{X_{Conv}}{X_{Dalitz}} = (X_{bp}+X_{air}^{eff})/X_{Dalitz} =
0.58/0.77 \approx 0.75$, in good agreement with fitted value of
0.73.

\pagebreak

\pagebreak

We obtain the electron contributions from photon conversions in
the Cocktail by multiplying the electron spectra from any given light
meson Dalitz decays with a constant factor. For $\pi^0$ this
factor is 0.73. For other light mesons, this factor is corrected
for ratio of the branching ratios BR($\gamma\gamma$)/BR(Dalitz) of
the meson, as compared to the same ratio for $\pi^0$. For the $\eta$
this ratio is 65.7 while for $\pi^0$ it is 82.5.

\subsubsection{Direct photons}

\begin{figure}[h]
\centering
\epsfig{figure=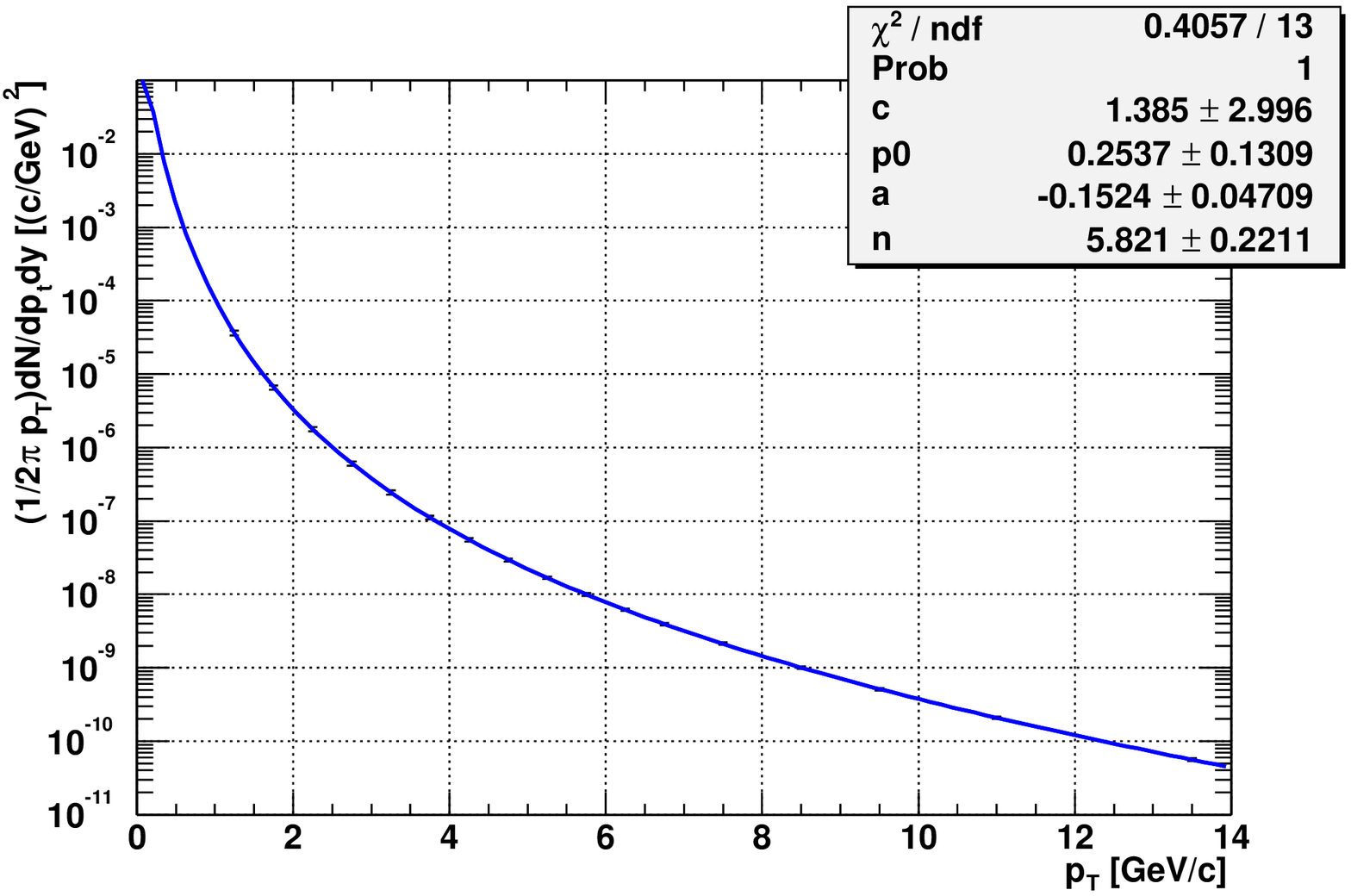,width=0.7\linewidth,clip}
\caption{\label{fig:ch4.direct_NLO} Modified power law fit to
direct photon yield NLO prediction~\cite{Vogelsang}.}
\end{figure}

Direct photon conversion contributions can also be a significant
source of background electrons. Direct photon radiation was measured
in PHENIX~\cite{ppg049,ana325} and agrees within systematic error with
NLO pQCD predictions~\cite{Vogelsang}. We use the NLO crossection as
an input for the analysis. Fig.~\ref{fig:ch4.direct_NLO} shows
modified power law fit (Eq.~\ref{eq:ch4.direct_fit_form}) to the direct
photon invariant yield~\cite{Vogelsang}. The fit function and
parameters are listed below:

\begin{equation}
E \frac{d\sigma}{dp_T^3} = \frac{c}{(e^{-a\cdot p_T} + p_T/p_0)^n}
\label{eq:ch4.direct_fit_form}
\end{equation}

where $c= 1.385$, $p_0= 0.25$, $a= -0.15$, $n= 5.82$.\\

\pagebreak

\subsubsection{Kaon $K_{e3}$ decay}

\begin{figure}[h]
\centering
\epsfig{figure=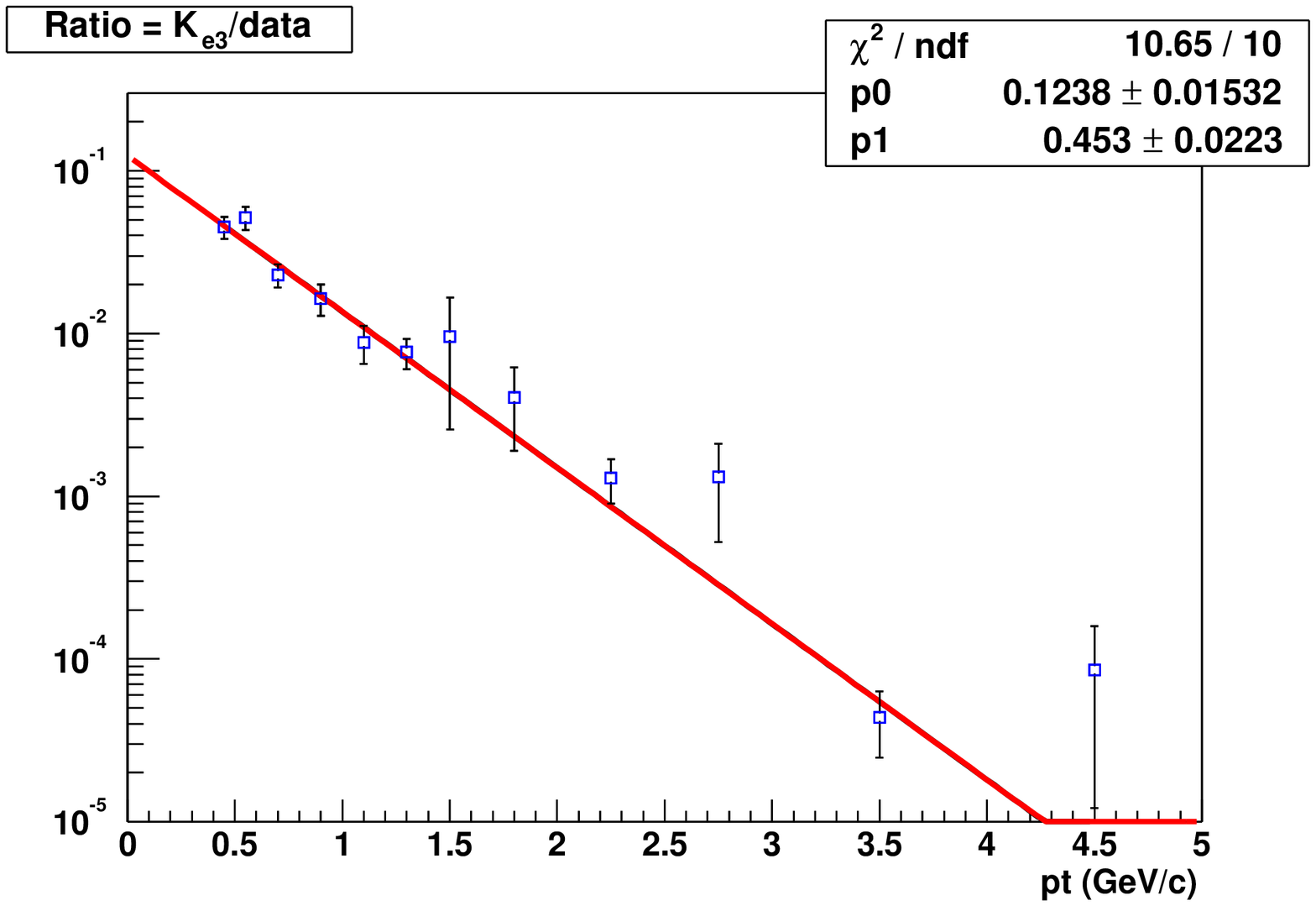,width=0.45\linewidth,clip}
\caption{\label{fig:ch4.Kaon_electron_ratio} Ratio of $K_{e3}$
decay electrons from simulation to inclusive electrons.}
\end{figure}

Kaons have a semi-leptonic decay mode ($K \rightarrow \pi e \nu_e$)
usually refereed to as the $K_{e3}$ decay. The decay length of the
different Kaon states and the $K_e3$ Branching Ratio are summarized in
Table.~\ref{tab:kaon_ke3} ~\cite{PDG}. Kaon decay is $\bold{not}$
"photon related" decay and in principle should not be evaluated as
"Photonic" electron contributor. Any time we talk about "Photonic"
contribution we need to mention whether Kaon decay is included or
not. To be consistent with the main idea of the Cocktail -
simulate explainable "background" electron level, $K_e3$ decay is
$\it{artificially}$ added into the mix (EXODUS does not simulate
Kaon decays!).
\begin{table}[h]
\caption{ $K_{e3}$ decay branching ratio and Decay Length for
different Kaon species.}
\begin{center}
\begin{tabular}[b]{|c|c|c|}
\hline  Decay mode& Decay Length [cm] & Branching Ratio [\%]\\
\hline
$K^{+} \rightarrow \pi^{0} e^{+}
\nu_e$& 371.3&$(4.82\pm 0.06)$ \\
$K^{0}_S \rightarrow \pi^{\pm} e^{\mp}
\nu_e$& 2.679&$(7.2\pm 1.4)\cdot 10^{-4}$ \\
$K^{0}_L \rightarrow \pi^{\pm} e^{\mp}
\nu_e$& 1551&$(38.78\pm 0.28)$ \\
\hline
\end{tabular} \label{tab:kaon_ke3} \end{center}
\end{table}
One can see that $K^{\pm},\ K^0_S$ and $K^0_L$ have completely
different decay rates and need to be treated separately. We used
the full PISA simulation of charged Kaons, $K^0_S$ and $K^0_L$ to
estimate the final electron rate. The spectra were normalized to
the PHENIX Kaon measurements in Run03 $p+p$ at $\sqs = 200$
GeV~\cite{ppg029}. The resulting electron crossection was compared
to the inclusive electron crossection
(see.~\ref{fig:ch4.final_inclusive}). The ratio of $K_{e3}$ decay
electrons to the inclusive is shown in
Fig.~\ref{fig:ch4.Kaon_electron_ratio}. The contribution is
exponentially falling and only relevant at low $p_T < 1$ Gev/c.

\subsection{Final Electron Cocktail}

\begin{figure}[h]
\centering
\epsfig{figure=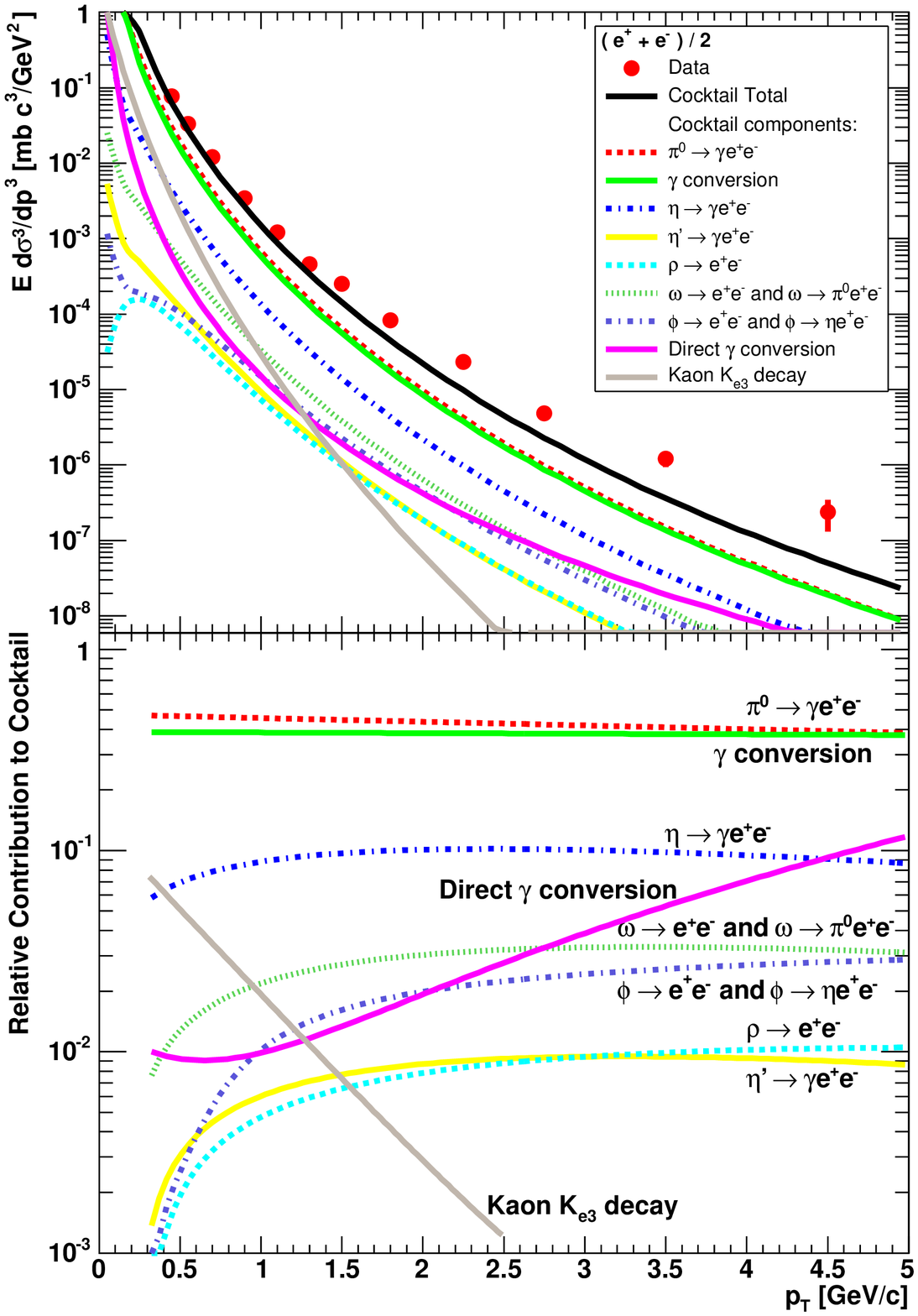,width=0.75\linewidth,clip}
\caption{\label{fig:ch4.final_cocktail} Final electron
``Cocktail'' broken into separate contributions overlaid with the
inclusive electron data. Bottom plot shows the relative
contributions of each Cocktail component to the total.}
\end{figure}

The final electron Cocktail and the breakdown into its
contributions are shown in Fig.~\ref{fig:ch4.final_cocktail}.

The Cocktail was calculated using the $p_T$-dependent BBC trigger
bias corrected charged pion spectrum (see
Section~\ref{sec:ch4.trig_bias}).

\section{"Non-photonic" Electron Crossection}\label{sec:ch4.Subtraction}

This section is devoted to the results of the cocktail subtraction.
Now we have all the ingredients to derive the electron component that
can not be described by leptonic decays of light mesons or photon
conversions in the apparatus material. The subtraction of the Cocktail
"background" was done bin by bin, using the same binning for the
cocktail as for the inclusive data. To take into account the BBC
trigger bias effect, the realistic BBC trigger bias was applied to the
input pion and kaon spectra. Then if we subtract from the non
bias-corrected crossection the non-biased cocktail we remove the
trigger bias effect for "Photonic" electrons. As was mentioned before,
the remaining non-photonic electrons originated primarily from heavy
flavor semi-leptonic decays should have a constant trigger bias as
appropriate for hard partonic interactions. Thus, applying this
correction to the subtraction results, we obtain bias-corrected
non-photonic electron invariant crossection.

\begin{figure}[h]
\centering
\epsfig{figure=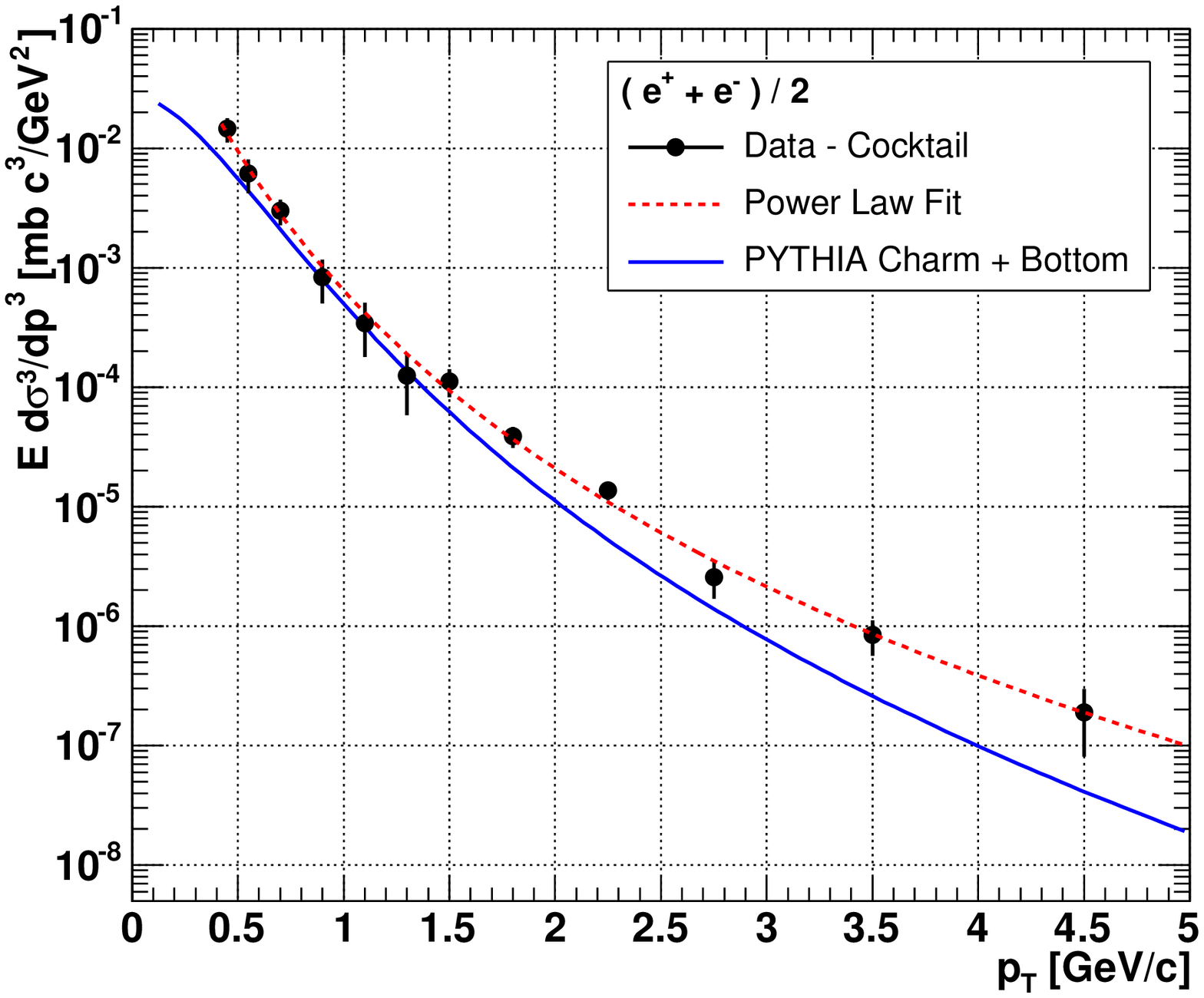,width=0.63\linewidth,clip}
\caption{\label{fig:ch4.final_nonphotonic} Final bin-width
corrected "Non-photonic" electron invariant crossection fitted
with modified power law function $\frac{A}{(p_0+p_{T})^{n}}$
overlaid with PYTHIA default~\cite{ppg011} prediction for Open
Charm + Bottom electron crossection.}
\end{figure}

The fully corrected "Non-photonic" electron invariant crossection is
shown in Fig.~\ref{fig:ch4.final_nonphotonic}. For comparison
 purposes the default PYTHIA prediction for Open
Charm + Bottom decay electrons (see
Section~\ref{sec:ch5.PYTHIA_comp}) is plotted. The final data points
are listed in Table~\ref{tab:final_nonphotonic}. The modified power
law fit $\frac{A}{(p_0+p_{T})^{n}}$ parameters are listed below:
\begin{itemize}
\item$A = (1.287\pm 1.714)\cdot 10^{-2}$ \item$p_0 = 0.547\pm
0.203$ \item$n = 6.87\pm 0.80$
\end{itemize}

\newpage
\begin{table}[h]
\caption{ Non-photonic electron invariant crossection.} \centering
\begin{tabular}[b]{|c|c|c|c|c|}
\hline $p_{T}$ bin & Non-photonic & Non-photonic & Non-photonic & Relative\\
$[GeV/c]$ & crossection&stat. error& syst. error &stat. \\
& $\cunit$ &$\cunit$&$\cunit$&error [\%]\\ \hline
0.4-0.5 &1.453e-02 &3.322e-03 &1.140e-02& 22.9 \\
0.5-0.6 &6.142e-03 &1.923e-03 &5.006e-03& 31.3 \\
0.6-0.8 &2.986e-03 &7.183e-04 &1.829e-03& 24.1 \\
0.8-1.0 &8.341e-04 &3.340e-04 &5.339e-04& 40.0 \\
1.0-1.2 &3.441e-04 &1.659e-04 &1.895e-04& 48.2 \\
1.2-1.4 &1.251e-04 &6.671e-05 &7.211e-05& 53.3 \\
1.4-1.6 &1.119e-04 &2.952e-05 &3.735e-05& 26.4 \\
1.6-2.0 &3.875e-05 &7.555e-06 &1.230e-05& 19.5 \\
2.0-2.5 &1.371e-05 &2.380e-06 &3.428e-06& 17.4 \\
2.5-3.0 &2.581e-06 &8.948e-07 &7.133e-07& 34.7 \\
3.0-4.0 &8.401e-07 &2.732e-07 &1.779e-07& 32.5 \\
4.0-5.0 &1.895e-07 &1.088e-07 &3.535e-08& 57.4 \\
\hline \end{tabular} \label{tab:final_nonphotonic}
\end{table}

\pagebreak
\section{Converter Subtraction Method}\label{sec:ch4.Converter}

The common alternative to a cocktail-subtraction analysis is the
convertor-subtraction analysis.  A converter subtraction analysis
was previously and successfully used to obtain the non-photonic
electron crossection in \linebreak Au+Au
~\cite{ana158,ana305,ana324}, d+Au~\cite{ana259}, and Run03
p+p~\cite{ana321} collisions. This method provides accurate
results, but is limited by the statistics of the conversion
sample. The main idea is to obtain the photon related electron
component by adding additional photon converter material with
$\bold{known\ radiation\ length}$ \linebreak ($X_C = 1.67$ \%)
close to the PHENIX interaction point for a portion of the Run
period.

By subtracting the inclusive electron crossection of the
Non-converter run period from the one from Converter run period we
directly obtain the "Converter" electron component, due to the
additional material in the acceptance. Since the amount of
material in PHENIX aperture ($X_{PHENIX}$) can be estimated, we
can scale the "Converter" component to measure the "Photonic"
component of the electron crossection and use this in place of the
Cocktail prediction. The steps required for convertor analysis
are:
\begin{itemize}
\item{We measure $\bold{biased}$ crossection in Converter and
Non-converter run periods (denote it as $N^C$ and $N^{NC}$).
\begin{eqnarray}
N^{NC} &=& N^{NC}_{P} + N^{NC}_{NP}\\ \nonumber
 N^{C} &=& R\cdot N^{C}_{P} +
N^{C}_{NP} \label{eq:ch4.conv_non_conv}
\end{eqnarray}
where index $P$ refers to "Photonic" component of crossection,
$NP$ - to "Non-photonic" component and $R$ is a factor,
representing the additional amount of converting material due to
photon converter installation}
\item{We account for the differences of BBC trigger bias for "Photonic" and "Non-photonic" electrons,
denoted as $\epsilon_P$ and $\epsilon_NP$ as:
from~\ref{eq:ch4.conv_non_conv}
\begin{eqnarray}
N^{unbiased}_P &=& \frac{1}{\epsilon_P}N_P =
\frac{N^{C}-N^{NC}}{\epsilon_P\cdot (R-1)}
\\
\nonumber
 N^{unbiased}_{NP} &=& \frac{1}{\epsilon_N}N_{NP} =
\frac{R\cdot N^{NC}-N^{C}}{\epsilon_N\cdot (R-1)}
\label{eq:ch4.ph_nph}
\end{eqnarray}}
\end{itemize}

The statistics of the Converter run is a limiting factor for this
analysis and unfortunately for Run02 the total Converter run
statistics(see Table~\ref{tab:runqatable}) does not allow us to make
an accurate measurement of the "Non-photonic" component.  Nonetheless,
we can use this analysis for a consistency crosscheck of the Cocktail
subtraction results.

\subsection{Converter and Non-converter
run group acceptance comparison}

\begin{figure}[hb]
\centering
\epsfig{figure=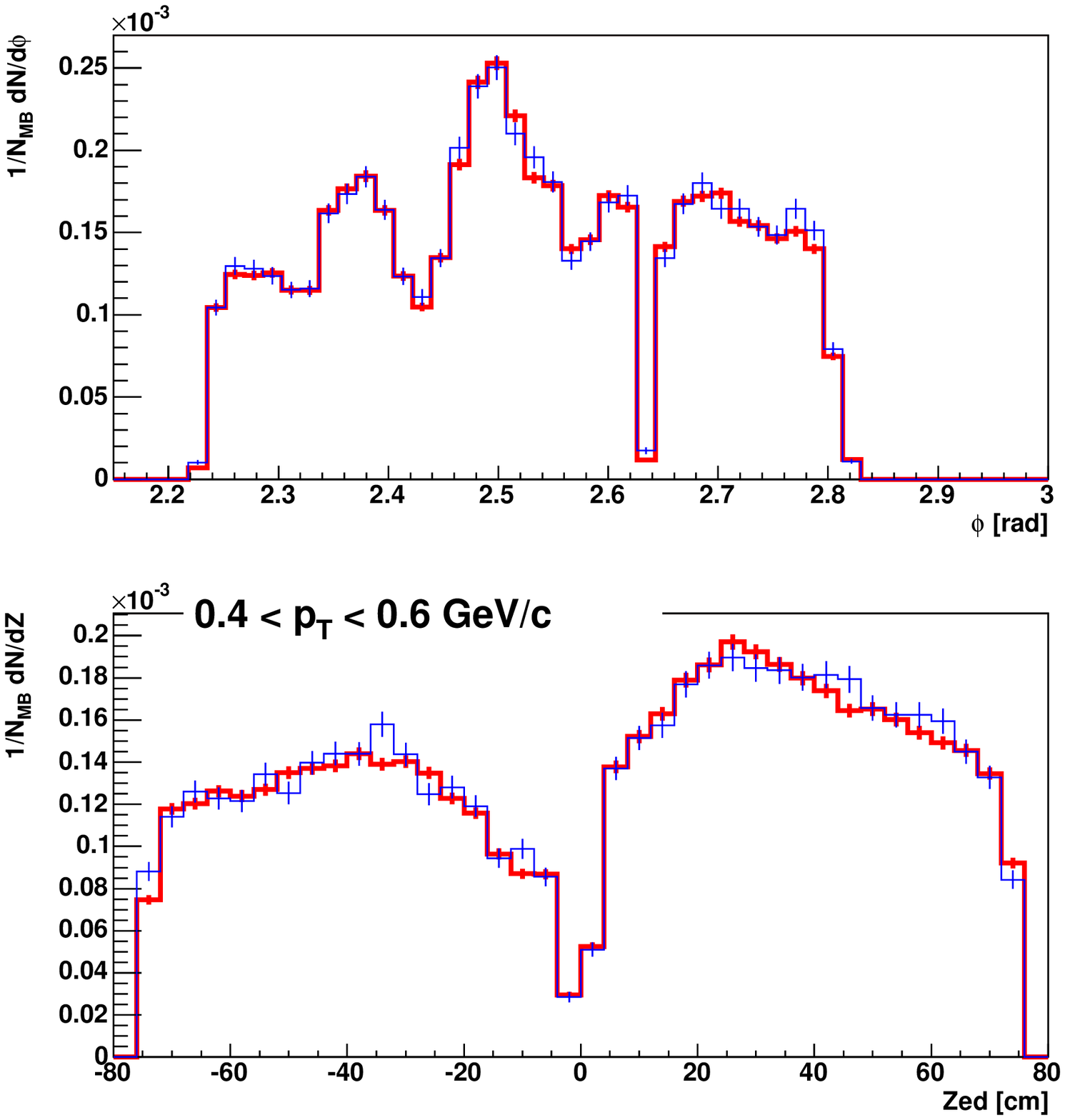,width=0.48\linewidth,clip,trim =
0in 0in 0.5in 0.2in}
\epsfig{figure=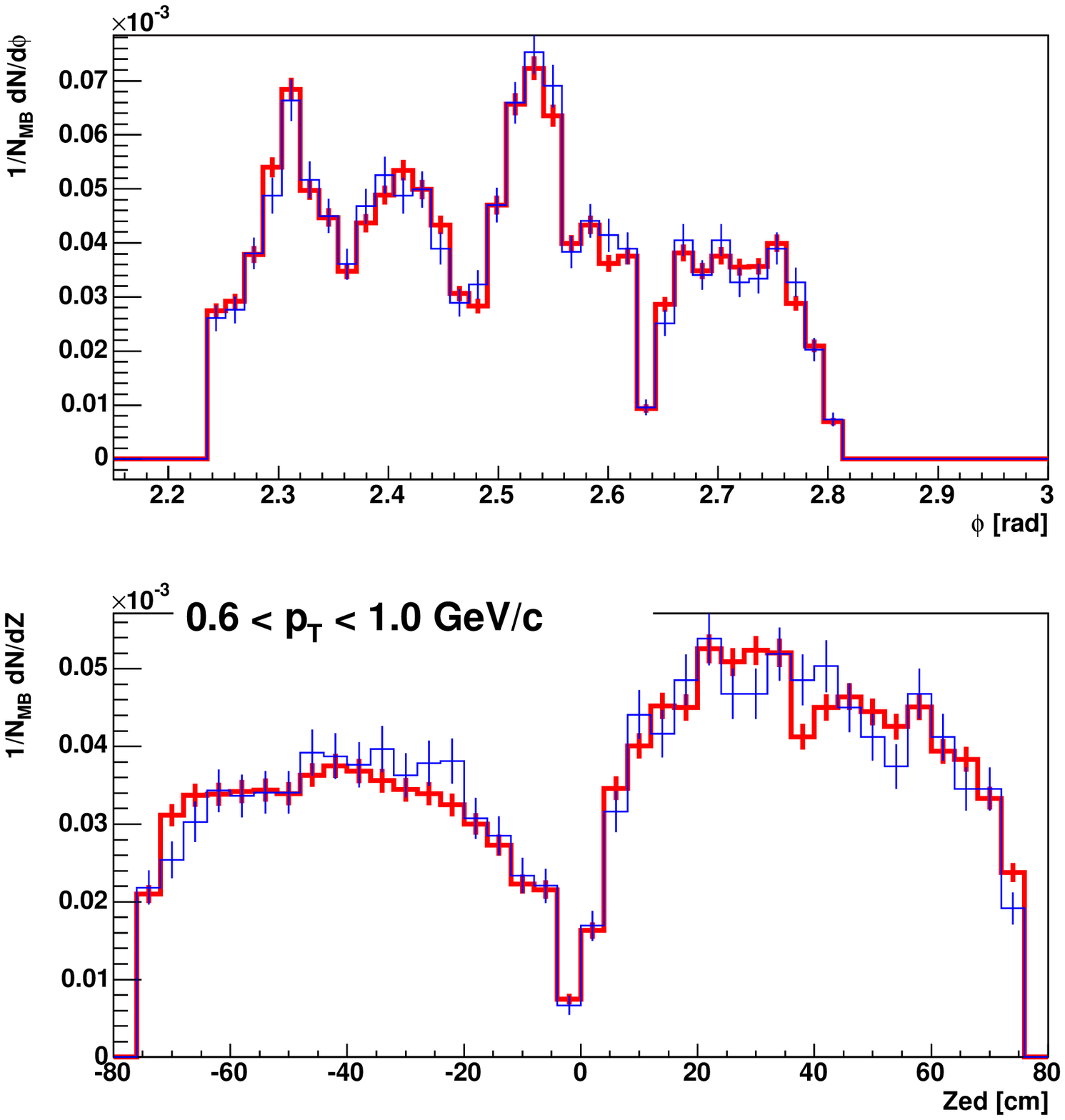,width=0.48\linewidth,clip,trim =
0in 0in 0.5in 0.2in}
\epsfig{figure=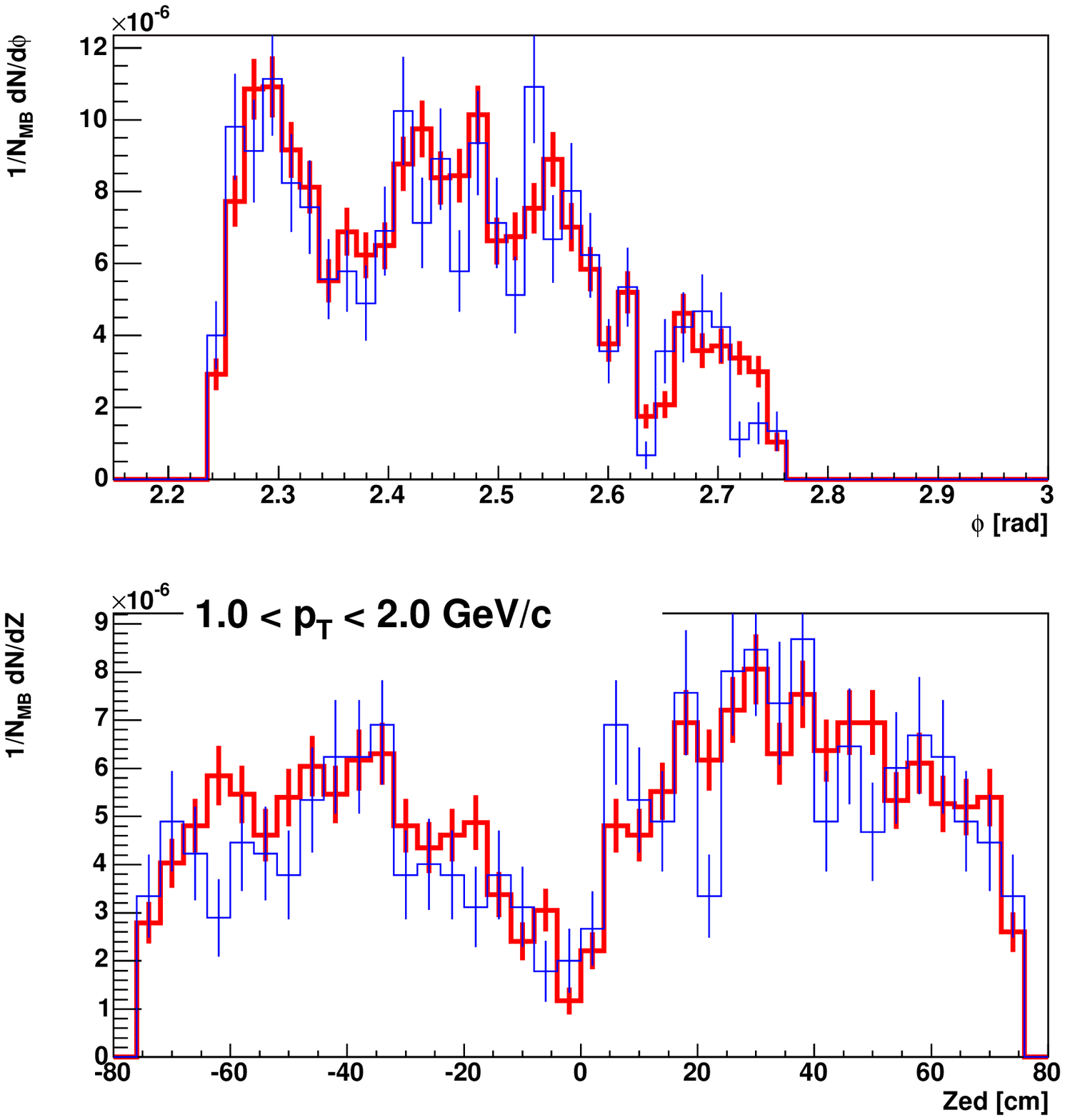,width=0.48\linewidth,clip,trim =
0in 0in 0.5in 0.2in}
\epsfig{figure=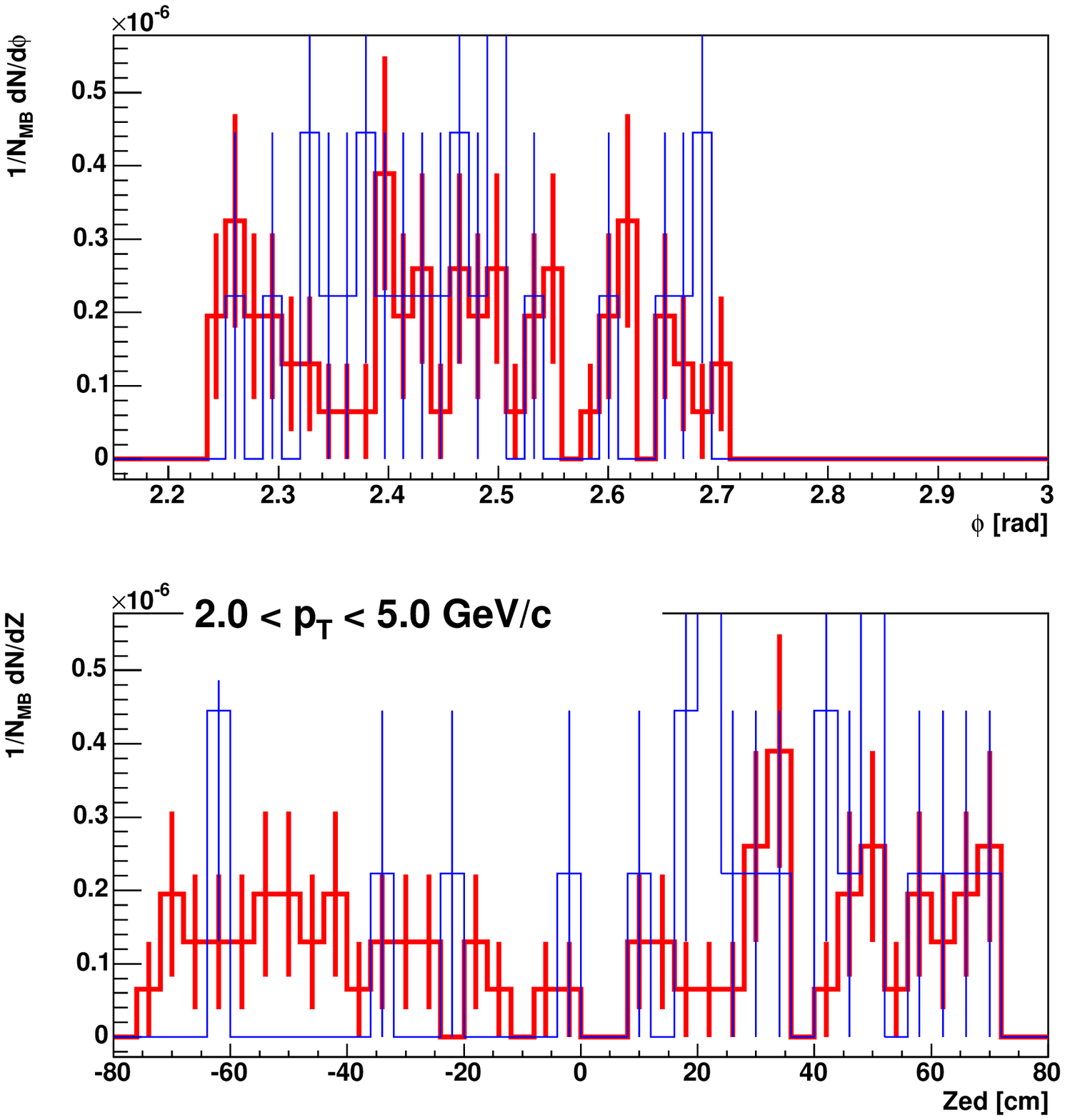,width=0.48\linewidth,clip,trim =
0in 0in 0.5in 0.2in} \caption{\label{fig:ch4.comp_conv_nonconv}
Comparison of $\phi$ and $Z$ acceptance of charged tracks in
Converter (thin) and Non-converter (thick) run period for
different $p_T$ bins. Minimum Bias sample, full electron ID cuts
(except $n0>1$).}
\end{figure}

Acceptance of Converter and Non-converter run period was done by
comparing $\frac{1}{N_{MB}}\frac{dN}{d\phi}$,
$\frac{1}{N_{MB}}\frac{dN}{dZ}$ distributions (normalized by the
number of MB events) for all charged tracks with standard eID cuts
(except $n0>1$). The results are shown in
Fig.~\ref{fig:ch4.comp_conv_nonconv} for different $p_{T}$ ranges.
One can see very nice agreement in the acceptance of both run
groups.

\begin{figure}[hb]
\centering
\epsfig{figure=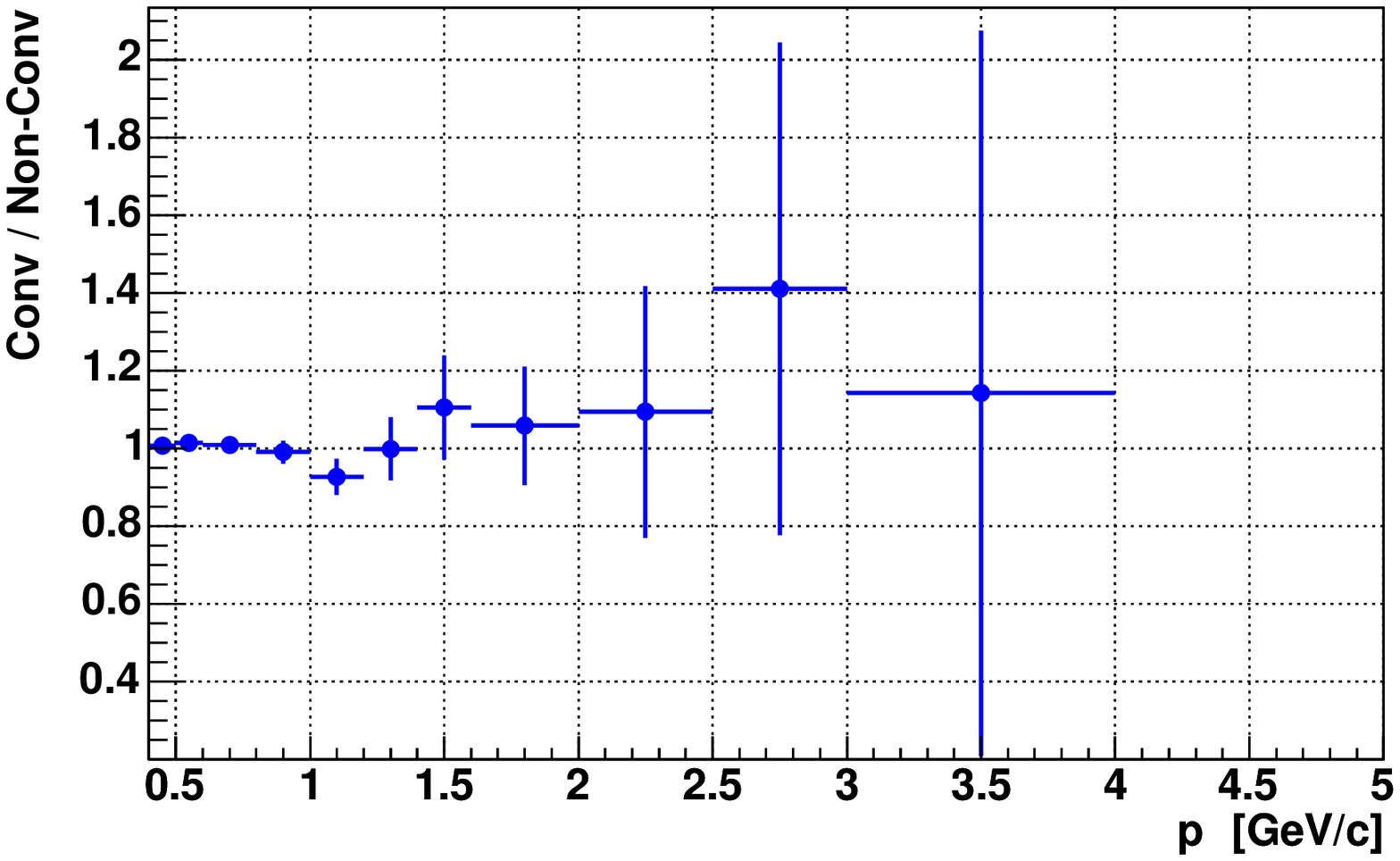,width=0.7\linewidth,clip,trim = 0in
0in 0.5in 0.2in}
 \caption{\label{fig:ch4.comp_conv_nonconv_pt}
Ratio of "raw" $\frac{1}{N_{MB}} \frac{dN}{dp_{T}}$ distributions
for Converter and Non-Converter run groups. Charged tracks, full
electron ID cuts (except $n0>1$).}
\end{figure}

We also compared $\frac{1}{N_{MB}} \frac{dN}{dp_{T}}$ distribution
for two run groups. Fig.~\ref{fig:ch4.comp_conv_nonconv_pt} shows
the ratio of $\frac{1}{N_{MB}} \frac{dN}{dp_{T}}$ distribution of
Converter run to the same distribution of Non-converter run. Ratio
is consistent with one within the statistical error. The increase
at high $p_{T}$ values may indicate that the relative contribution
of the electrons to the charged particles becomes significant and
we observe the increase of the electron production due to the
additional photon conversions in photon converter material.

\subsection{Inclusive electron crossection in Converter run}

\begin{figure}[t]
\centering
\epsfig{figure=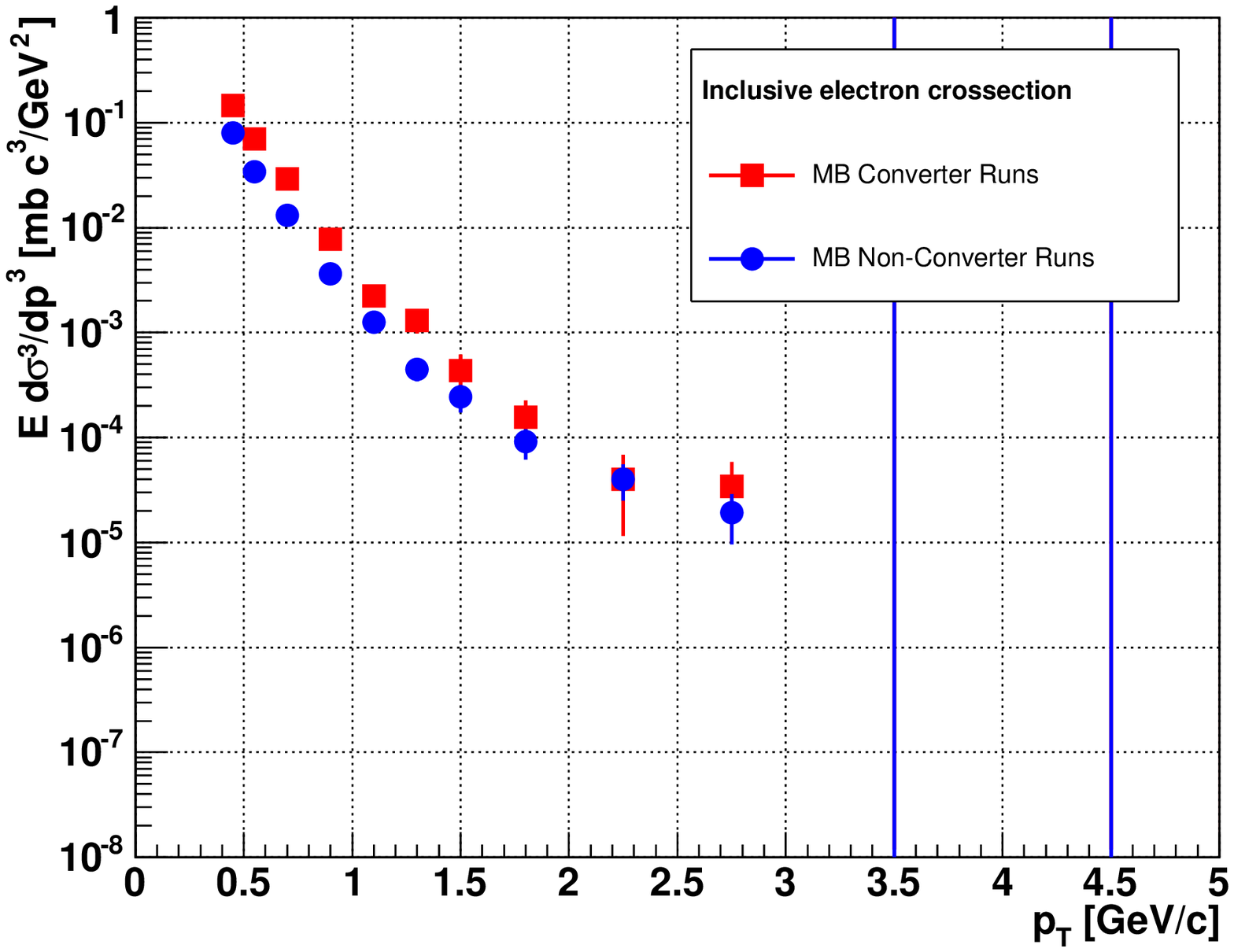,width=0.48\linewidth,clip,trim
= 0in 0in 0.5in 0.2in}
\epsfig{figure=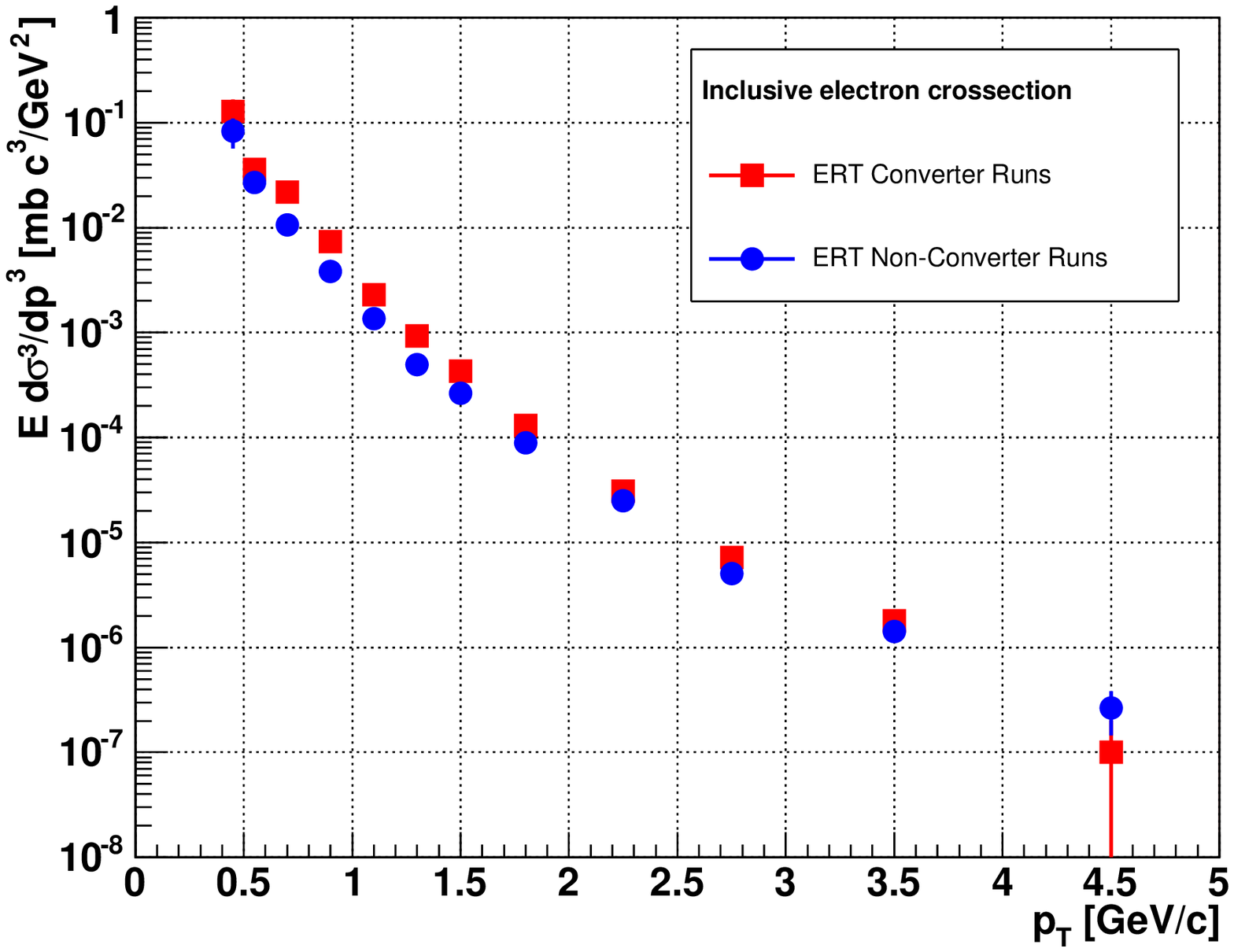,width=0.48\linewidth,clip,trim
= 0in 0in 0.5in 0.2in}
 \caption{\label{fig:ch4.incl_conv_nonconv}
Inclusive electron crossection for MB (left) and ERT trigger
(right) for two run groups: Converter (squares) and Non-converter
(circles). ERT sample have only statistical errors applied. }
\end{figure}

Inclusive electron crossection for MB and ERT sample for Converter
runs was obtained the same way as for Non-converter run analysis
(see Section~\ref{sec:ch4.Inclusive}). Subtraction of the
converter component should be performed separately for MB and ERT
sample in order not to double-count the ERT trigger efficiency
systematic error \footnote{Here we use an assumption that we only
do not know the absolute shape of ERT trigger efficiency and apply
the systematic error to the results of subtraction. This is a
valid approximation as acceptance of both run groups is almost
identical.} and resulted photonic and non-photonic electron
crossection should be combined using standard averaging formula
from Eq.~\ref{eq:ch4.weight_avg}.
Fig.~\ref{fig:ch4.incl_conv_nonconv} shows the inclusive electron
invariant crossection for Converter and Non-Converter run group.
As we expected, the electron yield in Converter run is higher due
to additional conversion material.

\subsection{Calculation of $R$}

In order to obtain the "Photonic" and "Non-photonic" crossection
using \linebreak Eq.~\ref{eq:ch4.ph_nph} we need to estimate
factor $R$ Eq.~\ref{eq:ch4.conv_non_conv}. This factor accounts
for the additional electrons produced through conversions in
Photon Converter material. Usually, this ratio was calculated
through full PISA simulation with photon converter implemented. We
will try to get the best estimation for this variable from the
first principles. From previous $Au+Au$ analysis
~\cite{ana305,ppg035} we know the radiation length of the
Converter material $X_C = (1.67\pm 0.02) \%$. The "$\it{effective\
radiation\ length}$"\footnote{Effective amount of material that
need to be added to create the same conversion electron yield as
from Dalitz $\pi^0$ decay.~\cite{PHENIXCDR}} of the $\pi^0$ Dalitz
decay can be approximated for $E > 1.0 $ GeV photons as
$X_{Dalitz}^{\pi^{0}} =0.6\% \cdot 9/7 = 0.77 \%$
~\cite{PHENIXCDR,Tsai}. Then we can derive $R$ the following way:

\begin{equation}
R = \frac{X_C +
X_{Dalitz}^{\pi^{0}}+X_{Convers}}{X_{Dalitz}^{\pi^0}+X_{Convers}}
= 1 + \frac{X_C}{X_{Dalitz}^{\pi^0}\cdot
(1+\frac{X_{Convers}}{X_{Dalitz}^{\pi^0}})}
\\
\label{eq:ch4.R}
\end{equation}

\begin{figure}[h]
\begin{tabular}{lr}
\begin{minipage}{0.5\linewidth}\centering
\epsfig{figure=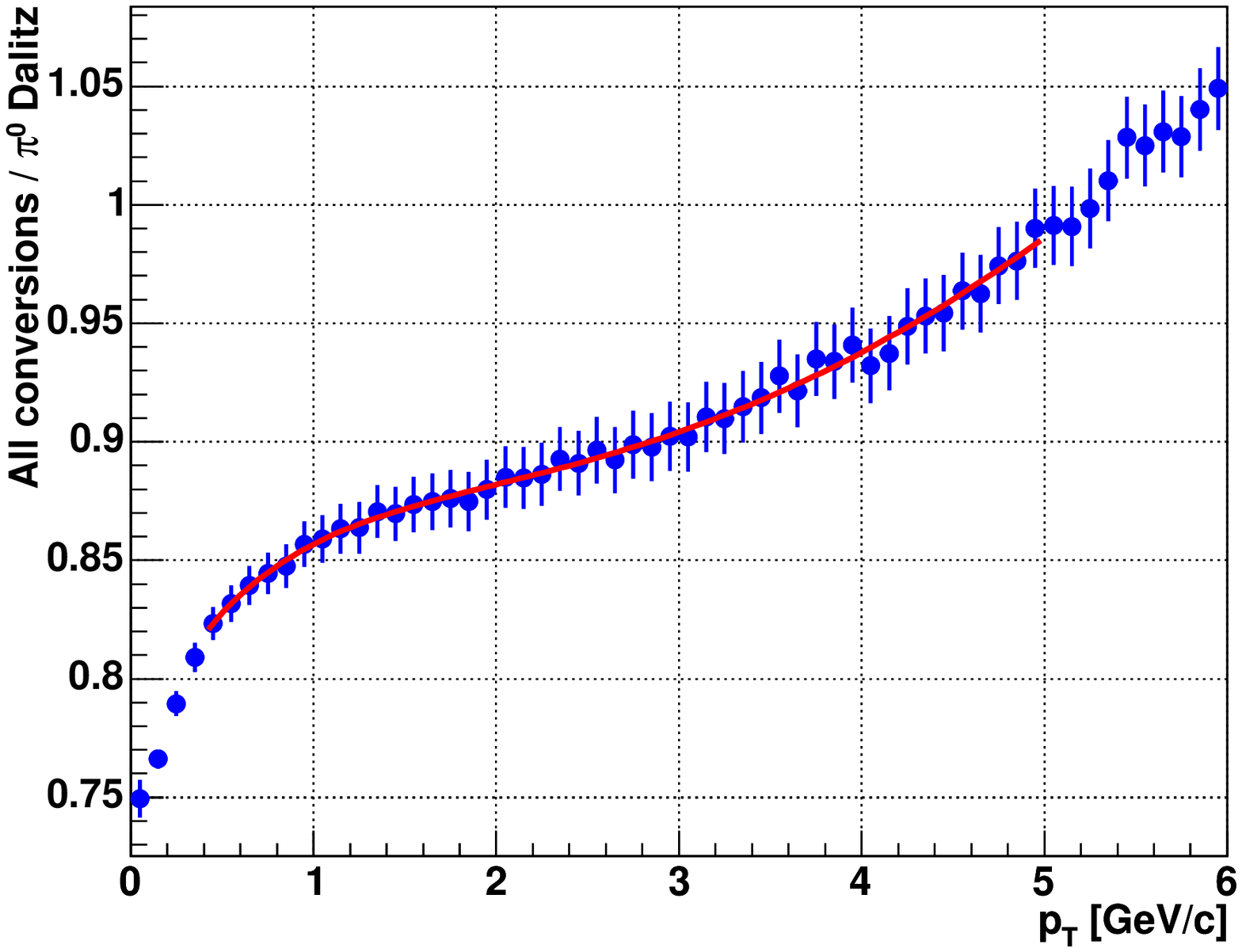,width=1\linewidth,clip,trim =
0in 0in 0.5in 0.2in}
 \caption{\label{fig:ch4.all_conv_dalitz}
Ratio of EXODUS Cocktail electrons from all conversions sources to
$\pi0$ Dalitz electrons. Fit to the data by second order
polynomial - exponential.}
\end{minipage}
&
\begin{minipage}{0.5\linewidth} \centering \epsfig{figure=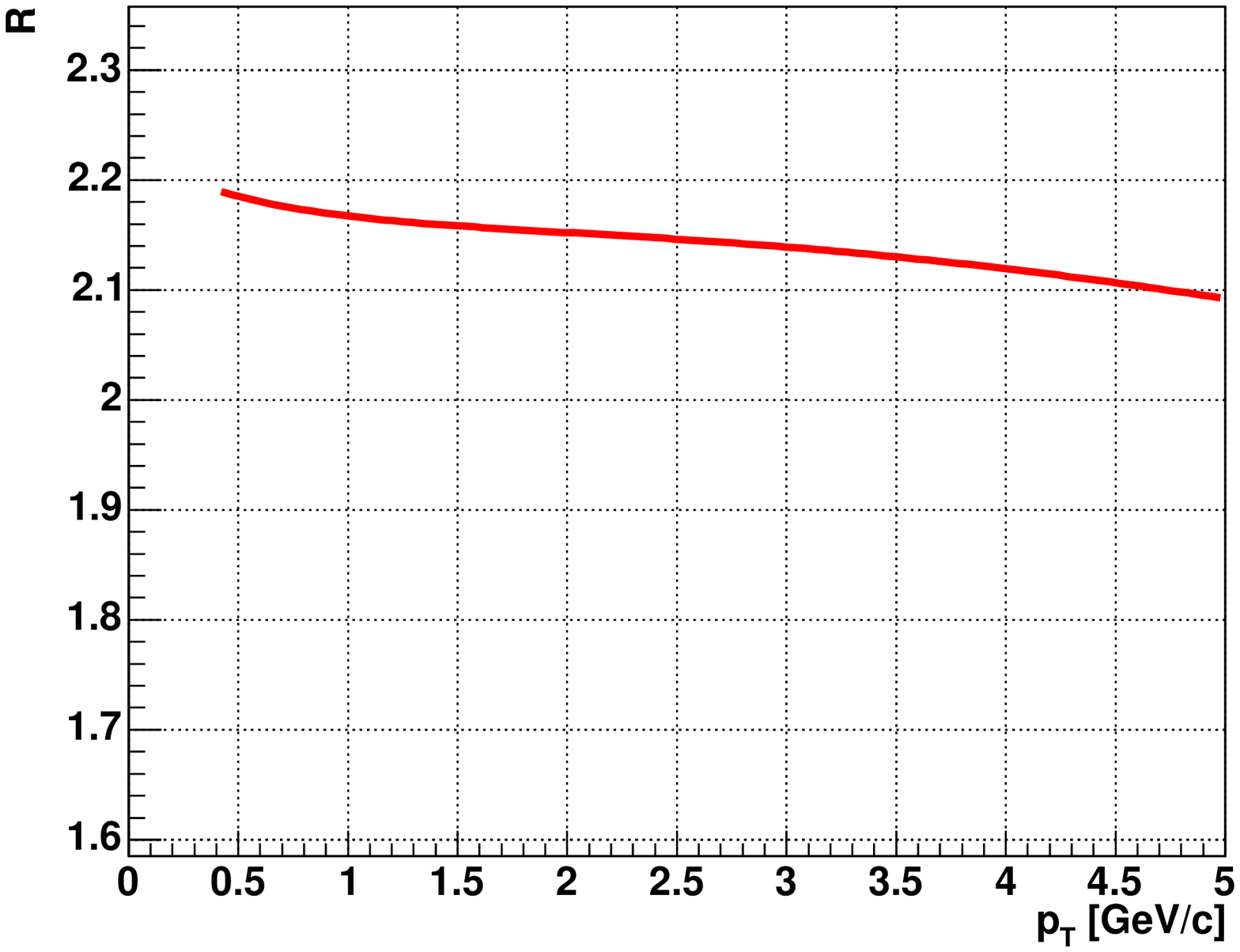,width=1\linewidth,clip,trim
= 0in 0in 0.5in 0.2in}
 \caption{\label{fig:ch4.R}
Ratio of photonic electron component with photon converter added
to photonic electron component without the converter.}
\end{minipage}
\end{tabular}
\end{figure}

Relative contribution of conversion electrons to $\pi^0$ Dalitz
electrons $\frac{X_{Convers}}{X_{Dalitz}^{\pi^0}}$ can be obtained
from Cocktail simulation. The ratio of all conversions to Dalitz
is shown in Fig.~\ref{fig:ch4.all_conv_dalitz} fitted with
arbitrary function.One can see that the ratio is higher then 0.73
value from Section~\ref{sec:ch4.Cocktail} as it also includes
photon conversions of $\eta$ and other light mesons.

Upon substitution the fit results of
$\frac{X_{Convers}}{X_{Dalitz}^{\pi^0}}$ into Eq.~\ref{eq:ch4.R},
obtain we the expression for $R$ as a function of electron $p_T$
shown in Fig.\ref{fig:ch4.R}.

\begin{equation}
R = 1 + \frac{1.67 \%}{0.77 \%\cdot (1.9 - 2.3\cdot
10^{-2}p_{T}+7.9\cdot 10^{-3} p_{T}^2 -e^{-2.0-1.5\cdot p_{T}})}
\\
\label{eq:ch4.R_sub}
\end{equation}

\subsection{"Photonic" and "Non-photonic" electron component from Converter subtraction method}

\begin{figure}[ht]
\centering
\epsfig{figure=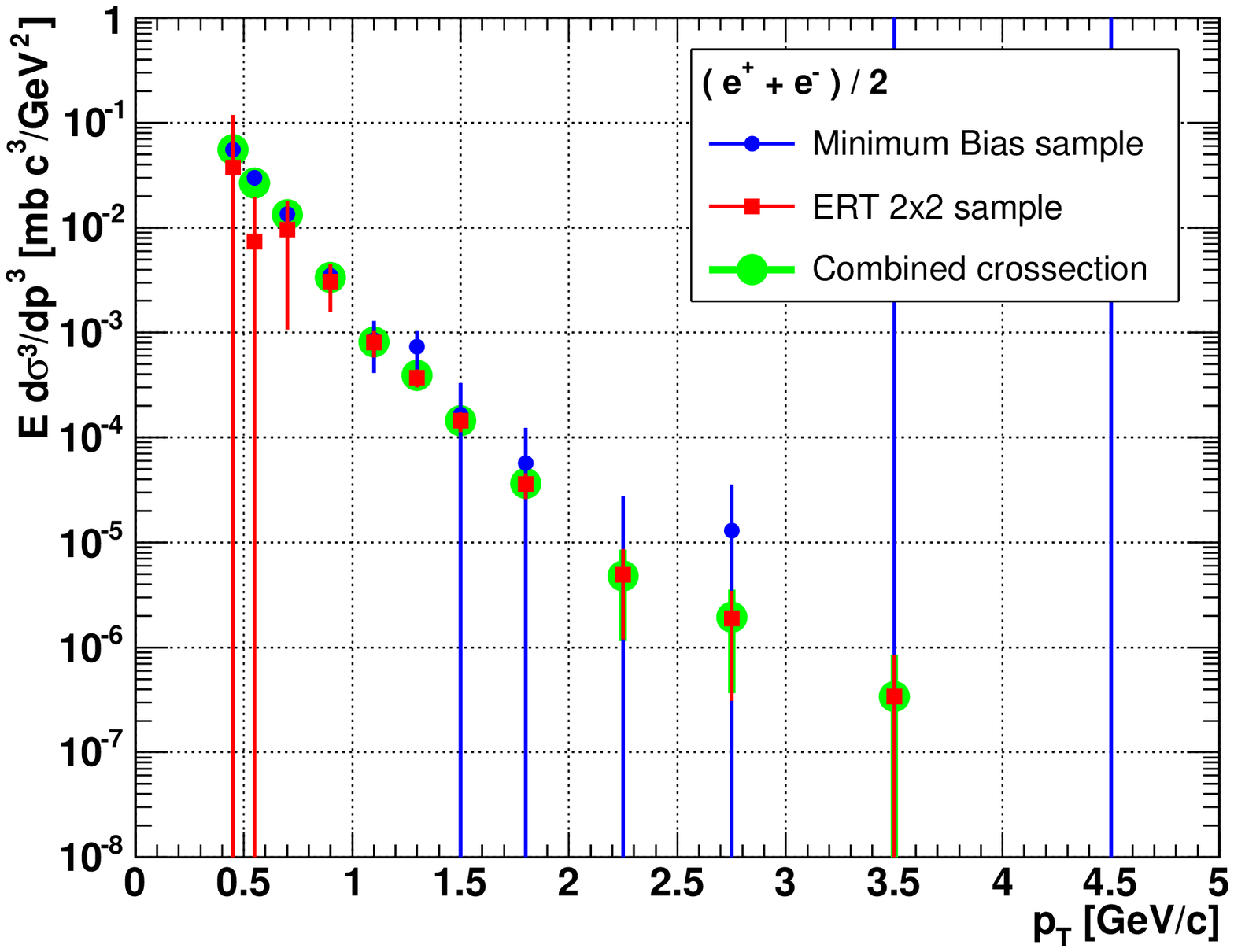,width=0.45\linewidth,clip,trim
= 0in 0.15in 0in 0.5in} \caption{\label{fig:ch4.phot_conv_mb}
"Photonic" electron invariant crossection for Minimum Bias
(circle) and ERT data sample (squares) (ERT statistical errors
includes the systematic error due to ERT trigger efficiency).
Combined by Eq.~\ref{eq:ch4.weight_avg} inclusive electron
invariant crossection (large circle). } \centering
\epsfig{figure=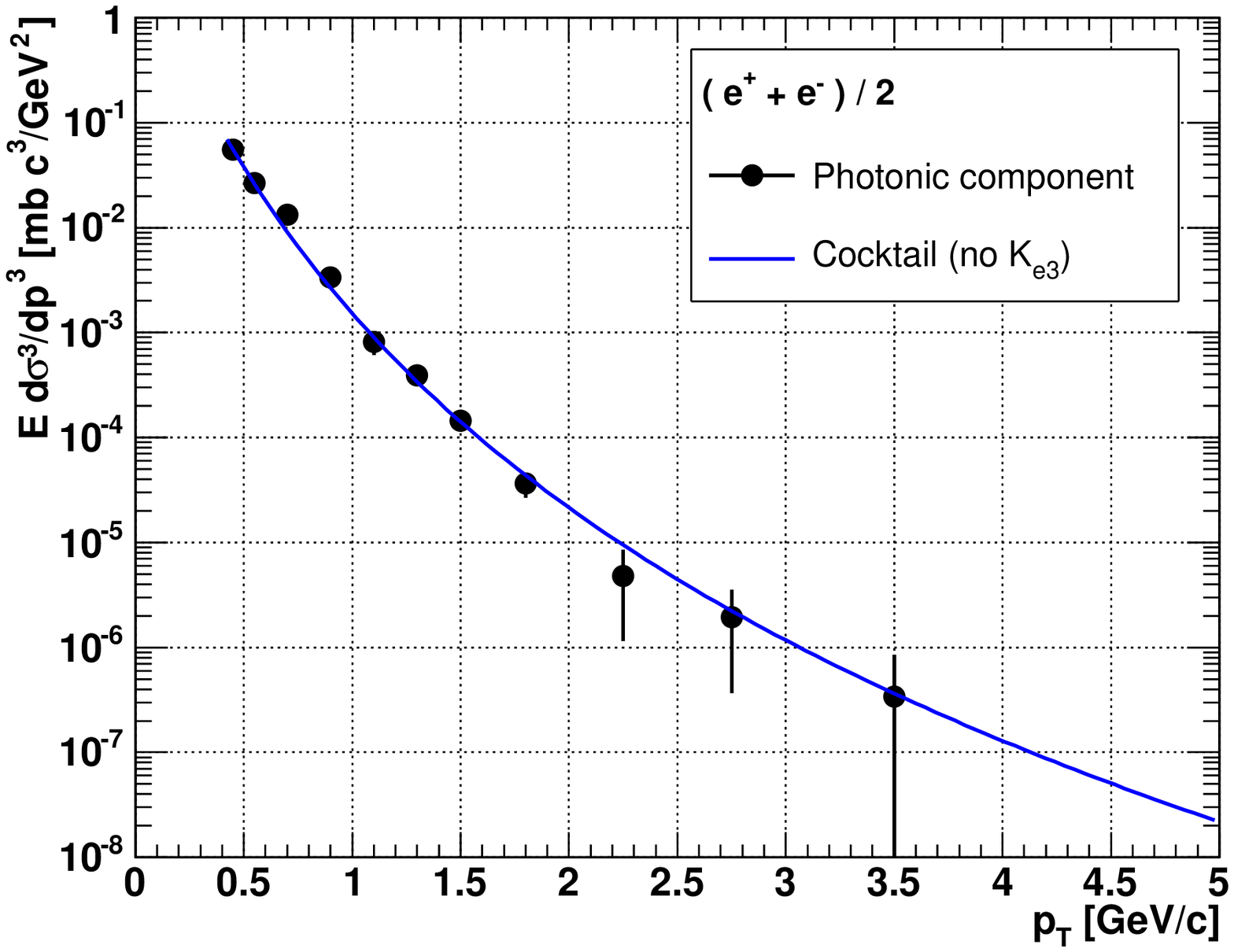,width=0.45\linewidth,clip,trim
= 0in 0in 0in 0.2in} \caption{\label{fig:ch4.phot_cocktail}
"Photonic" electron invariant crossection comparison to Cocktail
"Photonic" prediction. } \centering
\epsfig{figure=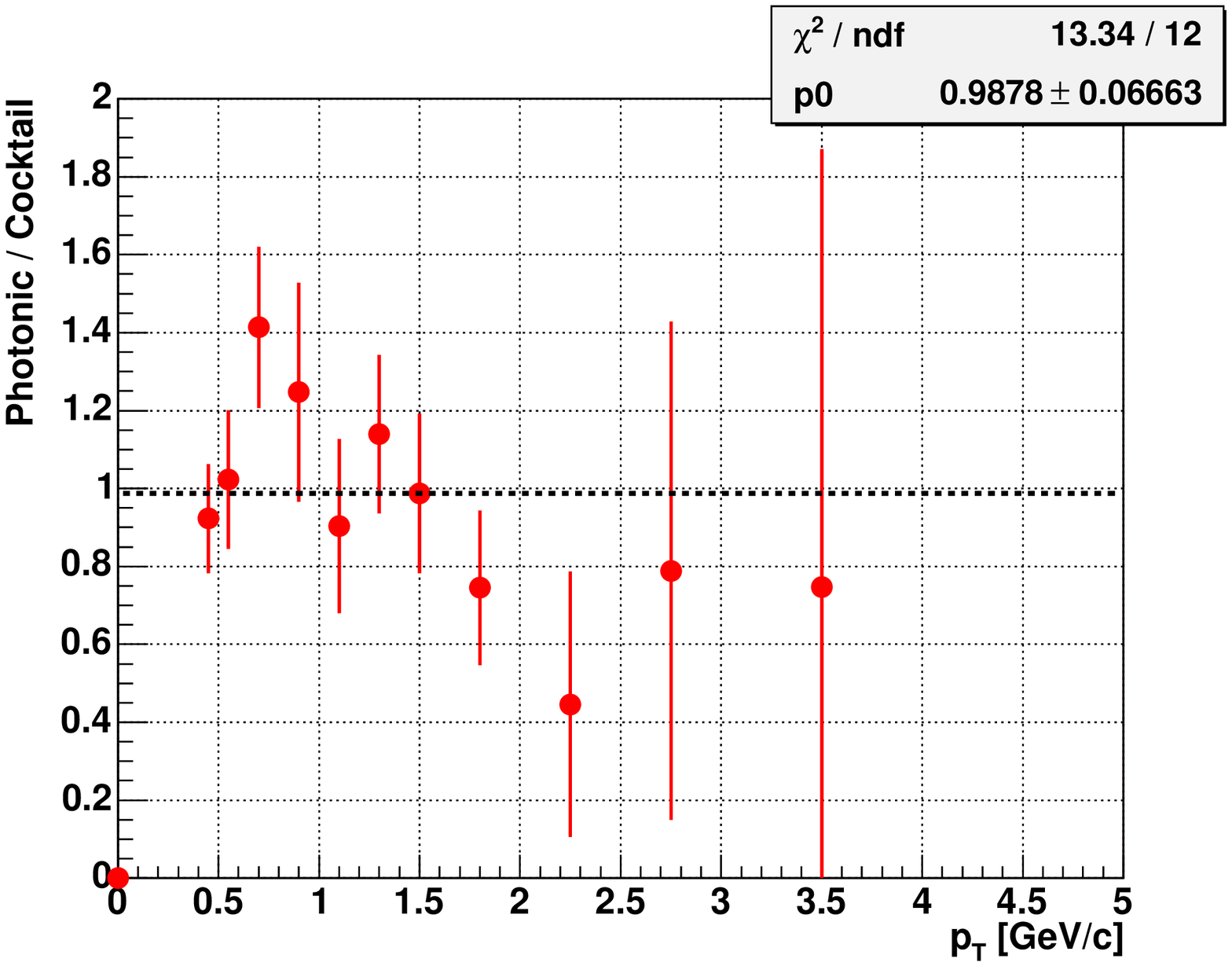,width=0.45\linewidth,clip}
\caption{\label{fig:ch4.ratio_phot_cocktail} Ratio of "Photonic"
electron invariant crossection to Cocktail photonic prediction. }
\end{figure}

Now we have all the ingredients to calculate "Photonic" and
"Non-photonic" electron crossection using Eq.~\ref{eq:ch4.ph_nph}.
The photonic electron crossection is using weighted average of
subtracted MB and ERT component. ERT trigger efficiency
uncertainty is added in quadrature to statistical error of ERT
trigger subtracted distribution. Bin width correction is applied
on the combined distribution. Fig.~\ref{fig:ch4.phot_conv_mb}
shows "Photonic" component of electron crossection for MB and ERT
sample and combined average. Fig.~\ref{fig:ch4.phot_cocktail}
presents the comparison of  final "Photonic" electron crossection
with Cocktail prediction for the photonic electron background from
all known sources (with exception of $K_{e3}$ decay which is not
"photon related" process). The ratio of data to Cocktail is shown
on Fig.~\ref{fig:ch4.ratio_phot_cocktail}.

\pagebreak

 One can see that the photonic component, obtained from
the data agrees with the Cocktail prediction within statistical
errors. The main error source is a low statistics of the converter
run period (especially in MB sample) which does not allow us to
make an accurate measurement of the "photon related" electrons.
Nevertheless, this agreement is very important
$\bold{independent}$ confirmation result for more accurate
Cocktail subtraction analysis.
\begin{figure}[b]
\centering
\epsfig{figure=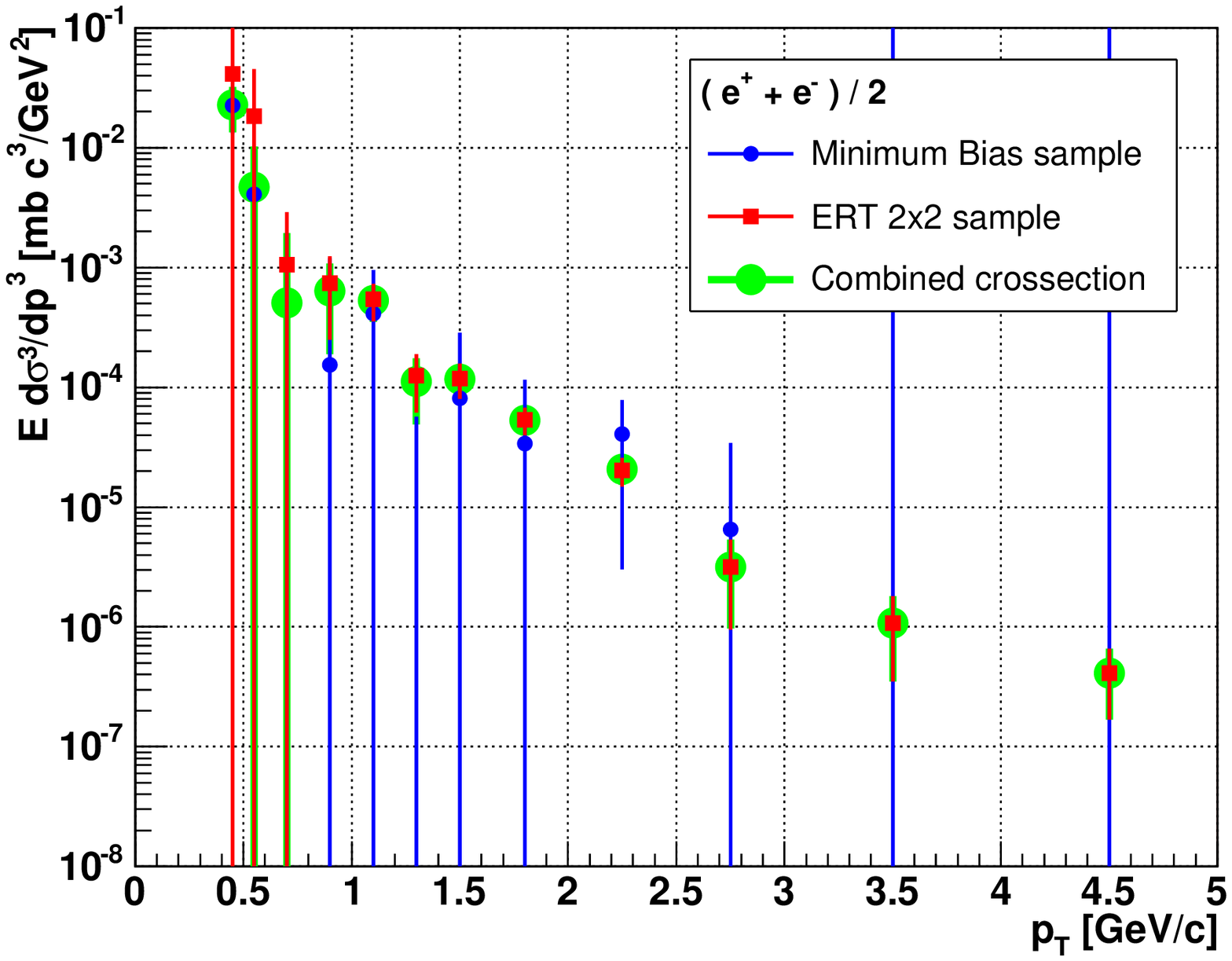,width=0.5\linewidth,clip,trim
= 0in 0.15in 0in 0.5in} \caption{\label{fig:ch4.non_phot_conv_mb}
"Non-photonic" electron invariant crossection for Minimum Bias
(circle) and ERT data sample (squares) (ERT statistical errors
includes the systematic error due to ERT trigger efficiency).
Combined by Eq.~\ref{eq:ch4.weight_avg} inclusive electron
invariant crossection (large circle). }
\epsfig{figure=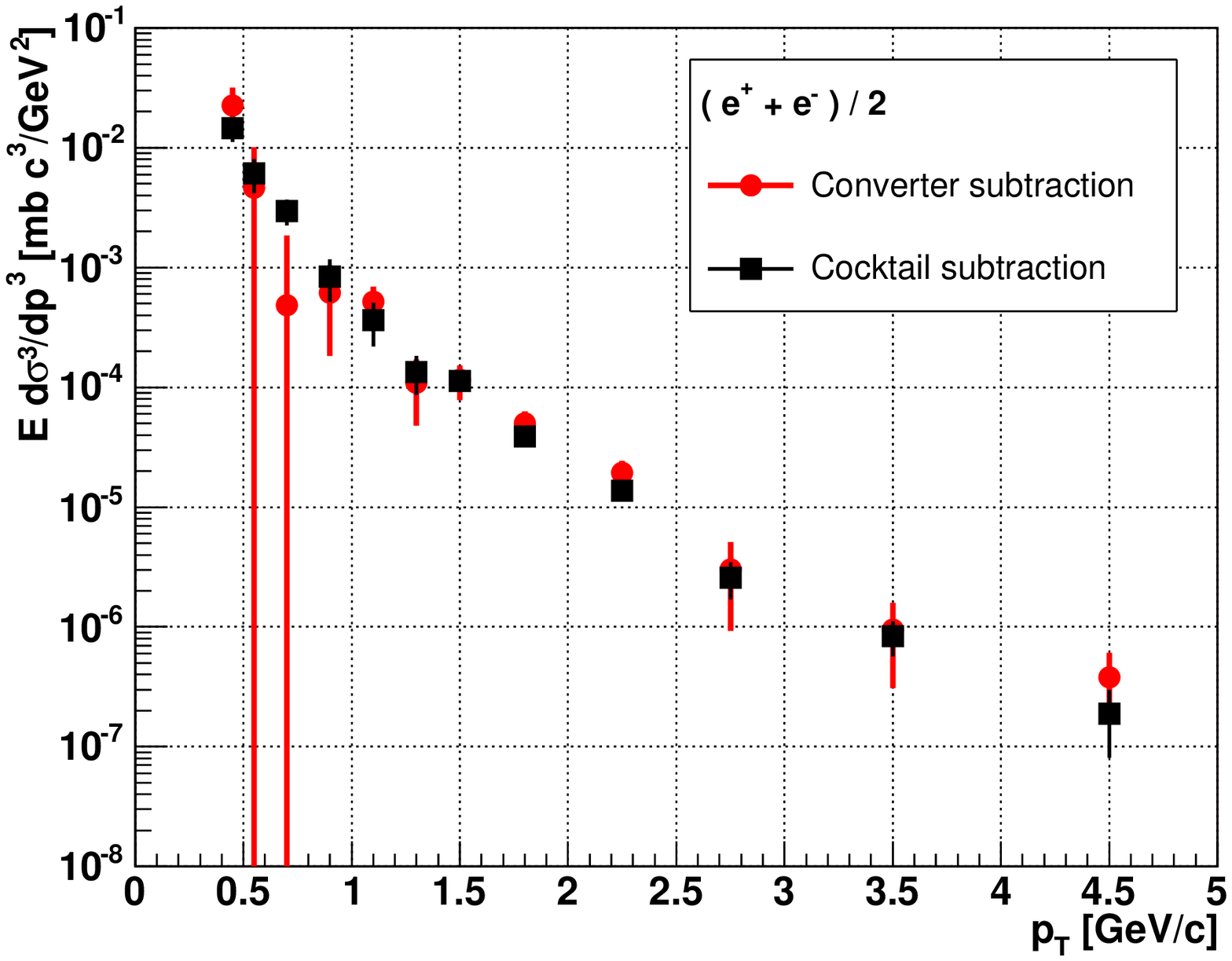,width=0.5\linewidth,clip,trim
= 0in 0in 0in 0.2in} \caption{\label{fig:ch4.non_phot_final}
"Non-Photonic" electron invariant crossection from Converter
subtraction analysis (circles) comparison to Cocktail subtracted
"Non-photonic" crossection (squares). }
\end{figure}

"Non-Photonic" electron component is calculated for similar way
both MB and ERT trigger data set using Eq.~\ref{eq:ch4.ph_nph}.
Combined crossection is calculated as weighted average of both
trigger samples. Fig.~\ref{fig:ch4.non_phot_conv_mb} shows
"Photonic" component of electron crossection for MB and ERT sample
and combined average.Fig.~\ref{fig:ch4.non_phot_final} shows the
final "Non-photonic" electron invariant crossection from the
converter subtraction method overlaid with Cocktail subtracted
Non-photonic results (Table.~\ref{tab:final_nonphotonic}). The
results of two independent analysis are in good agreement with
each other.

\begin{figure}[hb]
\centering \epsfig{figure=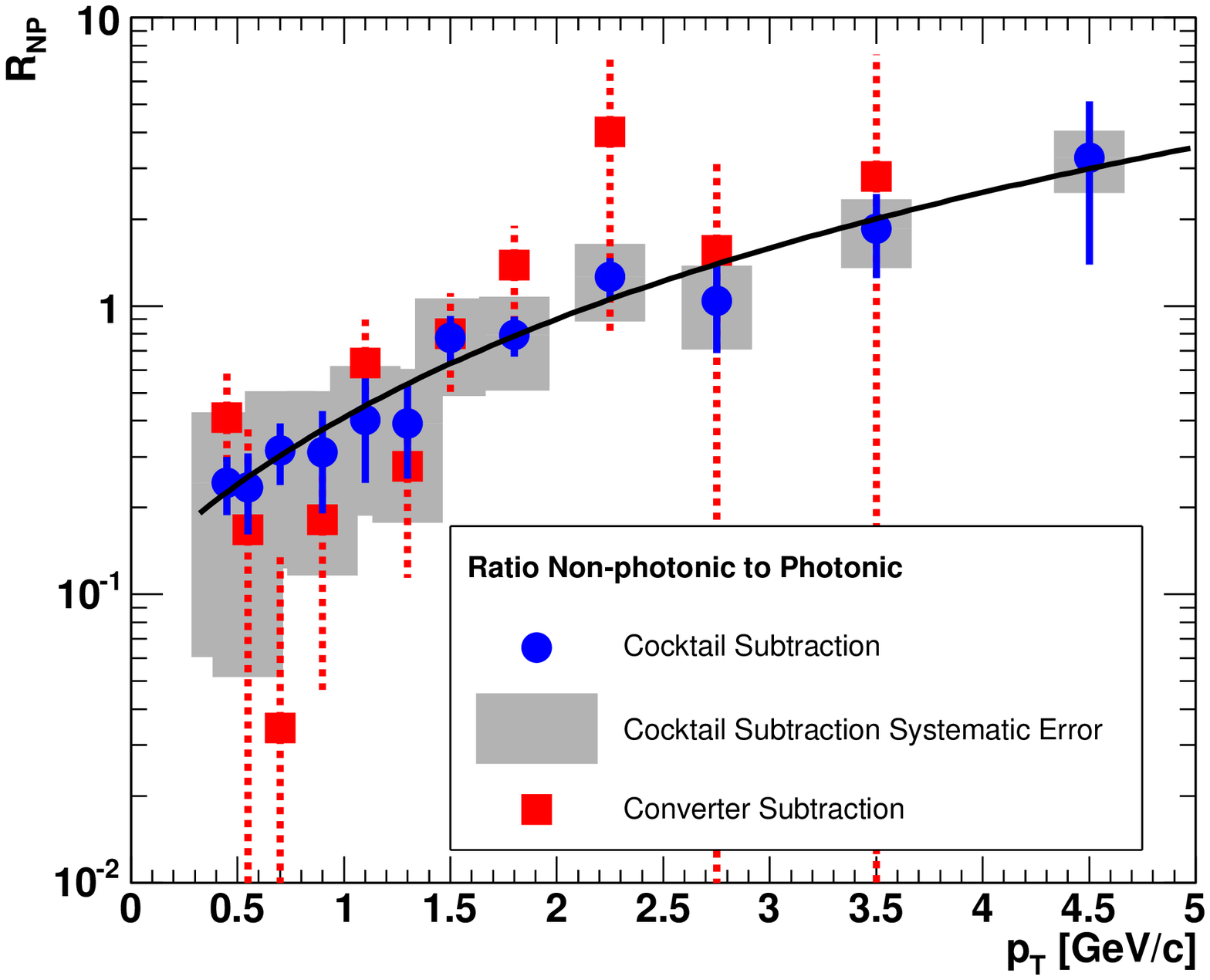,width=1\linewidth,clip}
\caption{\label{fig:ch4.ratio_np} Ratio of "Non-photonic" to
"Photonic" electron invariant crossection for cocktail subtraction
and converter subtraction methods. }
\end{figure}

The ratio of the Photonic component to Non-Photonic component for
Cocktail and Converter subtraction are plotted on
Fig.~\ref{fig:ch4.ratio_np}.

\newpage

\section{Systematic error analysis}\label{sec:ch4.Systematics}

Systematic error analysis have always been the most critical issue
for the experimentalist. The knowledge of the systematic error is
based on a priory assumptions to the detector performance and
quality of simulation. Systematic errors are the most important
ingredient for current analysis as it uses the results of two
completely uncorrelated measurements: (1) inclusive electron
crossection measurement and (2) neutral and charged pion
measurements for the cocktail input. The systematic errors in this
case will be completely independent and need to be treated
separately for the data and for the Cocktail.
Section~\ref{sec:ch4.Inclusive_Systematics} summarize the total
systematic error for the inclusive electron crossection,
section~\ref{sec:ch4.Cocktail_Systematics} describes the
contributions to the systematic error on the Cocktail prediction.
Finally section~\ref{sec:ch4.Subtracted_Systematics} gives the
final systematic error on the subtracted "Non-photonic" electron
crossection.

\subsection{Systematic error of the inclusive crossection}\label{sec:ch4.Inclusive_Systematics}

For the inclusive electron crossection we need to evaluate the
following contributions to the systematic error:
\begin{itemize}
\item eID cut error \subitem EMC matching cut systematic error
\subitem $E/p$ cut systematic error \subitem  Acceptance cut
systematic error \item Hadronic background systematic error \item
Correction function shape error \item Momentum scale systematic
error \item Momentum resolution systematic error \item Acceptance
MC$\leftrightarrow$Data systematic error \item ERT efficiency
systematic error
\end{itemize}

\subsubsection{eID cut systematic errors}

In order to estimate the systematic error on electron ID cuts,
each eID cut was varied within reasonably large limits. Complete
analysis was done using the modified cuts and the ratio of the new
inclusive electron crossection to the reference one was treated as
an estimator for the systematic error. This type of systematic
errors estimation suffers in case of low statistics when statistic
and systematic error can not be distinguished from each other. In
order to avoid the double-counting of the statistical error, we
always assume that relative systematic error is independent on
$p_T$ and is derived as a systematic trend at low-mediate $p_T$
range.

\subsubsection{EMC matching cut systematic error}

\begin{figure}[h]
\centering
\epsfig{figure=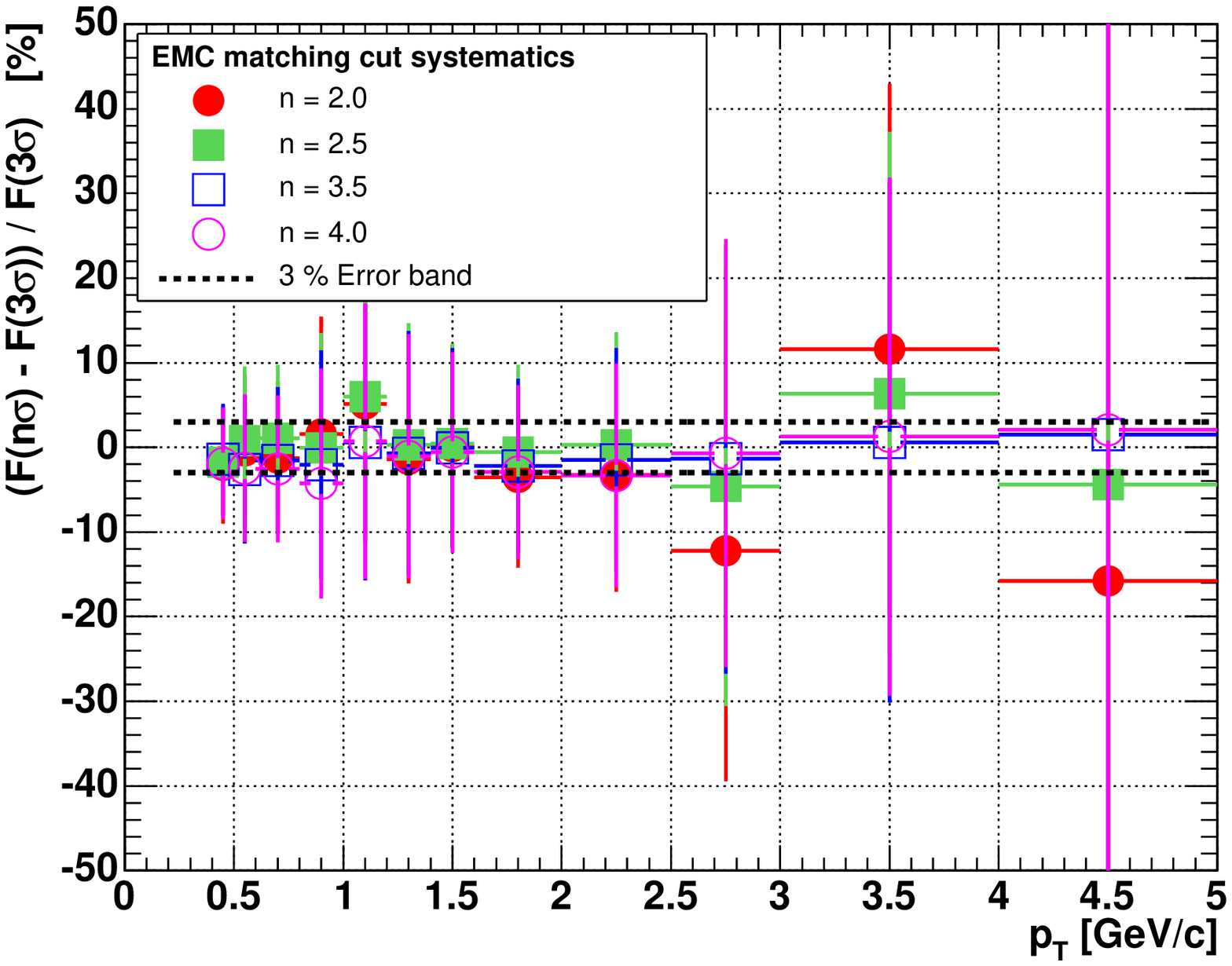,width=0.5\linewidth,clip,trim
= 0in 0in 0.5in 0.2in}
 \caption{\label{fig:ch4.matching_sys}
Variation of inclusive electron crossection for different
$d_{EMC}$ matching cut $|d_{EMC}| < 2.0,\ 2.5,\ 3.5,\ 4.0$.}
\end{figure}

EMC matching cut was changed from target value of $|d_{EMC}| <
3.0$ to $|d_{EMC}| < 2.0,\ 2.5,\ 3.5,\ 4.0$ level. Correction
functions, Hadronic background and inclusive crossection were
recalculated appropriately for each of the new matching cuts. The
ratio of $\bold{modified}$ inclusive electron distribution for
different matching cut to the reference is shown in
Fig.~\ref{fig:ch4.matching_sys}. One can see that most of the
points at low-mediate $p_T$ are covered by 3\% systematic error
band. At $p_T > 3$ GeV/c we can not make a conclusion due to
limited statistics.

\pagebreak
\subsubsection{$E/p$ cut systematic error}

To study the effect of $E/p$ cut we varied the cut value from
$3.0\sigma$ to $2.0\sigma,\ 2.5\sigma,\ 3.5\sigma,\ 4.0\sigma$.
Full analysis was repeated for each of the new cuts. The ratio of
modified inclusive electron distribution for different $E/p$ cut
to the reference is shown in Fig.~\ref{fig:ch4.Ep_sys}. One can
see that most of the points at low-mediate $p_T$ covered by 3\%
systematic error band.

\subsubsection{$Z$ cut systematic error}

\begin{figure}[ht]
\begin{tabular}{lr}
\begin{minipage}{0.5\linewidth}
\begin{flushleft}
\epsfig{figure=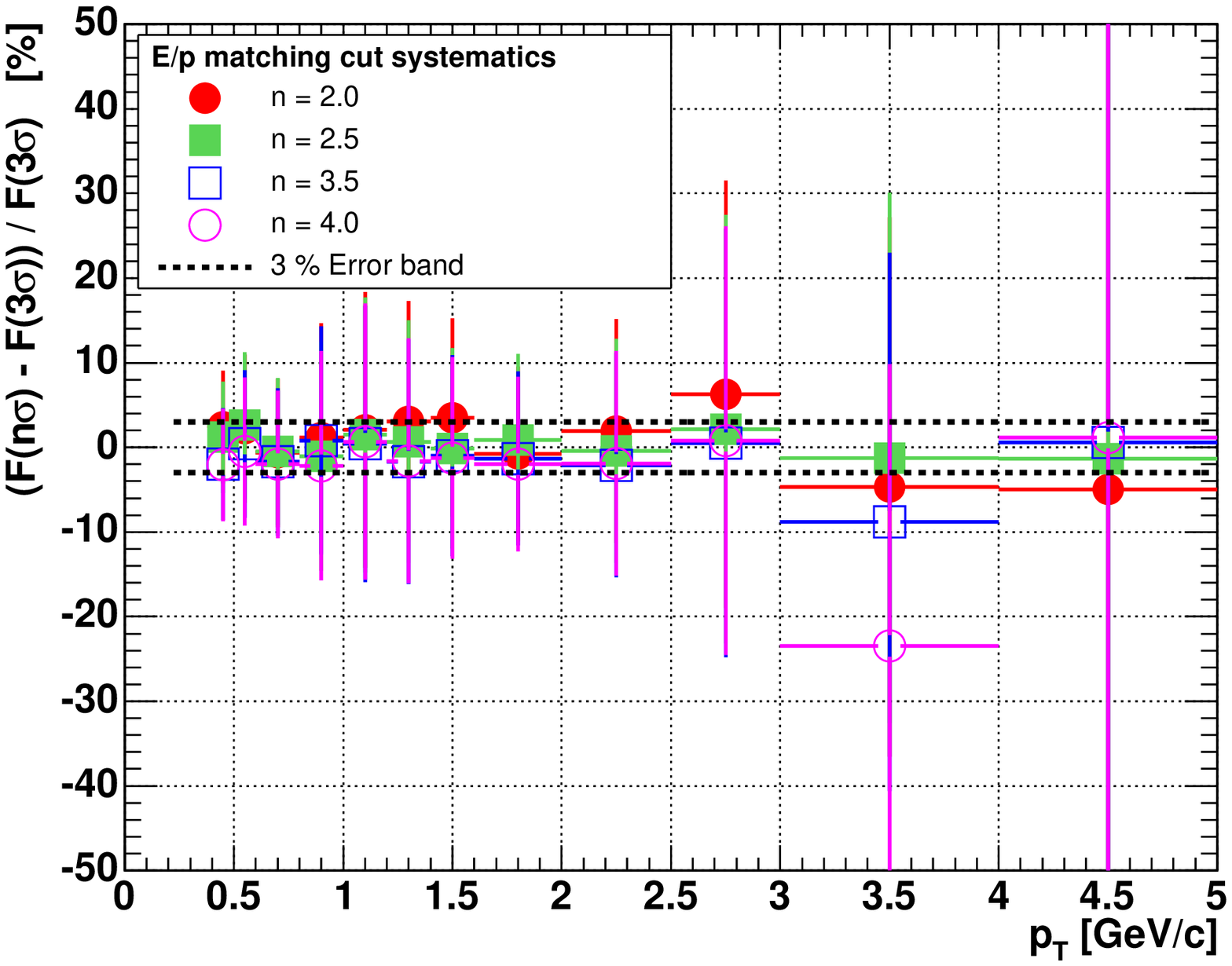,width=1\linewidth,clip,trim
= 0in 0in 0.5in 0.2in}
 \caption{\label{fig:ch4.Ep_sys}
Variation of inclusive electron crossection for different $E/p$
matching cut $|E/p| < 2.0\sigma,\ 2.5\sigma,\ 3.5\sigma,\
4.0\sigma$.}
\end{flushleft}
\end{minipage}
&
\begin{minipage}{0.5\linewidth}
\begin{flushright}
\epsfig{figure=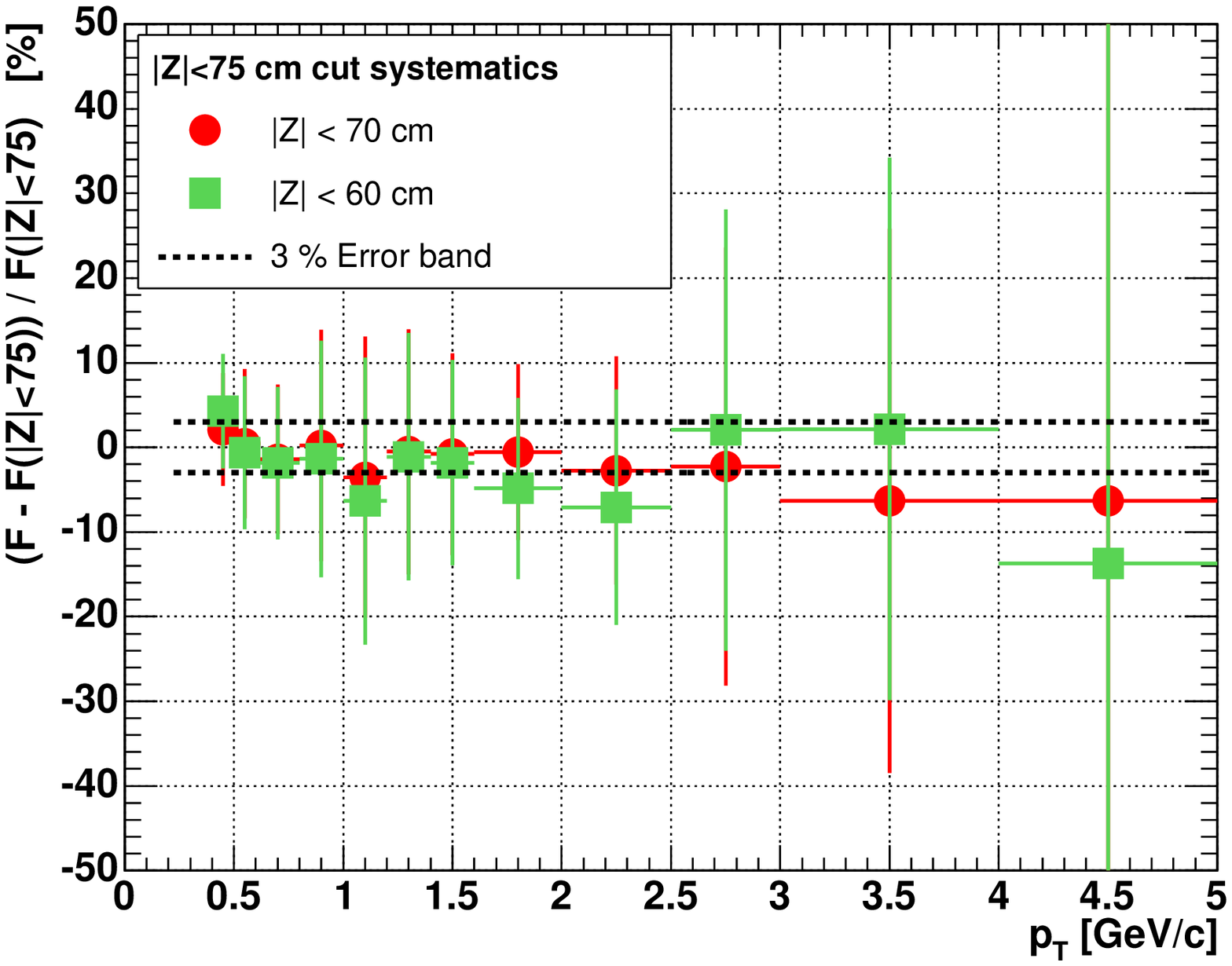,width=1\linewidth,clip,trim
= 0in 0in 0.5in 0.2in}
 \caption{\label{fig:ch4.zed_sys}
Variation of inclusive electron crossection for different $Z$
acceptance cuts $|Z| < 70,\ 60 cm$.}
\end{flushright}
\end{minipage}
\end{tabular}
\end{figure}

In order to estimate the error due to $|Z| < 75$ cm acceptance
cut, $Z$ cut value was changed from 75 cm to 70 cm and 60 cm. The
ratio of modified inclusive electron distribution for $Z$ cut of
70, 60 cm to the reference is shown in Fig.~\ref{fig:ch4.zed_sys}.
It is tough to estimate the trend of the ratio at high $p_T$, but
assuming that it should be $p_T$ independent, 3\% level of
systematic error seem to be a good estimate. It was also checked
that other acceptance cuts (TZR and NTC "shadow" cuts
Section~\ref{sec:ch4.acc_cuts}) does not seem to add any
significant systematic error to the inclusive crossection.

\pagebreak

\subsubsection{Hadronic background systematic errors}

\begin{figure}[h]
\centering
\epsfig{figure=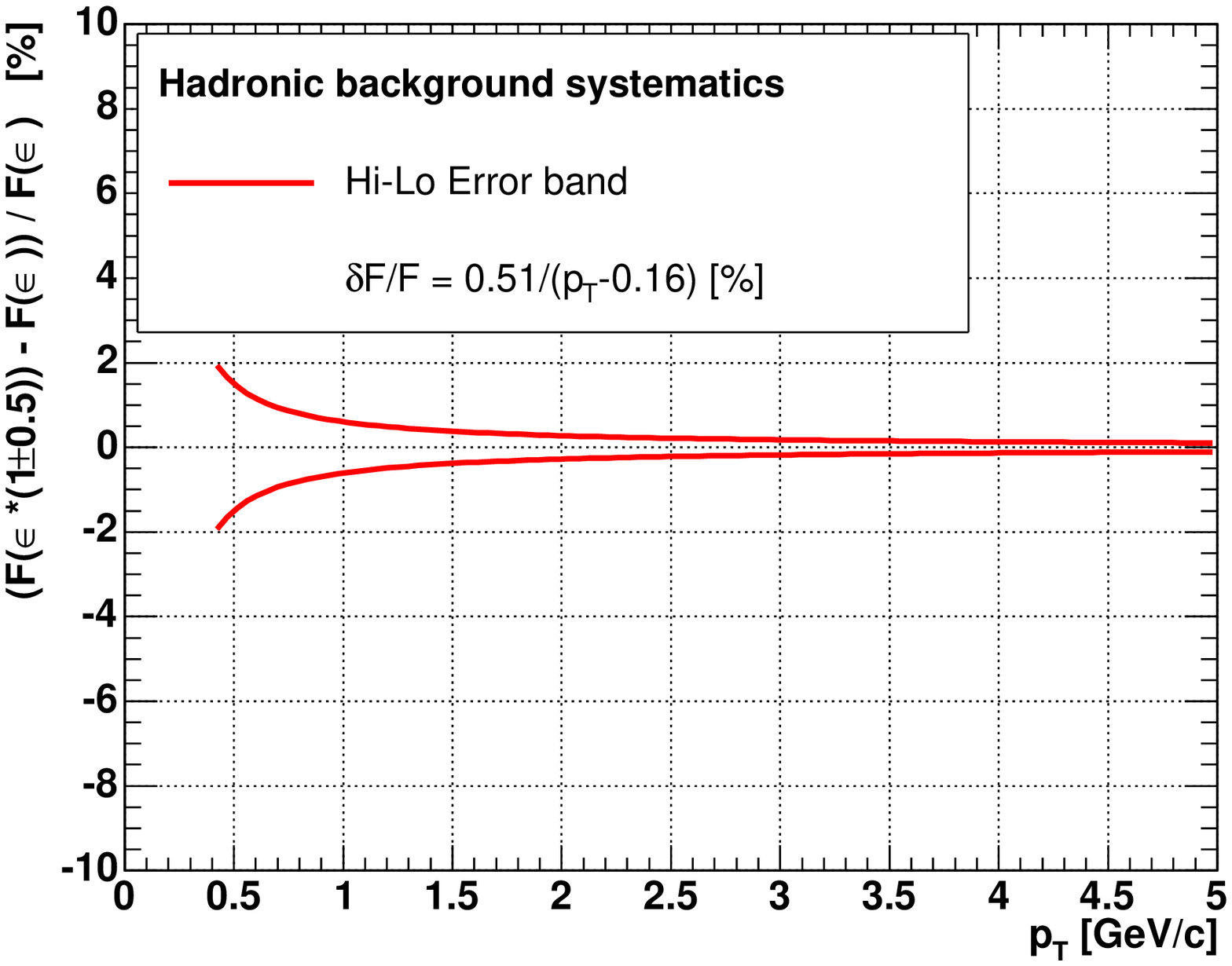,width=1\linewidth,clip,trim
= 0in 0in 0.5in 0.2in}
 \caption{\label{fig:ch4.hadr_sys}
Variation of inclusive electron crossection for different Random
Hadronic background association efficiency ($\pm1 \sigma$).}
\end{figure}

Hadronic background contribution to electron crossection is
calculated under assumption that RICH has certain probability to
"fire" on hadron (see Chapter~\ref{sec:ch4.hadr})
Fig.~\ref{fig:ch4.ep_hadrons} shows the $E/p$ distribution for
inclusive electrons, all charged scaled by  $\epsilon_{rand} =
(3\pm1.5(sys))\cdot10^{-4}$ and "swapped" background. Substantial
systematic error of 50\% was applied to this efficiency to take
into account low statistics of random background and possible
$p_T$ dependence of this parameter is applied to the random
association efficiency. The effect of the efficiency variation
($\pm1 \sigma$) to the final inclusive electron crossection is
shown in Fig.~\ref{fig:ch4.hadr_sys}. The systematic error band
can be fitted by functional form $\delta F / F =
0.51/(p_{T}-0.16)$ [\%].

\subsubsection{Correction function shape systematic errors}

Correction function shape (see Section~\ref{sec:ch4.cf}) is
limited by the total statistics of PISA single electron simulation
sample. We need to make a reasonable assumption how the shape of
this function looks like in case of high enough statistics. For
this purpose we use $\bold{Fast\ Acceptance\ Simulator}$, simple
MC code that simulates the $\bold{ideal}$ PHENIX acceptance by
assuming straight tracks and realistic pattern of acceptance
holes.

The major advantage of Fast Simulator is that it enables as to
calculate the correction function $\epsilon_{reco\ Fast}(p_T)$
without being limited by statistics. For the fast simulation we
used $100\cdot 10^6$ simulated $e^{\pm}$ generated with the same
initial assumption as in PISA single electron Simulation (this
gives us a factor of 20 more statistics than PISA sample).

\begin{itemize}
\item Uniform azimuthal angle distribution $0<\phi<2\pi$ \item
Uniform vertex $Z_{vtx}$ distribution $|Z_{vtx}|<25$ cm\item
Uniform rapidity distribution $-0.6 <y< 0.6$ units \item Uniform
$p_{T}$ distribution $0.0 <p_{T}< 5.0$ GeV/c
\end{itemize}

\begin{figure}[ht]
\begin{tabular}{lr}
\begin{minipage}{0.5\linewidth}
\begin{flushleft}
\epsfig{figure=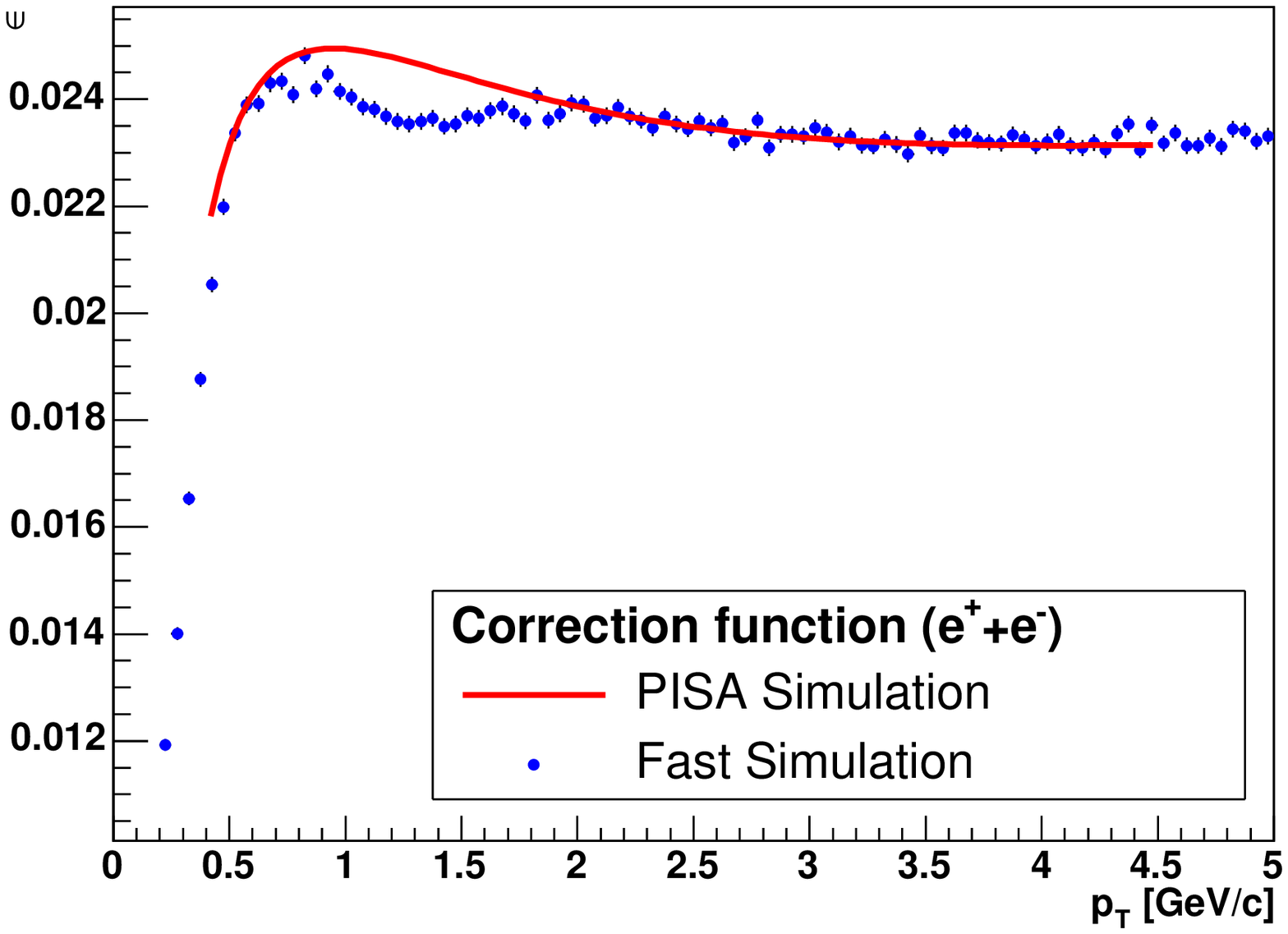,width=1\linewidth,clip,trim
= 0in 0in 0.5in 0.2in}
 \caption{\label{fig:ch4.fast_cf}
Comparison of correction function for Fast Simulator (points) and
full PISA Simulation (line) for $e^{+} +e^{-}$.}
\end{flushleft}
\end{minipage}
&
\begin{minipage}{0.5\linewidth}
\begin{flushright}
\epsfig{figure=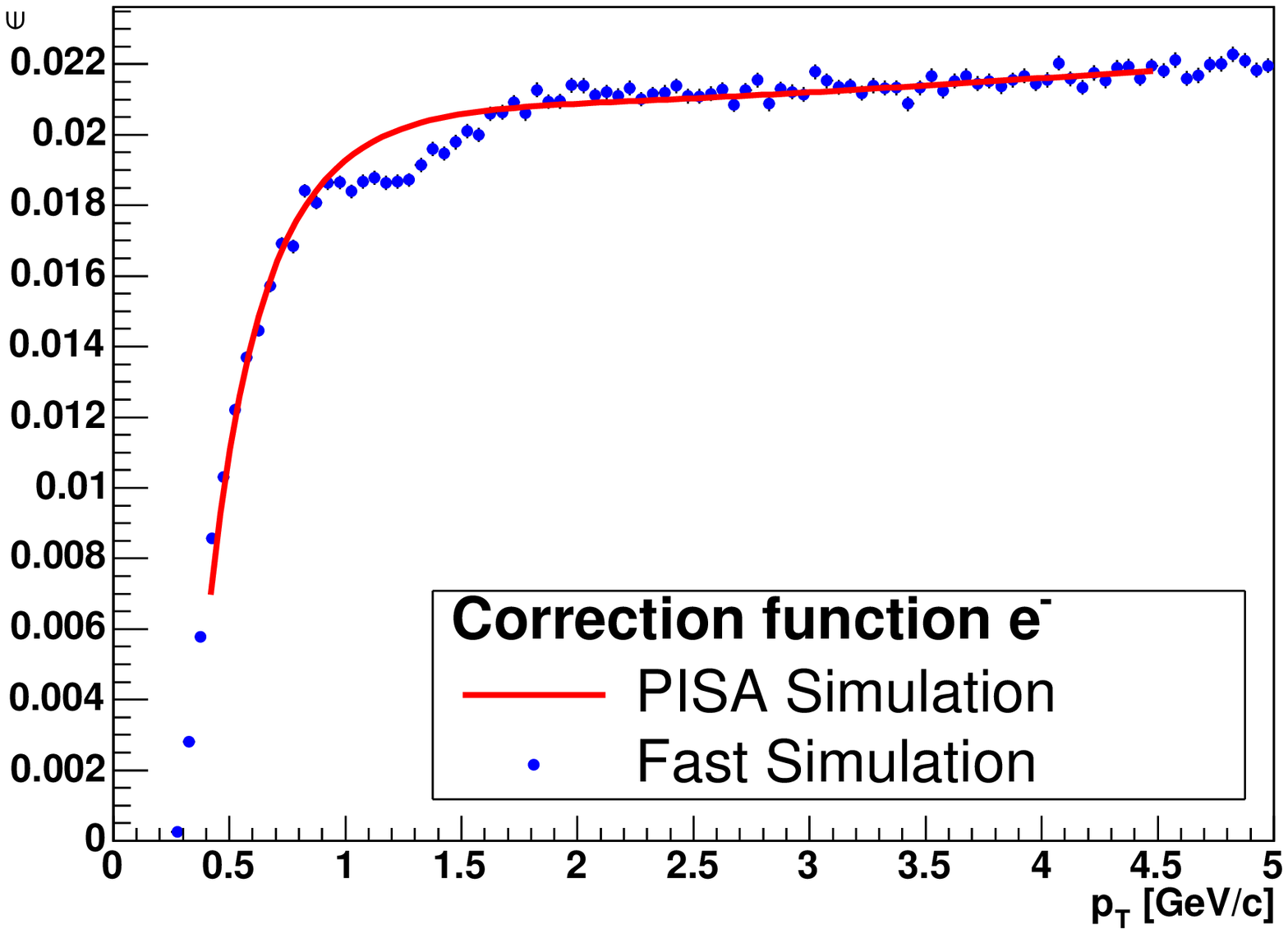,width=1\linewidth,clip,trim
= 0in 0in 0.5in 0.2in}
 \caption{\label{fig:ch4.fast_cf_electrons}
Comparison of correction function for Fast Simulator (points) and
full PISA Simulation (line) for electrons.}
\end{flushright}
\end{minipage}
\end{tabular}

\begin{tabular}{lr}
\begin{minipage}{0.5\linewidth}
\begin{flushleft}
\epsfig{figure=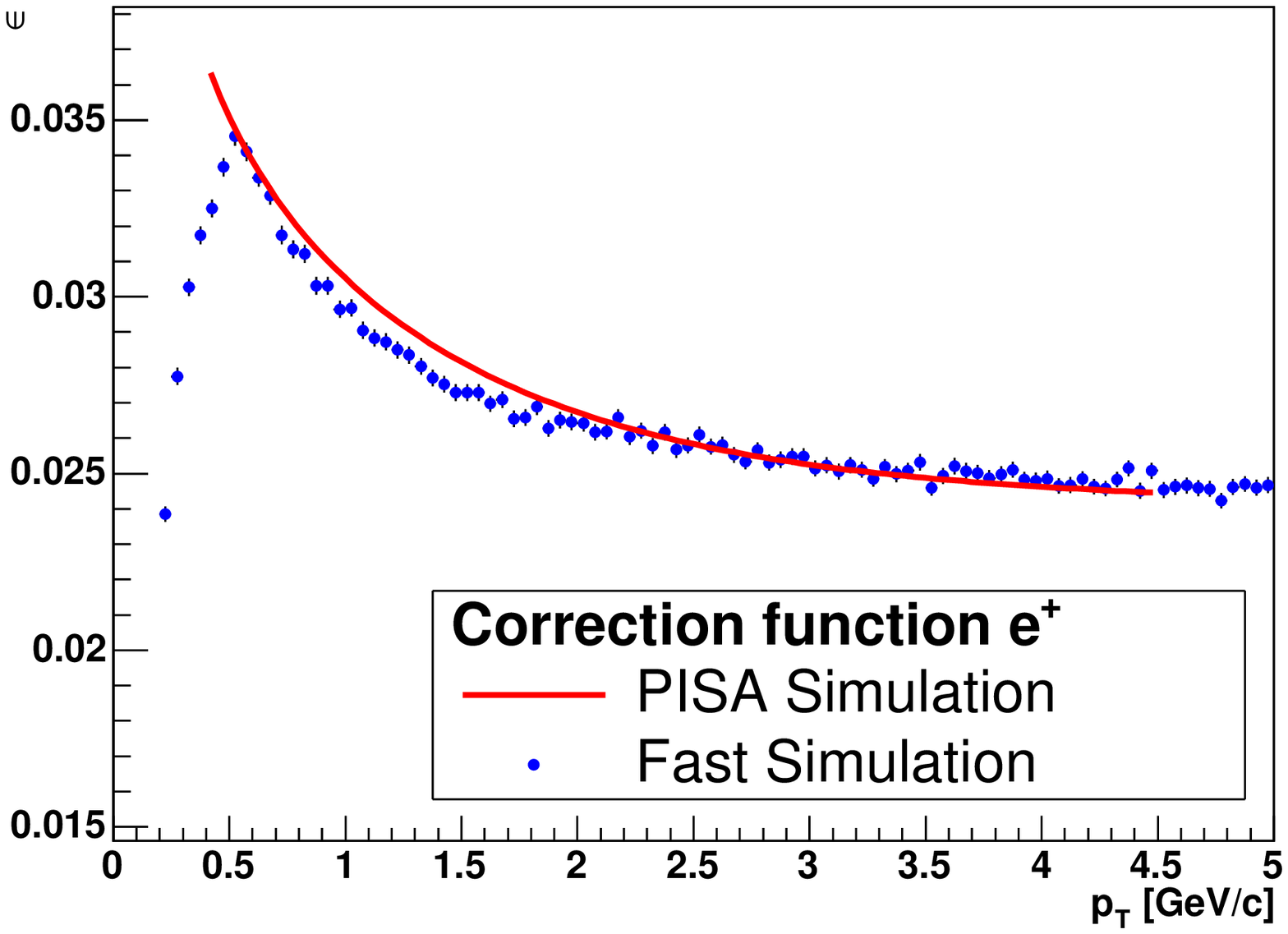,width=1\linewidth,clip,trim
= 0in 0in 0.5in 0.2in} \caption{\label{fig:ch4.fast_cf_positrons}
Comparison of correction function for Fast Simulator (points) and
full PISA Simulation (line) for positrons.}
\end{flushleft}
\end{minipage}
&
\begin{minipage}{0.5\linewidth}
\begin{flushright}
\epsfig{figure=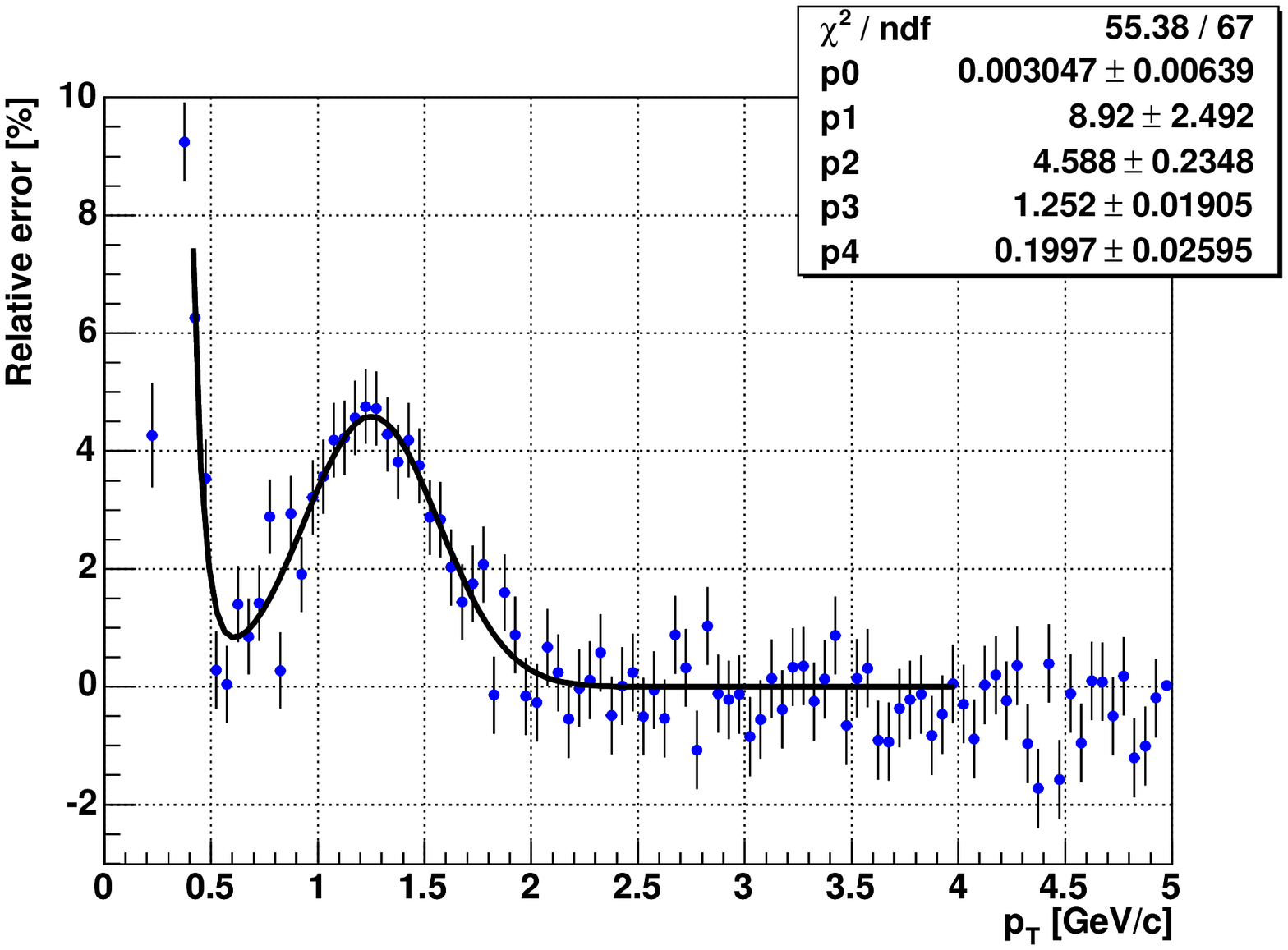,width=1\linewidth,clip}
 \caption{\label{fig:ch4.cf_sys}Relative difference between Fast Simulator and PISA simulation
correction functions from Fig.~\ref{fig:ch4.fast_cf}.}
\end{flushright}
\end{minipage}
\end{tabular}
\end{figure}

Full simulation, of cause, will be different from the ideal Fast
Simulation by the reconstruction efficiency. We can assume the
reconstruction efficiency to be $p_T$ independent to the first
order. By calculating the correction function for Fast Simulator
$\epsilon_{reco\ Fast}(p_T)$, scaling it by a constant
reconstruction efficiency ($\epsilon_{MC\ rec} = 0.853$ see below)
and comparing it with the PISA simulation of correction function
$\epsilon_{reco}(p_T)$ prediction for electrons, positrons and
total
(Fig.~\ref{fig:ch4.corr_function}~\ref{fig:ch4.corr_function_electrons}~\ref{fig:ch4.corr_function_positrons})
we can obtain the systematic error on the shape of the correction
function.
Fig.~\ref{fig:ch4.fast_cf}~\ref{fig:ch4.fast_cf_electrons}~\ref{fig:ch4.fast_cf_positrons}
shows the comparison of the correction function shape for Fast
simulation and PISA simulation. There is a ~ 5 \% deviation of
Fast simulator Correction function shape at $1 <p_T <2$ GeV/c
region. Other then that, the shape of the correction seems to be
in good agreement between two completely independent simulations.
The difference between the correction functions is assigned as a
systematic error on the correction function shape and shown in
Fig.~\ref{fig:ch4.cf_sys}.

\pagebreak

\subsubsection{Momentum resolution and momentum scale systematic errors}

\begin{figure}[ht]
\begin{tabular}{lr}
\begin{minipage}{0.5\linewidth}
\begin{flushleft}
\epsfig{figure=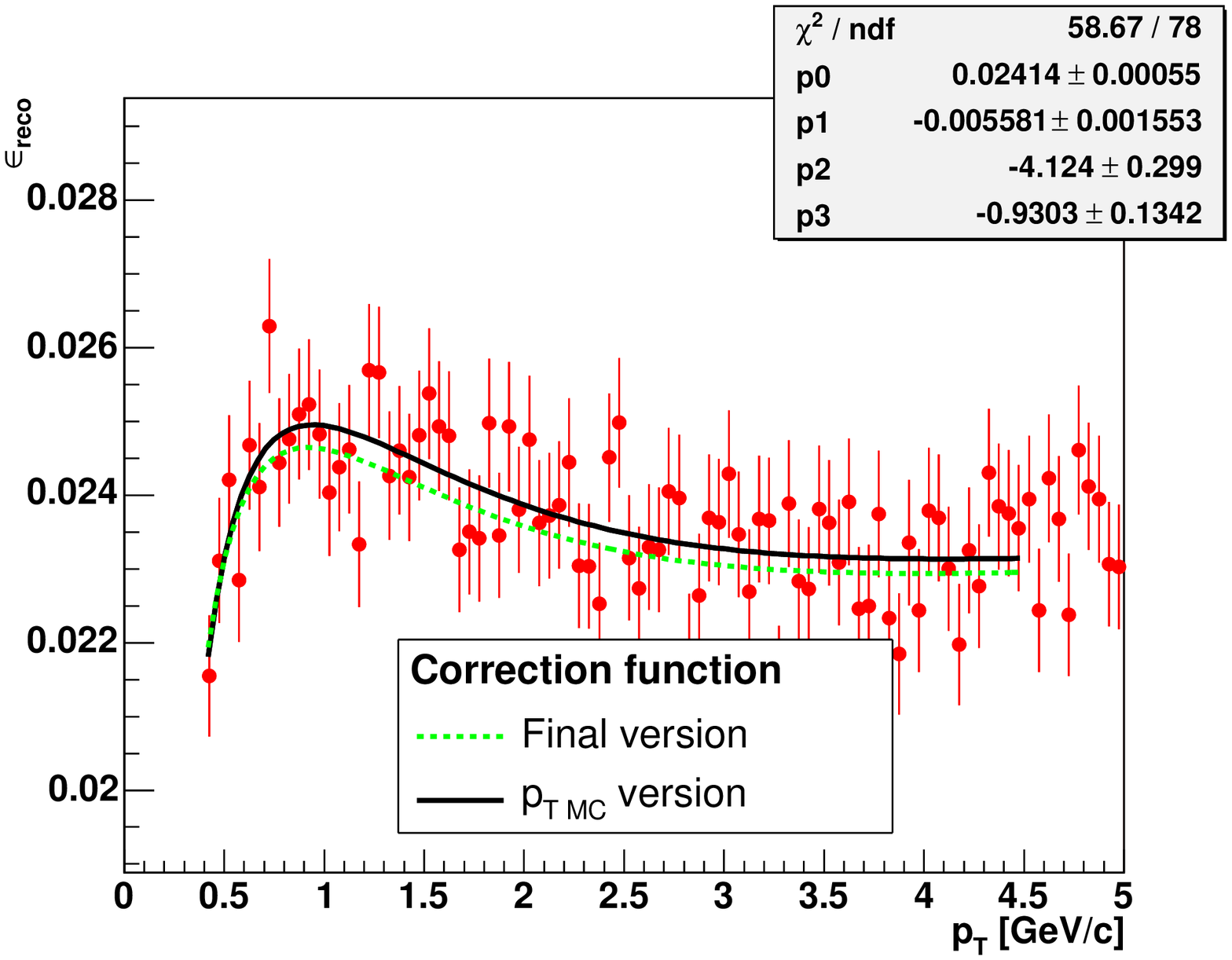,width=1\linewidth,clip}
 \caption{\label{fig:ch4.cf_no_resol}
Correction function calculated under assumption of ideal momentum
measurement (solid curve) compared with final Correction function
$\epsilon_{reco}(p_T)$ used in the analysis.}
\end{flushleft}
\end{minipage}
&
\begin{minipage}{0.5\linewidth}
\begin{flushright}
\epsfig{figure=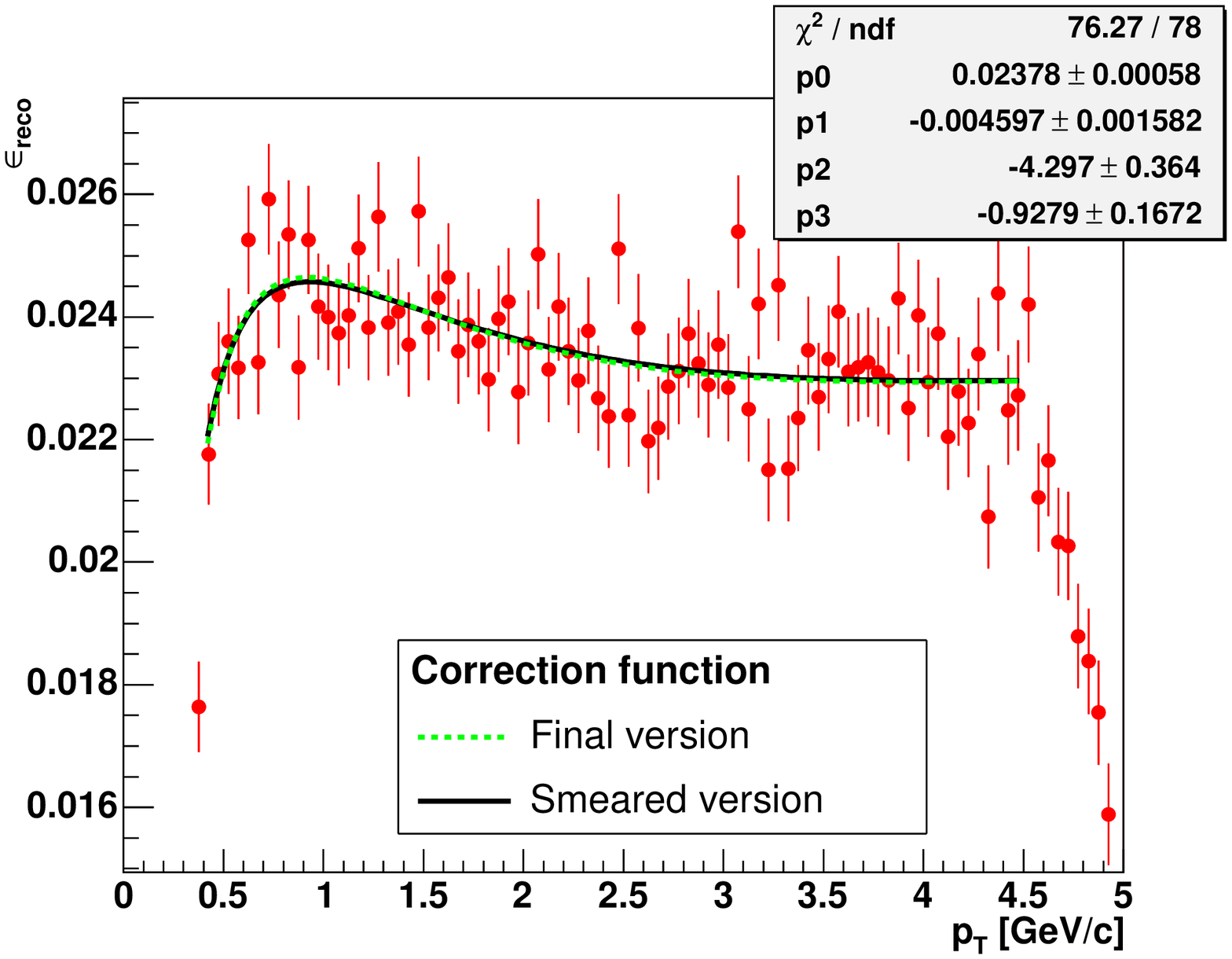,width=1\linewidth,clip}
 \caption{\label{fig:ch4.cf_data_resol}
Variation of the Correction function due to smearing of the
momentum resolution from $\sigma_{DCH} = 0.74 \%$ to $\sigma_{DCH}
= 1.48 \%$ compared with final Correction function
$\epsilon_{reco}(p_T)$ used in the analysis.}
\end{flushright}
\end{minipage}
\end{tabular}
\end{figure}

Momentum scale systematic error accounts for possible difference
of momentum resolution in Simulation and Data, it also includes
the uncertainty of the absolute momentum measurement. Effect of
the momentum resolution in MC Simulation and real data have
already been discussed in Chapter~\ref{sec:ch4.cf}.
Fig.~\ref{fig:ch4.cf_no_resol},~\ref{fig:ch4.cf_data_resol} shows
the variation of the correction function due to the momentum
resolution being varied in wide limits from 0\% to 1.48\%. The
total extent of the variation gives a systematic error of 0.9 \%.

\pagebreak

\begin{figure}[h]
\centering
\epsfig{figure=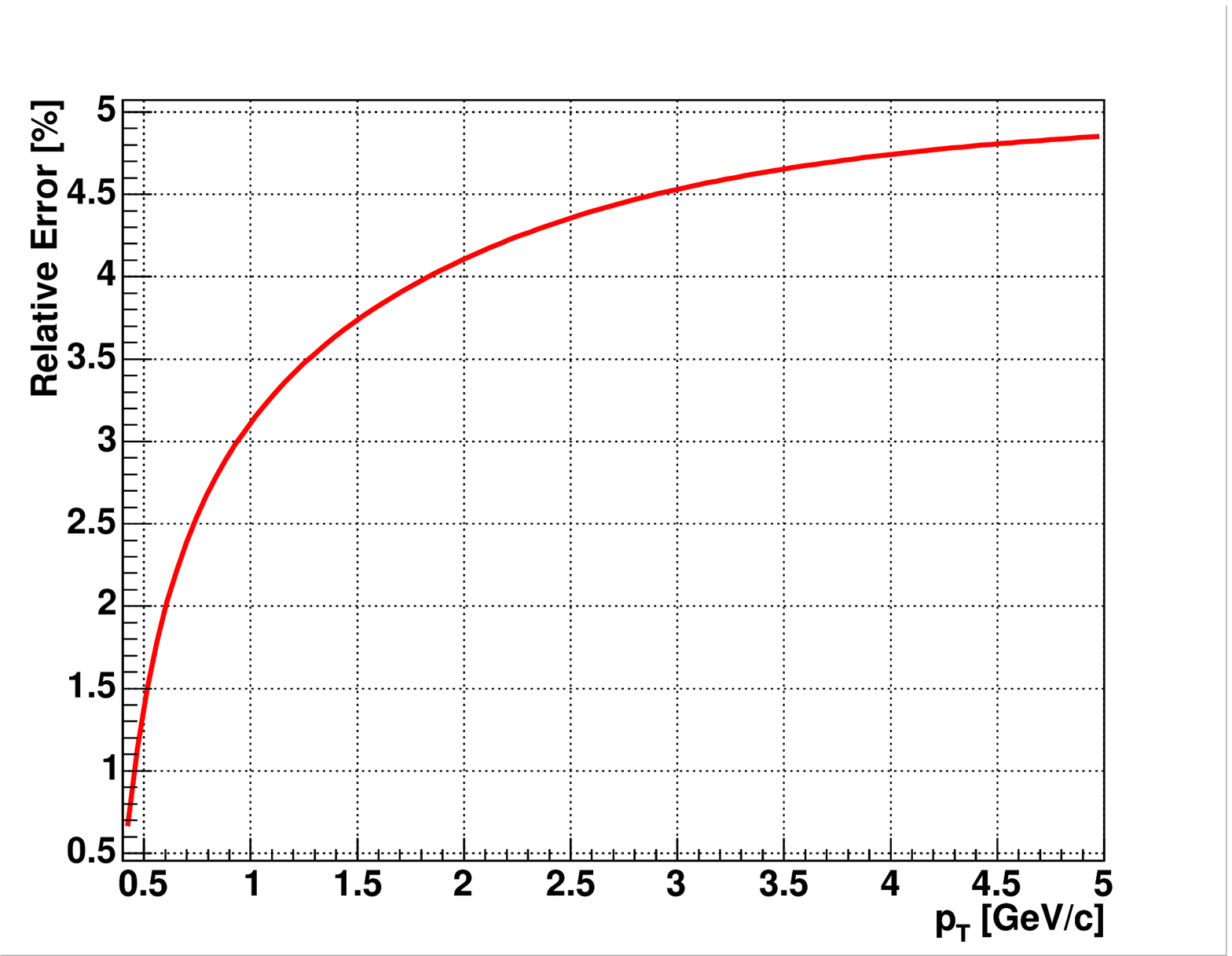,width=0.5\linewidth,clip,trim
= 0in 0.05in 0.5in 0.2in}
 \caption{\label{fig:ch4.mom_scale_sys}
Variation of inclusive electron crossection due to 1\% absolute
momentum scale uncertainty.}
\end{figure}

Momentum scale systematic is much harder to evaluate, we need to
make a reasonable assumption of how accurate we know the momentum
of the track. This value depends on the accuracy of the magnetic
field map representation used in reconstruction, quality of DCH
calibration constants and alignment of the beam with respect to
the center of the PHENIX coordinate system. The best estimate of
the absolute momentum scale accuracy can be obtained from studying
the mass of the proton peak in TOF acceptance. If we use the
results for the mass of proton $m^2$~\cite{ana172}, there is a
residual shift of the proton mass on the order of 1\% that is
primarily due to the remaining momentum scale error. Assuming the
accuracy of the momentum measurement to be 1\%, we can vary the
momentum scale of the PISA Simulation to obtain the new correction
functions corresponding to the mis-measured momentum. The ratio of
this correction function to the original correction function will
give as the systematic error associated with the momentum scale
uncertainty shown in Fig.~\ref{fig:ch4.mom_scale_sys}.

\pagebreak

\subsubsection{Reconstruction efficiency systematic error}

Reconstruction efficiency of Simulation and Data can be different,
we need to study pure efficiency of tracking, $n0$ cut, EMC
matching and $E/p$ cuts separately. As a reference we can select
Fast Simulator prediction which simulate the ideal detector
response\footnote{This assumption strongly relies on the exact
match of the acceptances in full PISA Monte Carlo and Fast
Simulator}.

\begin{figure}[ht]
\begin{tabular}{lr}
\begin{minipage}{0.5\linewidth}
\begin{flushleft}
\epsfig{figure=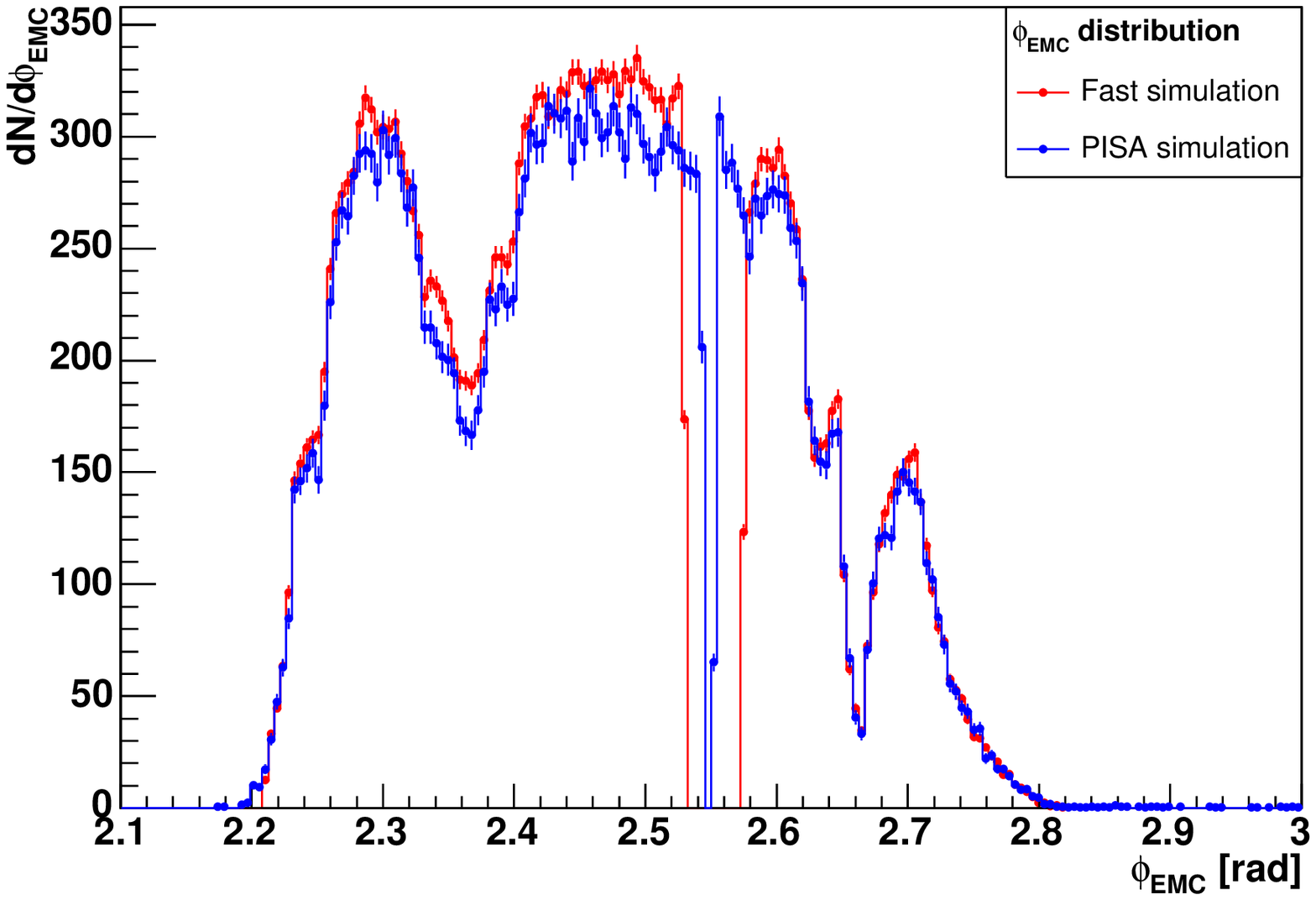,width=1\linewidth,clip}
 \caption{\label{fig:ch4.EMC_match_comp}
Comparison of $\phi_{EMC}$ distribution for Fast simulation and
PISA without any matching to outer detectors.}
\end{flushleft}
\end{minipage}
&
\begin{minipage}{0.5\linewidth}
\begin{flushright}
\epsfig{figure=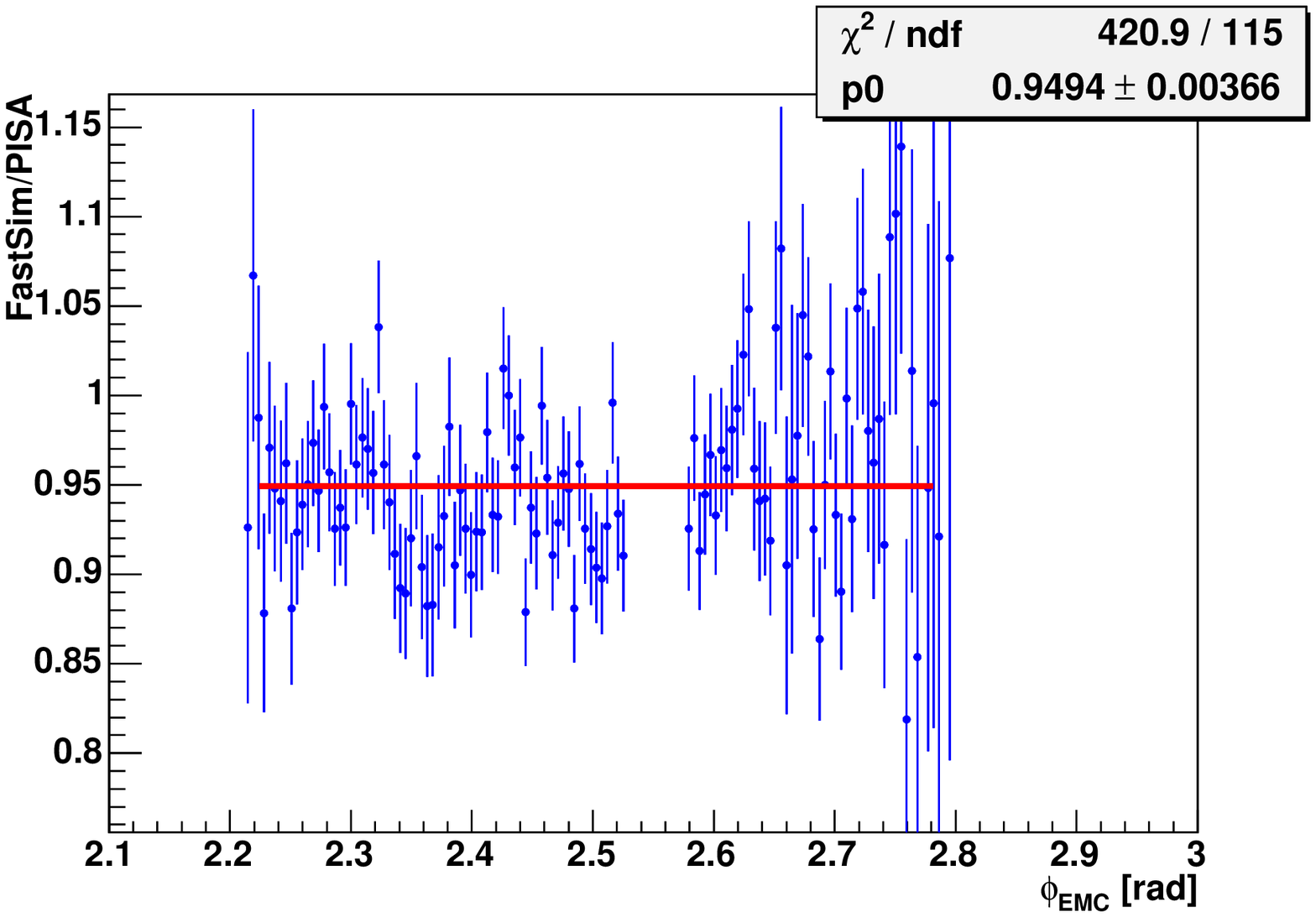,width=1\linewidth,clip}
 \caption{\label{fig:ch4.ratio_EMC_match}
Ratio of $\phi_{EMC}$ distributions from
Fig.~\ref{fig:ch4.EMC_match_comp} fitted with constant. This ratio
estimates tracking efficiency in Simulation.}
\end{flushright}
\end{minipage}
\end{tabular}
\centering
\epsfig{figure=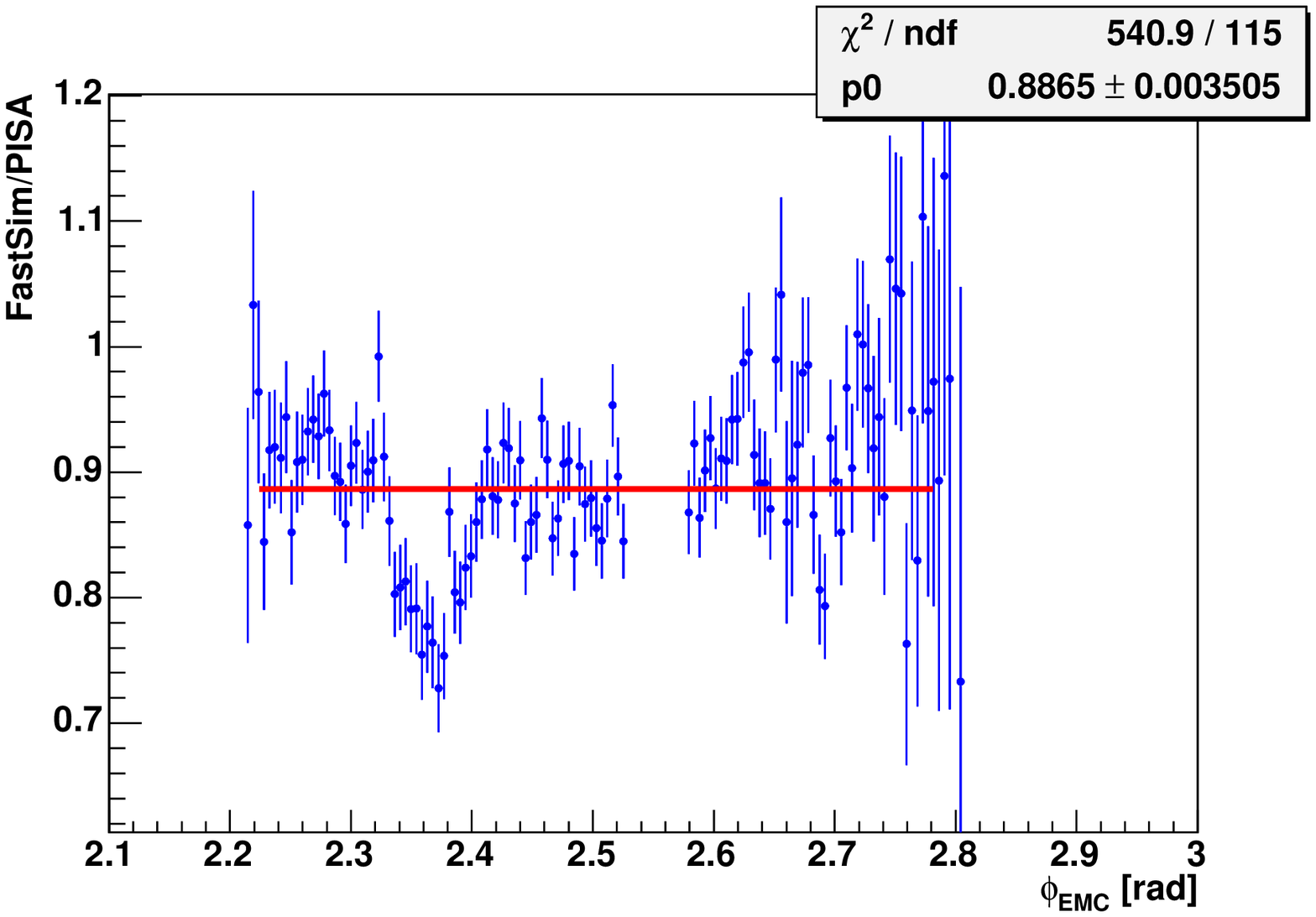,width=0.5\linewidth,clip}
 \caption{\label{fig:ch4.ratio_n0_match}
Ratio of $\phi_{EMC}$ distributions after $n0 >1$ cut is applied
fitted with constant. }
\end{figure}

First, we look at the DCH track acceptance, requiring no matching
to outer detectors (this is not exactly true as PISA simulation
have an internal requirement to have at least one MC hit in EMC)
our MC tracking efficiency will be slightly biased by EMC
matching. We can compare $\phi_{EMC}$ distribution for DCH tracks
which is shown in Fig.~\ref{fig:ch4.EMC_match_comp}. One can see
that the efficiency that except for the area of EMC sector
junction the shape of the EMC are in perfect agreement. The ratio
of the FastSim/PISA is shown in Fig.~\ref{fig:ch4.ratio_EMC_match}
with a constant fit in good agreement region. From this fit we can
conclude that tracking efficiency in Simulation is $\epsilon_{MC\
track} = (0.949\pm0.004)$. From  Run02 $p+p$ charged hadron
analysis~\cite{ana276} we obtain an estimate for the tracking
efficiency in Data $\epsilon_{Data track} = 0.982$. There may be
an effect of tighter Z cut for the charged analysis ($|Z| < 50cm$)
increasing the efficiency if it is \linebreak Z dependent but we
neglect it and take the worst case scenario. The difference
between those numbers is assigned to systematic error on tracking
efficiency. $\delta\epsilon_{track} = 3.3\%$.

\pagebreak

$n0$ cut efficiency can be studied the same way.
Fig.~\ref{fig:ch4.ratio_n0_match} shows the ratio of Fast
Simulation to PISA  $\phi_{EMC}$ distributions when $n0>1$ cut is
applied. The fit to the ratio is equal to  $\epsilon_{MC\
n0>1}\cdot \epsilon_{MC\ track} = (0.886\pm 0.004)$ which gives us
$\epsilon_{MC\ n0>1} = 0.934$. From $J/\psi$ Run02 $p+p$
analysis~\cite{ana139} we have an estimate for $n0>1$ cut
efficiency in Data $\epsilon_{Data\ n0>1} = (0.975\pm0.015)$.
There is a 3.8\% difference between Data and MC which is applied
as systematic error due to $n0$ cut efficiency.

EMC matching and E/p cuts lower the reconstruction efficiency by
another 3\% producing final $\epsilon_{MC\ rec} = 0.853$ but the
systematic error of those cuts was already estimated. 3\% loss
agrees with the amount of background removed by two $\pm3\sigma$
cuts on normally distributed variables.

\subsubsection{Acceptance systematic error}
\begin{figure}[h]
\centering
\epsfig{figure=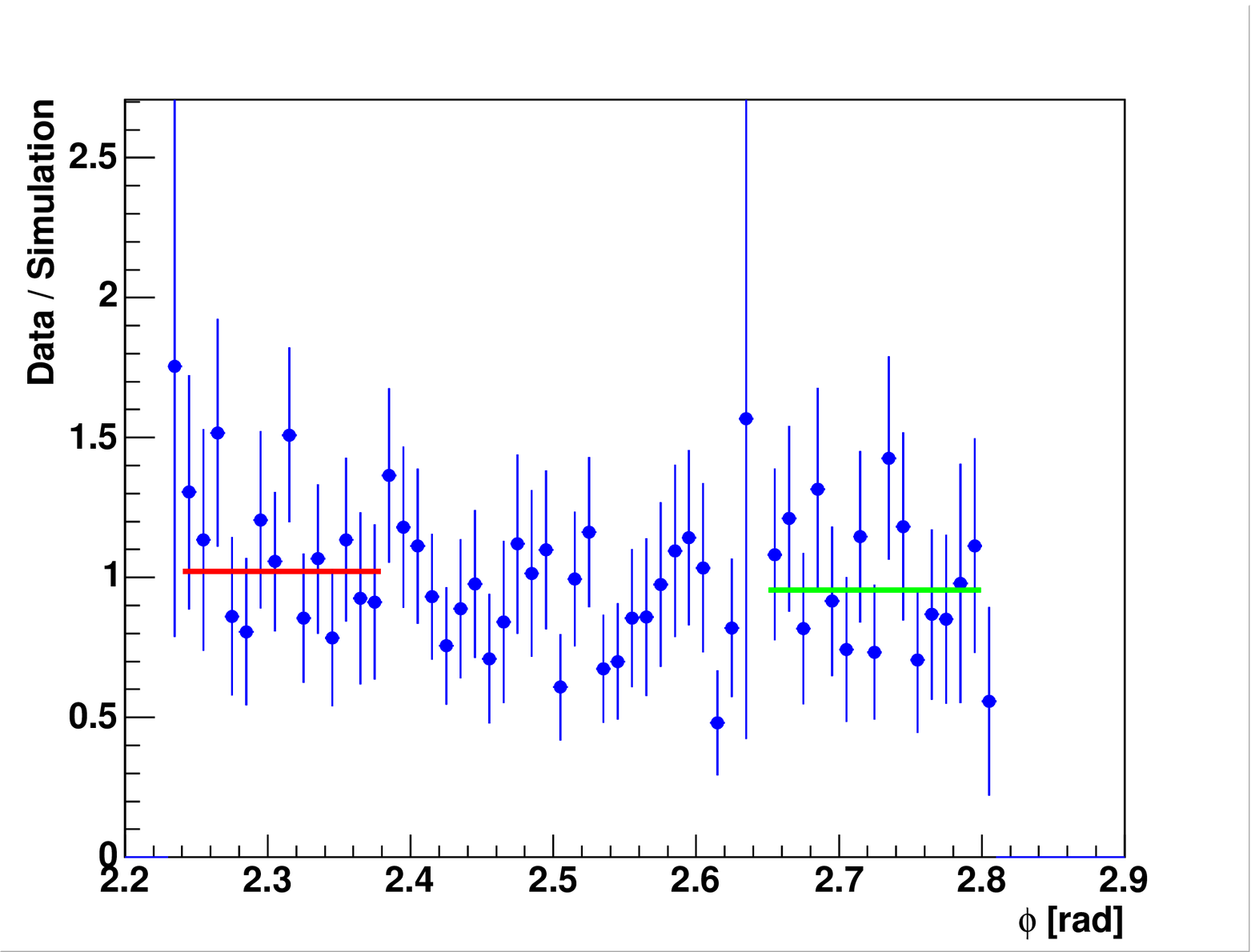,width=1\linewidth,clip,trim
= 0in 0.05in 0.5in 0.2in}
 \caption{\label{fig:ch4.ratio_acc_mb}
Ratio of $\phi$ acceptance of Minimum Bias electrons and PISA
simulation fitted in two $\phi$ regions shown in the plot.}
\end{figure}

 Acceptance systematic error takes into account possible
difference between Simulation and Data representations of the
acceptance. The $\phi$ acceptance of Simulation and MB Data sample
was compared (shown in Fig.~\ref{fig:ch4.comp_acc_mb}). The ratio
of $\phi$ acceptances presented on
Fig.~\ref{fig:ch4.ratio_acc_mb}.

Two different regions in $\phi$ (shown in
Fig.~\ref{fig:ch4.ratio_acc_mb}) were selected and the difference
of the average ratios in those regions was assumed to be a
systematic error $\delta_{acc} \approx 7\%$ on the acceptance
variation between MC and Data. This value probably includes a
significant amount of statistical error.

\pagebreak

\subsubsection{Run by bun systematic error}

Run by run systematic error was evaluated by dividing the run
period into two run groups with roughly equal statistics.

\begin{itemize}
\item {Run group I :  $Run \# < 40100$} \item  {Run group II: $Run
\# > 40100$}
\end{itemize}

\begin{figure}[h]
\centering
\epsfig{figure=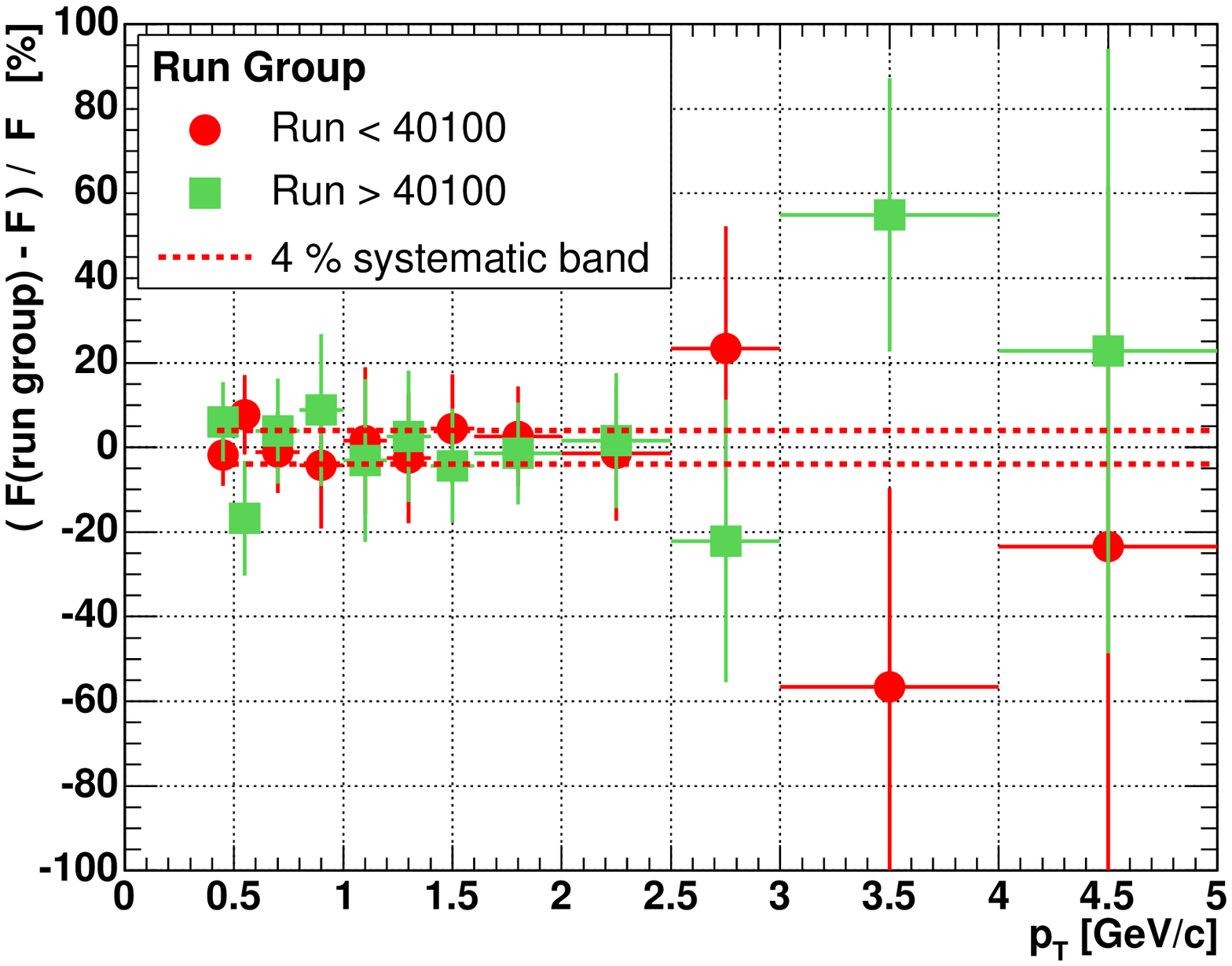,width=0.7\linewidth,clip}
\caption{\label{fig:ch4.ratio_run} Variation of inclusive electron
crossection for two separate run groups compared to total run
statistics.}
\end{figure}

The ratio of final inclusive crossection for Run Group to a total
is shown in Fig.~\ref{fig:ch4.ratio_run}. The systematic error of
4\% covers the ratio (at $p_T > 2.5$ GeV/c we run out of
statistics and can not make any conclusion).

\subsubsection{Final systematic error to the inclusive
crossection}

\begin{figure}[t]
\centering
\epsfig{figure=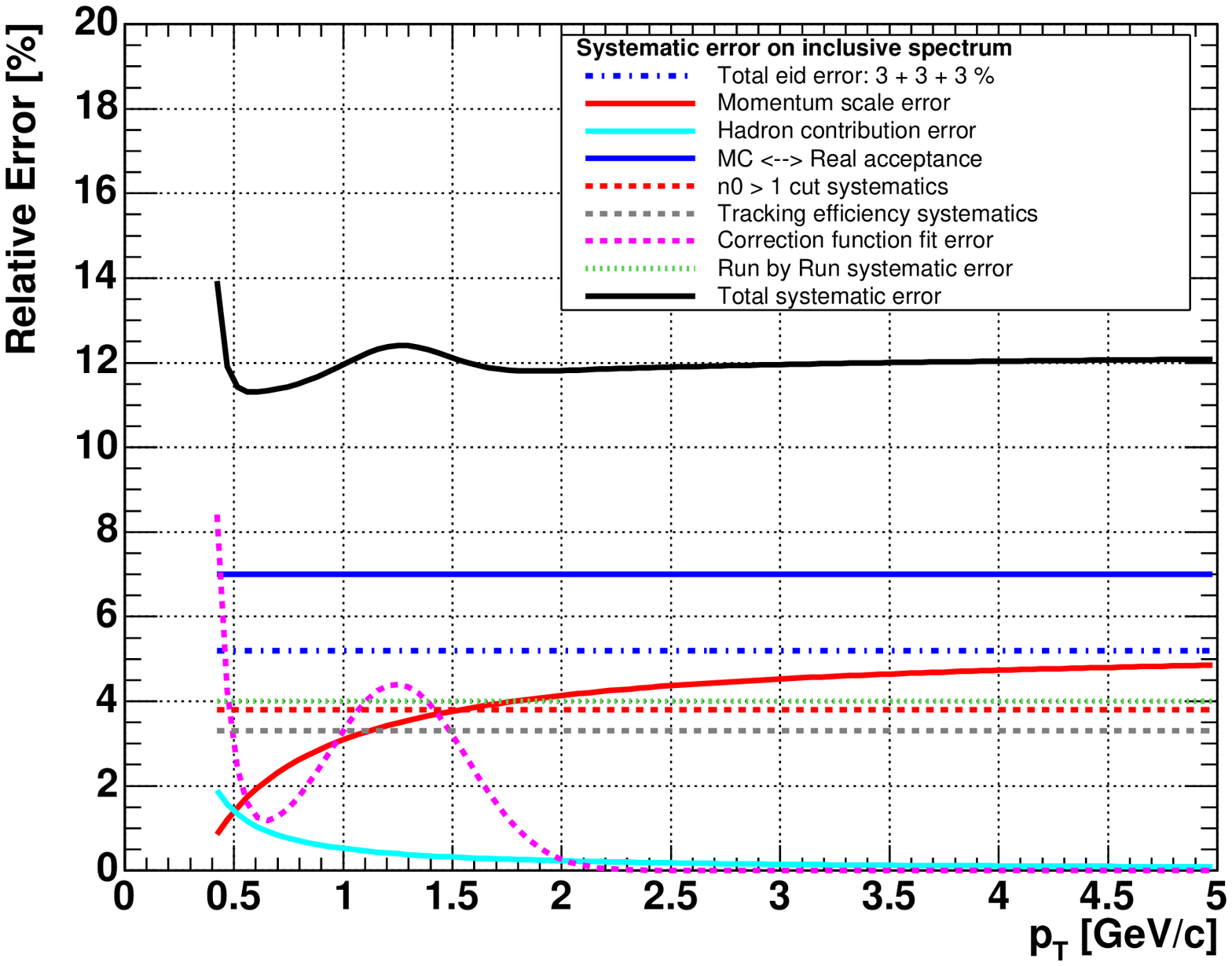,width=0.85\linewidth,clip,trim
= 0in 0.05in 0.5in 0.5in} \caption{\label{fig:ch4.final_inc_sys}
Total systematic error on the inclusive electron crossection.
Separate contributions to the systematic shown in the plot. Total
systematic is quadratic sum of individual components.}
\end{figure}

Final systematic error, the squared sum of all previously listed
components, is shown in Fig.~\ref{fig:ch4.final_inc_sys} and
summarized in Table~\ref{tab:final_inc_sys}. One can see that in
the total systematic error is almost constant for measured $p_T$
range and can be approximately taken as a $\bold{constant\ 12\%}$
throughout the whole $p_T$ range.

\begin{table}[h]
\caption{ Components of the total systematic error on the
inclusive crossection.} \centering
\begin{tabular}[b]{|c|c|c|}
\hline Error source & Value [\%]& Figure\\
\hline &&\\
Total eID cut& $3.0\oplus 3.0\oplus3.0$&~\ref{fig:ch4.matching_sys}-~\ref{fig:ch4.zed_sys}\\
Momentum Scale&$3.09+1.921\cdot ln(p_T)/p_T^{0.3485}$&~\ref{fig:ch4.mom_scale_sys}\\
Hadronic contribution& $0.51/(p_T-0.16)$& ~\ref{fig:ch4.hadr_sys}\\
MC to Data acceptance & 7.0& ~\ref{fig:ch4.ratio_acc_mb}\\
$n0>1$ cut & 3.8 & \\
Tracking efficiency & 3.0& \\
Correction function& $0.003047/p_T^{8.92}+4.588e^{-(p_T-1.252)^2/0.2} $& ~\ref{fig:ch4.fast_cf}\\
Run-by-run variation & 4.0 & ~\ref{fig:ch4.ratio_run}\\
&&\\
\hline
Total systematics & $\approx$ 12.0 & ~\ref{fig:ch4.final_inc_sys}\\

\hline \end{tabular} \label{tab:final_inc_sys}
\end{table}

\subsection{Systematic error of the Cocktail}\label{sec:ch4.Cocktail_Systematics}

Systematic error on the Cocktail electron prediction consists of
following contributions (errors listed from the most to the least
significant):
\begin{itemize}
    \item Systematic error on the input pion crossection
    \item Systematic error on Conversion/Dalitz ratio
    \item Systematic error on the meson/$\pi^0$ ratios
    \item Systematic error on Direct photon contribution
    \item Systematic error on Kaon $K_{e3}$ contribution
\end{itemize}

Total cocktail systematic is the quadratic sum of all separate
contributions.

\subsubsection{Input pion systematic error}

\begin{figure}[h]
\centering
\epsfig{figure=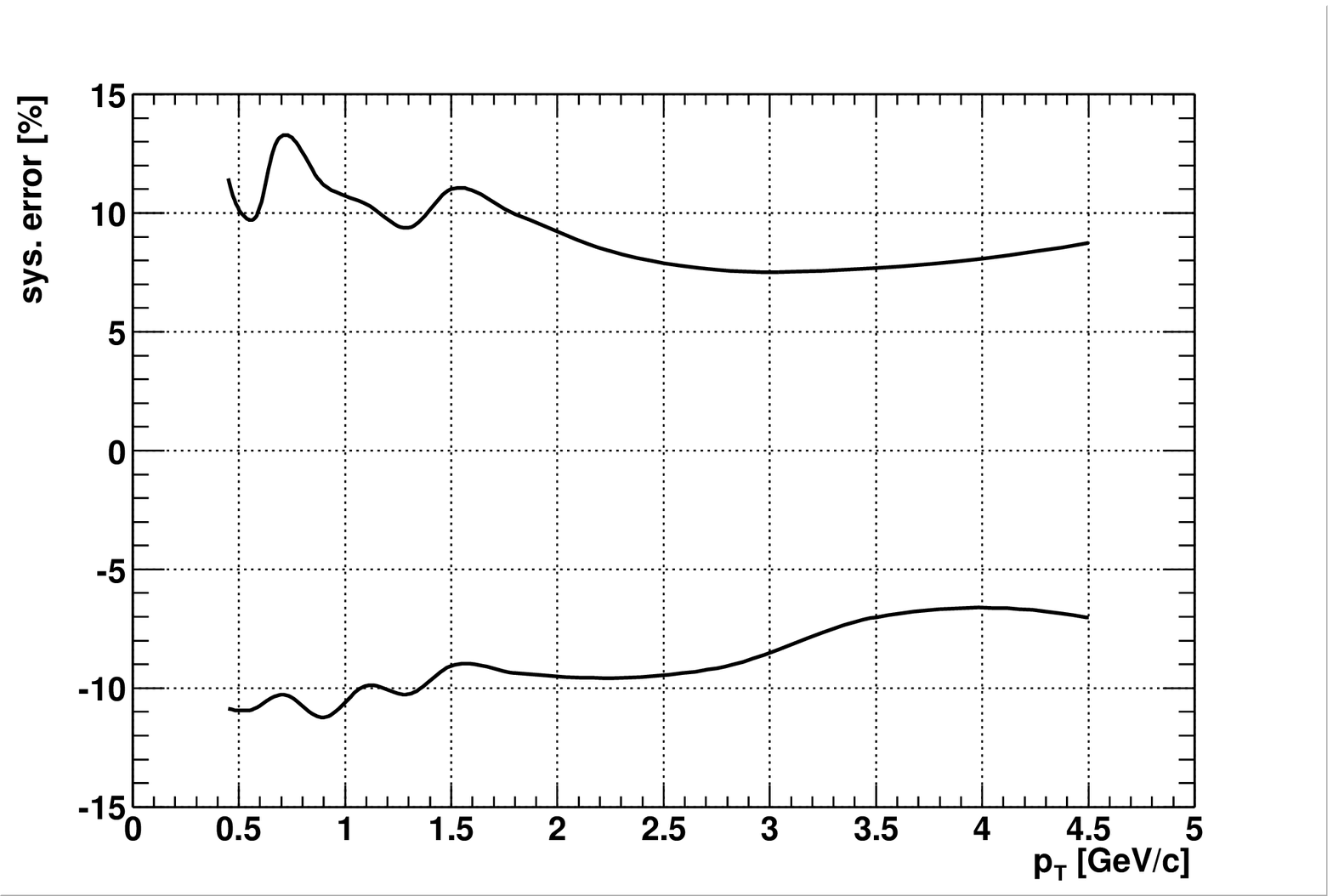,width=0.6\linewidth,clip,trim
= 0in 0.1in 0.1in 0in} \caption{\label{fig:ch4.pion_sys}
Systematic uncertainty of the cocktail due to the uncertainty in
the pion input spectra.}
\end{figure}

To quantify the systematic error of the pion input (see
Fig.~\ref{fig:ch4.pion_fit}) we move all neutral and charged pion
data points up (down) by their individual systematic error and
repeat the fit to obtain 1$\sigma$ error band for the pion input
spectra. Full cocktails are calculated with the upper (lower)
bounds of the pion input, and the difference to the optimum input
crossection provides an estimate for the pion input related
systematic uncertainty. Fig.~\ref{fig:ch4.pion_sys} shows the
systematic error hi-lo band due to the uncertainty of the pion
input.

\subsubsection{Other light mesons systematic error}

\begin{figure}[h]
\begin{tabular}{lr}
\begin{minipage}{0.5\linewidth}\centering
\epsfig{figure=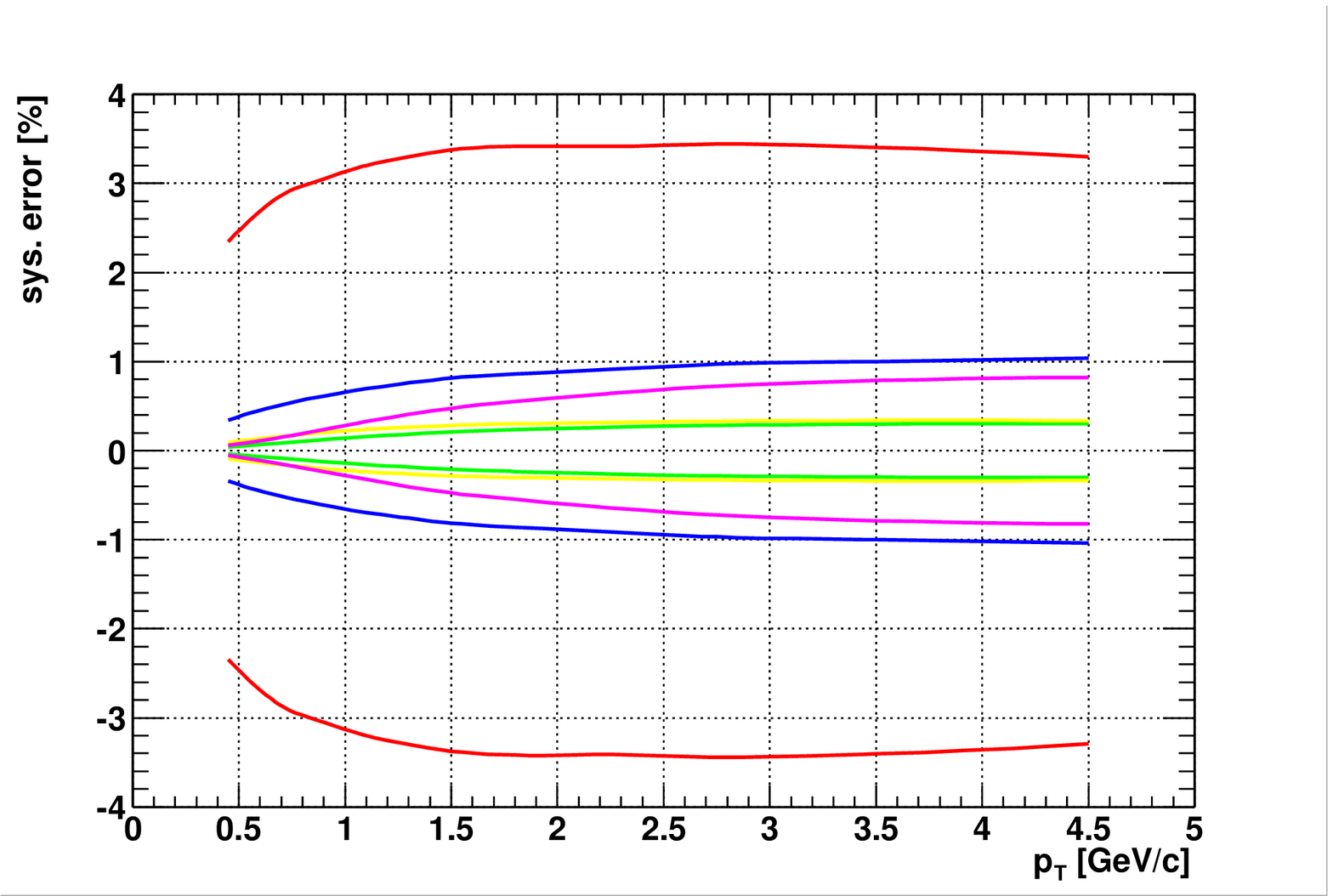,width=1\linewidth,clip,trim
= 0in 0.1in 0.1in 0in} \caption{\label{fig:ch4.meson_sys}
Systematic uncertainty of the cocktail due to the uncertainty in
the meson/$\pi^0$ ratios(see Table.~\ref{tab:part_ratios}).
$\eta$, $\eta'$, $\omega$, $\rho$, $\phi$ contributions shown in
different colors.}
\end{minipage}
&
\begin{minipage}{0.5\linewidth} \centering \epsfig{figure=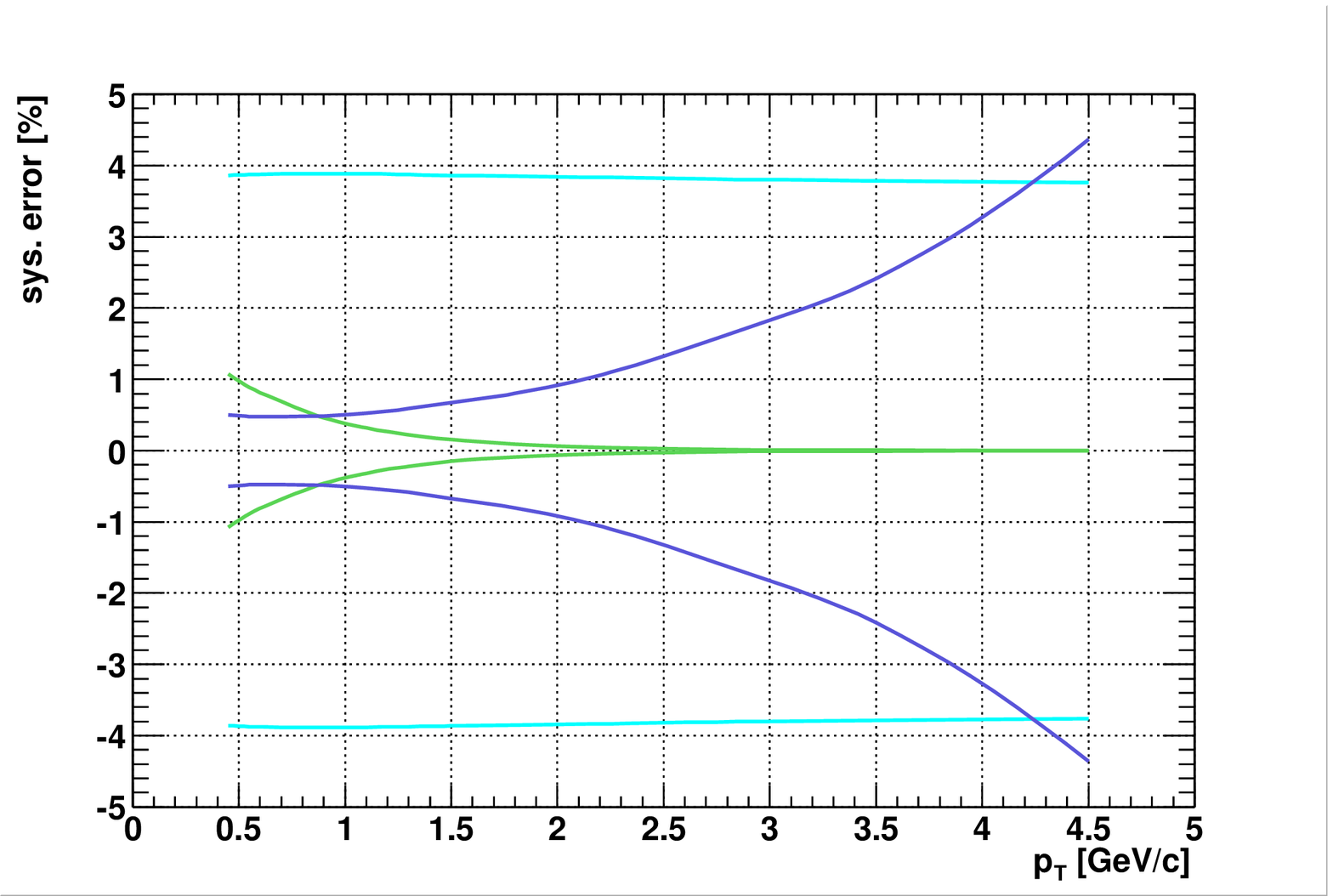,width=1\linewidth,clip,trim
= 0in 0.1in 0.1in 0in} \caption{\label{fig:ch4.sys_uns_small}
Systematic uncertainty of the cocktail due to the uncertainty in
the contribution from photon conversions (light blue curve at
about 4 \%), from weak kaon decays (green curve), and direct
photons (dark blue curve)}
\end{minipage}
\end{tabular}
\end{figure}

The systematic uncertainties of all cocktail ingredients were
discussed in the sections above (see
Table.~\ref{tab:part_ratios}). Just as for the pion input, the
ratio of each meson to $\pi^0$ was evaluated at hi-lo band of the
systematic uncertainty and the resulting ratio of modified
cocktail to the final one gives a systematic error due to each
contributing meson (shown in Fig.~\ref{fig:ch4.meson_sys}). The
largest systematic error of $\approx 3.5\%$ corresponds to
$\eta$-meson contribution. Contributions from the $\eta'$,
$\omega$, $\rho$, $\phi$ mesons are smaller then 1\%.

\subsubsection{Conversion rate uncertainty}

The systematics of the representation of PHENIX material in the
simulation was studied in $\Au$ Run02~\cite{ppg035} by
reconstructing the conversion pairs by the orientation with
respect to the magnetic field. In our case this analysis seems
unfeasible due to very low statistics of the conversion background
in $\pp$ collisions. The conclusion from $\Au$ analysis is that
the systematic error on the converter component is on the order of
10 \% independent of $p_T$. this translates to the error on the
Cocktail of 4\% also independent of $p_T$ (shown in
Fig.~\ref{fig:ch4.sys_uns_small}).

\subsubsection{Kaon decay uncertainty}

The systematic uncertainty from weak kaon decays, in general, is
tiny and is relevant at low $p_{T}$ only, if at all, as shown in
Fig.~\ref{fig:ch4.sys_uns_small}.

\subsubsection{Direct photon uncertainty}

The systematic uncertainty from direct photons increases with
increasing $p_{T}$ and becomes significant only at the highest
$p_{T}$ covered in this measurement as shown in
Fig.~\ref{fig:ch4.sys_uns_small}.

\subsubsection{Total cocktail systematic uncertainty}

\begin{figure}[h]
\centering
\epsfig{figure=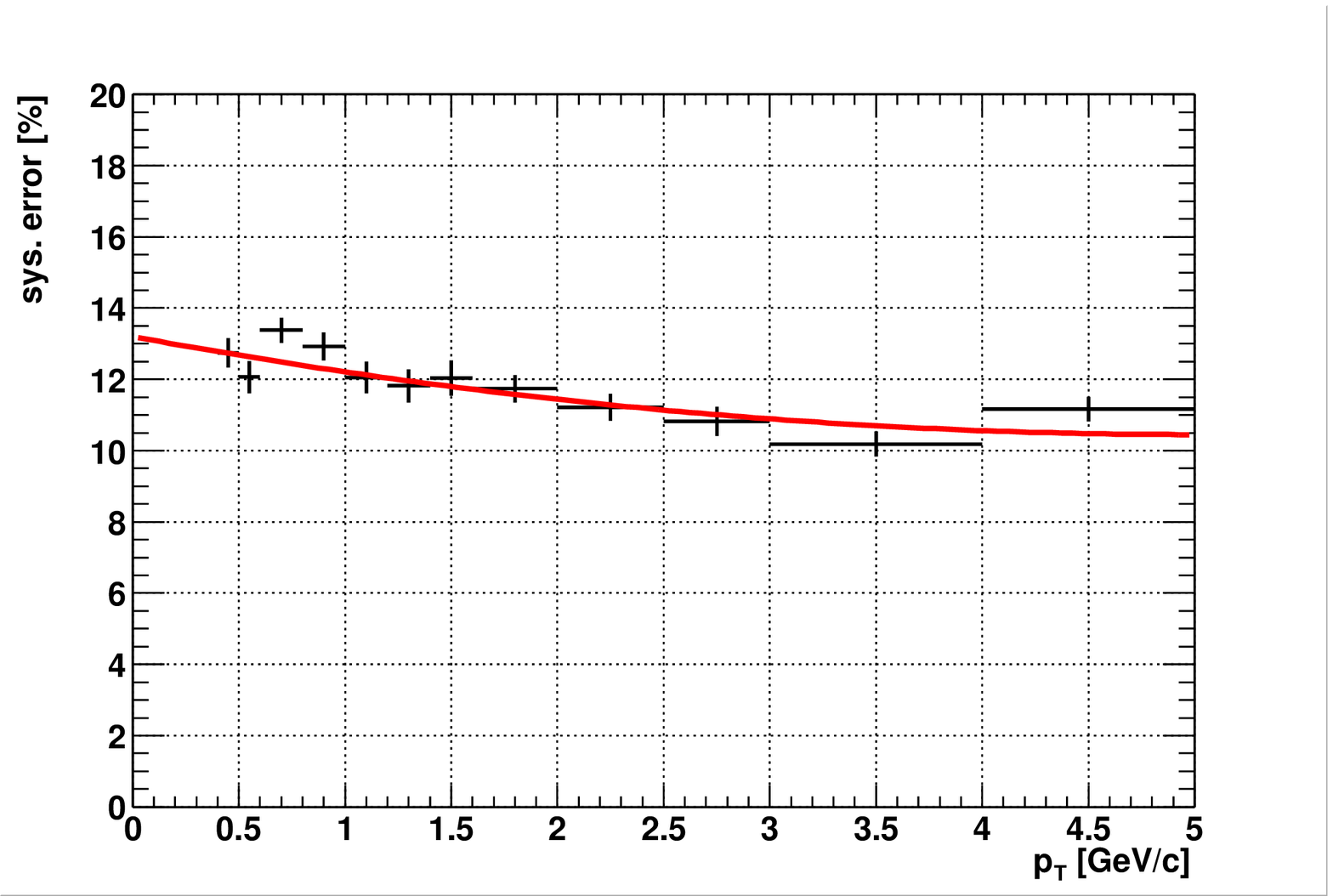,width=1\linewidth,clip,trim
= 0in 0.1in 0.1in 0in} \caption{\label{fig:ch4.sys_uns_total}
Total systematic error on the cocktail.}
\end{figure}

The total uncertainty on the cocktail is calculated as a quadratic
sum of all contributions and shown in
Fig.~\ref{fig:ch4.sys_uns_total}. One can see that it is slightly
falling with $p_T$ from $13 \%$ to $10 \%$.

\pagebreak

\subsection{Systematic error of the subtracted crossection}\label{sec:ch4.Subtracted_Systematics}

\begin{figure}[h]
\centering
\epsfig{figure=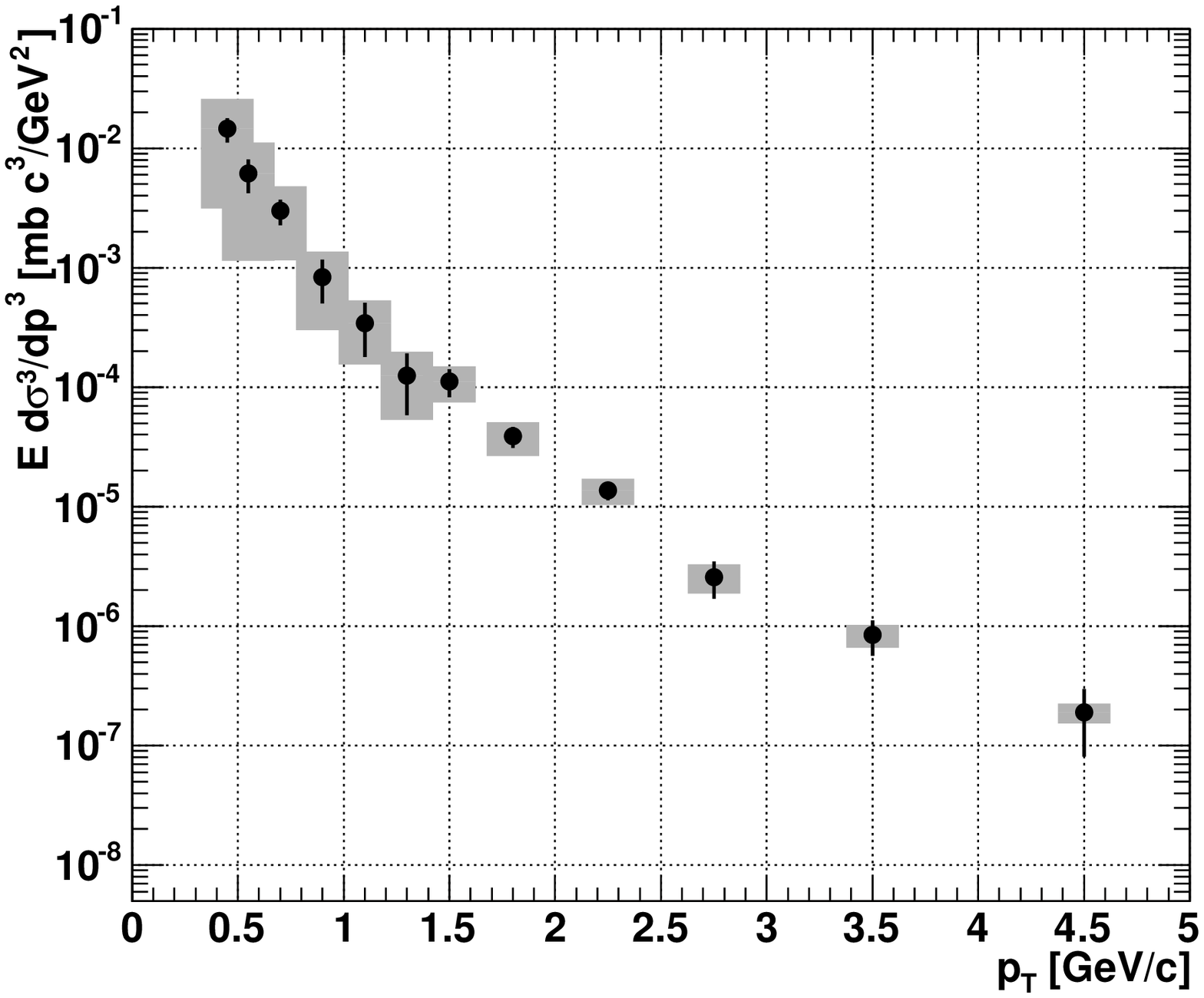,width=1\linewidth,clip,trim = 0in
0.1in 0.1in 0in} \caption{\label{fig:ch4.sys_final} Total
systematic error on the "Non-photonic" electron invariant
crossection.}
\end{figure}

Total systematic error on the data was calculated as a squared sum
of relative contributions to the inclusive and the cocktail. The
final "Non photonic" electron spectrum with the corresponding
systematic error band is shown in Fig.~\ref{fig:ch4.sys_final} and
summarized in Table~\ref{tab:final_nonphotonic}.

\chapter{Results}\label{sec:ch5}

In this chapter I present the comparison of various theoretical
predictions to the measured "Non-photonic" electron crossection
(see Section~\ref{sec:ch5.theory_comp}). The total Open Charm
crossection $\scc$ and single differential crossection
$\frac{d\scc}{dy}_{|y|<0.5}$ comparisons are described in
Section~\ref{sec:ch5.total_crossection}. We also compare the
result of our Open Charm measurement from $\pp$ collisions at
$\sqs = 200$ GeV with existing nucleus-nucleus measurements ($\dA$
and $\Au$) from PHENIX~\cite{ppg035} and STAR~\cite{STAR_se_pp}.

\section{Comparison with Theory}\label{sec:ch5.theory_comp}

Theory predictions of the Heavy Flavor crossection in $\pp$ collisions
(see Chapter~\ref{sec:ch2}) \textit{"remain at a rather primitive
level"}~\cite{whatK,phenom_cb,system_prod} due to the strong dependence of
the pQCD crossections on the input parameters and the simplistic
assumptions made for the fragmentation
mechanism~\cite{excess,saga,bottom_production}. In this section we
compare the measured electron crossection to the predictions of Open
Charm (Bottom) production as made by several theoretical models:

\begin{enumerate}
    \item Standard PYTHIA v.6.152~\cite{pythia} LO calculations
    \item Full NLO Monte Carlo code HVQLIB~\cite{hvqlib} for heavy
    quark production calculation + PYTHIA fragmentation and decay
    \item Theoretical prediction from FONLL by Matteo Cacciari.
\end{enumerate}

\pagebreak
\subsection{Comparison with PYTHIA}\label{sec:ch5.PYTHIA_comp}

PYTHIA~\cite{pythia} is a standard event generator widely used in
High Energy Physics to simulate particle production in $\pp$
collisions. PYTHIA uses Leading Order (LO) matrix elements and
the Lund string fragmentation model~\cite{LUND} to calculate Heavy
Flavor meson production. Keeping the full information of the
collision dynamics starting from the initial parton information to
the final decay products of the Open Charm particle enables us to
make a direct correspondence between "Non-photonic" electrons,
heavy mesons and the parent heavy quarks.

There is a long list of PYTHIA input parameters, however, the most
important ones are:
\begin{itemize}
    \item The mass of the heavy quark ($m_c$, $m_b$)
    \item The width of intrinsic parton $k_T$ smearing $\kt$~\cite{hq_prod}
    \item The assumed functional form of the $k_T$ smearing
    (gaussian is default)
    \item The $K-factor$~\cite{whatK} - \textbf{constant} factor,
    taking into account the ratio of NLO to LO heavy quark crossection
    \item The Parton Distribution Function choice from the PDFLIB
    package~\cite{PDFLIB}.
\end{itemize}

There is presently no good choice for the PHTHIA parameters that is
known to work well at all $\sqs$. We undertook the task of adjusting
the parameters within reasonable theorical and experimental limits so
as to produce the best fits to currently existent Heavy Flavor
measurements.  Following this, PYTHIA is then able to make a
``prediction'' for our measurements.

\subsubsection{Masses of heavy quarks}

From the PDG~\cite{PDG} the mass of Charm and Bottom quarks lies
within the limits:
\begin{itemize}
    \item $m_c \in (1.0 - 1.4)\ GeV/c^2$
    \item $m_b \in (4.0 - 4.5)\ GeV/c^2$
\end{itemize}

Those results are theoretical calculations from charmonium and D
meson masses of "running" mass $m_Q$ using scale $\mu=m_{Q}$ and
two-loop pQCD corrections to mass in the $\overline{MS}$ scheme.
\pagebreak

We adjusted the PYTHIA parameters so as to best describe the data from
existing $p+N$ and $\pi+N$ experiments ranging from lower energy fixed
target measurements at SPS to
FNAL~\cite{alves_e769,wa92,e791,ccrs,basile} results. These studies
~\cite{ana101} converged to the following parameter set, and provided
a good agreement with ($\sqs < 63$ GeV)world data:
\begin{itemize}
    \item Proton PDF $= CTEQ5L$
    \item $m_c = 1.25$ GeV/$c^2$, $m_b = 4.1$ GeV/$c^2$
    \item $\kt = 1.5$ GeV
    \item $K = 3.5$
\end{itemize}

This PYTHIA parametrization, referred to as the ``\textbf{standard}
PHENIX parametrization'', was used in our previous publications
~\cite{ppg011,ppg035} and will be used as a reference for comparison
with the current analysis.

\begin{figure}[h]
\centering
\epsfig{figure=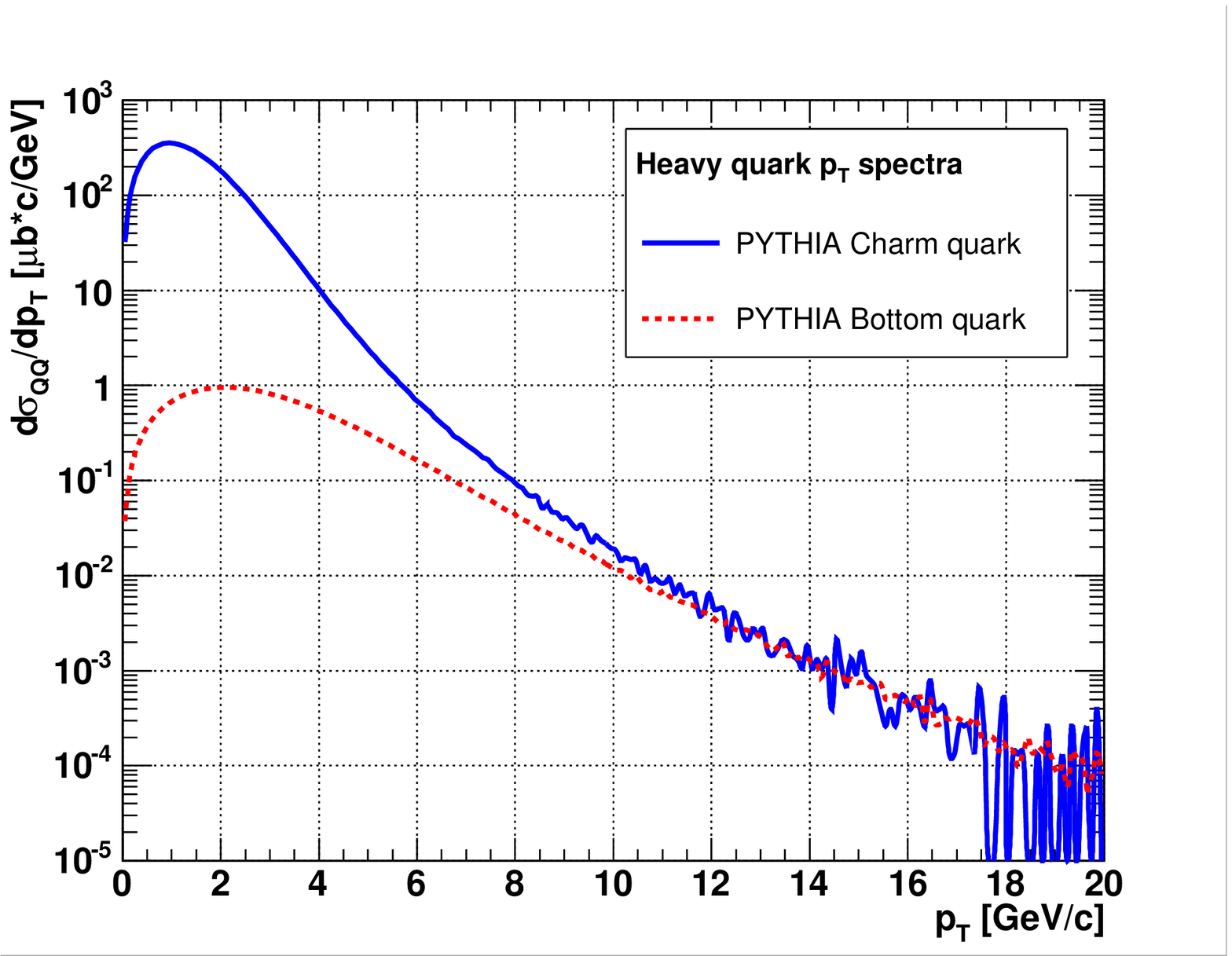,width=0.45\linewidth,clip,trim
= 0.in 0.1in 0.1in 0in}
\epsfig{figure=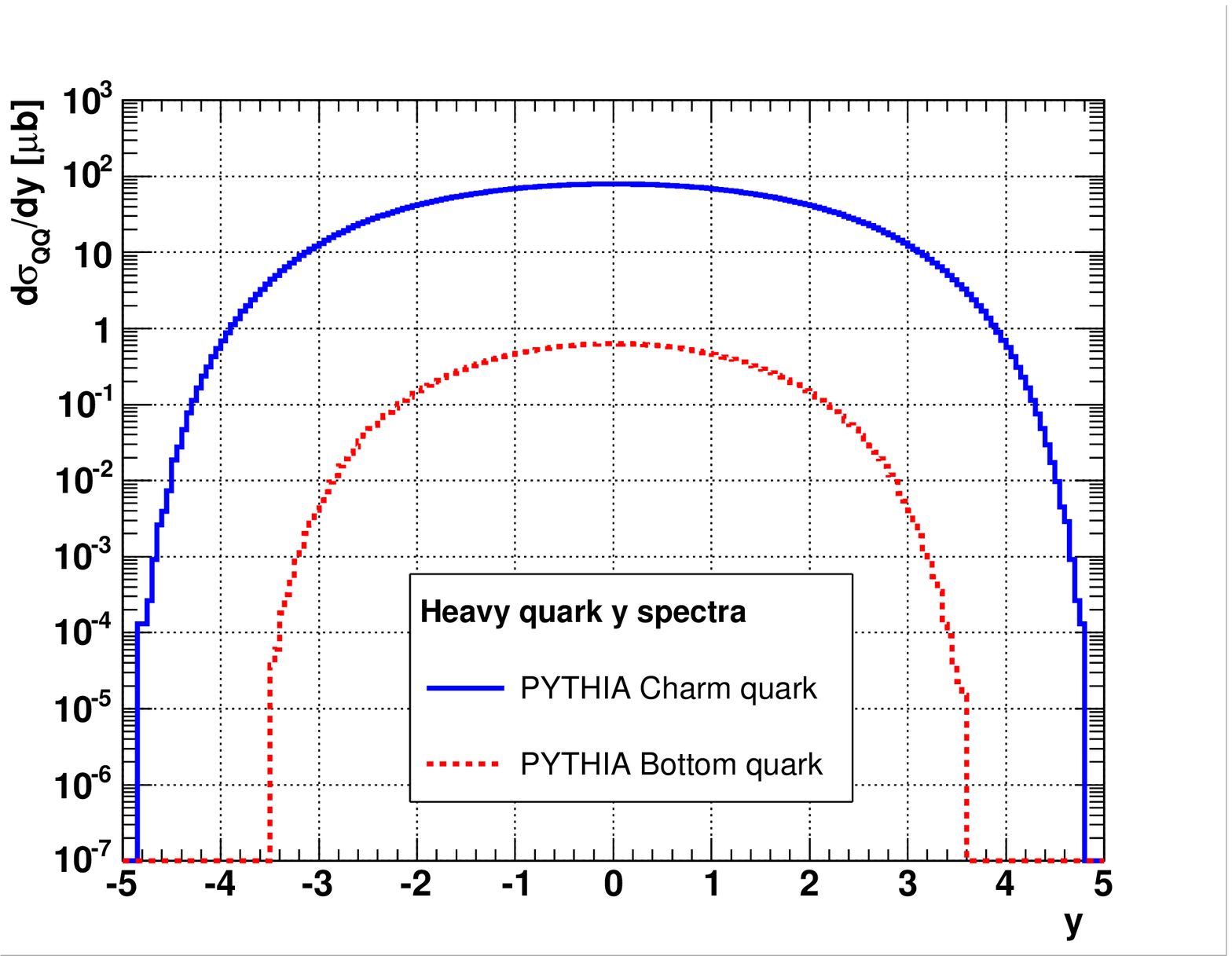,width=0.45\linewidth,clip,trim =
0.in 0.1in 0.1in 0in} \caption{\label{fig:ch5.pythia_c_b_pt_y}
Default PYTHIA $p_T$ (left panel) and rapidity (right panel)
distribution for charm (solid) and bottom (dashed) $Q\overline{Q}$
pairs.}

\epsfig{figure=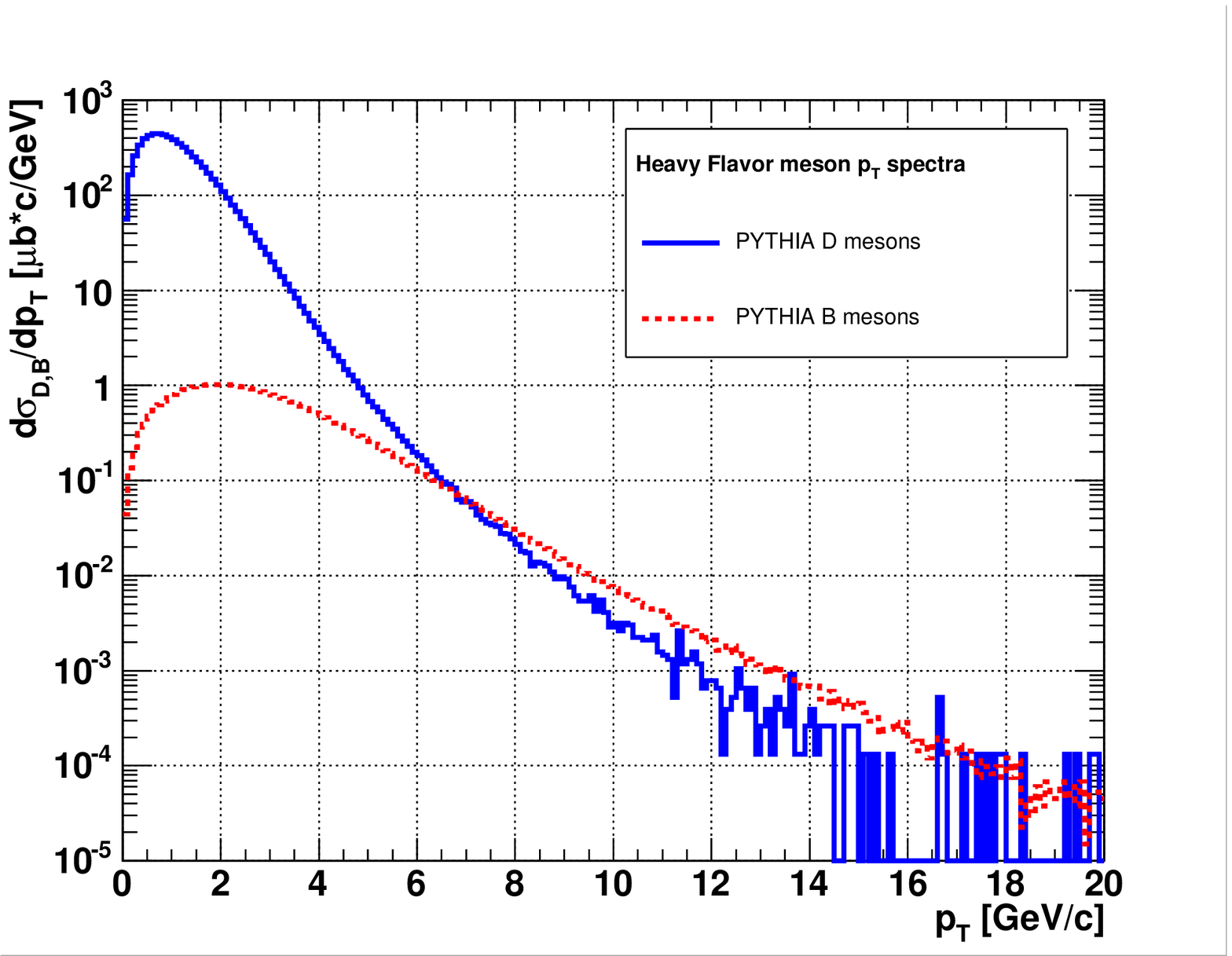,width=0.45\linewidth,clip,trim
= 0.in 0.1in 0.1in 0in}
\epsfig{figure=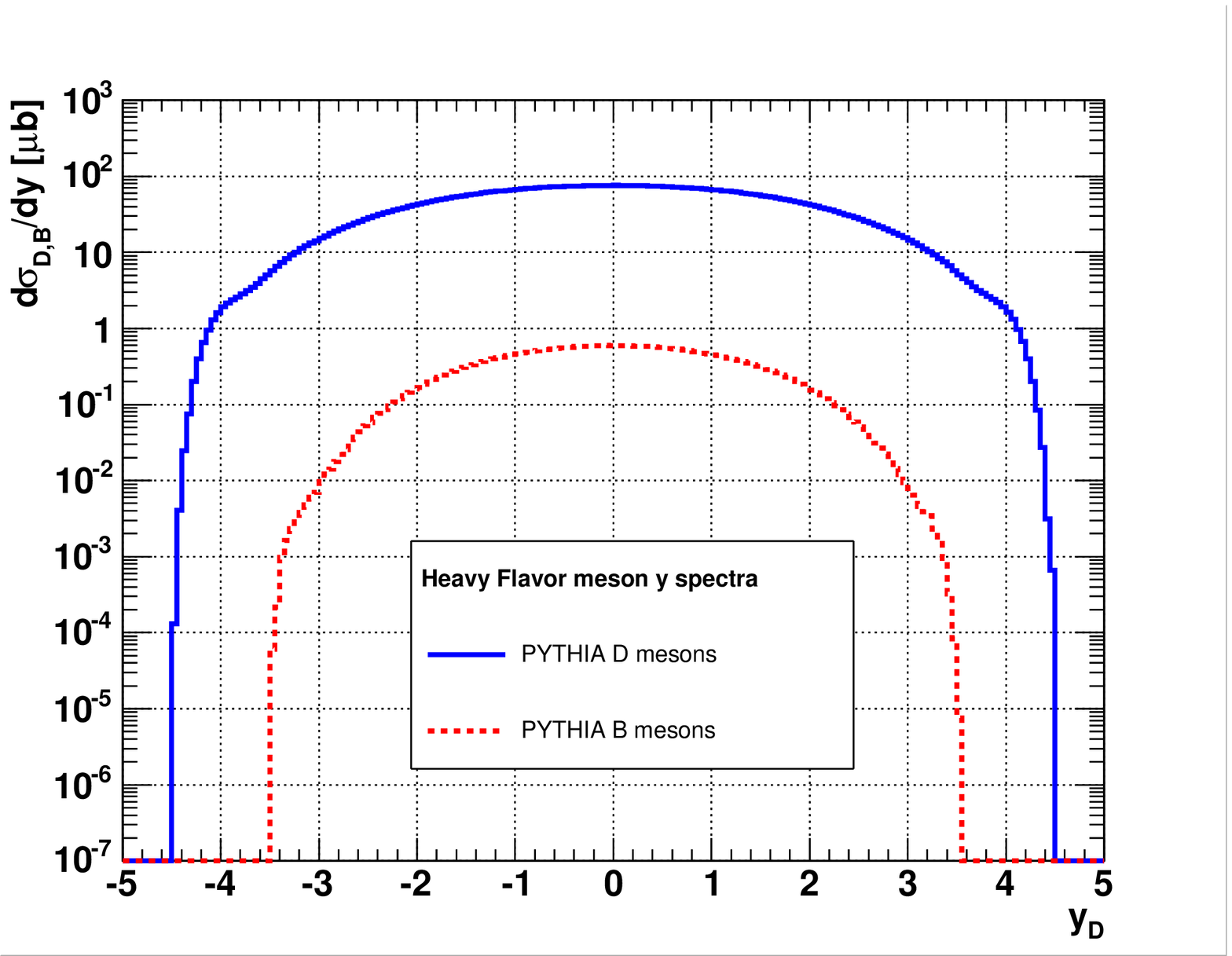,width=0.45\linewidth,clip,trim =
0.in 0.1in 0.1in 0in} \caption{\label{fig:ch5.pythia_D_B_pt_y}
Default PYTHIA $p_T$ (left panel) and rapidity (right panel)
distributions for Open Charm (solid) and Open Bottom (dashed)
particles.}
\epsfig{figure=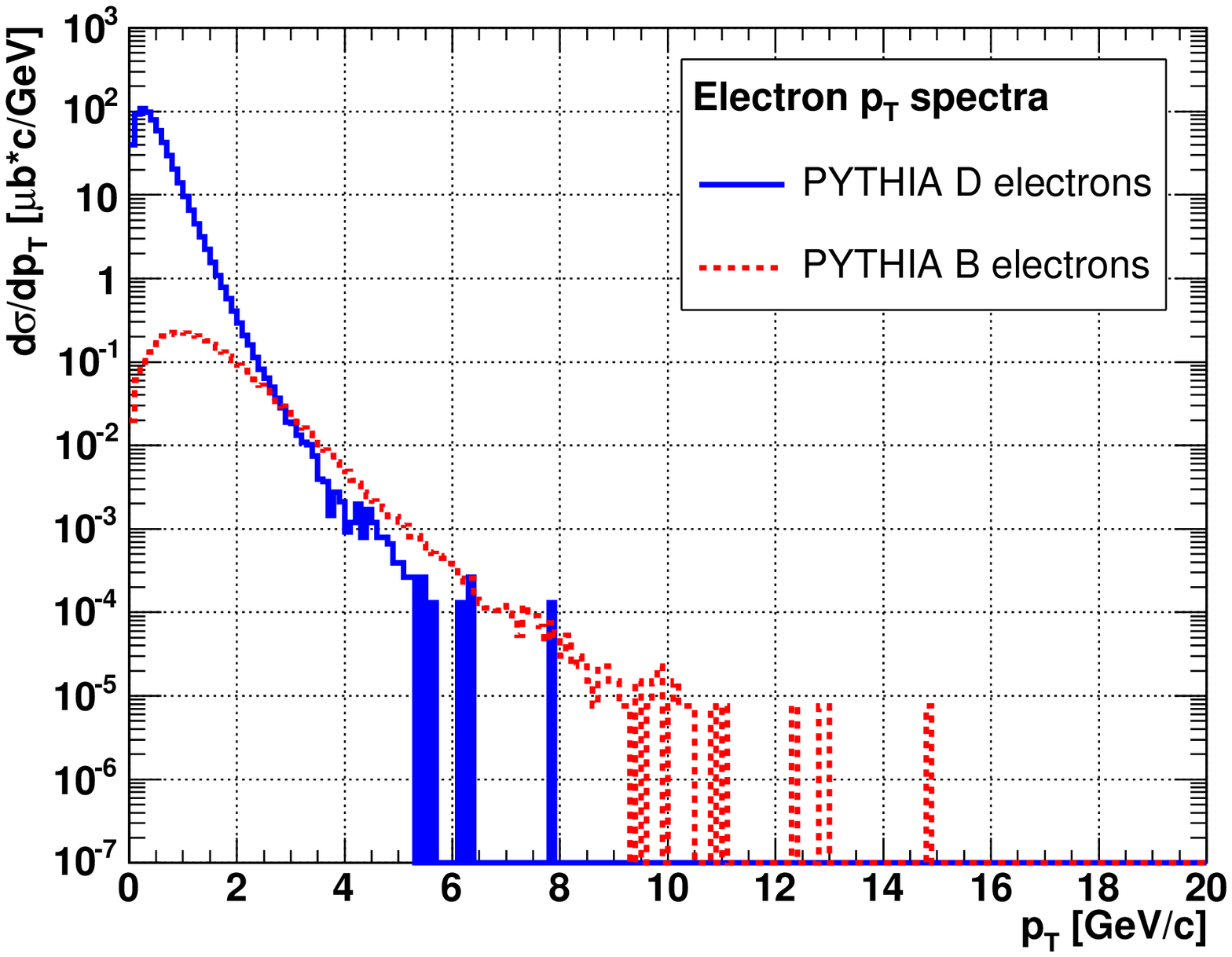,width=0.45\linewidth,clip,trim
= 0.in 0.1in 0.1in 0in}
\epsfig{figure=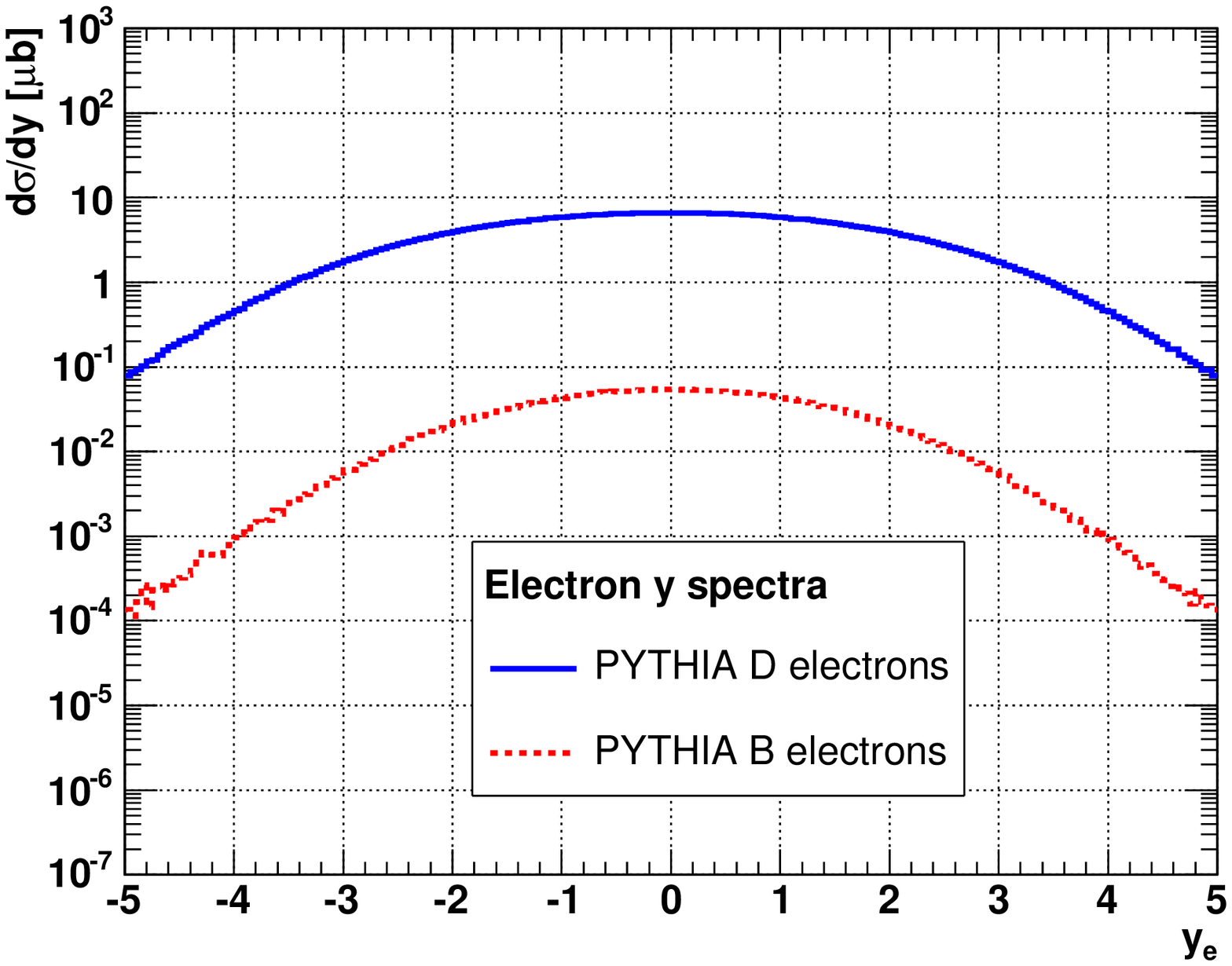,width=0.45\linewidth,clip,trim
= 0.in 0.1in 0.1in 0in} \caption{\label{fig:ch5.pythia_e_pt_y}
Default PYTHIA $p_T$ (left panel) and rapidity (right panel)
distributions for Open Charm (solid) and Open Bottom (dashed)
particle decay electrons.}
\end{figure}

Fig.~\ref{fig:ch5.pythia_c_b_pt_y} shows the Charm and Bottom quark
$\frac{d\sigma_{Q\overline{Q}}}{dp_T}$ and
$\frac{d\sigma_{Q\overline{Q}}}{dy}$ distributions from the default
PYTHIA parameters. Total heavy quark crossections derived from
integrals of these distributions are $\scc = 0.658\ mb$ and $\sbb = 3.77\
\mu b$.  Fig.~\ref{fig:ch5.pythia_D_B_pt_y} shows the $p_T$
distribution of produced $D$ and $B$ mesons and hadrons.

\begin{figure}[h]
\centering
\epsfig{figure=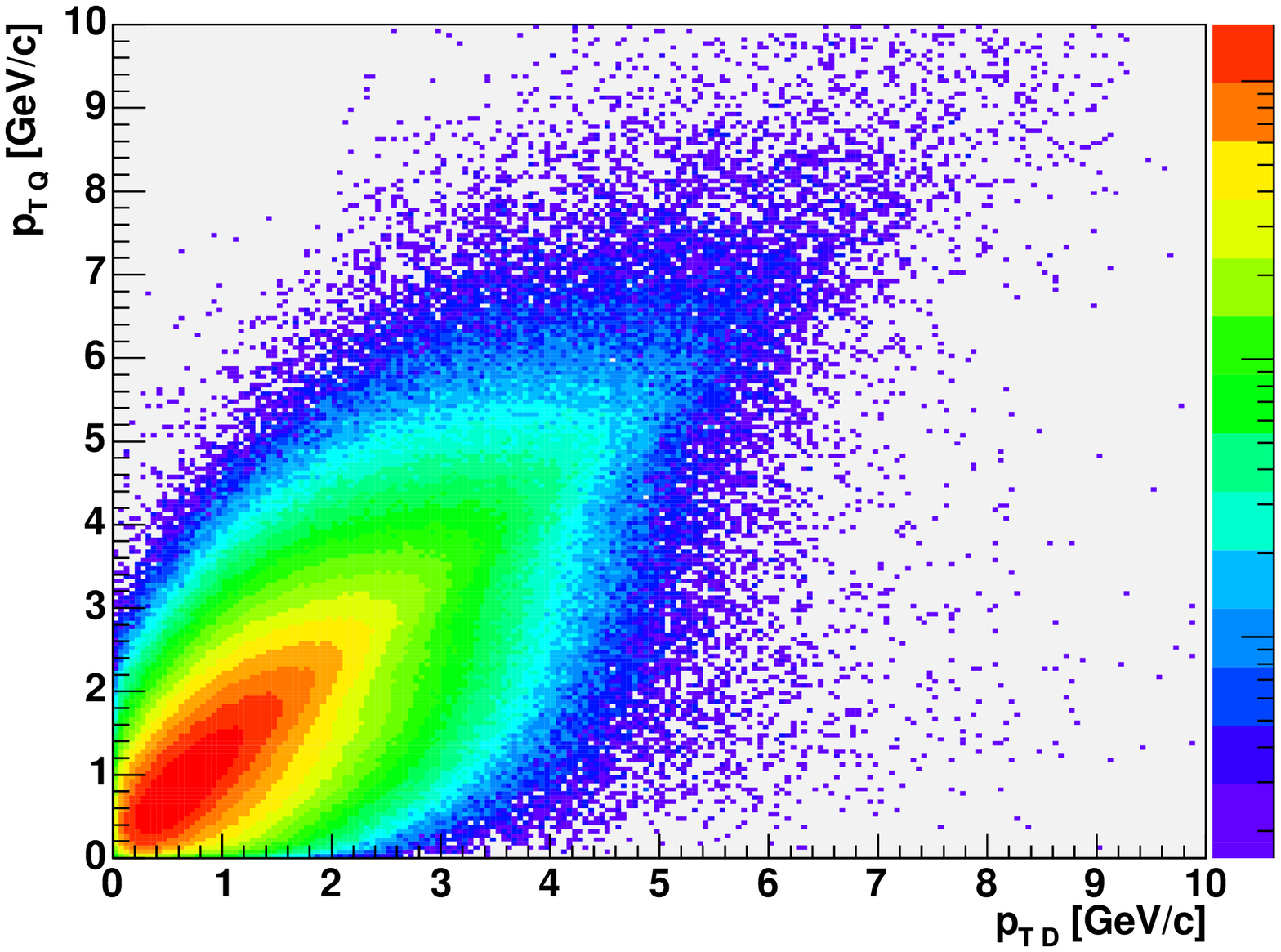,width=0.45\linewidth,clip,trim
= 0.in 0.1in 0.1in 0in}
\epsfig{figure=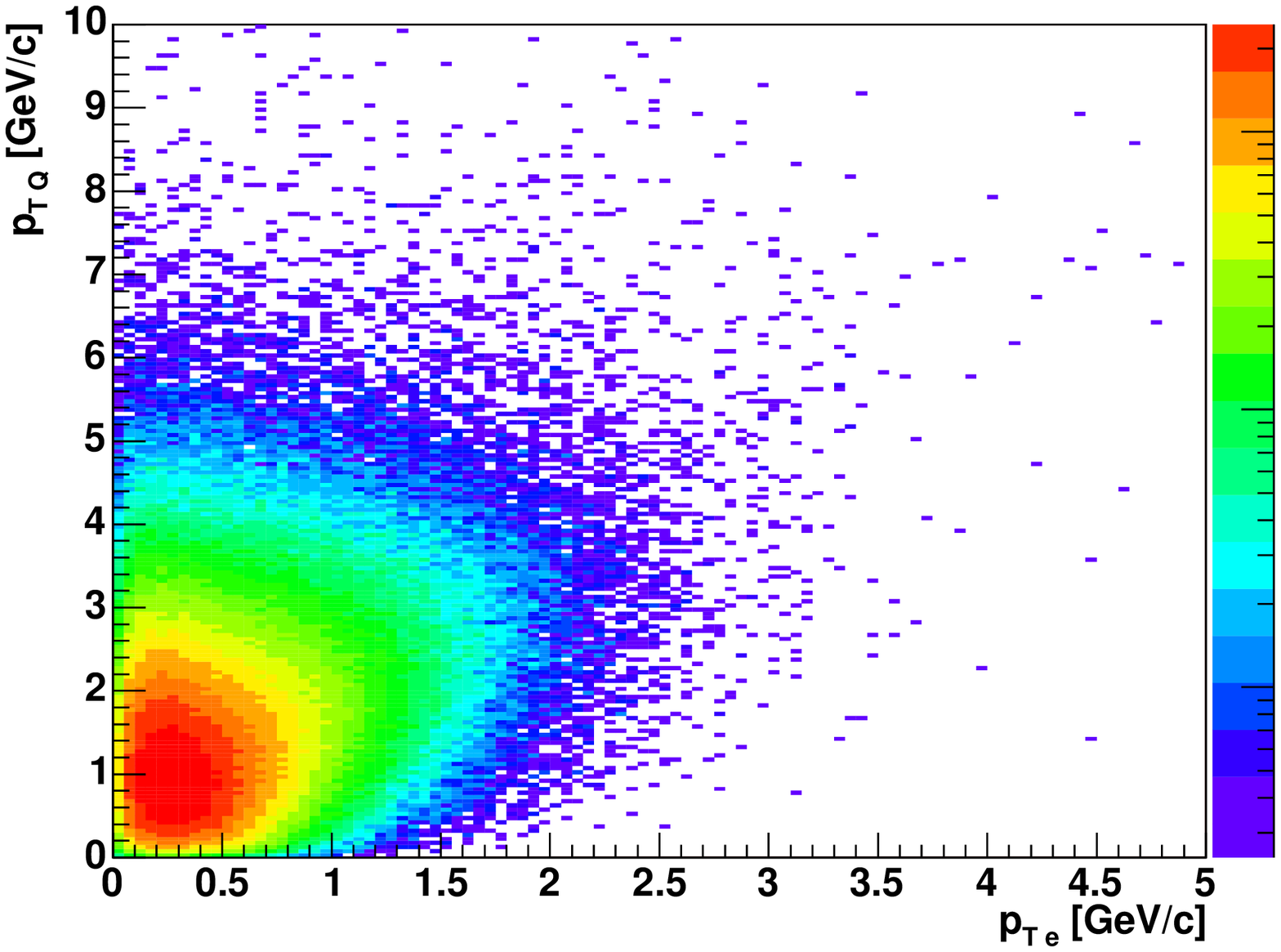,width=0.45\linewidth,clip,trim
= 0.in 0.1in 0.1in 0in} \caption{\label{fig:ch5.c_d_correlations}
Correlation between transverse momentum of charm quark and
daughter $D$-meson (left panel). Correlation between transverse
momentum of charm quark and $p_T$ of decay electron (right
panel).}
\end{figure}

The decay electron $p_T$ and rapidity distributions for charm and
bottom are shown in Fig.~\ref{fig:ch5.pythia_e_pt_y}. One can see that
due to weak decay kinematics, the electron rapidity and $p_T$
distribution's shape has a different slope than that of the input
quark. This effect is illustrated in
Fig.~\ref{fig:ch5.c_d_correlations} that shows the correlation between
the transverse momentum of the quark and the daughter $D-$meson.  One
can clearly see a nice quark-meson momentum correlation despite the
fact that the meson-electron correlation has almost vanished. This is
because of the very large q-value of the decay that can send the
electron in even the opposite direction as the parent meson.  This
fact is a \textbf{significant problem} for indirect Heavy Flavor
analysis through leptonic decay channel - it is very hard to make
direct correspondence of the measured electron signal to the original
quark distribution due to lost correlation in the decay.  On the other
hand, the large q-value of the decay is also responsible for
stiffening the electron spectrum and making these electrons distinct
from other sources at high $p_T$.  Thus, the large q-value of the
charm semi-leptonic decay is in a sense a double-edge sword.

\pagebreak

\subsubsection{Comparison to the "Non-photonic" electron crossection}

\begin{figure}[hb]
\centering
\epsfig{figure=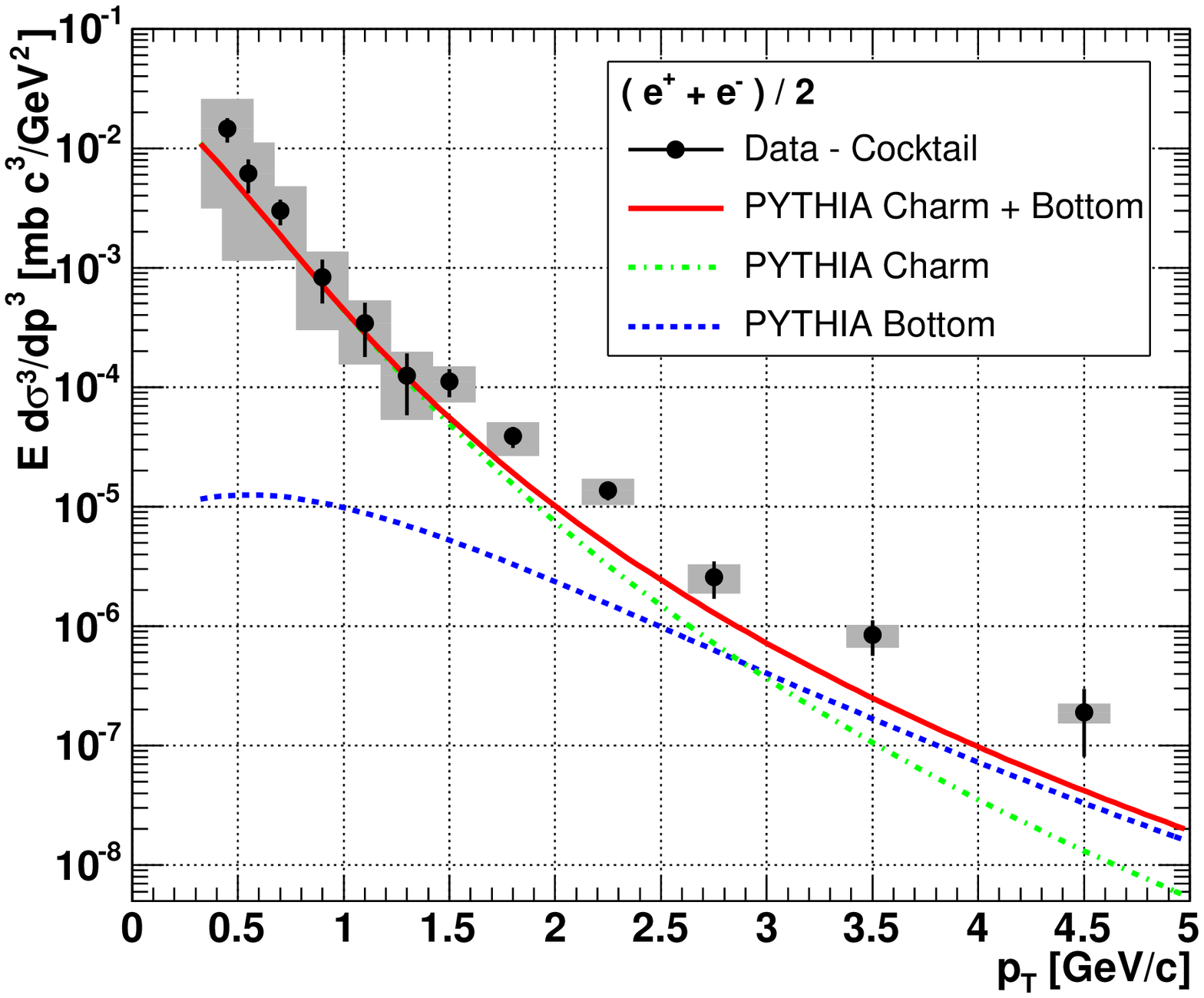,width=0.95\linewidth,clip,trim =
0.in 0.1in 0.1in 0in} \caption{\label{fig:ch5.comp_pythia}
Comparison of Non-Photonic electron crossection with default
PYTHIA expectation for Open Charm and Open Bottom semi-leptonic
decays.}
\end{figure}

 Now we have everything to compare the "Non-photonic electron spectrum
 with the default PYTHIA parametrization. The comparison is shown in
 Fig.~\ref{fig:ch5.comp_pythia}. One can see that default PYTHIA
 parametrization for Charm + Bottom \textbf{does not} do a good job in
 describing the electron invariant crossection and obviously can not
 be used as a reference for $\Au$ and $\dA$ nuclear modification
 effects. This is one of the most important results of this analysis.

There are several possible explanations for the observed discrepancy
 which we will try to investigate in the next sections on the paper:
\begin{enumerate}
\item Leading Order PYTHIA predictions for charm quark production
using a {\it $p_T$ independent} $K$ factor do not work.\\ Exact NLO
calculations predict $K$ factor to be {\it dependant} on the quark
momentum and rapidity~\cite{phenom_cb,whatK}. We need to perform
correct NLO calculations of the crossection. This was done in
Section~\ref{sec:ch5.NLO_comp}

\item The bottom crossection is actually higher then we expect and the
    shape difference at $p_T> 3$ GeV/c is described by increase of
    Open Bottom decay electron rate. \\ This argument agrees with the
    observations from $D0$ experiment measurements of Open Bottom
    crossection in single leptonic channel at $\sqs = 1.8$ TeV
    ~\cite{bottom_D0,bottom_CDF} which overpredicts NLO theory by a
    factor of $\sim 1.5 - 2.0$ ~\cite{excess,bottom_production}. So
    far there are several attempts to explain the discrepancy by
    adjusting the fragmentation function and a resummation
    technique~\cite{bottom_production}. So far there is still an
    excess of the experiment over theory on the order of $(1.7 \pm
    0.5(expt)\pm 0.5 (theory))$~\cite{bottom_production}.

\item The excess is due to higher order pQCD contributions to the
    charm crossection.\\ Unfortunately, the only attempt to calculate
    the Next-to-Next-Leading order (NNLO) charm and bottom
    crossections~\cite{NNLO} is only appropriate to use near the
    production threshold and is inapplicable to our
    measurements. \textit{"A full calculation of next-to-next-leading
    order QCD contributions, years ahead in the future, might finally
    also contribute the apparent
    discrepancy"}~\cite{bottom_production}.

\end{enumerate}

\newpage

\subsection{Comparison with NLO pQCD}\label{sec:ch5.NLO_comp}

As we already showed, the default PYTHIA LO prediction does not
describe experimental results so we attempt to utilize full NLO pQCD
calculations for Heavy Flavor production by using the standard fortran
Monte Carlo code ``HVQLIB''~\cite{nlo_charm,nlo2,hvqlib} (see
Section~\ref{sec:HQ_prod} for the main aspects of NLO pQCD).  HVQLIB
have been used previously by theorists to describe the total and
single, double differential crossection for charm and bottom
production with reasonable accuracyat many $\sqs$. The latest results
for NLO predictions for the Open Charm and Bottom production can de
found in~\cite{hq_prod,whatK,phenom_cb,saga,bottom_production}.

The comparson of such calculations to our data is complicated by the
fact that the pQCD calculations provide us with only the prediction
for the quark-antiquark pair production.  We must also include the
steps of hadronization and heavy flavor decay to make a prediction for
the single electron spectra. In order to produce the final electron
crossection prediction we used the following procedure:
\begin{itemize}
    \item Calculate $\frac{d^2N^Q}{dp^Q_Tdy^Q}$ distribution for heavy
    quarks from HVQLIB and PYTHIA.
    \item Weight each electron in PYTHIA with the weight $R(p^Q_T,y^Q)$ equal
    to:

\begin{equation}
    R(p^Q_T,y^Q)=
    \frac{\frac{d^2N^Q}{dp^Q_Tdy^Q}_{HVQLIB}}{\frac{d^2N^Q}{dp^Q_Tdy^Q}_{PYTHIA}}
    \end{equation}
\end{itemize}

This technique enables us to utilize the fragmentation and decay
machine of PYTHIA/JETSET while "artificially" changing the quark
density to NLO prediction. This simple model assumes that PYTHIA and
HVQLIB use the same or similar parton densities and quark masses.

The input parameters for NLO pQCD are the factorization and
renormalization scales $\mu_F$, $\mu_R$, respectively. Although these
scales are, in principle, independent, theoretical calculations
usually assume the scales to be identical $\mu = \mu_F = \mu_R$
~\cite{phenom_cb} because this assumption is inherent in the global
analysis of parton densities. We need to use NLO matrix elements in
these calculations. Although the use of LO PDFs is also possible in
HVQLIB, and produces similar results~\cite{whatK} for the same PDF
group, it is not recommended.

In order to make a fair comparison, the ratio of HVQLIB to PYTHIA
was calculated switching the intrinsic $k_t$ smearing off. In this
case we obtain the theoretical $K-factor$ $K(p_T^Q,y^Q)\equiv
\frac{\sigma_{NLO}}{\sigma_{LO}}$ ~\cite{whatK}.
\pagebreak

\begin{figure}[ht]
\begin{tabular}{lr}
\begin{minipage}{0.5\linewidth}\centering \epsfig{figure=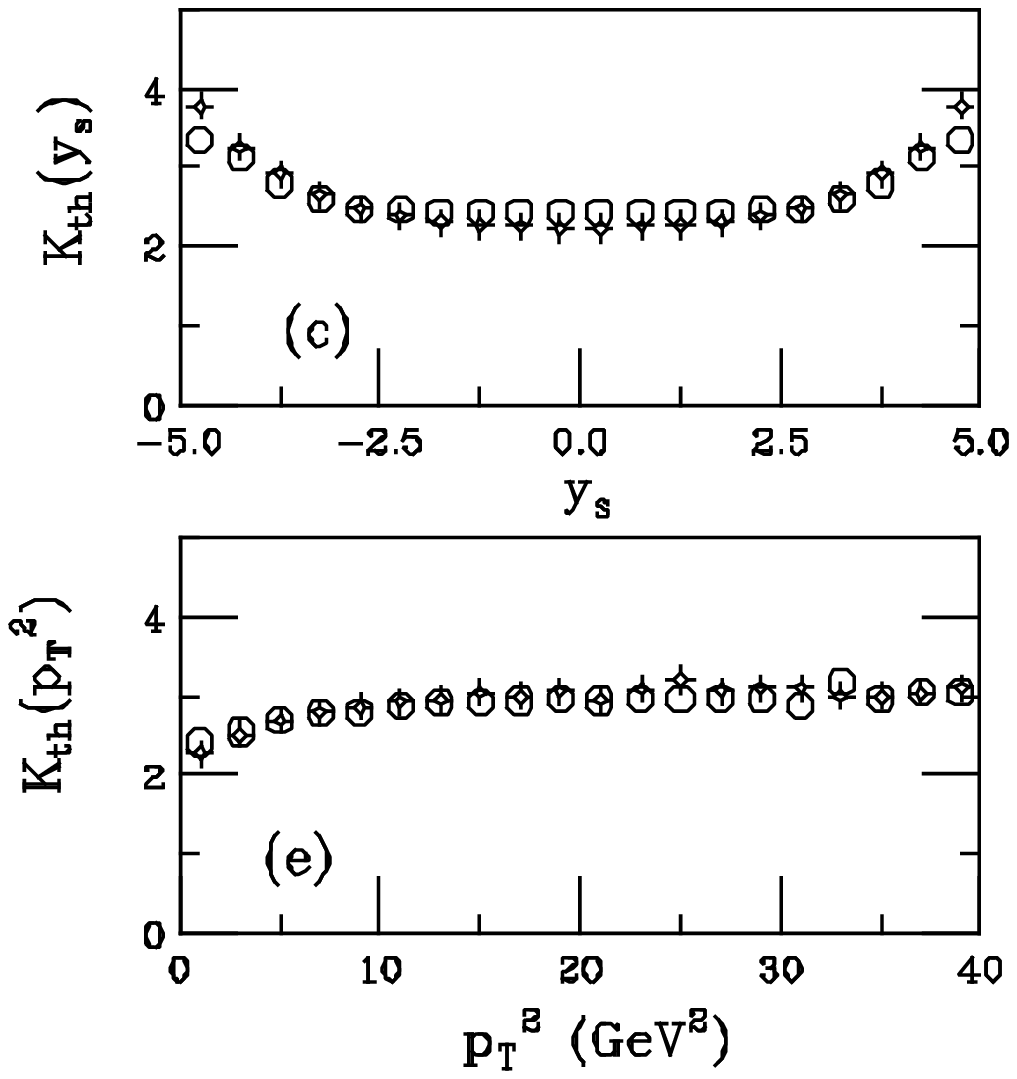,width=1\linewidth,clip}
\caption{\label{fig:ch5.K_init} K-factor for the charm quark as a
function of quark rapidity (top) and $p_t^2$ (bottom) for two sets
of PDFs: GRV HO (circles), MRS D-' (cross)~\cite{phenom_cb}.}
\end{minipage}
&
\begin{minipage}{0.5\linewidth} \centering \epsfig{figure=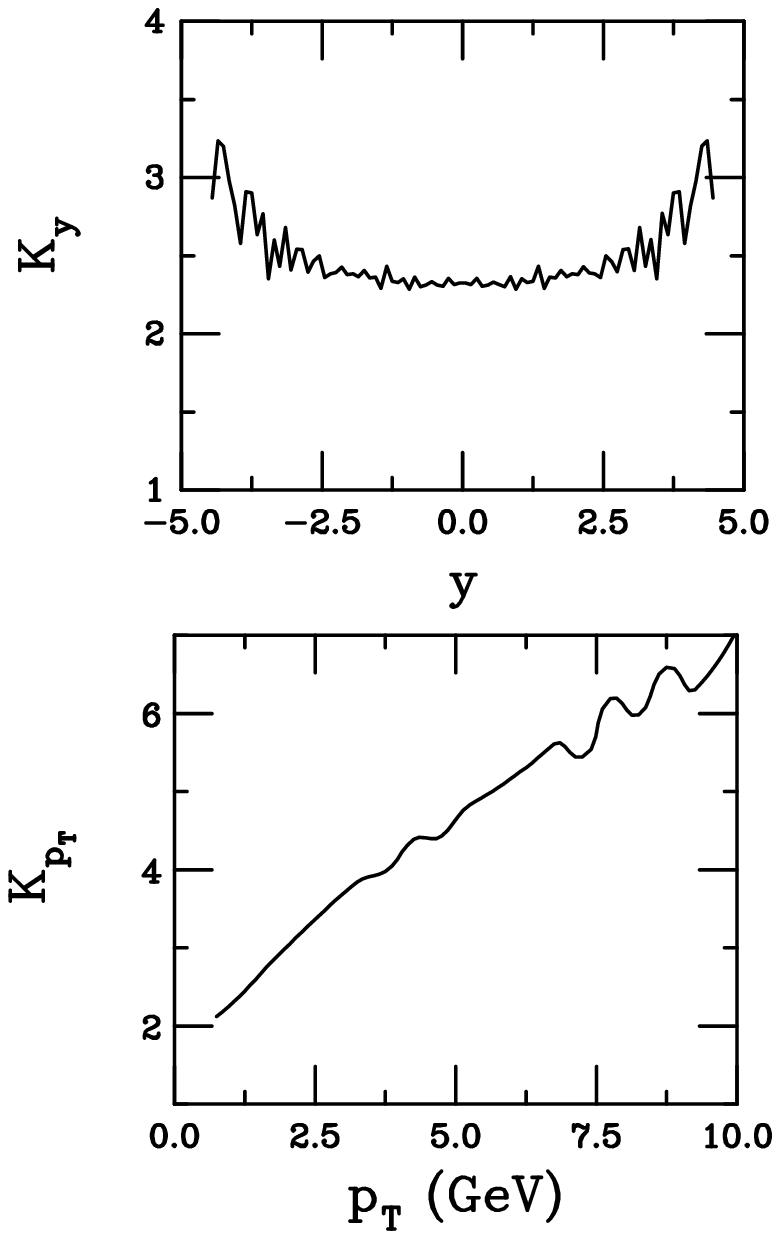,width=0.75\linewidth,clip}
\caption{\label{fig:ch5.K_new} K-factor for the charm quark as a
function of quark rapidity (top) and $p_t$ (bottom) for CTEQ5 PDF
set~\cite{whatK}.}
\end{minipage}
\end{tabular}
\end{figure}

 The
$K-factor$ in PYTHIA is a constant value of 3.5, there is a lot of
variety in the predictions for the $p_T$ dependence of this
variable. The original works of Ramona Vogt~\cite{phenom_cb} quote
a nearly constant $K-factor$ for the GRV HO and MRS D-' parton
distribution functions as shown in Fig.~\ref{fig:ch5.K_init}.
However, the most recent calculations of K-factor by the same
author predict a linearly growth with $p_T$ of the $K-factor$ as
shown in Fig.~\ref{fig:ch5.K_new} (CTEQ5 PDF set is used, $m_c =
1.2$ GeV, $\mu^2 = 4\,m_T^2$)~\cite{whatK}. The author claims that
the increase at high $p_T$ contributes to the significant amount
of "gluon splitting" contribution shown in
Fig.~\ref{fig:ch2.QQ_diag}e) $gg \rightarrow gg^* \rightarrow
gQ\overline{Q}$ where $Q\overline{Q}$ pair is opposite to a gluon
jet. Those contributions begin to be important when $p_T>m_Q$. The
paper also mention that $K-factor$ for the bottom quark has much
smaller $p_T$ dependence in the same range.

\begin{figure*}[ht]
\centering
\epsfig{figure=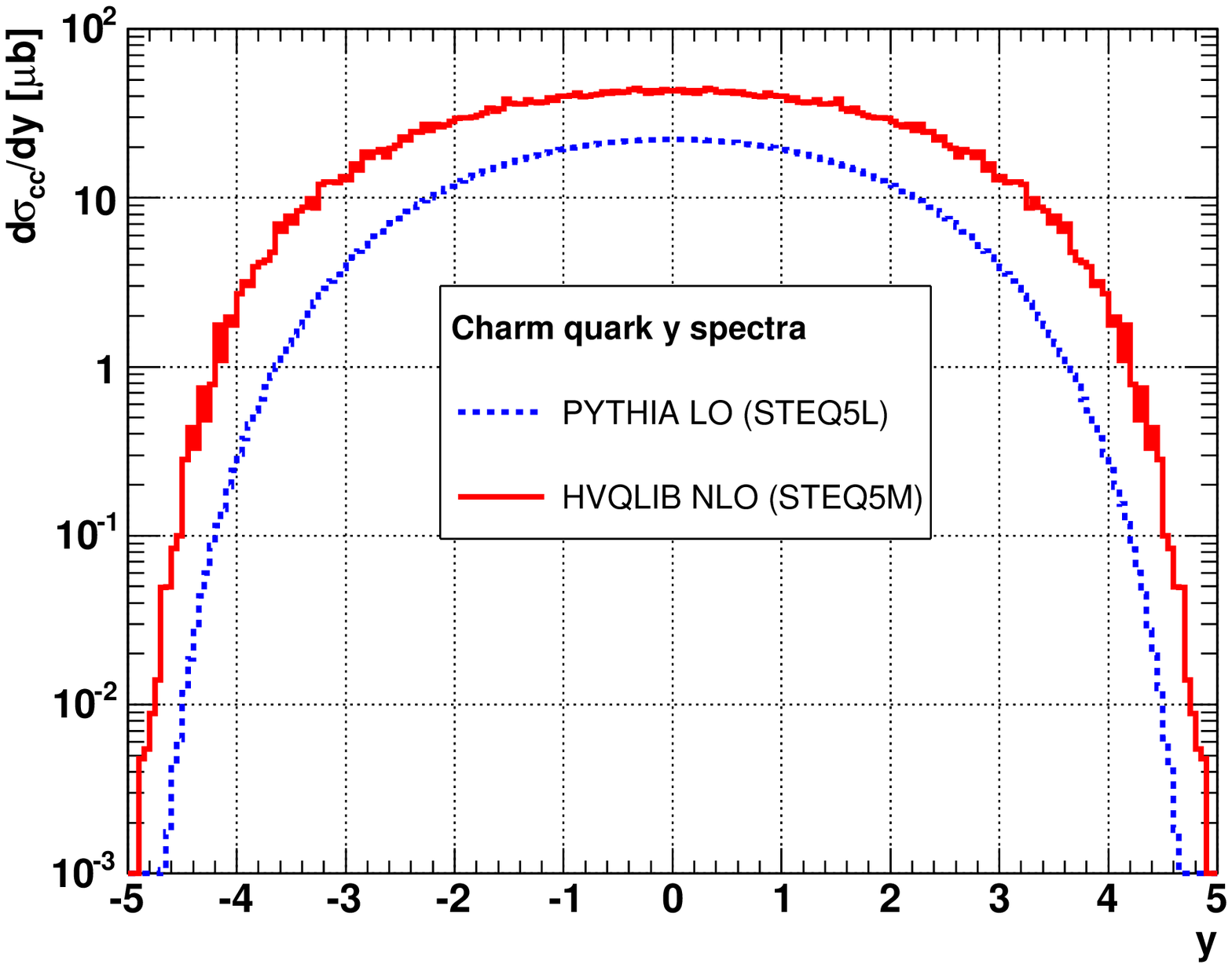,width=0.45\linewidth,clip,trim =
0.in 0.1in 0.1in 0in}
\epsfig{figure=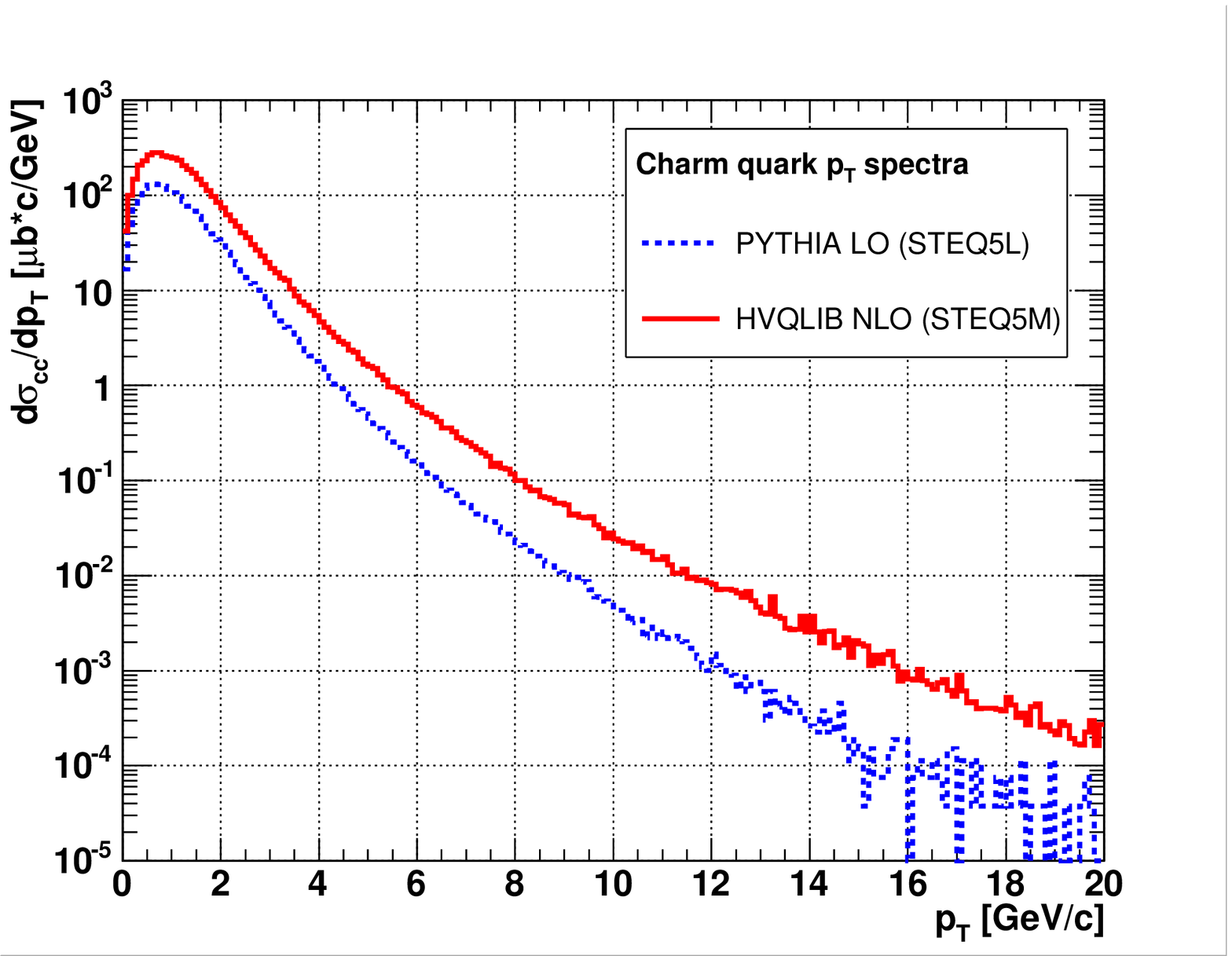,width=0.45\linewidth,clip,trim =
0.in 0.1in 0.1in 0in} \caption{\label{fig:ch5.quark_comp}
Comparison of rapidity and transversal momentum distributions for
NLO and LO charm quark production.}
\epsfig{figure=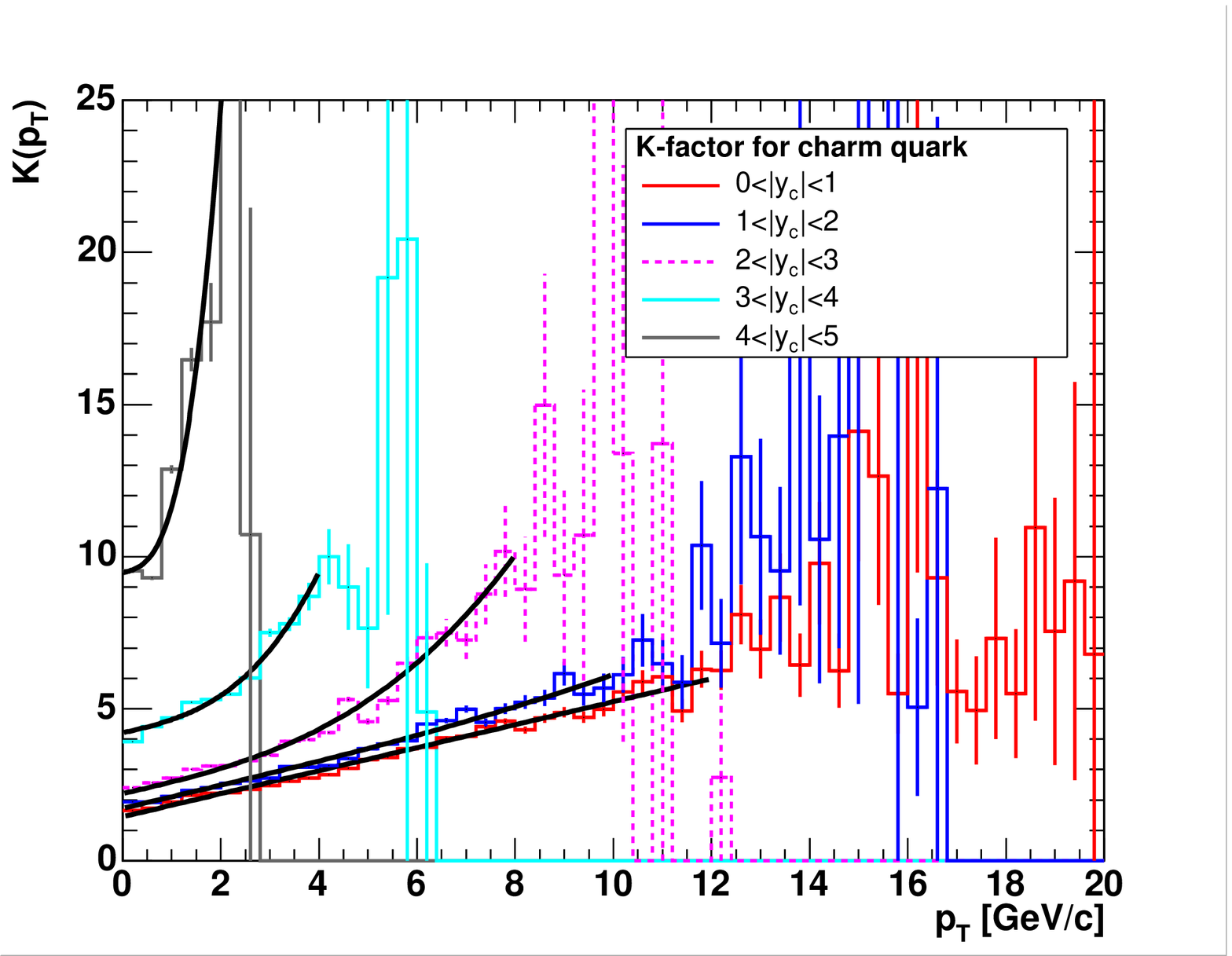,width=1\linewidth,clip,trim =
0.in 0.1in 0.1in 0in} \caption{\label{fig:ch5.ratio_NLO_LO}
Comparison of rapidity and transversal momentum distributions for
NLO and LO charm quark production. Fit functional form from
Eq.~\ref{eq:ch5.k_par}.}
\end{figure*}

In order to reproduce the results for the $K-factor$, a thorough
comparison of charm quark distribution in PYTHIA and HVQLIB was
performed using similar input theory parameters ($m_c = 1.25$ GeV,
CTEQ5 PDF set, $\mu_R^2 = \mu_F^2 = m_{T}^2$).
Fig.~\ref{fig:ch5.quark_comp} shows the comparison of the LO
PYTHIA and NLO HVQLIB charm quark $p_T$ and rapidity
distributions.

In the next step, we took a ratio of the $p_T$ distributions in NLO to
PYTHIA LO prediction for different slices in rapidity. This ratio was
used to parameterize the $K-factor$ as a function of $y$ and
$p_T$. Fig.~\ref{fig:ch5.ratio_NLO_LO} shows the partial $K(p_T)$ for
different rapidity ranges as denoted in the plot. \footnote{The error
bars on this plot may be not exactly correct as HVQLIB does not
calculate statistical error on the crossection, so the statistical
error, assigned to NLO prediction is equal to the statistical error of
the PYTHIA statistical error in given $p_T$ bin.} One can see that
$K-factor$ calculated by this procedure is in good agreement with
latest predictions by Ramona Vogt (see Fig.~\ref{fig:ch5.K_new}).

Using the best fit to the shape of the ratio, we obtain the
following phenomenological parametrization for the $K-factor$:
\begin{equation}
K(p_T,y) = 1.31 +e^{-2.52+1.08\cdot |y|}+0.378\cdot p_T
+\left(e^{-4.3+1.14\cdot |y|}\cdot p_T\right)^3
\label{eq:ch5.k_par}
\end{equation}

\begin{figure}[h]
\centering
\epsfig{figure=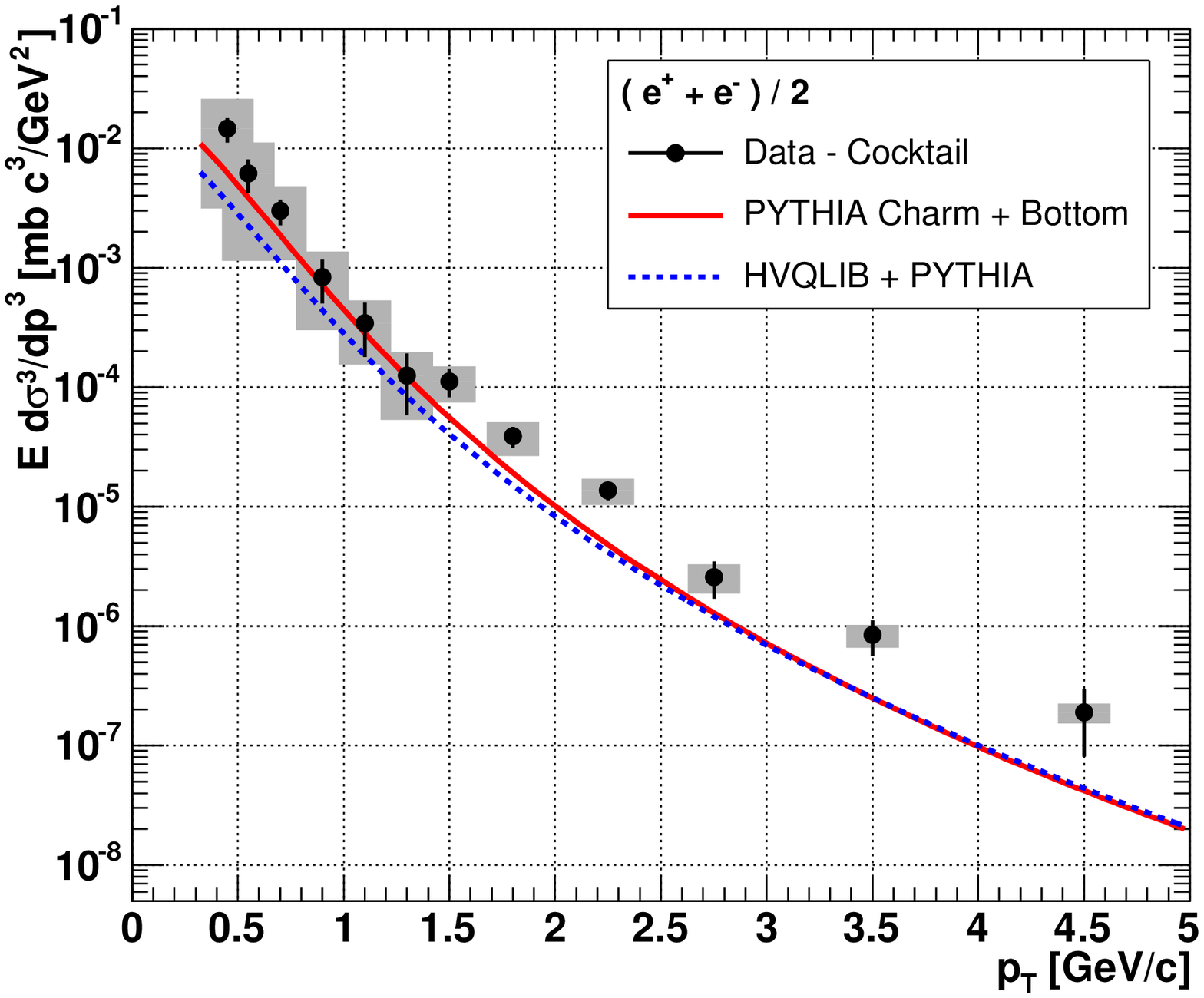,width=1\linewidth,clip,trim =
0.in 0.1in 0.1in 0in} \caption{\label{fig:ch5.comp_tuned_pythia}
Comparison of Non-Photonic electron crossection with default
PYTHIA expectation for Charm + Bottom (solid curve) and HVQLIB
tuned PYTHIA expectation for Charm + default PYTHIA bottom (dashed
curve).}
\end{figure}

This factor was applied as a weight to the charm quarks and the
corresponding decay electrons in default PYTHIA Open Charm
calculations. the resulting electron crossection, compared to the the
data and standard PYTHIA prediction is shown on
Fig.~\ref{fig:ch5.comp_tuned_pythia}. One can see that tuned PYTHIA
Open Charm crossection became \textit{"harder"} and much better
describes the shape of the data indicating that the NLO processes in
Open Charm production significantly change the initial $p_T$
distribution of the produced $\QQ$ pairs.  It may also be true that
NNLO processes are also required to fully explain the shape of the
Heavy Flavor invariant crossection.

We also have a recent theory prediction from Matteo Cacciari of heavy
flavor crossection calculations using the FONNL (\textbf{Fixed Order
Next Leading Logarithm}) resummation technique~\cite{pt_spectrum} and
the improved fragmentation functions~\cite{excess}. The results of
FONLL calculations repeat the shape of our own ``poor man's'' NLO
predictions and also indicate significant excess of the Data over
Theory.

\begin{figure}[h]
\centering
\epsfig{figure=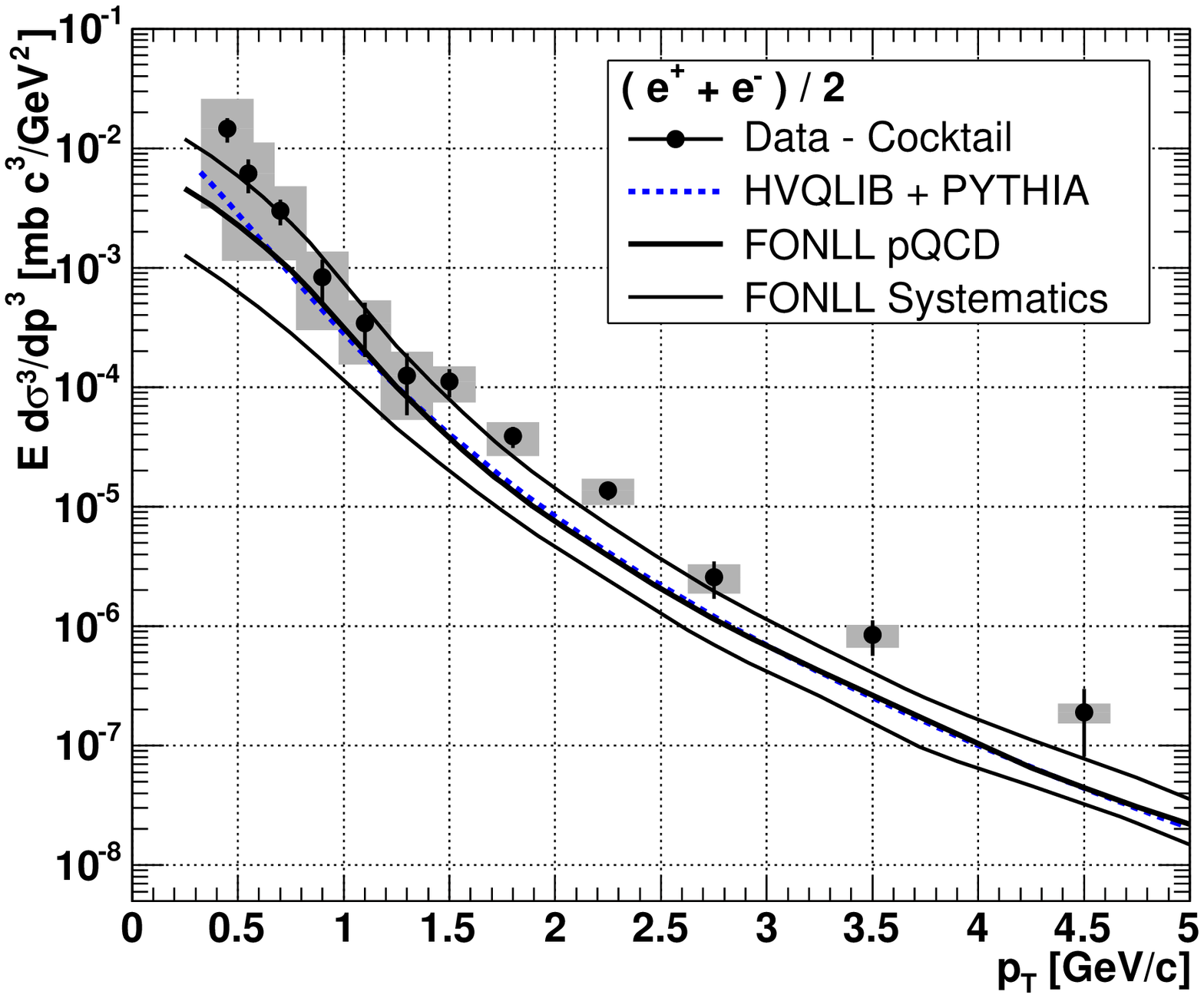,width=0.9\linewidth,clip,trim
= 0.in 0.1in 0.1in 0in}
\epsfig{figure=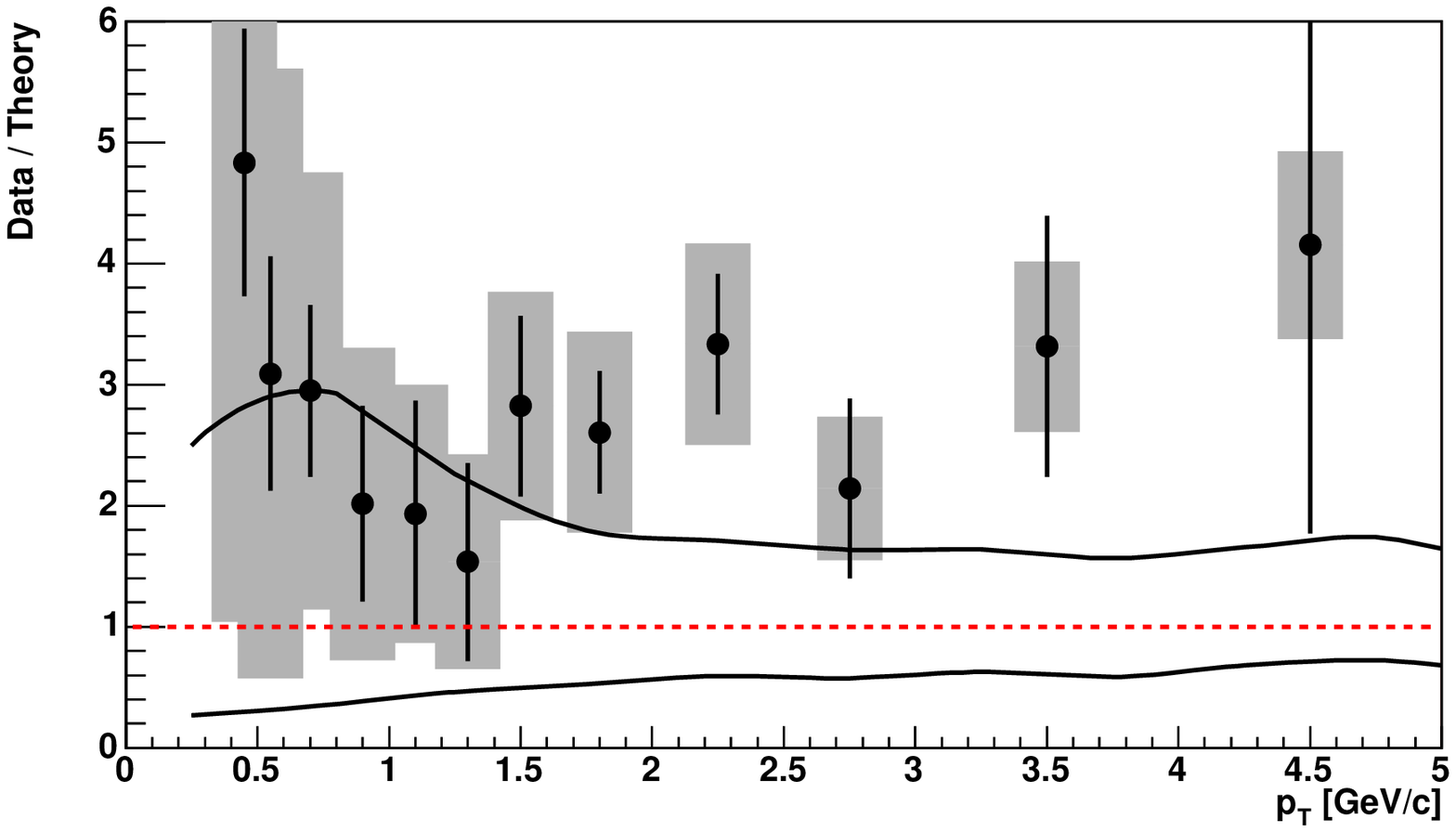,width=0.9\linewidth,clip,trim =
0.in 0.1in 0.1in 0in} \caption{\label{fig:ch5.comp_matteo}
Comparison of Non-Photonic electron invariant crossection with
HVQLIB tuned PYTHIA (dashed curve) and theoretical prediction from
FONLL~\cite{matteo}.}
\end{figure}

Fig.\ref{fig:ch5.comp_matteo} shows the comparison of the data to
the NLO predictions for the invariant electron crossection. The
ratio of Data to Theory central value is shown on the bottom
inlet.

One can clearly see that \textbf{theory underpredicts the data by
a factor of 3} although describing the shape of the invariant
crossection. This is the other major result from this analysis.
\pagebreak

\pagebreak

\pagebreak

\pagebreak

\section{Estimation of total Open Charm crossection}\label{sec:ch5.total_crossection}

In previous section we analyzed in detail the shape of the invariant
crossection of Heavy Flavor electrons. This chapter is devoted to
a calculation of the total Open charm crossection
$\sigma_{c\overline{c}}$.

In order to obtain the total crossection, we need to assume some shape
for the differential crossection in the low $p_T$ region. One can see
that the total crossection is dominated by the low momentum region
where the measurement is most difficult and the error bars are the
largest.  Nonetheless we shall use the PYTHIA spectral shape to
extrapolate down to $p_T$=0.  At high $p_T$, the shape clearly
deviates from standard PYTHIA prediction which can therefore not be
used for the extrapolation to $p_T\rightarrow\infty$.  However, we can
produce a reasonable spectral shape from the overall PYTHIA outputs by
scaling the bottom crossection upward until the prediction matches the
data. Note that the variation of the bottom contribution does not
effect the low $p_T$ part of the invariant crossection where it is
negligibly small.  We can obtain the total charm crossection by
fitting the spectrum with $f(p_T)=w_c\cdot f_c(p_T) + w_b\cdot
f_b(p_T)$, letting parameters $w_c$ and $w_b$ vary so as to best
describe the spectral shape. This in no way means that we can measure
the bottom crossection via $w_b$!! At high $p_T$ the variation of the
shape can be explained both by charm and by bottom contributions.  The
mechanical adjustment of the bottom rate is just a means to a smooth
fit and not a bottom quark determination.  Presently we have no way to
disentangle those components.  Such an analysis will be made easy with
upgrades to the existing PHENIX apparatus as described in Chapter 6.

\begin{figure}[h]
\centering
\epsfig{figure=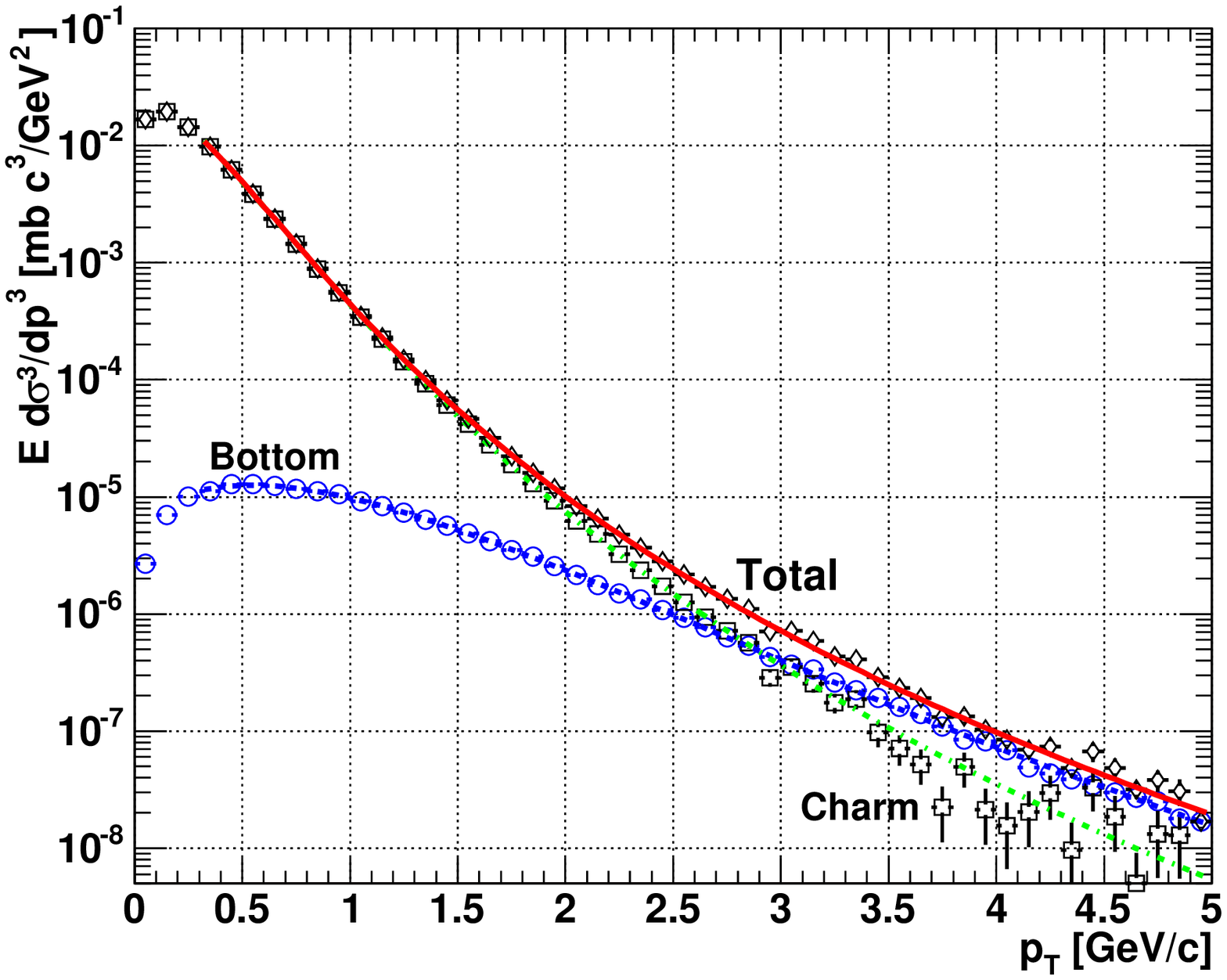,width=0.7\linewidth,clip,trim
= 0.in 0.1in 0.1in 0in} \caption{\label{fig:ch5.fits_c_b} Results
of the best fit to the shape of the charm, bottom and charm+bottom
electron crossections from default PYTHIA.}
\end{figure}

To begin with we need to parameterize the shape of the default
PYTHIA invariant crossections for Charm and Bottom related
electrons.  Fig.~\ref{fig:ch5.fits_c_b} shows the fits to PYTHIA electron
crossection for charm, bottom and charm+bottom contribution. The fit
functionalorm used is is purely arbitrary, obtained from the best
match to the data:
\begin{eqnarray}
    f_c(p_T)&=& \frac {3.59\cdot10^{-2}}{(1+p_T/1.773 + p_T^2)^{4.657}}\\
    f_b(p_T)&=& e^{-11.55-1.311\cdot p_T} + e^{-8.983-2.857\cdot p_T}\cdot
    p_T^2 \nonumber
    \label{eq:ch5.fits_c_b}
\end{eqnarray}
\pagebreak

 The procedure used for the crossection derivation is
described below:

\begin{itemize}
    \item Fit the data with $f(p_T)=w_c\cdot f_c(p_T) +
w_b\cdot f_b(p_T)$ assuming $w_c$ and $w_b$ to be a fit parameters
and shape for the charm and bottom is fixed by
Eq.~\ref{eq:ch5.fits_c_b}.
    \item Assuming some value for the low $p_T$ cut-off $p_{T\ cut-off}$,
    integrate double differential crossection to obtain
    $\frac{d\sigma}{dy}(p_T>p_{T\ cut-off})$.
\begin{equation}
    \frac{d\sigma}{dy}(p_T>p_{T\ cut-off}) = \int\limits_{p_{T\ cut-off}}^{5.0} p_T\cdot
    f_c(p_T)dp_T
    \label{eq:ch5.dsig_dy}
\end{equation}

    \item From PYTHIA, calculate the correction factor to
    translate the differential crossection to full $p_T$ range
\begin{equation}
    R_p = \frac{d\sigma}{dy}(p_T>p_{T\
    cut-off})/\frac{d\sigma}{dy}.
    \label{eq:ch5.R_p}
\end{equation}
    \item From PYTHIA, calculate the portion of the
    crossection that falls into our measurement mid-rapidity range $|y|<0.5$ compared to full
    rapidity range
\begin{equation}
    R_y = \frac{\sigma_{c\bar{c}\ |y|<0.5}}{\sigma_{c\bar{c}}}
    \label{eq:ch5.R_y}
\end{equation}

    \item Compute the total $c\overline{c}$ crossection as:
\begin{eqnarray}
    \sigma_{c\bar{c}} &=& \frac{\frac{d\sigma}{dy}(p_T>p_{T\
    cut-off})}{R_p\cdot R_y \cdot BR(c\rightarrow eX)}
    \label{eq:ch5.sigma}\\
    \frac{d\sigma_{c\bar{c}}}{dy} &=& \frac{\frac{d\sigma}{dy}(p_T>p_{T\
    cut-off})}{R_p \cdot BR(c\rightarrow eX)}
    \label{eq:ch5.dsigma}
\end{eqnarray}
    where $BR(c\rightarrow eX)$ is a branching ratio for Open
    Charm meson semi-leptonic decay.
\end{itemize}

The branching ratio $c\rightarrow eX$ must be evaluated both for
PYTHIA and for the data. The ratio is very sensitive to the different
Open Charm meson species relative ratios~\cite{shuryak,STAR_se_pp} due
to the fact that neutral $D^0$ meson has a measured branching ratio of
($6.75\pm 0.29\ \%$)~\cite{PDG}, quite different from the charged
$D^{\pm}$ mesons ($17.2\pm 1.9\ \%$).

\begin{figure}[t]
\centering
\epsfig{figure=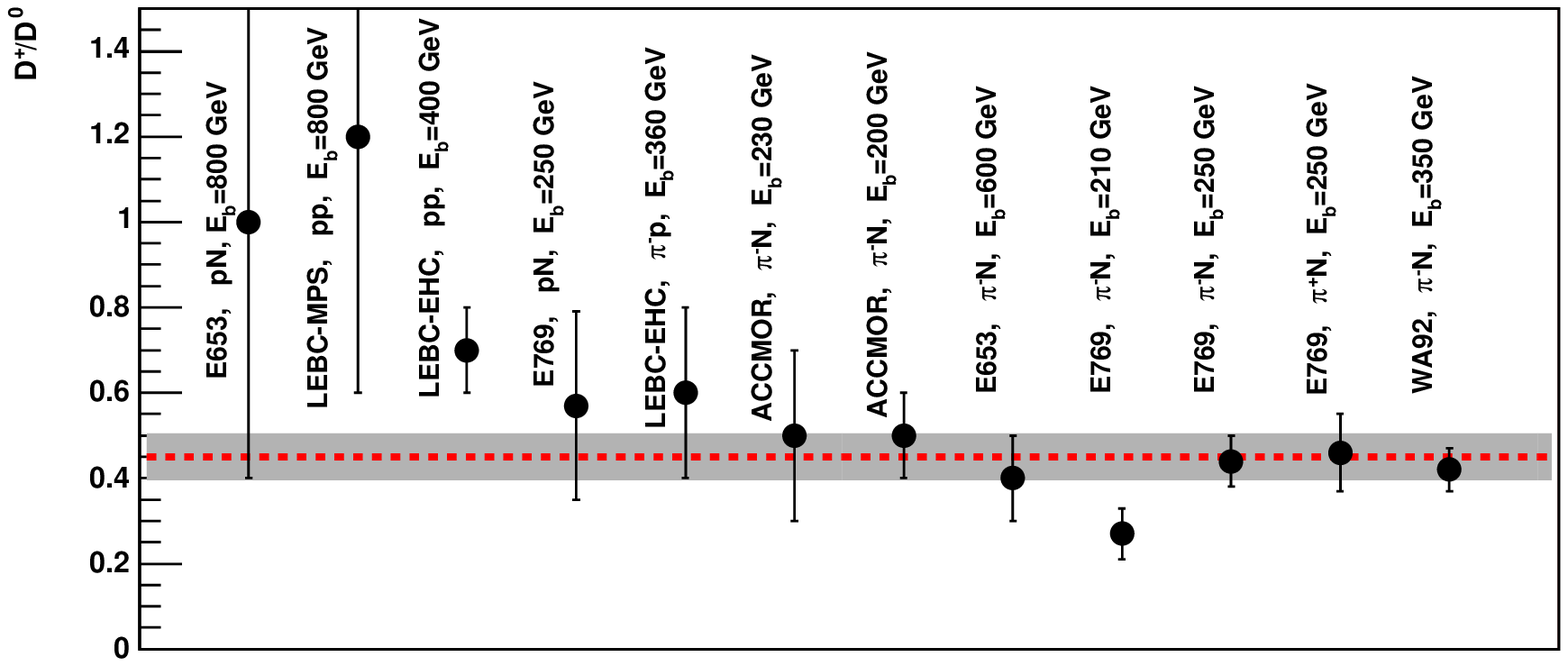,width=0.8\linewidth,clip,trim
= 0.in 0.1in 0.1in 0in} \caption{\label{fig:ch5.D_ratios} World
data compilation for the ratio of charged to neutral $D-$mesons
$R_{D^\pm/D^0}$.}
\end{figure}

\pagebreak The ratio of the charged to neutral D-mesons was
recently measured measured by many experiments and the compilation
of the most recent results from fixed target experiments presented
in Fig.~\ref{fig:ch5.D_ratios}. A theoretical prediction for this
ratio at $\sqs =200$ GeV $R_{D^\pm/D^0}=0.40$~\cite{shuryak} is
very close to World Data average in fixed target experiments. For
this analysis we used $R_{D^\pm/D^0}=0.45\pm 0.05$. \pagebreak

Another contributor to $D$ particle mixture comes from $D_s$
mesons and the $\Lambda_c$ baryon.  The values we used for the
current analysis are presented in Table~\ref{tab:D_ratios}. The
total branching ratio for the particle mixture was calculated as
the weighted average of the individual semi-leptonic branching
ratios. It is necessary to note that PYTHIA 6.152, uses the PDG92
value~\cite{pythia} for $BR(D^0\rightarrow eX)$ which is
significantly higher than the current World average result.
Finally, although the contribution of charged $D^\pm$ mesons is
underpredicted in PYTHIA, the overall branching ratio is very
similar($9.55 \%$) to our expectations ($9.4 \%$). \nopagebreak

\begin{figure}[h]
\centering \epsfig{figure=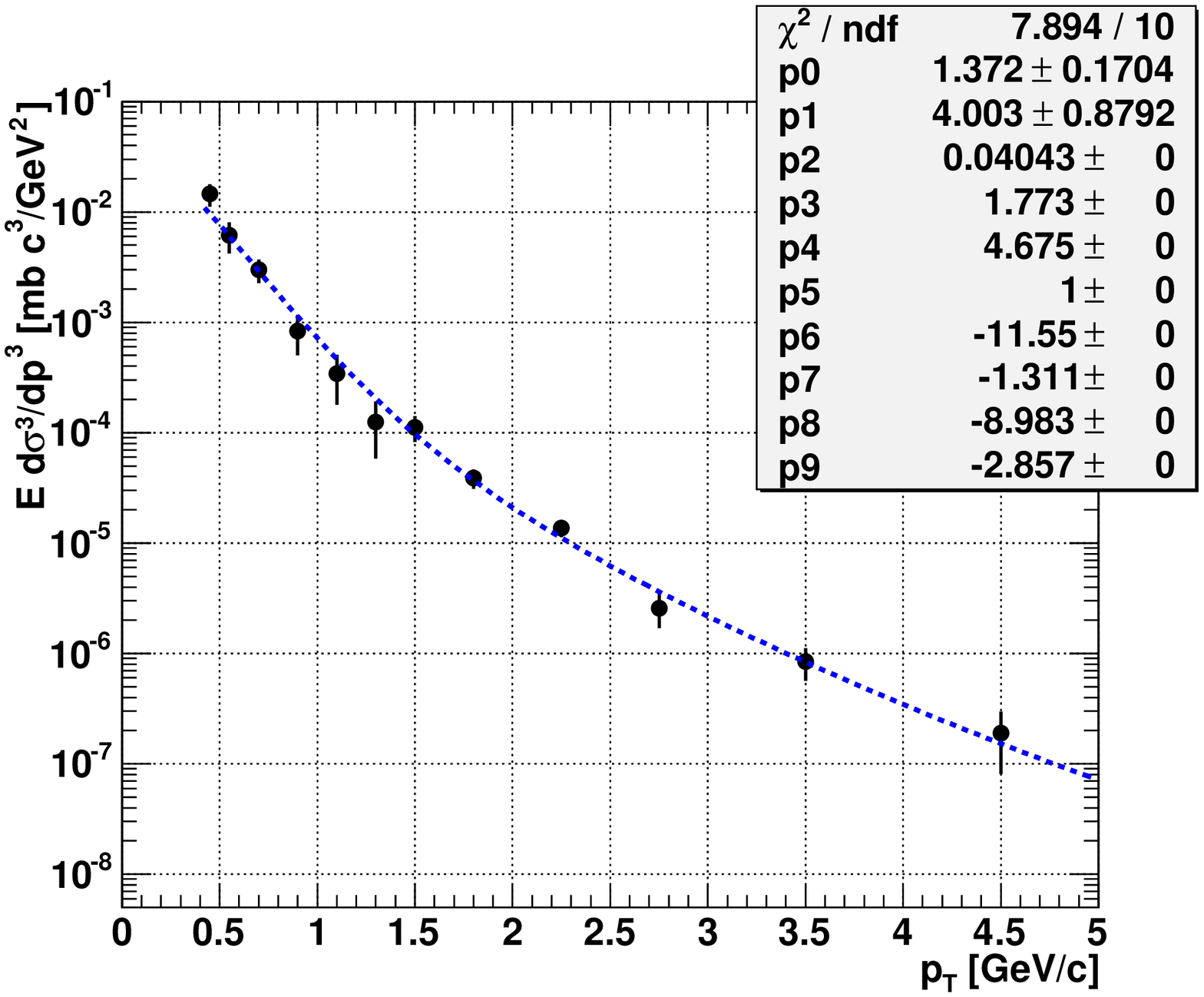,width=0.48\linewidth}
\epsfig{figure=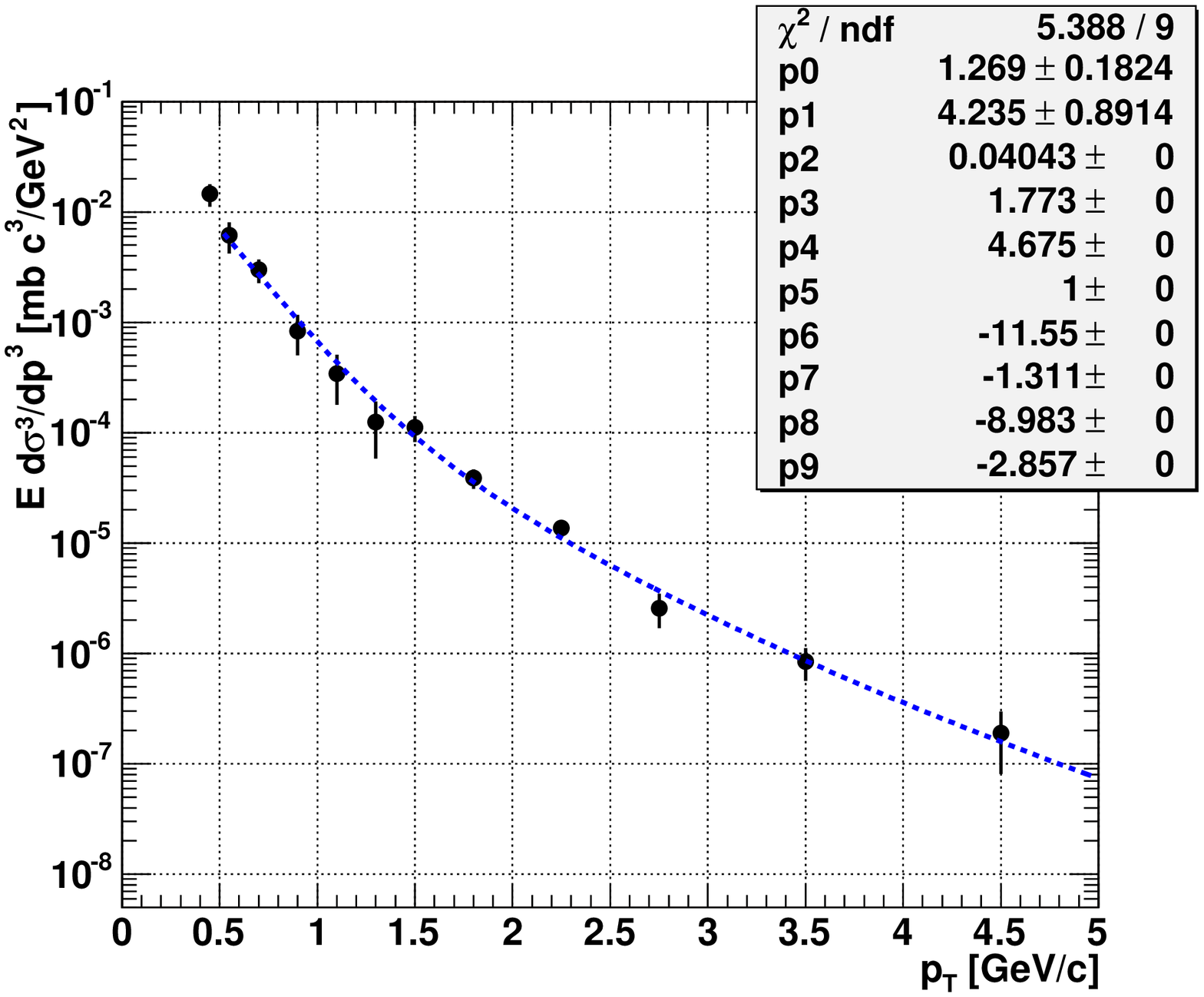,width=0.48\linewidth}

\epsfig{figure=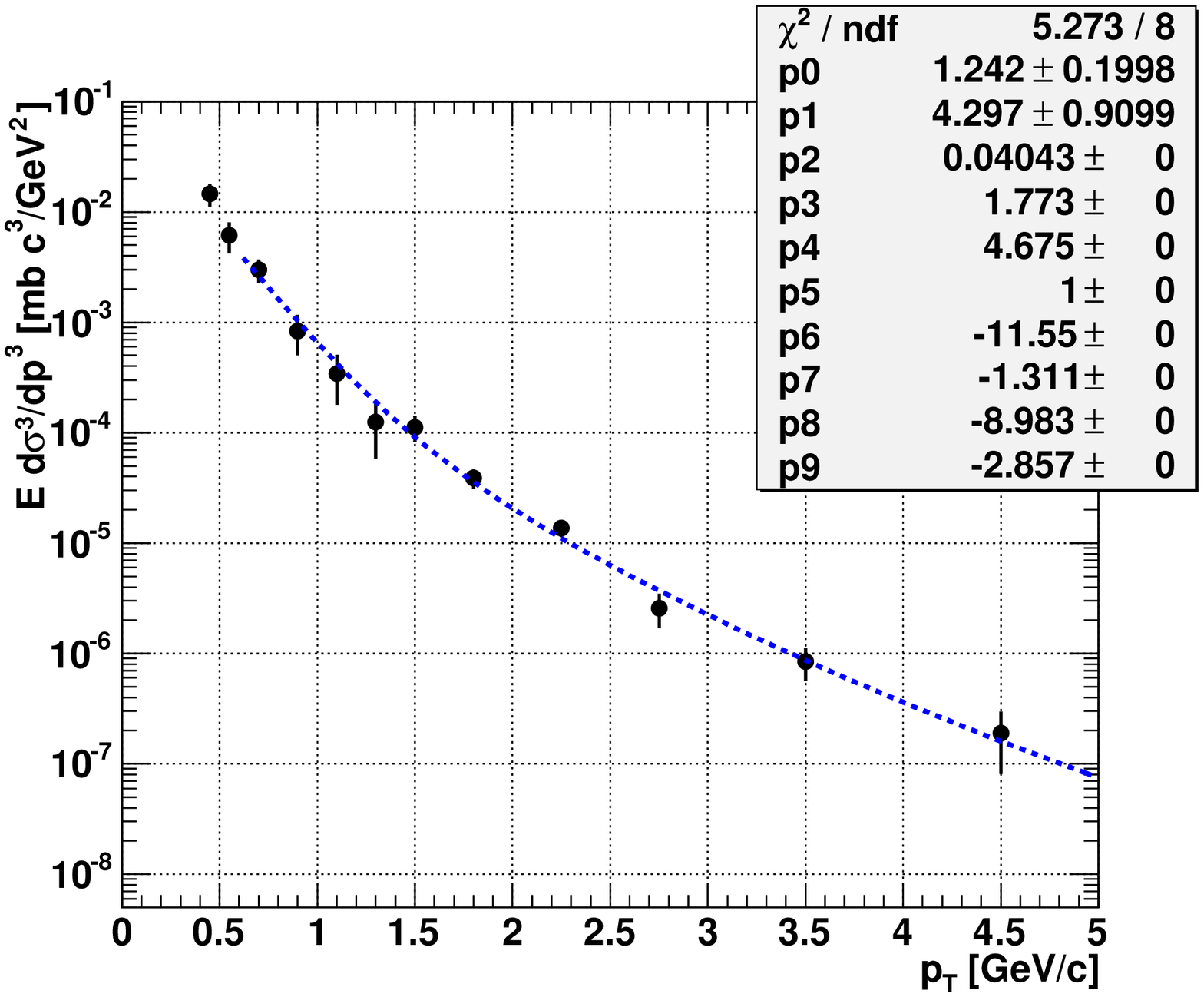,width=0.48\linewidth}
\epsfig{figure=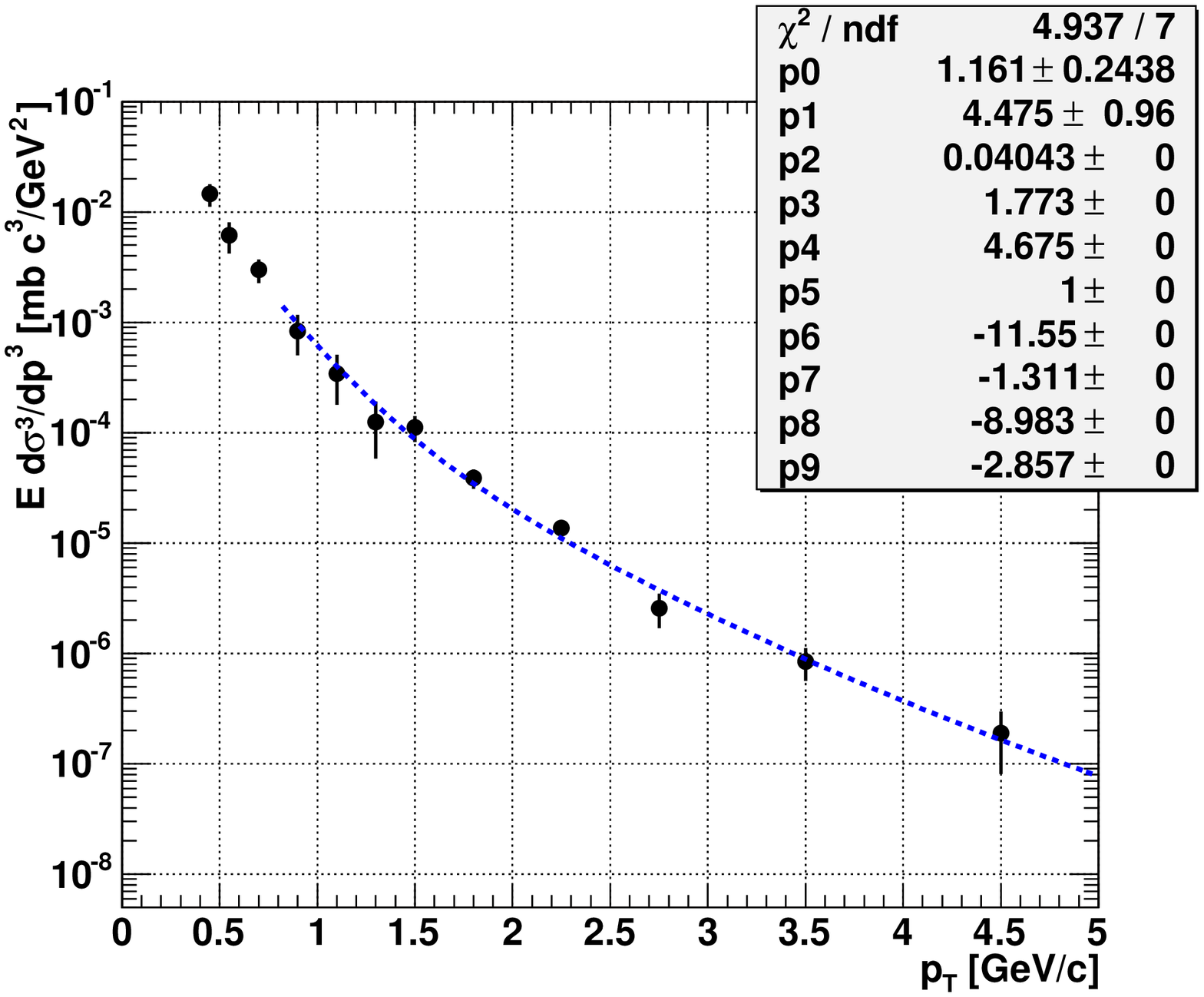,width=0.48\linewidth}
\caption{\label{fig:ch5.pt_cutoff} Fits to the "Non-photonic"
electron crossection using PYTHIA expectation for open charm and
bottom crossection shape. Different low $\pt$ cut-off values (0.4,
0.5, 0.6, 0.8 GeV/c) used for the fits.}
\end{figure}

 Now we have all the ingredients to calculate the total and
differential crossection using
Eq.~\ref{eq:ch5.sigma},~\ref{eq:ch5.dsigma}. First of all we need to
obtain the dependence of the total crossection upon our arbitrary
chose low $p_T$ cut-off in the fit function. Four values of $p_{T\
cut-off}$ were tested: 0.4, 0.5, 0.6, 0.8 GeV/c. The fits to the data
using the floating charm and bottom shape parametrization are shown in
Fig.~\ref{fig:ch5.pt_cutoff}.

\begin{table}[t]
\label{tab:D_ratios} \caption{Ratio of the Open Charm particles to
$D^0$ and the corresponding semi-leptonic decay branching ratios
for World Data average and PYTHIA.}

\begin{center}
\begin{tabular}{|c|c|c|c|c|}
  \hline
  Open Charm &\multicolumn{2}{c|}{Ratio to $D^0$}&\multicolumn{2}{c|}{Branching Ratio [$\%$]} \\
  \cline{2-5}
  Particle & World average & PYTHIA & World average& PYTHIA \\
  \hline
  $D^0$ & 1 & 1 & $6.57\pm 0.28$& 7.8  \\
  $D^{\pm}$ & $0.45\pm 0.05$ &   0.3 & $17.2\pm 1.9$& 17.21 \\
  $D_s$ &$0.25\pm 0.05$&   0.17& $7.7\pm 5.0$& 7.7 \\
  $\Lambda_c$ & $0.1\pm 0.05$&  0.1 &$4.5\pm 1.7$& 4.5 \\
  \hline
  \hline
  $BR(c\rightarrow eX)$ &\multicolumn{2}{c|}{}&$9.4\pm0.4$&
  9.55\\
  \cline{1-1}\cline{4-5}
\end{tabular}
\end{center}
\end{table}

 The shape of the fit function is well determined, however, we
take some care in setting the normalization, by successively
fitting different portions of the data.  The high $p_T$ end of the
spectrum has little influence on the integral and so out
successive sets of fits will be defined by the choice of low end
cutoff.  When corrected to the full integral range, we would hope
that the full integral is independent of the low end cutofff of
the fit range.



\begin{figure}[h]
\centering
\epsfig{figure=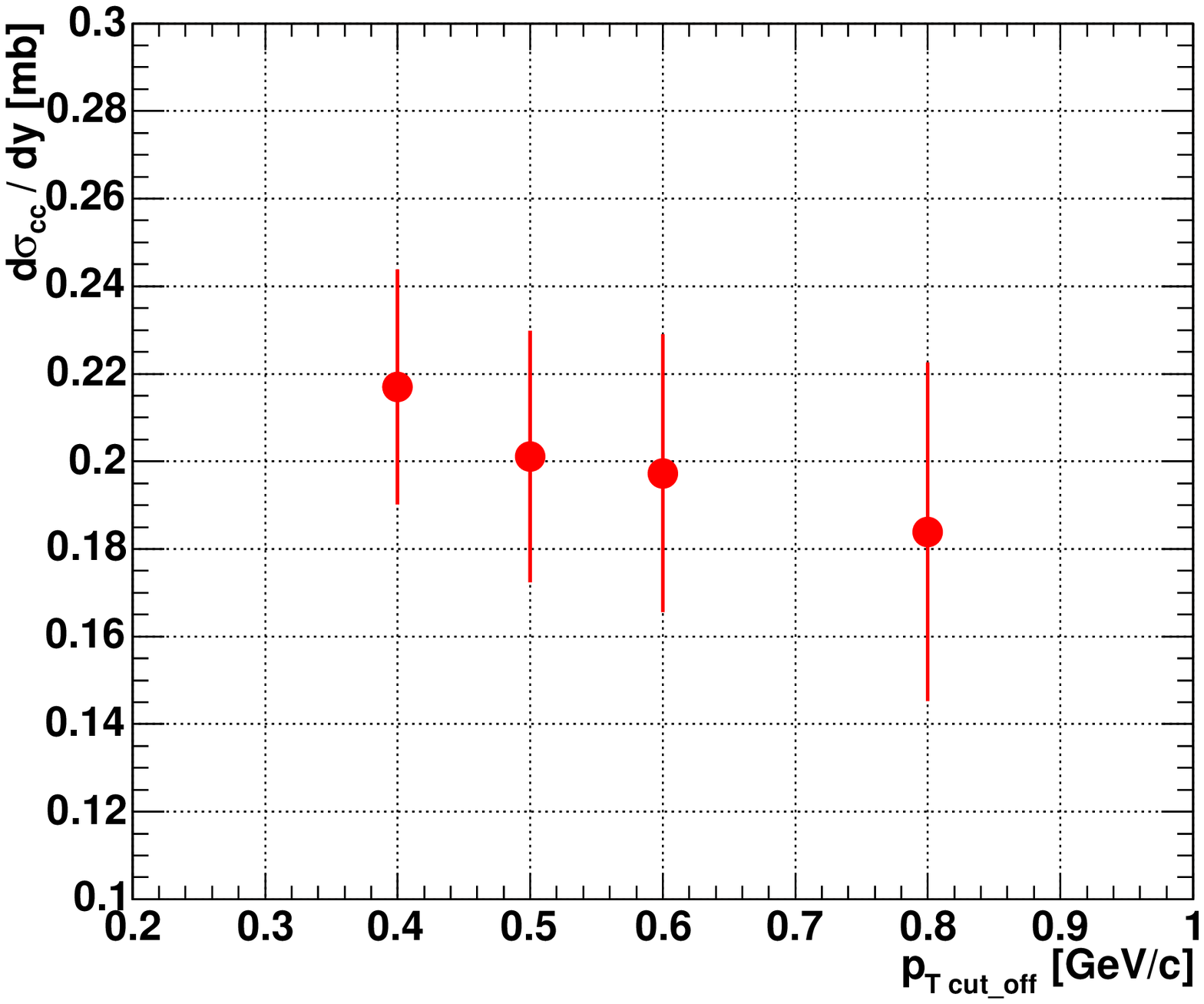,width=0.5\linewidth,clip,trim =
0.in 0.1in 0.1in 0in} \caption{\label{fig:ch5.dsigma_dy}
$\frac{d\sigma_{c\bar{c}}}{dy}$ as a function of low $p_T$
cut-off.} \centering
\epsfig{figure=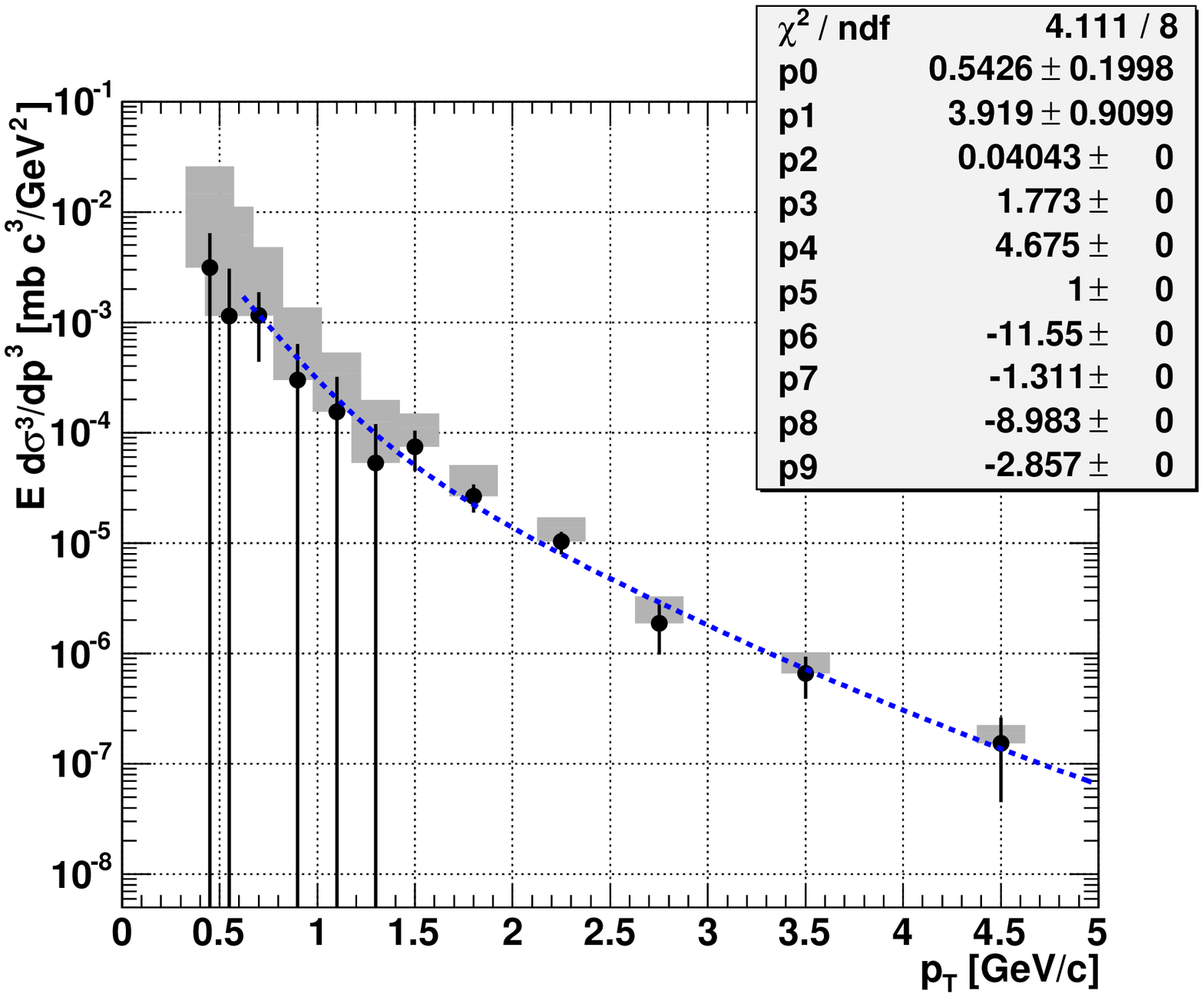,width=0.45\linewidth,clip}
\epsfig{figure=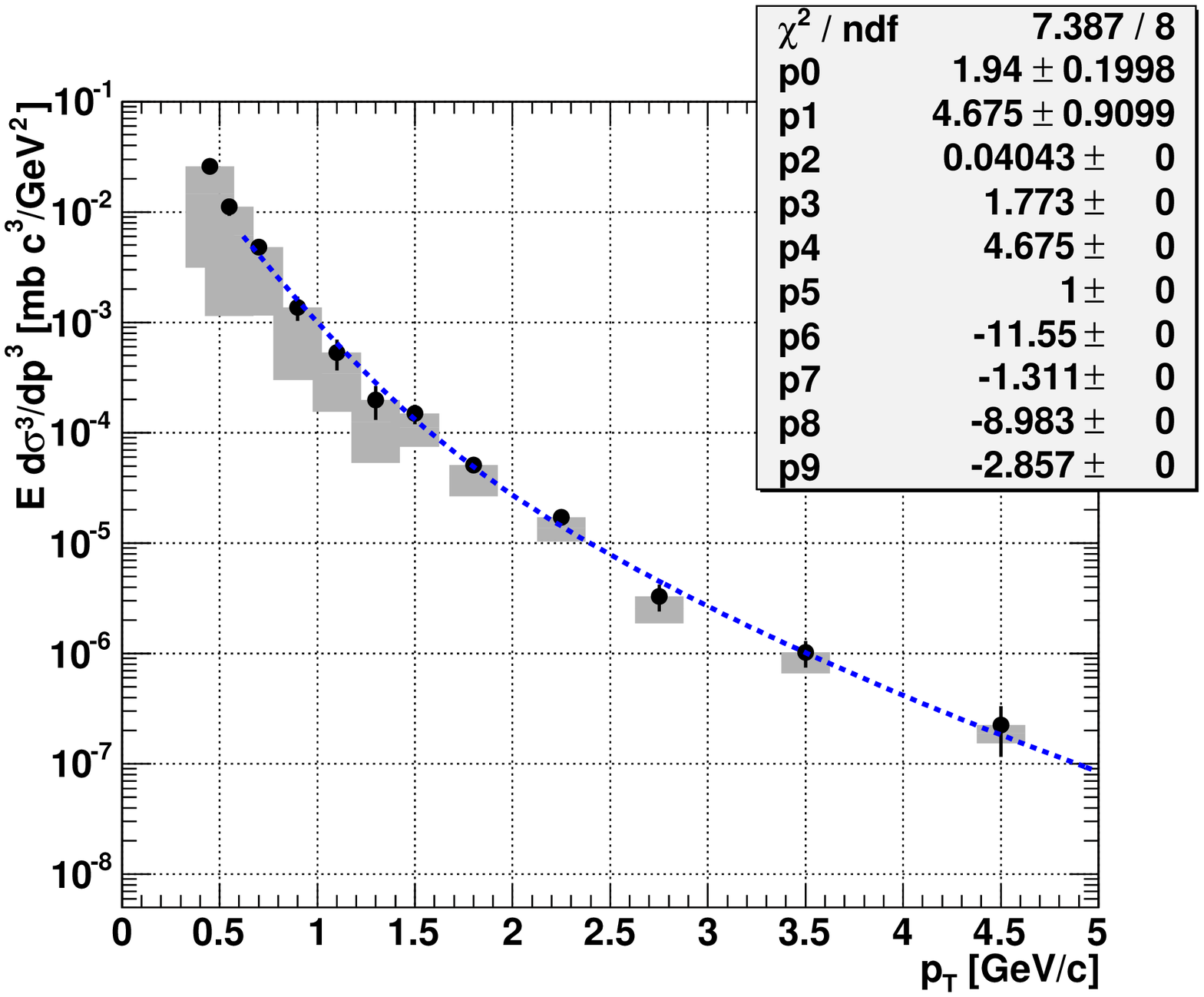,width=0.45\linewidth,clip}
\caption{\label{fig:ch5.fit_hi_lo} Variation of the charm
crossection due to the systematic error of the "Non-photonic"
crossection.}
\end{figure}

Now calculate $\frac{d\sigma_{c\bar{c}}}{dy}$ as a function of $p_{T\
cut-off}$ using Eq.~\ref{eq:ch5.dsigma}, the results are shown in
Fig.~\ref{fig:ch5.dsigma_dy}. One can see that there is no significant
sensitivity of the single differential crossection on the low $p_T$
cut-off limit. We arbitrarily chose $p_T>0.6$ GeV/c as a low limit for
the total crossection calculations.

\pagebreak

 What is left now is the translation of the mid-rapidity
to the full rapidity range. From PYTHIA prediction we determined
the fraction of open charm within one unit of rapidity to be $R_y
= 21.3 \%$. We then applied $\frac{1}{R_y}$ as a scaling factor to
determine the total crossection for the Open Charm:

\begin{itemize}
    \item $\sigma_{c\bar{c}}= (0.920 \pm 0.148(stat))$ mb
    \item $\frac{d\sigma_{c\bar{c}}}{dy} = (0.196 \pm 0.031(stat))$ mb
\end{itemize}

Systematic error on the crossection consists of the following
components:

\begin{itemize}
    \item  Total systematic error of the "Non-photonic" electron
    crossection.
    \item  Systematic error due to PYTHIA parameterization.
    \item    Uncertainty from $d\sigma/dy$ extrapolation to full rapidity
    range.
    \item   Branching ratio systematic error.
\end{itemize}

Each component of the systematic error can be calculated
separately. First we moved the data up and down to the full extent
of the systematic error on the "Non-photonic" electron
crossection. Fit results to the shifted data are shown on
Fig.~\ref{fig:ch5.fit_hi_lo}. Resulting systematic error is equal
to 56.3\%, out of which 34.4 \% comes from the Cocktail
uncertainty, 43.6 \% from inclusive electron crossection
uncertainty and 10 \% comes from absolute normalization.

The error on the shape of the invariant crossection was tested by
varying the $\kt$ within reasonable values, $1.0 - 2.0$ GeV. This
caused a 10.2 \% variation of the crossection. The rapidity dependence
was varied by using a different PDFs for the PYTHIA predictions. The
amount of variation in this case was 6.4 \%~\cite{ppg035}. The error
on the branching ratio is estimated in Table~\ref{tab:D_ratios} and
adds another 4 \% systematic error.

Combining all the values together in quadrature we obtain the
total value for the total crossection systematics of 57.7 \% which
produce the following results:

\begin{itemize}
    \item $\sigma_{c\bar{c}}= (0.920 \pm 0.148(stat)\pm 0.524(sys))$ mb
    \item $\frac{d\sigma_{c\bar{c}}}{dy} = (0.196 \pm 0.031(stat)\pm 0.117(sys))$ mb
\end{itemize}

\begin{figure}[h]
\centering
\epsfig{figure=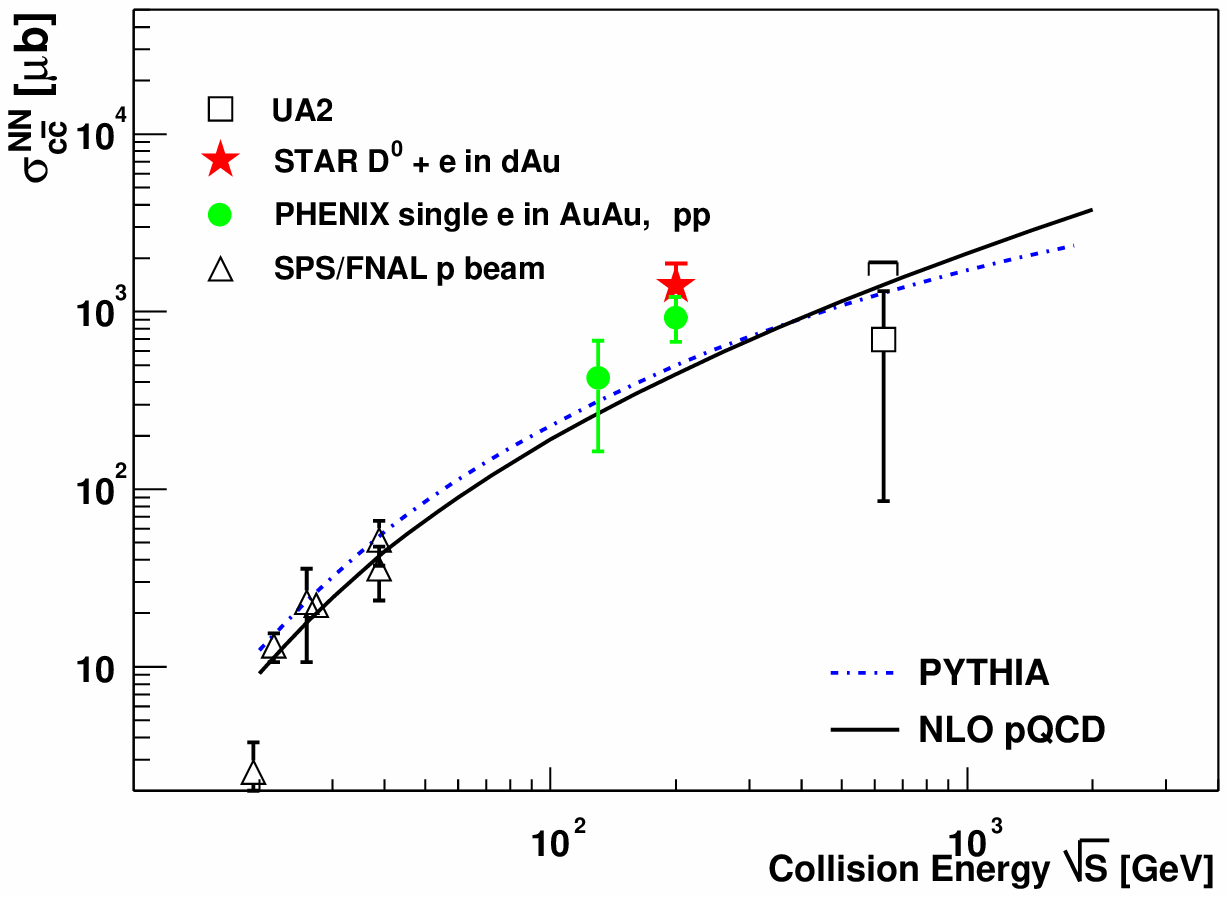,width=0.8\linewidth,clip}
\caption{\label{fig:ch5.sigma_sqs} World Data compilation for Open
Charm crossection as a function of $\sqs$. Modified version of the
plot from~\cite{STAR_se_pp}.}
\end{figure}

 Fig.~\ref{fig:ch5.sigma_sqs} shows the current World
Data for the total charm crossection as a function of $\sqs$ from
different experiments. The PHENIX and STAR results are in quite
good agreement with each other within statistical and systematic
errors.  However, one can easily see that both these measurements
are roughly factor of $2-3$ higher than NLO prediction for the
crossection. Our data seem to indicate a steeper rise in charm
production than either the PYTHIA or the NLO calculation would
predict. It is important to note that the UA2 point was included
in the process of tuning PYTHIA's parameters and may have driven
the predictions downward.

\chapter{Summary}\label{sec:ch6}

In the summary I would like to emphasize the most important
results of the Thesis. It is obvious that PHENIX has excellent
capabilities to perform single electron analysis of the Heavy
Flavor decay component in collisions ranging from $\pp$ to $\Au$.
We have measured for the first time the total charm production
crossection at $\sqs = 200$ GeV to be $\sigma_{c\bar{c}}= (0.920
\pm 0.148(stat)\pm 0.524(sys))$ mb.

Two completely different and absolutely independent analysis
techniques were performed: (1) Cocktail subtraction, (2) Converter
subtraction and yield identical results. The systematic error on
the total crossection is large \linebreak (57.7 \%).  The reason
for this large error is simple and natural.  Our signal is the
difference between the inclusive electron yield and the electron
yield due to the decays of light vector mesons.  Even though each
of these two measurements have reasonably small systematic errors,
their difference has a large percentage error particularly at the
low $p_T$ end (the range in the spectrum with the most influence
on the integrated yield).  This trouble is driven primarily by the
small signal to background ratio.

The limitations of the two techniques are quite different. For the
Cocktail analysis, the limitation is the systematic error
(normalization of the Cocktail) and will not be significantly
improved with higher statistics.  However, the limitation of the
convertor technique comes primarily from the statistics of the
convertor run itself.  Future PHENIX data could improve over the
precision of the present results if a significantly longer period
of convertor running is allowed. \pagebreak

 pQCD NLO calculations
underpredict the data by a factor of $2-3$ possibly indicating
that higher order interaction processes are important in
determining the total heavy flavor cross section. The theoretical
$K-factor$ has a significant $p_T$ dependence due to the "gluon
splitting" contributions. pQCD theory can describe the data only
by reducing the quark mass to $m_c=1.2$ GeV and the scale to
$\mu^2=\frac{m_c^2}{4}$. With these modifications perturbative
theory predictions becomes unreliable as $\alpha_s \sim 1$. The
fragmentation of the charm quarks also require additional
theoretical consideration.

The $\pp$ data set, collected by PHENIX during Run03-05 will
definitely provide us enough statistics to reach a $p_T$ range of up
to 10 GeV/c, where we expect to be completely dominated by Open
Bottom.  The experimental analysis of that data set will present new
challenges since at this momentum pions exceed the RICH threshold
($p_T>4.8$ GeV/c). One of the possible solutions for precise
estimation of RICH efficiency to pions is to use the newly installed
Transition Radiation Detector (TRD) which can effectively separate
clean high momentum electron sample.

Additionally, PHENIX plans to improve measurements of heavy flavor
decays by explicitly tagging off-vertex sources of electrons. This
tagging will be possible after the PHENIX upgrades that include
the installation of fine granularity silicon vertex detector
surrounding the collision point. The device will be able to
directly measure off-vertex decays whose distance exceeds ~100-200
$\mu m$.  The difference in the lifetimes of charm and bottom
($c\tau_{D^{\pm}} = 315\ \mu m;\ c\tau_{D^0} =123\ \mu m;\
c\tau_{B^{\pm}} = 503\ \mu m;\ c\tau_{B^0} =462\ \mu m$).  These
differences will allow PHENIX to distinguish charm and bottom
quarks at high $p_T$.  The principal drawback of this detector is
that it will contribute $\approx 6 - 10 \%$ radiation length into
the acceptance of PHENIX which will increase conversion background
by a factor of 10, however, to some level it will also tag the
background it produces.

As I already mentioned in the introduction, it is also very important
to have robust and reliable results for the single muon measurements
using the Muon Arm detectors. Ongoing single muon analysis seem to
converge to a final result during the writing of this work.

Another possible avenue for the PHENIX Open Charm program can be
the study of one of several hadronic decay modes. There are
currently preliminary results for the expected $D^0\rightarrow
K\pi$ reconstruction rate which seems to be too small to measure,
but there are also other channels that might be kinematically
profitable in case of 2 arm PHENIX acceptance (for example
$D^0\rightarrow \pi^+ \pi^- \pi^0$ with $BR = (1.6 \pm 1.1) \%$).
\pagebreak

At this point I would like to end the Thesis and hope to leave the
reader with the feeling of confidence that the long and tedious
analysis of a baseline Open Charm measurement in $\pp$ collisions have
come to a logical and correct conclusion. I look forward to
participating in the upcoming $\Au$ Run04 analysis that will definitely
provide enough statistics for accurate measurements of the $R_{AA}$ of
the heavy quarks.

One more time, I would like to express my deepest gratitude to my
colleagues from PPG037, Matteo Cacciari for investing so much time
into theoretical predictions for this analysis and to everyone who
believed in me.

\appendix

\newpage
\begin{thebibliography}{99}
\bibitem{asp_qcd} A.~Pich, Lectures at the ICTP Summer School in
Particle Physics (Triest, 1999)
\bibitem{nlo_charm} S.~Frixione, M.~L.~Mangano, P.~Nason and G.~Ridolfi,
\Journal{\NPB}{431}{453}{1994}.
\bibitem{qcd_hq1} M.~L.~Mangano, S.~Dawson and R.K.~Ellis,
\Journal{\NPB}{303}{607}{1988}.
\bibitem{qcd_hq2} M.~L.~Mangano, S.~Dawson and R.K.~Ellis,
\Journal{\NPB}{327}{49}{1989}.
\bibitem{nlo1} M.~L.~Mangano, P.~Nason and G.~Ridolfi,
\Journal{\NPB}{373}{295}{1992}.
\bibitem{nlo2} M.~L.~Mangano, P.~Nason and G.~Ridolfi,
\Journal{\NPB}{405}{507}{1993}.
\bibitem{hvqlib} M.~L.~Mangano, P.~Nason and G.~Ridolfi,
\Journal{\NPB}{405}{507}{1993}: HVQLIB FORTRAN code for heavy
quark NLO calculations.

\bibitem{gell_mann} M.~Gell-Mann,
\Journal{\PL}{8}{214}{1964}.
\bibitem{PDG} Review of Particle
Physics \Journal{\PRD}{66}{1}{2002}.
\bibitem{han} M.~Han, Y.~Nambu,
\Journal{\PRB}{139}{1006}{1965}.
\bibitem{Bjorken} J.~D.~Bjorken
\Journal{\PRB}{179}{1547}{1969}.
\bibitem{Gross} D.~J.~Gross, F.~Wilczek,
\Journal{\PRL}{30}{1343}{1973}.
\bibitem{MS_scheme} W.~A.~Bardeen, A.~J.~Buras, D~.W.Duke,
T.~Muta, \Journal{\PRD}{18}{3998}{1978}.
\bibitem{beta_dec} T.~van~Ritbergen, J.~A.~M.~Vermaseren and S.~A.~Larin,
\Journal{\PRB}{400}{379}{1997}.
\bibitem{bethke} S.Bethke, \textit{"Jet Physics at LEP and World summary of
$\alpha_s$"}, [ArXiv:\textsf{hep-ex/9812026}]

\bibitem{intrinsic_flavor} S.~J.~Brodsky and C.~Peterson,
\Journal{\PRD}{23}{2745}{1981}.
\bibitem{Vogt_intrinsic} R.~Vogt and S.~J.~Brodsky,
\Journal{\NPB}{438}{261}{1995}.
\bibitem{PShower} E.~Norrbin, T.~Sj\"{o}strand,
\Journal{\EPJ}{17}{137}{2000}.
\bibitem{whatK} R.~Vogt,
\Journal{\HIP}{17}{75}{2003}.
\bibitem{Vogt_eta} N.~Kidonakis, E.~Laenen, S.~Moch and R.~Vogt,
\Journal{\PRD}{64}{114001}{2001}.
\bibitem{PDFLIB} H.Plothow-Besch, \textit{"PDFLIB: Proton,Pion and Photon Parton Dencity Function of
 the Nucleos and $\alpha_s$ calculations"} - Version 8.04, W5051
 PDFLIB, CERN-PPE.
\bibitem{CTEQ5} H.~L.~Lai, J.~Huston, S.~Kuhlmann,
J.~Morfin, F.~Olness, J. F.~Owens, J.~Pumplin, W. K.~Tung
\Journal{\EPJ}{12}{375}{2000}.
\bibitem{gribov} V.N. Gribov, L.N. Lipatov, Sov. J. Nucl. Phys. {\bf 15} 438 (1972).
\bibitem{altarelli} G. Altarelli, G. Parisi,\Journal{\NPB}{126}{298}{1977}.
\bibitem{dokshitzer} Yu.L. Dokshitzer, Sov. Phys. JETP {\bf 46} (1977) 641.
\bibitem{begel} M.~Begel for the E706 Collaboration, Proceedings of
DPF99, [\textsf{http://home.fnal.gov/~begel/dpf99.html}].
\bibitem{LUND} B.~Anderson, G.~Gustavson, G.~Ingelman and
T.~Sj\"{o}strand, \Journal{\PR} {97}{31}{1983}.
\bibitem{bowler} M.~G.~Bowler,
\Journal{\ZPC} {11}{169}{1981}.
\bibitem{peterson} C.~Peterson, D.~Shlatter, I.~Schmitt and P.~M.~Zerwas,
\Journal{\PRD} {27}{105}{1983}.
\bibitem{pet_eps} M.~Cacciari, M.~Greco,
\Journal{\PRD} {55}{7134}{1997}.
\bibitem{chrin} J.~Chrin,
\Journal{\ZPC} {36}{163}{1987}.
\bibitem{two_lectures} M.~Mangano, \textit{"Two Lectures on Heavy Quark
Production in Hadronic collisions"},
[ArXiv:\textsf{hep-ex/9711337}]
\bibitem{saga} M.~Mangano, \textit{"The saga of bottom production in proton-antiproton collisions"}, Presented at the 2004 Hadron Collider Physics Workshop, East Lansing, MI, June
2004, [ArXiv:\textsf{hep-ex/0411020}].
\bibitem{excess} M.~Cacciari, P.~Nason,
\Journal{\PRL} {89}{122003}{2002}.
\bibitem{pt_spectrum} M.~Cacciari, M.~Greco, P.~Nason,
\Journal{\JHEP} {05}{007}{1998}.
\bibitem{Ellis} K.~Ellis, \textit{"Heavy Quark Production: Phenomenological Review"}, CTEQ School presentation, June
2004,\-
[\textsf{http://www.phys.psu.edu/~cteq/schools/summer04/ellis}].


\bibitem{PHENIXCDR} PHENIX Conceptional Design Report.
\bibitem{zdcmickey} M.~Chiu {\it et~al.},
\Journal{\PRL}{89}{012302}{2002}.
\bibitem{zdcnim} C.~Adler {\it et~al.},
\Journal{\NIM}{470}{488}{2001}.
\bibitem{PHENIXNIM} K.~Adcox {\it et al.}, \Journal{\NIMA}{499}{469}{2003}.
\bibitem{garf} R.~Veenhof, GARFIELD Fortran code for gaseouse
detector simulation, Version 7.0, W5050 GARFIELD,
[\textsf{http://garfield.web.cern.ch/garfield/}]
\bibitem{jjiathesis} J.~Jia, Ph.D thesis, SUNY at Stony Brook, Sep. 2003.
\bibitem{tracknim} K.~Adcox {\it et al.},
\Journal{\NIMA}{499}{489}{2003}; J.T.~Mitchell {\it et al.},
\Journal{\NIMA}{482}{491}{2002}.
\bibitem{sashathesis} A.~Milov, Ph.D thesis, Weizmann Institute,
Rohovot Israel, May 2002.
\bibitem{richnim} M.~Aizawa {\it et al.}, \Journal{\NIMA}{499}{508}{2003}.


\bibitem{ana143} A.~Basilevsky,{\it et al.}, PHENIX internal analysis note 143.
\bibitem{pp_pi0} S.~S.~Adler,{\it et al.},
\Journal{\PRL}{91}{182301}{2003}.
\bibitem{ana224} A.~Basilevsky,{\it et al.}, PHENIX internal analysis note 224.
\bibitem{jpsi} S.~S.~Adler,{\it et al.},
\Journal{\PRC}{69}{014901}{2004}.
\bibitem{ppg011} K.~Adcox,{\it et al.},
\Journal{\PRL}{88}{082301}{2002}.
\bibitem{ppg035} S.~S.~Adler,{\it et al.}, \textit{"Centrality Dependence of Charm Production from single Electrons
at $\sqrt{s_{NN}} = 200$ GeV"}, to be published in PRL,
[ArXiv:\textsf{nucl-ex/0409028}].
\bibitem{ana148} S.~Belikov,{\it et al.}, PHENIX internal analysis note 148.
\bibitem{ana139} W.~Xie,{\it et al.}, PHENIX internal analysis note 139.
\bibitem{EXODUS} EXODUS lepton-pair generator, C++
code created and maintained by Ralf Averbeck.
\bibitem{PISA} PHENIX Integrated Simulation Code, [\textsf{http://vpac17.phy.vanderbilt.edu/index.html}].
\bibitem{ana073} K.~Reygers,{\it et al.}, PHENIX internal analysis note 073.
\bibitem{Kroll_Wada} N.~Kroll, W.~Wada,
\Journal{\PR}{98}{1355}{1955}.
\bibitem{ppg029} F.~Matethias,{\it et al.}, PHENIX internal PPG029, in preparation for publication.
\bibitem{ana333} H.~Hiejima,{\it et al.}, PHENIX internal analysis note 333.
\bibitem{ana337} M.~Kaufman,{\it et al.}, PHENIX internal analysis note 337.
\bibitem{ana305} T.~Hachiya,{\it et al.}, PHENIX internal analysis note
305. Supplimentary note for~\cite{ppg035}.
\bibitem{Tsai} Y.-S.~Tsai,
\Journal{\RMP}{46}{815}{1974}.
\bibitem{ppg049} M.~Tannenbaum,{\it et al.}, PHENIX internal PPG049, in preparation for publication.
\bibitem{ana325} K.~Okada,{\it et al.}, PHENIX internal analysis note
325. Supplimentary note for~\cite{ppg049}.
\bibitem{Vogelsang} B.~Jader, A.~Schafer, M.~Stratmann, W.~Vogelsang,
\Journal{\PRD}{67}{054005}{2003}.
\bibitem{ana158} Y.~Akiba,{\it et al.}, PHENIX internal analysis note
158. Supplimentary note for~\cite{ppg035}.
\bibitem{ana324} T.~Hachiya, Y.~Akiba, PHENIX internal analysis note 324.
\bibitem{ana259} Y.~Akiba,{\it et al.}, PHENIX internal analysis note 259.
\bibitem{ana321} M.~Togava,{\it et al.}, PHENIX internal analysis note 321.
\bibitem{ana172} A.~K.~Purwar,{\it et al.}, PHENIX internal analysis note 172.
\bibitem{ana276} K.~Okada,{\it et al.}, PHENIX internal analysis note 276.


\bibitem{STAR_se_pp} A.~Tai ,
\Journal{\JPG}{30}{S809-S818}{2004}.
\bibitem{phenom_cb} R.~Vogt ,
\Journal{\ZPC}{71}{475}{1996}.
\bibitem{system_prod} R.~Vogt , Proc. 18th Winter Workshop on
Nuclear Dynamics (2002), [ArXiv:\textsf{hep-ph/0203151}].
\bibitem{bottom_production} P.~Nason,{\it et al.} \textit{"1999 CERN Workshop on SM physics (and more) at the LHC"},
[ArXiv:\textsf{hep-ph/0003142}].
\bibitem{pythia} T.~Sj\"{o}strand ,
\Journal{\CPC}{82}{74}{1994}; H.-U.~Bengtsson, T.~Sj\"{o}strand ,
\Journal{\CPC}{46}{43}{1987}; PYTHIA 6.2 Physics and Manual
[ArXiv:\textsf{hep-ph/0108264}].
\bibitem{hq_prod} R.~V.~Gavai,{\it et al.},
\Journal{\IJMPA}{10}{2999}{1995}.
\bibitem{alves_e769} G.~A.~Alves,{\it et al.}, E769 Collaboration,
\Journal{\PRL}{77}{2388}{1996}.
\bibitem{wa92} M.~I.~Adamovich,{\it et al.}, WA92 Collaboration,
\Journal{\NPB}{495}{3}{1997}.
\bibitem{e791} E.~M.~Aitala,{\it et al.}, E791 Collaboration,
\Journal{\PLB}{462}{401}{1999}.
\bibitem{ccrs} F.~W.~Busser,{\it et al.}, CCRS Collaboration,
\Journal{\NPB}{113}{189}{1976}.
\bibitem{basile} M.~Basile,{\it et al.}, SPIRES Collaboration,
\Journal{\NPB}{65}{421}{1981}.
\bibitem{ana101} R.~Averbeck,{\it et al.}, PHENIX internal analysis note
101. Supplimentary note for~\cite{ppg011}.
\bibitem{bottom_CDF} F.~Abe,{\it et al.}, CDF Collaboration,
\Journal{\PRL}{71}{500}{1993};\Journal{\PRL}{71}{2396}{1993};
\Journal{\PRL}{71}{2537}{1993}.
\bibitem{bottom_D0} S.~Abachi,{\it et al.}, D0 Collaboration,
\Journal{\PRL}{74}{1451}{1995}.
\bibitem{NNLO} N.~Kidonakis, R.~Vogt,
\Journal{\EPJ}{36}{201}{2004}.
\bibitem{matteo} M.~Cacciari. Private communictaions.
\bibitem{shuryak} R.~Rapp, E.~V.~Shuryak,
\Journal{\PRD}{67}{074036}{2003}.

\end{thebibliography}
\end{document}